\documentclass[a4paper]{article}
\pdfoutput=1

\usepackage{macros}
\usepackage{booktabs}
\usepackage{multirow,makecell}
\usepackage{import}
\usepackage{tikz}
\usepackage{standalone}
\usepackage[boxsize=0.4em]{ytableau}
\usepackage{asymptote}
\newcommand{\arrowHeadPosition}{0.7}
\newcommand{\fra}{\ensuremath{\mathfrak{a}}}

\newcommand{\PT}{\ensuremath{\operatorname{PT}}}
\preprint{ }
\title{Three-dimensional spiked quiver theory and the tetrahedron Pandharipande--Thomas vertex}

\author[a]{Jiaqun Jiang,}
\author[b,c]{Taro Kimura,}
\author[d]{Satoshi Nawata,}
\author[e]{Go Noshita,}
\author[d]{Jiahao Zheng,}

\affiliation[a]{Center for Geometry and Physics, Institute for Basic Science (IBS), Pohang 37673, Republic of Korea}
\affiliation[b]{Université Bourgogne Europe, CNRS, IMB UMR 5584, 21000 Dijon, France}
\affiliation[c]{Institut Universitaire de France (IUF), France}
\affiliation[d]{Department of Physics and Center for Field Theory \& Particle Physics, Fudan University, 2005 Songhu Road, 200438 Shanghai, China}
\affiliation[e]{Center for High Energy Physics, Peking University, Beijing, China}

\emailAdd{jiangjiaqun@gmail.com}
\emailAdd{taro.kimura@ube.fr}
\emailAdd{snawata@gmail.com}
\emailAdd{gnoshita969hep@gmail.com}
\emailAdd{azjh1997@gmail.com}
\abstract{We study the half-index of the spiked quiver, a 3d $\mathcal{N}=2$ gauge theory on D2-branes coupled to D6-branes. For a single D6-brane, the normalized vortex contribution reproduces the one-leg K-theoretic Pandharipande--Thomas (PT) vertex for $\mathbb{C}^{3}$. We propose that the vortex contributions give higher-rank one-leg PT vertices for parallel D6-branes and one-leg tetrahedron PT vertices for intersecting D6-branes, and verified these identifications.

We also relate the factorization of the twisted index on
$\mathbb{S}^{1}\times\mathbb{S}^{2}$ to the gluing of these vertices along $\mathbb{P}^1$, with the adjoint $R$-charges fixing the degrees of the normal line bundles. We test the resulting relation between twisted-index and PT generating functions in explicit calculations for several local curves and brane configurations, and compare with known plethystic formulas where they are available.}

\begin{document}
\allowdisplaybreaks

\maketitle

\section{Introduction and summary}\label{sec:intro}
The half-index of a 3d \(\mathcal N=2\) gauge theory
is a supersymmetric partition function on \(\mathbb S^1\times D\),
with a supersymmetric boundary condition imposed on
\(\mathbb S^1\times\partial D\) \cite{Yoshida:2014ssa}.
Holomorphic blocks associated with massive vacua provide local
factors for supersymmetric partition functions on closed
three-manifolds \cite{Pasquetti:2011fj,Beem:2012mb}.
In Higgs-branch localization, the block at a given vacuum
contains a vortex series, normalized by the zero-flux
contribution \cite{Fujitsuka:2013fga}.
For \(T[SU(N)]\), the vortex series and the twisted index are
related to equivariant indices of local and global Laumon spaces,
respectively~\cite{Crew:2020jyf}.
More generally, equivariant K-theoretic quasimap vertex functions of Nakajima quiver varieties provide a geometric framework for
vortex partition functions, with a gauge-theoretic realization
in the \(q\)-Langlands correspondence
\cite{Okounkov:2015spn,Aganagic:2016aa,AganagicFrenkelOkounkov:2018}.

For the 3d $\mathcal N=4$ ADHM theory with one fundamental hypermultiplet, the hemisphere partition function with an exceptional Dirichlet boundary condition gives an $A$-shifted block whose vortex series agrees with the bare quasimap vertex of $\operatorname{Hilb}^{N_4}(\mathbb C^2)$ at the corresponding torus fixed point \cite{Crew:2020psc}. Under the correspondence between quasimaps to the Hilbert
scheme and stable pairs on \(\mathbb C^3\), the same Young diagram
labels the asymptotic data of the K-theoretic
Pandharipande--Thomas (PT) vertex
\cite{Pandharipande:2007sq,Okounkov:2018crossroads}.

The 3d \(\mathcal N=4\) ADHM theory admits a
D2--D6 brane realization in type IIA string theory on
\(\mathbb C^4\times\mathbb R\times\mathbb S^1\), where
\(\mathbb C^4=\mathbb C_1\times\mathbb C_2\times
\mathbb C_3\times\mathbb C_4\). In this realization, \(N_4\) D2-branes wrap
\(\mathbb C_4\times\mathbb S^1\), and a stack of \(N\) D6-branes
wraps \(\mathbb C^3_{124}\times\mathbb S^1\), where
\(\mathbb C^3_{abc}=\mathbb C_a\times\mathbb C_b\times\mathbb C_c\). This brane realization suggests a natural extension of the correspondence between vortex series and one-leg PT vertices to systems with intersecting D6-branes. Keeping the D2-branes fixed, we add stacks of $L$ and $M$ D6-branes wrapping $\mathbb{C}^3_{234}\times \mathbb{S}^1$ and $\mathbb{C}^3_{134}\times \mathbb{S}^1$, respectively. The theory on the D2-branes becomes a 3d
\(\mathcal N=2\) quiver gauge theory with gauge group \(U(N_4)\), which we call the \emph{spiked quiver}.
The matter consists of three adjoint chirals and a fundamental--antifundamental pair for each D6-brane. The cubic superpotential \(W\) couples the three adjoints to each other and couples each fundamental--antifundamental pair to the adjoint transverse to its D6-brane \cite{Benini:2009qs,Rapcak:2018nsl}.

At positive Fayet--Iliopoulos (FI) parameter, the Higgs branch of the
spiked quiver is the moduli space of spiked instantons on $\mathbb{C}^{2}_{23},\mathbb{C}^{2}_{31},\mathbb{C}^{2}_{12}$, where $\mathbb{C}^{2}_{ab}=\mathbb{C}_{a}\times\mathbb{C}_{b}$, in its critical-locus formulation. It is
obtained as the GIT quotient of the stable critical
locus of \(W\)~\cite{Arbesfeld:2026zrf}. This identification has a brane interpretation. After compactifying
and T-dualizing the two real directions of \(\mathbb C_4\), the
D2-branes become D0-branes and the D6-brane stacks become D4-branes
on the three coordinate two-planes. The resulting D0--D4 system
realizes the spiked instantons of gauge origami within $\mathbb{C}^{3}$ \cite{Nekrasov:2015wsu,Nekrasov:2016ydq,Nekrasov:2016gud}.
The name spiked quiver comes from this connection to spiked
instantons.
When only one D6-brane stack is present, the theory has
\(\mathcal N=4\) supersymmetry and its Higgs branch is the ADHM
moduli space of framed instantons on \(\mathbb C^2\).
The torus-fixed Higgs vacua are labeled by tuples \(\underline{\vec\lambda}\)
with one Young diagram per D6-brane and total size
\(|\underline{\vec\lambda}|=N_4\).

To describe vortices, we return to the original D2--D6
configuration. The coupling of the D2 worldvolume magnetic flux
to the Ramond--Ramond one-form gives the vortices D0-brane charge.
This motivates a comparison with the D6--D2--D0 quantum mechanics,
in which the D2-branes fix the asymptotic data along
\(\mathbb C_4\). In the absence of D2-branes, a single D6-brane gives the Donaldson--Thomas (DT) theory of \(\mathbb C^3\), while intersecting
D6-branes give the tetrahedron instantons~\cite{Pomoni:2021hkn,Pomoni:2023nlf}. The latter admit a geometric vertex formalism based on Quot schemes~\cite{Fasola:2023ypx}. With D2-branes present, the asymptotic Young diagrams specify
nontrivial legs. The two chambers of the associated supersymmetric quantum mechanics give the DT and PT vertices for a single D6-brane and their
tetrahedron counterparts for intersecting D6-branes
\cite{Kimura:2025lfo}. We refer to the vertices associated with general D6-brane configurations as the \emph{tetrahedron DT and PT vertices}. The vortex series of the D2-brane theory with a single D6-brane reproduces the one-leg PT vertex associated with a Young diagram \cite{Crew:2020psc}.

To study the gluing of PT vertices, we replace the common line \(\mathbb C_4\)
by \(\mathbb P^1\) and compute the twisted index on
\(\mathbb S^1\times\mathbb S^2\)
\cite{Benini:2015noa,Closset:2016arn}.
The adjoint $R$-charges \(r_i\) determine the relation between the
equivariant weights at the north and south poles of $\mathbb{P}^1$. With \(q_4\) the tangent
weight at the north pole, the south-pole weights are
\(\tilde q_4=q_4^{-1}\) and \(\tilde q_i=q_iq_4^{r_i}\) for
\(i=1,2,3\). These are the transition rules for the equivariant
weights of the local curve
\bea
Z=\operatorname{Tot}\!\left(
\mathcal O(-r_1)\oplus\mathcal O(-r_2)\oplus\mathcal O(-r_3)
\longrightarrow\mathbb P^1\right),
\qquad r_1+r_2+r_3=2.
\eea
Here $r_i$ is the $R$-charge of the adjoint chiral $\Phi_i$, and $-r_i$ is the degree of the corresponding normal line bundle. The condition $r_1+r_2+r_3=2$ gives the cubic superpotential $R$-charge two and makes $Z$ a Calabi--Yau fourfold. Let $z_i$ be the fiber coordinate of $\mathcal O(-r_i)$. The D2-branes wrap the zero section $C=\{z_1=z_2=z_3=0\}\simeq\mathbb P^1$, while the stacks of $L$, $M$, and $N$ D6-branes wrap the divisors $z_1=0$, $z_2=0$, and $z_3=0$, respectively. Each D6-brane thus wraps $C$ together with two of its three normal directions. All branes also wrap the circle $\mathbb S^1$.

\medskip
\noindent We now summarize the main results of this paper and state the scope of each.
\paragraph{Half-index and vortex partition function}
At a torus-fixed Higgs vacuum \(\underline{\vec\lambda}\) of the spiked quiver, we impose Dirichlet boundary conditions on the vector multiplet, following the construction of \cite{Crew:2020psc}, and compute the half-index on \(\mathbb S^1\times D\). Normalized by its zero-flux contribution, the vortex series takes the triality-invariant closed form \eqref{eq:exceptional-vortex-triality}. It is a sum over integer fluxes \(k_\mathsf{a}\), one for each box \(\mathsf a\in\underline{\vec\lambda}\), weighted by \(\mathfrak q^{\sum_\mathsf{a}k_\mathsf{a}}\) and by ratios of \(q_4\)-Pochhammer symbols. For generic framing fugacities, the flux sum is supported on reverse plane partitions with bases \(\underline{\vec\lambda}\) \eqref{eq:colored-rpp-condition}, and the total flux is the D0-brane charge. The contour-integral representation with a Neumann vector multiplet \eqref{eq:halfindex-integral} gives the same series, and for a single D6-brane the formula reduces to that of \cite{Crew:2020psc}.

The boundary condition \eqref{eq:boundary-conditions} used in the main text is not itself an exceptional Dirichlet boundary condition: with several framings, its boundary locus can contain directions along the complex gauge orbit of the vacuum. For \(N\) parallel D6-branes, Appendix~\ref{app:rankNADHM_bcd} removes these directions and constructs a rank-\(N\) exceptional Dirichlet boundary condition with the same vortex factor. The analogous construction for intersecting D6-branes is left open. Other choices of chiral boundary conditions can introduce flux-dependent signs, but for our choice they cancel in each flux sector.

\paragraph{Vortex--PT correspondence}
We propose that, at every Higgs vacuum \(\underline{\vec\lambda}\), the normalized vortex series equals the one-leg PT generating function with asymptotic Young diagrams \(\underline{\vec\lambda}\) along \(\mathbb C_4\),
\bea
\mathcal{Z}_{\mathrm{vortex}}^{\underline{\vec\lambda}}(\mathfrak q;q_1,q_2,q_3,q_4)
=\mathsf{PT}_{\underline{\vec\lambda}}[\mathfrak q;q_1,q_2,q_3,q_4],
\eea
where the right-hand side is defined by the D6--D2--D0 quantum mechanics in the PT chamber (Section~\ref{sec:DTPT-vortexQM}). For parallel D6-branes, \(\mathsf{PT}_{\underline{\vec\lambda}}\) is the higher-rank one-leg PT vertex; for intersecting D6-branes, it is the one-leg K-theoretic tetrahedron PT vertex. We further propose that the charge-\(k\) vortex moduli space is equivariantly identified with the moduli space of PT-stable D6--D2--D0 quiver representations with the same leg.

For a single D6-brane, the vortex--PT correspondence reduces to the result of~\cite{Crew:2020psc}. For higher-rank cases, we have confirmed the identity \eqref{eq:higher-rank-vortex-PT-proposal} in various configurations and present explicit flux-by-flux comparisons for $(L,M,N)=(0,0,1)$, $(0,0,2)$, and $(1,1,0)$ in Section~\ref{sec:DTPT-vortexQM}. A direct derivation of the proposed moduli-space identification from the 3d vortex equations is left for future work.

\paragraph{Twisted index and gluing of PT vertices}
On \(\mathbb S^1\times\mathbb S^2\), the twisted index factorizes into holomorphic blocks at the two poles of \(\mathbb P^1\). Under the vortex--PT correspondence, the normalized north and south vortex factors give the two one-leg PT vertices, the common Higgs vacuum \(\underline{\vec\lambda}\) specifies the asymptotic Young diagrams along the compact edge, and the zero-flux factors give the edge weight. Suppressing the classical weights and orientation signs, we may write the gluing formula schematically as
\bea
\sum_{N_4\geq0}Q^{N_4}\mathcal I_{N_4}
&=\sum_{\underline{\vec\lambda}}Q^{|\underline{\vec\lambda}|}\,
\mathcal I_{\mathrm{edge}}^{\underline{\vec\lambda}}\,
\mathcal Z_{\mathrm{vortex}}^{\mathrm N,\underline{\vec\lambda}}\,
\mathcal Z_{\mathrm{vortex}}^{\mathrm S,\underline{\vec\lambda}} \\
&=\sum_{\underline{\vec\lambda}}Q^{|\underline{\vec\lambda}|}\,
\mathsf E_{\underline{\vec\lambda}}\,
\mathsf V_{\underline{\vec\lambda}}^{\mathrm N}\,
\mathsf V_{\underline{\vec\lambda}}^{\mathrm S}
=\mathsf{PT}[\mathfrak q,Q].
\eea
Here \(Q\) weights the gauge rank \(N_4=|\underline{\vec\lambda}|\), corresponding to the D2-brane charge, and \(\mathfrak q\) is the FI fugacity counting D0-brane charge. The correspondence is summarized by
\bea\label{tab:index-PT-correspondence}
&\begin{array}{c|c}
\text{Twisted index} & \text{PT gluing} \\
\hline
\text{Normalized vortex factors} & \text{One-leg PT vertices} \\
\text{Classical and one-loop factors} & \text{Edge weight} \\
\text{Higgs vacuum }\underline{\vec\lambda} & \text{Edge Young diagrams} \\
\text{Gauge rank }N_4 & \text{D2-brane charge} \\
\text{FI fugacity }\mathfrak q & \text{D0-brane fugacity} \\
\text{Rank fugacity }Q & \text{Exponentiated K\"ahler parameter}
\end{array}
\eea

For $N$ parallel D6-branes, we take $L=M=0$ and $r_3=0$. The D6-branes wrap the local threefold $X=\{z_3=0\}=\operatorname{Tot}(\mathcal O(-r_1)\oplus\mathcal O(-r_2)\to \mathbb{P}^1)$, and the right-hand side is the rank-$N$ PT partition function of $X$. At $N=1$ and $(r_1,r_2,r_3)=(1,1,0)$, this recovers the resolved-conifold correspondence between the $A$-twisted ADHM index and refined vertex gluing~\cite{Crew:2020psc}. We extend this correspondence to arbitrary rank $N$ and, by varying the adjoint $R$-charges, to all normal-bundle degrees $-r_1,-r_2$ with $r_1+r_2=2$ (Section~\ref{sec:rankone-twistedindex}).

For intersecting D6-branes, we propose a \emph{tetrahedron PT vertex formalism} in which gluing two one-leg tetrahedron PT vertices along $\mathbb{P}^1$ defines the \emph{tetrahedron PT invariants of the local curve} $Z$ (Appendix~\ref{app:tetrahedronDTPTvertex}). We propose that their generating function coincides with the twisted-index generating function of the spiked quiver and have confirmed this correspondence in explicit examples (Section~\ref{sec:tetrahedron-twisted}).

We also study the D2--D8--$\overline{\D8}$ system, with the D8-branes and antibranes wrapping $Z$ and the D2-branes wrapping $C$. We compare its twisted index with the gluing of one-leg fourfold PT vertices~\cite{Piazzalunga:2023qik,Cao:2023lon,Kimura:2024xpr} and specialize the D8--$\overline{\D8}$ masses in the corresponding plethystic formulas~\cite{CaoKoolMonavari:2019,CaoKoolMonavari:2023CRC,Monavari:2025csu}, as described in Section~\ref{sec:D2-D8}. For the twist $(1,1,0)$, this specialization recovers the rank-$N$ resolved-conifold formula. For the twist $(2,0,0)$, it yields a new plethystic formula \eqref{eq:local-tetra-PE} for the tetrahedron PT partition function of two intersecting D6-brane stacks of flavor rank $(0,M,N)$.

\paragraph{Organization of the paper}
The paper is organized as follows. The brane configuration and the resulting 3d spiked quiver are introduced in Section~\ref{sec:3dADHMflavor}, together with a description of the Higgs branch and its torus fixed points. In Section~\ref{sec:halfindex}, we compare the vortex contribution to the half-index with the one-leg PT vertex computed in D6--D2--D0 quantum mechanics, and give a free-field realization of the vortex sums in terms of screening charges. The relation between twisted-index factorization and PT vertex gluing is developed in Section~\ref{sec:twisted-gluing} for parallel and intersecting D6-branes and for the D2--D8 system. We conclude in Section~\ref{sec:conclusion} with a summary of the results and a discussion of open problems.

The Young-diagram conventions used throughout the paper are
summarized in Appendix~\ref{app:notations}. For the rank-$N$ ADHM
theory with parallel D6-branes, we construct an exceptional
Dirichlet boundary condition in Appendix~\ref{app:rankNADHM_bcd}.
Appendix~\ref{app:JK} reviews Jeffrey--Kirwan residues, and
Appendix~\ref{app:one-leg-detail} gives the superpotential and the
\(J\)- and \(E\)-terms of the one-leg PT quantum mechanics.

In Appendix~\ref{app:gluing}, we review the threefold and fourfold
PT gluing rules, propose their tetrahedron counterpart, and collect
plethystic formulas for local curves. Appendix~\ref{app:edge-comparison} compares the
glued perturbative factors with the PT edge factors, including their
relative signs.
Finally, Appendix~\ref{app:orbifold-topstring} relates higher-rank
DT counting to the type IIA topological string on
\(\mathbb C^2/\mathbb Z_k\) times a Calabi--Yau threefold.

\section{Three-dimensional spiked quiver theory}\label{sec:3dADHMflavor}
In this section, we introduce the 3d gauge theory on D2-branes that intersect D6-branes in type IIA string theory. We take the 10d spacetime to be $\mathbb{R}\times \mathbb{S}^{1}\times \mathbb{R}^{8}$ and write it as $\mathbb{R}\times\mathbb{S}^{1}\times \mathbb{C}^{4}$ upon choosing a complex structure on $\mathbb{R}^{8}$. The coordinates are assigned as
\bea\label{eq:string-background}
\renewcommand{\arraystretch}{1.3}
\begin{tabular}{|c|c|c|c|c|c|c|c|c|c|}
\hline
 \multicolumn{2}{|c|}{$\mathbb{C}_{1}$} & \multicolumn{2}{c|}{$\mathbb{C}_{2}$} & \multicolumn{2}{c|}{$\mathbb{C}_{3}$} & \multicolumn{2}{c|}{$\mathbb{C}_{4}$} & \multicolumn{2}{c|}{$\mathbb{R}\times\mathbb{S}^{1}$} \\
\cline{1-10}   1 & 2 & 3 & 4& 5 & 6 & 7 & 8 & 9& 0\\
\cline{1-10} $x^{1}$ & $x^{2}$ & $x^{3}$ & $x^{4}$& $x^{5}$ & $x^{6}$& $x^{7}$ &$x^{8}$ & $x^{9}$& $x^{0}$\\
\hline \multicolumn{2}{|c|}{$z_{1}$} & \multicolumn{2}{c|}{$z_{2}$} & \multicolumn{2}{c|}{$z_{3}$} & \multicolumn{2}{c|}{$z_{4}$} & \multicolumn{2}{c|}{$x,t$} \\
\hline
\end{tabular}
\eea
where $\{x^{i}\}_{i=0}^{9}$ are the real coordinates and $\{z_{a}\}_{a=1}^{4}$, with $z_{a}=x^{2a-1}+ix^{2a}$, are the complex ones. 

We fix the following conventions for the rest of the paper. The index sets labeling the four complex lines and the four complex three-planes inside $\mathbb{C}^{4}$ are denoted by $\four=\{1,2,3,4\}$ and $\four^{\vee}=\{123,124,134,234\}$, respectively. For $a\in\four$, the complement $\bar{a}\in\four^{\vee}$ is the triple obtained by deleting $a$; for instance, $a=4$ gives $\bar{a}=123$. We write $\mathbb{C}_{a}$ for the complex line labeled by $a\in\four$, and $\mathbb{C}^{3}_{\bar{a}}$ for the complex three-plane labeled by $\bar{a}\in\four^{\vee}$. In later sections, we will also use the complex two-planes, which we denote $\mathbb{C}^{2}_{A}$ ($A\in\six=\{12,14,23,24,31,34\}$). We write $\mathbb{C}^{2}_{\bar A}$ for the complementary coordinate two-plane. The planes contained in $\mathbb{C}^{3}_{123}$ are labeled in cyclic order by $\three=\{12,23,31\}$, and those containing $\mathbb{C}_4$ by $\three^{\vee}=\{14,24,34\}$. Thus, $\six=\three\cup\three^{\vee}$.

The 3d gauge theory of interest arises as the low-energy effective theory on the following D-brane configuration:
\bea\label{eq:DbraneADHMflavor}
\renewcommand{\arraystretch}{1.05}
\begin{tabular}{|c|c|c|c|c|c|c|c|c|c|c|}
\hline
& \multicolumn{2}{c|}{$\mathbb{C}_{1}$} & \multicolumn{2}{c|}{$\mathbb{C}_{2}$} & \multicolumn{2}{c|}{$\mathbb{C}_{3}$} & \multicolumn{2}{c|}{$\mathbb{C}_{4}$} & \multicolumn{2}{c|}{$\mathbb{R}^{1}\times \mathbb{S}^{1}$} \\
\cline{2-11} \raisebox{-0mm}{}  & 1 & 2 & 3 & 4& 5 & 6 & 7 & 8 & 9& 0\\[-2pt]
\hline
    \raisebox{0mm}{$\textcolor{red!60!yellow}{\D2_{4}}$} & $\bullet$ & $\bullet$ & $\bullet$ & $\bullet$ & $\bullet$ & $\bullet$ & $-$ & $-$ & $\bullet$ & $-$ \\
\hline
\raisebox{-0.5mm}{$\textcolor{blue}{{\D6}_{124}} $}& $-$ & $-$& $-$ & $-$  & $\bullet$ & $\bullet$ & $-$ & $-$ & $\bullet$ & $-$ \\
\hline
\raisebox{-0.5mm}{$\textcolor{red}{{\D6}_{134}} $}& $-$ & $-$  & $\bullet$ & $\bullet$ & $-$ & $-$& $-$ & $-$ & $\bullet$ & $-$ \\
\hline \raisebox{-0.5mm}{$\textcolor{green!50!black}{{\D6}_{234}}$} & $\bullet$ & $\bullet$ & $-$ & $-$ & $-$ & $-$ & $-$ & $-$ & $\bullet$ & $-$ \\
\hline
\end{tabular}
\eea
where the subscript on each brane records the subspace of $\mathbb{C}^{4}$ that it spans. Concretely, we wrap $N_{4}$ D2$_{4}$-branes on $\mathbb{C}_{4}\times \mathbb{S}^{1}$ and $N_{\bar{a}}$ D6$_{\bar{a}}$-branes on $\mathbb{C}^{3}_{\bar{a}}\times\mathbb{S}^{1}$ for $a=1,2,3$. For the three stacks of D6-branes, we use throughout the shorthand
\bea\label{eq:LMN-def}
L=N_{\bar{1}},\qquad M=N_{\bar{2}},\qquad N=N_{\bar{3}},
\eea
and retain the notation $N_{\bar{a}}$ only when the stacks are treated simultaneously.

The configuration \eqref{eq:DbraneADHMflavor} preserves four
supercharges and supports a 3d $\mathcal{N}=2$ $U(N_{4})$ gauge
theory on the D2-branes, while the D6-branes provide flavor
symmetries.  We denote the vector multiplet by
$(A_{\mu},\sigma,\lambda,D)$ and a chiral multiplet by
$(\varphi,\eta,F)$.  We use the standard kinetic terms and a
holomorphic superpotential $W$ of $R$-charge two.  The
Chern--Simons level is zero throughout.

The central $U(1)\subset U(N_{4})$ has a topological symmetry
$U(1)_{J}$ with current
$J^{\mu}=\frac{1}{4\pi}\epsilon^{\mu\nu\rho}\Tr F_{\nu\rho}$.
We denote its fugacity by $\fq$; it counts the magnetic flux and will
be identified with the vortex counting parameter below.

The deformations relevant to the Higgs branch are the real masses and
the FI parameter.  A real mass $m_{R}$ is introduced as a constant
background value of the scalar in a flavor vector multiplet and acts
on the scalar fields by the shift
\bea\label{eq:real-mass-shift}
\sigma
\longrightarrow
\sigma+m_{R}.
\eea
A complex mass for a pair of chiral multiplets in conjugate
representations may instead be introduced through the superpotential,
$W\supset m_{C}\widetilde{\Phi}\Phi$.  The real FI parameter
$\xi_{R}$ for an abelian factor is introduced by
\bea\label{eq:FI-term}
\xi_{R}\int\mathrm{d}^{4}\theta\,V_{\mathrm{abelian}},
\eea
which contributes $-\xi_{R}\Tr D$ to the component Lagrangian.  In
the theory considered below, we turn it on for the central
$U(1)\subset U(N_{4})$.

The auxiliary fields $F$ and $D$ enter algebraically and can be
eliminated.  Denoting by $\varphi_{a}$ the scalars of the chiral
multiplets, transforming in representations $R_{a}$, the scalar
potential is
\bea\label{eq:3dN2-scalar-potential}
V(\varphi,\sigma)
&=
\sum_{a}
\left|
\frac{\partial W}{\partial\varphi_{a}}
\right|^{2}
+
\sum_{a}
\left|
(\sigma+m_{R})\cdot\varphi_{a}
\right|^{2}
+
\frac{g^{2}}{2}
\sum_{A}
\left(
\mu_{\mathbb{R}}^{A}(\varphi)
-\xi_{R}\delta^{A,\mathrm{ab}}
\right)^{2},
\eea
where
$\mu_{\mathbb{R}}^{A}(\varphi)
=\sum_{a}\varphi_{a}^{\dagger}T^{A}\varphi_{a}$
is the real moment map.  Both $\sigma$ and $T^{A}$ act in the
representation $R_{a}$; in particular,
$\sigma\cdot\varphi_{a}=[\sigma,\varphi_{a}]$ for an adjoint chiral
multiplet, while $\delta^{A,\mathrm{ab}}$ restricts the FI coupling
to the abelian direction.

Since the three terms in
\eqref{eq:3dN2-scalar-potential} are non-negative, a supersymmetric
vacuum satisfies
\bea\label{eq:general-vacuum-equations}
\frac{\partial W}{\partial\varphi_{a}}
&=0,
&
\mu_{\mathbb{R}}^{A}(\varphi)
&=
\xi_{R}\delta^{A,\mathrm{ab}},
&
(\sigma+m_{R})\cdot\varphi_{a}
&=0.
\eea
At vanishing real masses, the Higgs branch is obtained by setting
$\sigma=0$ and solving the first two equations modulo $U(N_{4})$.
When generic background masses are retained, the last equation is
instead the equivariant fixed-point condition; in that case
$\sigma$ need not vanish.

For the D2--D6 system \eqref{eq:DbraneADHMflavor}, the matter content arising from open strings stretched between the branes is as follows.
\begin{itemize}
    \item \textbf{D2$_4$--D2$_{4}$ strings}: On their own, the D2-branes support 3d $\mathcal{N}=8$ $U(N_{4})$ super Yang--Mills, preserving sixteen supercharges. In the eight-supercharge language, this decomposes into an $\mathcal{N}=4$ vector multiplet and an $\mathcal{N}=4$ adjoint hypermultiplet; in the four-supercharge language, into an $\mathcal{N}=2$ vector multiplet and three $\mathcal{N}=2$ adjoint chiral multiplets $\Phi_{1,2,3}$. Geometrically, $\Phi_{i}$ parametrizes the fluctuation of the D2-branes along the transverse plane $\mathbb{C}_{i}\subset\mathbb{C}^{3}_{123}$.  
    \item \textbf{D2$_{4}$--D6$_{\bar{i}}\,\,(i=1,2,3)$ strings}: These give a 3d $\mathcal{N}=4$ hypermultiplet in the bifundamental representation of $U(N_{4})\times U(N_{\bar{i}})$. In $\mathcal{N}=2$ language, it splits into a chiral multiplet $p_{i}$ in the fundamental representation and a chiral multiplet $\tilde{p}_{i}$ in the antifundamental representation.
\end{itemize}
This matter content is encoded in a quiver diagram, in which the round node carries the $U(N_{4})$ gauge symmetry of the D2-branes and each square node a flavor symmetry $U(N_{\bar{a}})$ carried by a stack of D6-branes.
\bea\label{spiked-instanton-quiver}
\adjustbox{valign=c}{\begin{tikzpicture}[decoration={markings,mark=at position \arrowHeadPosition with {\arrow{latex}}}]
 \tikzset{
        box/.style={draw, minimum width=0.6cm, minimum height=0.6cm, text centered,thick},
        ->-/.style={decoration={
        markings,mark=at position #1 with {\arrow[scale=1.5]{>}}},postaction={decorate},line width=0.5mm},
        -<-/.style={decoration={
        markings,
        mark=at position #1 with {\arrow[scale=1.5]{<}}},postaction={decorate},line width=0.5mm}    
    }
    
\begin{scope}[xshift=0cm]
\node[green!60!black,anchor=east] at (-0.18,0.85) {$p_{1}$};
\node[green!60!black,anchor=west] at ( 0.18,0.85) {$\widetilde p_{1}$};

\node[red,anchor=south west] at ( 0.72,-0.18) {$\widetilde{p}_{2}$};
\node[red,anchor=north west] at ( 0.52,-0.55) {$ p_{2}$};

\node[blue,anchor=south east] at (-0.72,-0.23) {$\widetilde{p}_{3}$};
\node[blue,anchor=north east] at (-0.6,-0.5) {$p_{3}$};

    \draw[red!60!yellow,thick,scale=1.3,postaction={decorate}] (0,0) arc(90:450:0.13 and 0.5) ;
    \draw[red!60!yellow,thick,scale=1.3,postaction={decorate}, rotate=-120] (0,0) arc(90:450:0.13 and 0.5) ;
    \draw[red!60!yellow,thick,scale=1.3,postaction={decorate},rotate=120] (0,0) arc(90:450:0.13 and 0.5) ;
   
    \draw[red!60!yellow,thick,postaction={decorate}] (-1.02+0.95,1.2)--(-1.02+0.95,0.3);
     \draw[red!60!yellow,thick,postaction={decorate}](-0.88+0.95,0.3)--(-0.88+0.95,1.2);

    \draw[red!60!yellow,thick,postaction={decorate}] (0,0.1)--(-2.2+0.95,-0.42);
    \draw[red!60!yellow,thick,postaction={decorate}] (-2.2+0.95,-0.42-0.1)--(0,0.1-0.1);

    \draw[red!60!yellow,thick,postaction={decorate}] (0,0.1)--(1.25,-0.42);
    \draw[red!60!yellow,thick,postaction={decorate}] (1.25,-0.42-0.1)--(0,0.1-0.1);

     \draw[red!60!yellow,fill=white,thick](0,0) circle(0.3cm);
    \node[red!60!yellow] at (0,0) {$N_{4}$};

    \node[green!50!black,fill=white,box] at (0,1.5){$L$};
    \node[blue,fill=white,box] at (-1.55,-0.7){$N$};
    \node[red,fill=white,box] at (1.55,-0.7){$M$};

     \node[right] at (1.1,0.7){$\textcolor{blue}{\Phi_{3}}$};
      \node[left] at (-1.1,0.7){$\textcolor{red}{\Phi_{2}}$};
      \node[below] at (0,-1.25){$\textcolor{green!50!black}{\Phi_{1}}$};
    
\end{scope}
\end{tikzpicture}}\qquad {W}=\Tr\Phi_{1}[\Phi_{2},\Phi_{3}]+\sum_{i=1,2,3}\Tr\, \tilde{p}_{i}\Phi_{i}p_{i}
\eea
 The $F$-term equations are
\bea\label{eq:spiked-Fterms}
\frac{\partial W}{\partial \Phi_1}&=[\Phi_{2},\Phi_{3}]+p_{1}\tilde{p}_{1}=0,\qquad \frac{\partial W}{\partial \Phi_2}=[\Phi_{3},\Phi_{1}]+p_{2}\tilde{p}_{2}=0,\qquad \frac{\partial W}{\partial \Phi_3}=[\Phi_{1},\Phi_{2}]+p_{3}\tilde{p}_{3}=0,\\
\frac{\partial W}{\partial p_{i}}&=\tilde{p}_{i}\Phi_i=0,\quad \frac{\partial W}{\partial \tilde{p}_{i}}=\Phi_ip_i=0,\quad i=1,2,3,
\eea
which we denote collectively by $\mu_{\mathbb{C}}$. Together with the $D$-term equations
\bea\label{eq:spikedD-terms}
\mu_{\mathbb{R}}=\sum_{i=1}^{3}[\Phi_i,\Phi_i^{\dagger}]+\sum_{i=1}^{3}p_{i}p_{i}^{\dagger}-\sum_{i=1}^{3}\tilde{p}_{i}^{\dagger}\tilde{p}_{i}=\xi,
\eea
they cut out the Higgs branch of the theory, the moduli space of the intersecting brane configuration \eqref{eq:DbraneADHMflavor}
\bea\label{eq:spikedmodulispace}
\mathcal{M}_{H}=\bigl(\mu_{\mathbb{R}}^{-1}(\xi)\cap \mu_{\mathbb{C}}^{-1}(0)\bigr)/U(N_4).
\eea

For $\xi>0$, when only one D6-brane stack is present, say
$(L,M,N)=(0,0,N)$, these equations reduce to the ordinary ADHM
equations, and $\mathcal{M}_{H}$ is the moduli space of $N_{4}$
instantons of rank $N$ on $\mathbb{C}^{2}_{12}$, understood as the
resolved ADHM moduli space of framed torsion-free sheaves. In the rank-one
case $N=1$, it is the Hilbert scheme
$\operatorname{Hilb}^{N_{4}}(\mathbb{C}^{2}_{12})$ of $N_{4}$ points~\cite[Chapter~2]{NakaBook}. For general $(L,M,N)$, the same equations define the corresponding
\textit{spiked-instanton} moduli space~\cite{Rapcak:2018nsl,Arbesfeld:2026zrf},
with instantons supported on the three
coordinate planes
$\mathbb{C}^{2}_{23}$, $\mathbb{C}^{2}_{31}$ and
$\mathbb{C}^{2}_{12}$.\footnote{This is the critical-locus version of
the moduli space. For its relation to Nekrasov's original gauge-origami
construction~\cite{Nekrasov:2016qym,Nekrasov:2016gud,Nekrasov:2016ydq},
including the distinction between the defining equations, see
\cite[Remark~2.8 and Section~2.5]{Arbesfeld:2026zrf}.}
In the brane picture, compactifying and T-dualizing along $\mathbb{C}_{4}$ maps the
D2-branes to D0-branes and the three D6-brane stacks to D4-branes
wrapping these planes.
Lifting the D4-branes to M5-branes and applying M-theory/type-IIB
duality to the toric $T^{2}$ fibration relates this configuration to
the Gaiotto--Rap\v{c}\'ak junction of D3-branes ending on a trivalent
$(p,q)$ fivebrane web~\cite[Section~1.3.3]{Rapcak:2018nsl}.
The multiplicities $(L,M,N)$ become the numbers of D3-branes in the three
sectors, whose junction supports the vertex algebra
$Y_{L,M,N}$~\cite{Gaiotto:2017euk}.
This relates the vertex algebra acting on the spiked-instanton states
to the junction algebra, although the generic instanton modules and
the specialized junction modules need not coincide~\cite[Remark~1.3.1]{Rapcak:2018nsl}.

\paragraph{Flavor charges}
The rotations of the three planes
$\mathbb{C}_{1,2,3}$ transverse to the D2-branes act on the quiver
\eqref{spiked-instanton-quiver} as a flavor symmetry.  We parametrize
this geometric torus by $t_{1,2,3}$, with the convention that the
adjoint chiral multiplet $\Phi_i$, describing the motion along
$\mathbb{C}_i$, has weight $t_i^{-1}$.  Invariance of the cubic term
$\Tr\Phi_1[\Phi_2,\Phi_3]$ then requires
\bea\label{eq:t-constraint}
t_1t_2t_3=1.
\eea
The geometric flavor torus is therefore two-dimensional. 

Each stack of D6-branes also carries a $U(N_{\bar{a}})$ flavor symmetry, whose Cartan fugacities we denote by $v_{\bar{a},\alpha}$, $\alpha=1,\ldots,N_{\bar a}$; with the notation \eqref{eq:spiked-Adictionary} below, we also write them as $v_{A4,\alpha}$. Under the geometric torus, the chiral multiplets transform as
\bea\label{eq:3dflavor}
\Phi_{i}\rightarrow t_{i}^{-1}\Phi_{i},\qquad p_{i}\rightarrow p_{i},\qquad \tilde{p}_{i}\rightarrow t_{i}\tilde{p}_{i},\qquad i=1,2,3.
\eea
The superpotential term
$\Tr\widetilde p_i\Phi_i p_i$ fixes only the product of the
geometric weights of the framing pair,
$\operatorname{wt}(p_i)\operatorname{wt}(\widetilde p_i)=t_i$.
Their individual weights may be shifted oppositely by the central
$U(1)$ of the corresponding D6-brane flavor group.  This freedom can
be absorbed into the framing fugacity $v_{A4,\alpha}$, and in
\eqref{eq:3dflavor} we have chosen $p_i$ to be neutral.

We denote the $R$-symmetry fugacity by $q$. The $R$-charges of the chiral multiplets can be chosen freely as long as the superpotential has $R$-charge two,
\bea
W\rightarrow q^{2}W.
\eea
In the main text, we use the assignment 
 \bea\label{eq:flavor_table_sym}
 \begin{tabular}{c|c c c }
  &   $U(1)_{t_{1}} $  & $U(1)_{t_{2}}$  & $U(1)_{R,q}$ \\\hline
     $\Phi_{1}$ & $-1$ & $0$  &  $r_{1}$ \\
    $\Phi_{2}$   & $0$ & $-1$& $r_{2}$ \\
    $\Phi_{3}$  & $1$  & $1$ & $r_{3}$\\
    $p_{1}$  & $0$& $0$ & $1$\\
    $p_{2}$  & $0$ & $0$  & $1$ \\
    $p_{3}$  & $0$  & $0$ & $1$ \\
    $\tilde{p}_{1}$  & $1$ & $0$ & $1-r_{1}$\\
    $\tilde{p}_{2}$  & $0$ & $1$ & $1-r_{2}$ \\
    $\tilde{p}_{3}$  &  $-1$ &  $-1$& $1-r_{3}$\\
 \end{tabular}
 \eea
 where the condition on the superpotential imposes
 \bea\label{eq:r-constraint}
r_{1}+r_{2}+r_{3}=2.
 \eea
These are trial $R$-charges used to define the equivariant index, not the infrared superconformal $R$-charges. We focus on the case in which all of them are integers $r_{i}\in\mathbb{Z}$, as required by the topological twist of Section~\ref{sec:twisted-gluing}. By relabeling the three coordinate directions together with their D6-brane stacks, we may assume the ordering 
\bea\label{eq:r-ordering}
r_1\geq r_2\geq r_3.
\eea
Where the D6-brane orientation is fixed, we keep the corresponding coordinate labels and impose only the conditions on the $r_i$ stated there.

It is convenient to combine the flavor charges and the $R$-charge into
\bea\label{eq:qi-def}
q=q_{4}^{1/2},\qquad q_{i}=t_{i}q^{-r_{i}}=t_{i}q_{4}^{-r_{i}/2},\qquad i=1,2,3 .
\eea
The two constraints \eqref{eq:t-constraint} and \eqref{eq:r-constraint} then combine into the single relation
\bea\label{eq:q-constraint}
q_{1}q_{2}q_{3}q_{4}=1
\eea
familiar from gauge origami on $\mathbb{C}^{4}$. Here $q_{4}$ generates the rotation of the plane $\mathbb{C}_{4}$, which every brane wraps. In this notation, the chiral fields carry the combined weights
\bea\label{eq:3dflavor2}
\Phi_{i}\rightarrow q_{i}^{-1}\Phi_{i},\qquad p_{i}\rightarrow q_{4}^{1/2}p_{i},\qquad \tilde{p}_{i}\rightarrow q_{i}q_{4}^{1/2}\,\tilde{p}_{i},\qquad i=1,2,3 .
\eea

\paragraph{Stability condition and the GIT quotient}
For $\xi>0$, the description \eqref{eq:spikedmodulispace} of the Higgs branch has an equivalent algebraic form: imposing the $D$-term equation \eqref{eq:spikedD-terms} and quotienting by the compact group $U(N_{4})$ is the same as imposing a \emph{stability condition} and quotienting by its complexification $GL(\bfK)$. To state it, assign a vector space to each node of the quiver \eqref{spiked-instanton-quiver},
\bea\label{eq:spiked-vectorspaces}
\bfK=\mathbb{C}^{N_{4}},\qquad \bfN_{\bar{i}}=\mathbb{C}^{N_{\bar{i}}}\quad (i=1,2,3),
\eea
with $(N_{\bar{1}},N_{\bar{2}},N_{\bar{3}})=(L,M,N)$ as in \eqref{eq:LMN-def}, and read the chiral multiplets as linear maps
\bea\label{eq:spiked-maps}
\Phi_{i}:\bfK\rightarrow \bfK,\qquad p_{i}:\bfN_{\bar{i}}\rightarrow \bfK,\qquad \tilde{p}_{i}:\bfK\rightarrow \bfN_{\bar{i}}.
\eea
The stability condition is the requirement that the only subspace $\bfK'\subseteq\bfK$ which contains all the images $p_{i}(\bfN_{\bar{i}})$ and is preserved by all the $\Phi_{i}$ is $\bfK$ itself. Equivalently, $\bfK$ is generated from the images of the $p_{i}$ by acting with the adjoint fields,
\bea\label{eq:spiked-stability}
\bfK=\sum_{i=1}^{3}\mathbb{C}\langle\Phi_{1},\Phi_{2},\Phi_{3}\rangle\,p_{i}(\bfN_{\bar{i}}),
\eea
where the brackets denote the algebra generated by all words in the adjoints.
On stable configurations, $GL(\bfK)$ acts freely, and the Higgs branch becomes the GIT quotient
\bea
\mathcal{M}_{H}\simeq\bigl\{(\vec{\Phi},\vec{p},\vec{\tilde{p}})\;\bigm|\;\text{$F$-terms \eqref{eq:spiked-Fterms}}\bigr\}/\!\!/\,GL(\bfK).\label{eq:spiked-GIT}
\eea

For each two-plane $A=ab\in\three$, let $c$ be the complementary index and write
\bea\label{eq:spiked-Adictionary}
A4=\bar{c},\qquad \bfN_{A4}\equiv\bfN_{\bar{c}},\qquad p_{A4}\equiv p_{c},\qquad N_{A4}\equiv N_{\bar{c}} .
\eea
Thus the stack D6$_{\bar{c}}$ spans $\mathbb{C}^{2}_{A}\times\mathbb{C}_{4}$ and is transverse to $\mathbb{C}_{c}$.

\paragraph{Torus fixed points}
The equivariant torus acts on $\mathcal{M}_{H}$ by \eqref{eq:3dflavor2}, with $q_{1,2,3}$ independent, and by the framing rotations.
At a fixed point, every matrix element of $\tilde p_c\Phi_{i_1}\cdots\Phi_{i_\ell}p_d$ vanishes.
For distinct framing labels its weight contains a ratio of independent framing fugacities.
For the same framing label its weight is $q_cq_4\prod_{j=1}^{\ell}q_{i_j}^{-1}$, which is nontrivial since $q_cq_4=(q_aq_b)^{-1}$ for $A=ab$.
Stability then implies $\tilde p_c=0$ for all $c$.
The $F$-terms make the adjoints commute, and $\Phi_cp_c=0$ restricts the subspace generated by $p_c$ to the two in-plane adjoints.
The independent framing weights give the direct sum
\bea\label{eq:spiked-Kdecomposition}
\bfK=\bigoplus_{A\in\three}\bfK_{A},\qquad \bfK_{A}=\mathbb{C}[\Phi_{a},\Phi_{b}]\,p_{A4}(\bfN_{A4}),\qquad A=ab .
\eea
Each summand is an ordinary rank-$N_{A4}$ ADHM fixed point for the plane $\mathbb{C}^{2}_{A}$. The fixed points are isolated and are labeled by one Young diagram per D6-brane,
\bea\label{eq:spiked-fixedpointlabels}
\vec{\lambda}_{A}=\bigl(\lambda_{A}^{(1)},\ldots,\lambda_{A}^{(N_{A4})}\bigr),\qquad
\lambda_{A}^{(\alpha)}=\bigl(\lambda_{A,1}^{(\alpha)}\geq\lambda_{A,2}^{(\alpha)}\geq\cdots\bigr),
\eea
subject to the constraint that the total number of boxes is the number of D2-branes,
\bea\label{eq:spiked-boxcount}
\sum_{A\in\three}\sum_{\alpha=1}^{N_{A4}}\bigl|\lambda_{A}^{(\alpha)}\bigr|=N_{4}.
\eea
Our Young-diagram conventions are collected in Appendix~\ref{app:notations}.

Writing $\bfN_{A4}=\bigoplus_{\alpha}\mathbb{C}e_{A}^{(\alpha)}$ for the fixed-point basis of the flavor space, the corresponding decomposition of the gauge space is
\bea\label{eq:spiked-Kbasis}
\bfK_{A}=\bigoplus_{\alpha=1}^{N_{A4}}\bfK_{A,\alpha},\qquad
\bfK_{A,\alpha}=\bigoplus_{(i,j)\in\lambda_{A}^{(\alpha)}}\mathbb{C}\;\Phi_{a}^{\,i-1}\Phi_{b}^{\,j-1}\,p_{A4}\,e_{A}^{(\alpha)},\qquad A=ab ,
\eea
where $(i-1,j-1)$ records the powers of the in-plane adjoints. The first index of a box always runs along $\mathbb{C}_{a}$ and the second along $\mathbb{C}_{b}$, with $ab$ taken in the cyclic order $12,23,31$; this fixes the orientation of each Young diagram. For later use, we denote the whole collection by
\bea\label{eq:spiked-alldiagrams}
\underline{\vec{\lambda}}=\bigl(\lambda_{A}^{(\alpha)}\bigr)_{A\in\three}^{\alpha=1,\ldots,N_{A4}} .
\eea

We label gauge indices by boxes $x=(A,\alpha;i,j)$ and write
$(\Phi_d)_{yx}$ for the component mapping the basis vector
labeled by $x$ to that labeled by $y$.
For $A=ab$, the fixed-point adjoint components are
\bea\label{eq:spiked-fixedpoint-components}
(\Phi_d)_{(B,\beta;k,\ell),(A,\alpha;i,j)}
=
\begin{cases}
1, & (B,\beta)=(A,\alpha),\quad d=a,\quad
     (k,\ell)=(i+1,j),\\
1, & (B,\beta)=(A,\alpha),\quad d=b,\quad
     (k,\ell)=(i,j+1),\\
0, & \text{otherwise},
\end{cases}
\eea
where $d=1,2,3$ and both indices label boxes in their respective diagrams. In particular, the transverse adjoint $\Phi_c$, with $A4=\bar c$, vanishes on $\bfK_{A,\alpha}$.

The framing components are
\bea\label{eq:spiked-fixedpoint-framing}
(p_{A4})_{(B,\beta;k,\ell),\alpha}
&=
\begin{cases}
1, & (B,\beta)=(A,\alpha),\quad (k,\ell)=(1,1),\\
0, & \text{otherwise},
\end{cases}
\\
(\tilde p_{A4})_{\alpha,(B,\beta;k,\ell)}&=0,
\eea
where $\tilde p_{A4}\equiv\tilde p_c$ for $A4=\bar c$.

\section{Vortex partition function and one-leg PT vertices}\label{sec:halfindex}
In this section, we relate the half-index of the spiked quiver
to one-leg K-theoretic PT vertices.
At a fixed Higgs vacuum $\underline{\vec{\lambda}}$, dividing
the half-index block by its zero-flux contribution isolates the
vortex partition function. On the PT side, the same tuple of
Young diagrams specifies the asymptotic data of the leg, and
the total vortex flux corresponds to the D0-brane charge.
A single D6-brane gives the ordinary one-leg PT vertex.
We propose that parallel D6-branes give its higher-rank generalization
and intersecting D6-brane stacks give the tetrahedron PT vertex.

In Section~\ref{sec:halfindex-def}, we compute the half-index
blocks and separate their vortex and zero-flux contributions.
In Section~\ref{sec:DTPT-vortexQM}, we construct the corresponding
vertices from D6--D2--D0 quantum mechanics and compare their
generating functions with the vortex partition functions.
Section~\ref{sec:BPSCFT} gives a free-field realization of the
vortex sums in terms of screening charges. The zero-flux factors will
enter the gluing of two such blocks in
Section~\ref{sec:twisted-gluing}.

\subsection{Half-index and vortex partition function}\label{sec:halfindex-def}
The gauge theory \eqref{spiked-instanton-quiver} is supported on the D2$_{4}$-brane worldvolume $\mathbb{S}^{1}\times \mathbb{C}_{4}$. We consider its \emph{half-index}, that is, the supersymmetric index of the theory on $\mathbb{S}^{1}\times D$ with a boundary condition imposed at the edge of the disk $D\simeq\mathbb{C}_{4}$ \cite{Yoshida:2014ssa,Costello:2020ndc}, 
\bea\label{eq:halfindex-def}
\mathcal{Z}=\Tr\left[(-1)^{F}\,\fq^{\,\mathfrak{J}}\,q_{4}^{\,J_{4}+\frac{R_{4}}{2}}\,v^{\,F_{v}}\right],
\eea
where $J_{4}$ generates the rotation of $\mathbb{C}_{4}$, $R_{4}$ is the $R$-charge, $\mathfrak{J}$ is the charge under the topological $U(1)_{J}$ of Section~\ref{sec:3dADHMflavor} whose fugacity $\fq$ counts vortex number, and $v$ collectively denotes the D6-brane flavor fugacities. The flavor fugacities $t_{1,2}$ of \eqref{eq:flavor_table_sym} combine with the $R$-symmetry fugacity into the equivariant parameters \eqref{eq:qi-def}, which satisfy \eqref{eq:q-constraint}.

We first list the one-loop determinants of the individual multiplets for each choice of boundary condition. We denote by $y_I$, $I=1,\ldots,N_4$, the fugacities
of the Cartan subgroup of $U(N_4)$ and by
$ k_I\in\mathbb Z$ the corresponding magnetic fluxes.

For each component of a chiral multiplet, let $w$ denote its
multiplicative equivariant weight, including its gauge, flavor and
$R$-symmetry fugacities. In the presence of magnetic flux, we write
\bea
 w_{k}
 &=
 q_4^{Q(k)}w ,
\eea
where $Q(k)$ is the gauge charge of the component evaluated
on $k$. For example, for an adjoint component with weight
$w_{IJ}$ and a fundamental component with weight $w_{I\alpha}$, we
have
\bea
 w_{IJ,k}
 &=
 q_4^{k_I-k_J}w_{IJ},
 &
 w_{I\alpha,k}
 &=
 q_4^{k_I}w_{I\alpha}.
\eea

The finite $q$-Pochhammer symbol $(a;q_4)_n$ is defined by
\begin{equation}\label{q-Pochhammer}
(a;q_4)_n =
\begin{dcases}
  \displaystyle \prod_{k=0}^{n-1} \left(1 - a q_4^k\right), & n \in \mathbb{Z}_{\geq 0}, \\[1.2em]
  \displaystyle \frac{1}{\prod_{k=1}^{-n} \left(1 - a q_4^{-k}\right)}, & n \in \mathbb{Z}_{< 0}.
\end{dcases}
\end{equation}
For $|q_4| < 1$, we define the infinite product
\begin{equation}
    (a;q_4)_\infty = \prod_{k=0}^\infty (1 - a q_4^{k}) .
\end{equation}
Furthermore, we introduce
\bea\label{eq:symm-q-Pochhammer}
{(x;q_{4})'}_\infty&=e^{-\mathcal{E}[-\log(x)]}(x;q_4)_{\infty},\quad
(z;q_4)'_n
=
\frac{(z;q_4)'_\infty}{(q_4^n z;q_4)'_\infty}
=
q_4^{-\frac{n(n-1)}4}z^{-\frac n2}(z;q_4)_n
\eea
for $n\in\mathbb Z$, where
\bea
\mathcal{E}[x]=-\frac{\eps_4}{24}-\frac{x}{4}-\frac{x^{2}}{4\eps_4}
\eea
comes from the zeta-function regularization of the zero-point energies.

\begin{itemize}

\item \textbf{FI term.} The classical FI factor is
\bea\label{eq:3dFI}
\prod_{I=1}^{N_4}y_I^\xi,\quad \xi=\frac{\log\fq}{\log q_4}.
\eea
In a magnetic sector $k=(k_I)$, the shift
$y_I\mapsto q_4^{k_I}y_I$ gives
\bea
\prod_{I=1}^{N_4}(q_4^{k_I}y_I)^\xi
=
\left(\prod_{I=1}^{N_4}y_I^\xi\right)
\fq^{\sum_{I=1}^{N_4}k_I}.
\eea

\item \textbf{$3d$ chiral multiplet with Neumann boundary condition.}
Its one-loop determinant is
\bea
 \mathcal{Z}_{\mathrm{chiral}}^{\mathrm N}
 \left(w,k\right)
 &=
 \frac{1}
 {\left(w_{k};q_4\right)'_\infty}.
\eea

\item \textbf{$3d$ chiral multiplet with Dirichlet boundary condition.}
Its one-loop determinant is
\bea
 \mathcal{Z}_{\mathrm{chiral}}^{\mathrm D}
 \left(w,k\right)
 &=
 \left(q_4w_{k}^{-1};q_4\right)'_\infty .
\eea

\item \textbf{$U(N_4)$ vector multiplet with Dirichlet boundary
condition.}
The gauge fugacities $y_I$ become fugacities for a boundary global
symmetry, and the contribution in the magnetic sector
$k=(k_1,\ldots,k_{N_4})$ is
\bea
 \mathcal{Z}_{\mathrm{vec}}^{\mathrm D}
 \left(\vec y,k\right)
 &=
 \frac{1}{{(q_4;q_4)'}^{N_{4}}_\infty}\prod_{\substack{I,J=1\\ I\neq J}}^{N_4}
 \frac{1}{
 \left(
 q_4^{1+k_I-k_J}
 \frac{y_I}{y_J};
 q_4
 \right)'_\infty
 }.
\eea

\item \textbf{$U(N_4)$ vector multiplet with Neumann boundary
condition.}
The gauge symmetry remains gauged at the boundary. Its contribution
to the integrand is
\bea
 \mathcal{Z}_{\mathrm{vec}}^{\mathrm N}
 \left(\vec y\right)
 &=
 {(q_4;q_4)'}^{N_{
 4}
 }_\infty
 \prod_{\substack{I,J=1\\ I\neq J}}^{N_4}
 \left(
 \frac{y_I}{y_J};
 q_4
 \right)'_\infty .
\eea
The corresponding gauge fugacities are integrated over with the
Haar measure
\bea
 \frac{1}{N_4!}
 \prod_{I=1}^{N_4}
 \frac{dy_I}{2\pi i\,y_I}.
\eea

\end{itemize}

\paragraph{Half-index with a Dirichlet vector multiplet}
Following Crew--Dorey--Zhang \cite{Crew:2020psc}, we impose
Dirichlet boundary conditions on the vector multiplet.
The gauge symmetry then becomes a global symmetry at the boundary,
whose Cartan fugacities are denoted by $y_\mathsf{a}$.
We use inverse gauge fugacities, so the gauge transformation
acts on the basis vector $e_\mathsf{a}$ with eigenvalue $y_\mathsf{a}^{-1}$.

Fix a Higgs vacuum \(\underline{\vec{\lambda}}\), and label its
boxes by \(\mathsf{a}=(A,\alpha;i_\mathsf{a},j_\mathsf{a})\), where
\(A=ab\in\{12,23,31\}\) and
\((i_\mathsf{a},j_\mathsf{a})\in\lambda_A^{(\alpha)}\). At zero magnetic flux, we specialize the gauge fugacities to
\bea\label{eq:boundary-fugacities}
y_\mathsf{a}=\tilde s_\mathsf{a}=q_4^{1/2}s_\mathsf{a},
\qquad
s_\mathsf{a}=v_{A4,\alpha}
q_{a}^{1-i_\mathsf{a}}q_{ b}^{1-j_\mathsf{a}}.
\eea
In the magnetic sector $k=(k_\mathsf{a})_{\mathsf{a}\in\underline{\vec{\lambda}}}$, we evaluate the one-loop determinants at $y_\mathsf{a}=q_4^{k_\mathsf{a}}\tilde s_\mathsf{a}$.

Let us denote the sets of Neumann and Dirichlet chiral components by
\(\mathscr N_{\underline{\vec{\lambda}}}\) and
\(\mathscr D_{\underline{\vec{\lambda}}}\), respectively. We keep these sets unspecified for now. For a chiral component $\chi$, let $w_\chi$ denote its zero-flux weight and $w_{\chi,k}$ its weight in the magnetic sector $k$. Explicitly,
\bea
\renewcommand{\arraystretch}{1.4}
\begin{array}{c|c|c}
\chi&w_\chi&w_{\chi,k}\\ \hline
(\Phi_{d})_{\mathsf{ba}}
&q_{d}^{-1}s_\mathsf{a}/s_\mathsf{b}
&q_4^{k_\mathsf{a}-k_\mathsf{b}}q_{ d}^{-1}s_\mathsf{a}/s_\mathsf{b}\\
(p_{A4})_{\mathsf{a}\alpha}
&v_{A4,\alpha}/s_\mathsf{a}
&q_4^{-k_\mathsf{a}}v_{A4,\alpha}/s_\mathsf{a}\\
(\tilde p_{A4})_{\alpha \mathsf{a}}
&q_A^{-1}s_\mathsf{a}/v_{A4,\alpha}
&q_4^{k_\mathsf{a}}q_A^{-1}s_\mathsf{a}/v_{A4,\alpha}
\end{array}
\eea
where the box indices $\mathsf{a},\mathsf{b}$ run over $\underline{\vec\lambda}$.

Using \eqref{eq:3dFI}, we obtain the half-index
\bea\label{eq:exceptional-half-index}
\widetilde{\mathcal Z}^{\underline{\vec\lambda}}_{
\mathbb S^1\times D}
={}&
\exp\left[
\frac{\log\mathfrak q}{\log q_4}
\sum_{\mathsf{a}}\log\tilde s_{\mathsf a}
\right]
\sum_{\{k_\mathsf{a}\in\mathbb Z\}}
\mathfrak q^{\sum_\mathsf{a} k_\mathsf{a}}
\prod_{\mathsf{a},\mathsf{b}}
\frac{1}{
(q_4^{1+k_\mathsf{a}-k_\mathsf{b}}s_\mathsf{a}/s_\mathsf{b};q_4)'_\infty
}
\\
&\times
\prod_{\chi\in\mathscr N_{\underline{\vec\lambda}}}
\frac{1}{(w_{\chi,k};q_4)'_\infty}
\prod_{\chi\in\mathscr D_{\underline{\vec\lambda}}}
(q_4w_{\chi,k}^{-1};q_4)'_\infty .
\eea
The product over $\mathsf{a},\mathsf{b}$ includes the diagonal terms,
which give the Cartan determinant of the Dirichlet vector. We combine the determinant factors at generic gauge fugacities
before taking the fixed-point limit. The normalization below assumes that the zero-flux contribution has a finite, nonzero limit.

Following \cite{Crew:2020psc}, we split the zero-flux contribution
into a classical factor, which contains the FI term and the
zero-point energies, and a one-loop factor written in terms of
ordinary $q$-Pochhammer symbols:
\bea
\widetilde{\mathcal Z}^{\underline{\vec\lambda}}_{
\mathbb S^1\times D}
=
\mathcal Z_{\mathrm{cl}}^{\underline{\vec\lambda}}
\mathcal Z_{\text{1-loop}}^{\underline{\vec\lambda}}
\mathcal Z_{\mathrm{vortex}}^{\underline{\vec\lambda}}.
\eea
Here the classical factor is
\bea\label{eq:exceptional-classical}
\mathcal Z_{\mathrm{cl}}^{\underline{\vec\lambda}}
=
\exp\biggl[
&\frac{\log\mathfrak q}{\log q_4}
\sum_\mathsf{a}\log\tilde s_\mathsf{a}
+\sum_{\mathsf a, \mathsf b}\mathcal E\left[-\log(q_4s_\mathsf{a}/s_\mathsf{b})\right]\\
&+\sum_{\chi\in\mathscr N_{\underline{\vec\lambda}}}
\mathcal E[-\log w_\chi]
-\sum_{\chi\in\mathscr D_{\underline{\vec\lambda}}}
\mathcal E[-\log(q_4w_\chi^{-1})]
\biggr],
\eea
and
\bea\label{eq:exceptional-one-loop}
\mathcal Z_{\text{1-loop}}^{\underline{\vec\lambda}}
=
\prod_{\mathsf a, \mathsf b}
\frac{1}{(q_4s_\mathsf{a}/s_\mathsf{b};q_4)_\infty}
\prod_{\chi\in\mathscr N_{\underline{\vec\lambda}}}
\frac{1}{(w_\chi;q_4)_\infty}
\prod_{\chi\in\mathscr D_{\underline{\vec\lambda}}}
(q_4w_\chi^{-1};q_4)_\infty .
\eea
The normalized vortex factor is obtained by dividing each
flux sector by its zero-flux determinant:
\bea\label{eq:exceptional-vortex}
\mathcal Z_{\mathrm{vortex}}^{\underline{\vec\lambda}}
={}&
\sum_{\{k_\mathsf{a}\in\mathbb Z\}}
\mathfrak q^{\sum_\mathsf{a} k_\mathsf{a}}
\prod_{\mathsf{a},\mathsf{b}}
\frac{
(q_4s_\mathsf{a}/s_\mathsf{b};q_4)'_\infty
}{
(q_4^{1+k_\mathsf{a}-k_\mathsf{b}}s_\mathsf{a}/s_\mathsf{b};q_4)'_\infty
}\prod_{\chi\in\mathscr N_{\underline{\vec\lambda}}}
\frac{(w_\chi;q_4)'_\infty}
     {(w_{\chi,k};q_4)'_\infty}
\prod_{\chi\in\mathscr D_{\underline{\vec\lambda}}}
\frac{(q_4w_{\chi,k}^{-1};q_4)'_\infty}
     {(q_4w_\chi^{-1};q_4)'_\infty}\\
    ={}&
\sum_{\{k_\mathsf{a}\in\mathbb Z\}}
\fq^{\sum_\mathsf{a} k_\mathsf{a}}
\prod_{\mathsf{a},\mathsf{b}}
(q_4s_\mathsf{a}/s_\mathsf{b};q_4)'_{k_\mathsf{a}-k_\mathsf{b}}
\prod_{\chi\in\mathscr N_{\underline{\vec\lambda}}}
(w_\chi;q_4)'_{d_\chi}
\prod_{\chi\in\mathscr D_{\underline{\vec\lambda}}}
\frac{1}{(q_4w_\chi^{-1};q_4)'_{-d_\chi}},
\eea
where $d_\chi\in\mathbb Z$ is the flux shift of the component $\chi$, defined by $w_{\chi,k}=q_4^{d_\chi}w_\chi$.

\paragraph{Boundary condition}
For the explicit computations, we use the boundary condition \eqref{eq:boundary-conditions} below. It is motivated by the construction of
Crew--Dorey--Zhang~\cite{Crew:2020psc}.\footnote{This boundary condition admits the fixed-point configuration
labeled by $\underline{\vec{\lambda}}$, and both the superpotential and
its variation along boundary-preserving deformations vanish.
For a single D6-brane, the local Higgs-branch analysis of
\cite{Crew:2020psc} applies directly. With several framings, the boundary
locus can contain directions along the complex gauge orbit of the fixed point,
even when the D6-branes are parallel. For \(N\) parallel D6-branes,
Appendix~\ref{app:rankNADHM_bcd} removes these directions with additional
tree constraints, which extends the exceptional Dirichlet boundary condition
of \cite{Crew:2020psc} to the rank-\(N\) \(3\mathrm d\ \mathcal N=4\)
ADHM theory.} For each Young diagram, we use the prescription of Crew--Dorey--Zhang for the two in-plane adjoints and fix the boundary values along the corresponding spanning tree. To assign boundary conditions to adjoint components connecting distinct diagrams, we choose a
total order on the framing labels. Changing the chiral boundary conditions can introduce
flux-dependent signs in the normalized vortex sum; for the assignment below, these signs cancel in each flux
sector. We leave a systematic study of boundary conditions for the spiked quiver to future work.

We label boxes by $\mathsf{a},\mathsf{b}$ and coordinate directions by
$ a, b, c,d\in\{1,2,3\}$.
We write $<$ for the chosen total order on the framing labels $(A,\alpha)$.
For a box $\mathsf{a}$, write $\beta(\mathsf{a})$ for its framing label, and $c(\mathsf{a})$ for the transverse coordinate label. Within a Young diagram $\lambda_A^{(\alpha)}$, with
$A=a b$ in the cyclic order $12,23,31$ and $\{a,b,c\}=\{1,2,3\}$,
write $\mathsf{a}=(i_\mathsf{a},j_\mathsf{a})$ and define
\bea
(\lambda_{A}^{(\alpha)})_{\mathrm B}
=
\left\{
(i,j)\in\lambda_A^{(\alpha)}
\,\middle|\,
i=(\lambda_A^{(\alpha)})^T_j
\right\}.
\eea
For an ordered pair $(\mathsf{a},\mathsf{b})$ in this Young diagram, let
\bea
\mathcal C(\mathsf{a},\mathsf{b}):
\qquad
\mathsf{b}\notin(\lambda_{A}^{(\alpha)})_{\mathrm B}
\quad\text{or}\quad
\left(
\mathsf{b}\in(\lambda_{A}^{(\alpha)})_{\mathrm B}
\ \text{and}\ i_\mathsf{a}>i_\mathsf{b}
\right).
\eea
With rows labeling targets and columns labeling sources,
the boundary assignments we use are
\bea\label{eq:boundary-conditions}
\renewcommand{\arraystretch}{1.5}
\begin{array}{c|c|c|c}
\text{component}
&\text{weight}
&\text{condition}
&\text{b.c.}
\\ \hline
U(N_4)\text{ vector}
&y_\mathsf{a}/y_\mathsf{b}
&\text{all components}
&\mathrm D
\\
(\Phi_{c(\mathsf{a})})_{\mathsf{b}\mathsf{a}}
&q_{ c(\mathsf{a})}^{-1}y_\mathsf{a}/y_\mathsf{b}
&\text{all }\mathsf{a},\mathsf{b}
&\mathrm D
\\
\bigl((\Phi_{ a})_{\mathsf{a}\mathsf{b}},(\Phi_{ b})_{\mathsf{b}\mathsf{a}}\bigr)
&\bigl(q_{ a}^{-1}y_\mathsf{b}/y_\mathsf{a},\
       q_{ b}^{-1}y_\mathsf{a}/y_\mathsf{b}\bigr)
&\substack{\beta(\mathsf{a})=\beta(\mathsf{b})\\ \mathcal C(\mathsf{a},\mathsf{b})}
&(\mathrm D,\mathrm N)
\\
\bigl((\Phi_{a})_{\mathsf{a}\mathsf{b}},(\Phi_{ b})_{\mathsf{b}\mathsf{a}}\bigr)
&\bigl(q_{ a}^{-1}y_\mathsf{b}/y_\mathsf{a}, q_{ b}^{-1}y_\mathsf{a}/y_\mathsf{b}\bigr)
&\substack{\beta(\mathsf{a})=\beta(\mathsf{b})\\ \neg\mathcal C(\mathsf{a},\mathsf{b})}
&(\mathrm N,\mathrm D)
\\
(\Phi_{d})_{\mathsf{b}\mathsf{a}}
&q_{ d}^{-1}y_\mathsf{a}/y_\mathsf{b}
&\substack{\beta(\mathsf{a})<\beta(\mathsf{b})\\  d\ne c(\mathsf{a})}
&\mathrm N
\\
(\Phi_{d})_{\mathsf{b}\mathsf{a}}
&q_{ d}^{-1}y_\mathsf{a}/y_\mathsf{b}
&\substack{\beta(\mathsf{a})>\beta(\mathsf{b})\\  d\ne c(\mathsf{a})}
&\mathrm D
\\
(p_{A4})_{\mathsf{a}\alpha}
&q_4^{1/2}v_{A4,\alpha}/y_\mathsf{a}
&\text{all components}
&\mathrm D
\\
(\tilde p_{A4})_{\alpha\mathsf{a}}
&q_{ c}q_4^{1/2}y_\mathsf{a}/v_{A4,\alpha}
&A4=\bar{c}
&\mathrm N
\end{array}
\eea
where $A=ab$ is the
plane of the common diagram, and $\mathsf{a}=\mathsf{b}$ is included.
The fundamental components are fixed to their vacuum values,
so $(p_{A4})_{(A,\alpha;1,1),\alpha}=1$ for each nonempty
diagram, with all other components zero.
In addition, we freeze the adjoint components along every
$a$-direction edge and every $ b$-direction
edge whose target belongs to $(\lambda_{A}^{(\alpha)})_{\mathrm B}$
to their fixed-point value $1$, overriding the assignments above.
These edges form a spanning tree in each nonempty diagram.
All remaining Dirichlet components vanish.

For the boundary condition chosen above, the normalized vortex
factor admits the manifestly triality-invariant expression
\bea\label{eq:exceptional-vortex-triality}
\mathcal Z_{\mathrm{vortex}}^{\underline{\vec\lambda}}
={}&
\sum_{\{k_\mathsf{a}\in\mathbb Z\}}
\mathfrak q^{\sum_\mathsf{a} k_\mathsf{a}}
\prod_{A\in\three}
\prod_{\alpha=1}^{N_{A4}}
\prod_{\mathsf{a}\in\underline{\vec{\lambda}}}
\frac{
(v_{A4,\alpha}/s_\mathsf{a};q_4)'_{-k_\mathsf{a}}
}{
(q_{A4}v_{A4,\alpha}/s_\mathsf{a};q_4)'_{-k_\mathsf{a}}
}\\
&\qquad \times 
\prod_{\mathsf{a}\ne\mathsf{b}}
\frac{
\displaystyle\prod_{ d=1}^{3}
(q_{d}^{-1}s_\mathsf{a}/s_\mathsf{b};q_4)'_{k_\mathsf{a}-k_\mathsf{b}}
}{
(s_\mathsf{a}/s_\mathsf{b};q_4)'_{k_\mathsf{a}-k_\mathsf{b}}
}.
\eea
Triality acts by simultaneously permuting the coordinate
parameters and the corresponding framing data. Using $q_1q_2q_3q_4=1$ and combining opposite matrix
components, we can equivalently write this in the form
of \cite{Crew:2020psc}:
\bea\label{eq:exceptional-vortex-2}
\mathcal Z_{\mathrm{vortex}}^{\underline{\vec\lambda}}
={}&
\sum_{\{k_\mathsf{a}\in\mathbb Z\}}
\mathfrak q_{\mathrm{eff}}^{\sum_\mathsf{a} k_\mathsf{a}}
\prod_{A\in\three}
\prod_{\alpha=1}^{N_{A4}}
\prod_{\mathsf{a}\in\underline{\vec{\lambda}}}
\frac{
(v_{A4,\alpha}/s_\mathsf{a};q_4)_{-k_\mathsf{a}}
}{
(q_{A4}v_{A4,\alpha}/s_\mathsf{a};q_4)_{-k_\mathsf{a}}
}\\
&\quad \times 
\prod_{\mathsf{a}\ne\mathsf{b}}
\frac{
(q_1^{-1}s_\mathsf{a}/s_\mathsf{b};q_4)_{k_\mathsf{a}-k_\mathsf{b}}
(q_2^{-1}s_\mathsf{a}/s_\mathsf{b};q_4)_{k_\mathsf{a}-k_\mathsf{b}}
}{
(s_\mathsf{a}/s_\mathsf{b};q_4)_{k_\mathsf{a}-k_\mathsf{b}}
(q_{12}^{-1}s_\mathsf{a}/s_\mathsf{b};q_4)_{k_\mathsf{a}-k_\mathsf{b}}
},
\eea
with
\bea\label{eq:effective-topological}
\mathfrak q_{\mathrm{eff}}
=
\mathfrak q
\prod_{A\in\three}q_{A4}^{-\frac{N_{A4}}2}
=
\mathfrak q
\prod_{c=1}^{3}
q_{c}^{\frac{N_{\bar{ c}}}2}.
\eea
Here $q_{A4}=q_Aq_4=q_{c}^{-1}$ for
$A4=\bar{c}$.
The regularization monomials cancel between the numerator
and denominator of the adjoint--vector contribution.
The remaining framing monomial is absorbed into
$\mathfrak q_{\mathrm{eff}}$.
The choice of directions $1,2$ is conventional, and any
other pair gives the same result.

For comparison with \cite{Crew:2020psc}, consider a single
D6-brane on the plane $\mathbb{C}^{3}_{124}$.
Their parameters are related to ours by
\bea\label{eq:crew-dictionary}
q_1^{-1}=uz,\qquad
q_2^{-1}=u/z,\qquad q_3^{-1}=q/u^2,\qquad 
q_4=q,\qquad
u=t^{1/2}q^{1/4},
\eea
with $\mathfrak q=\zeta$, or equivalently
\bea
\mathfrak q_{\mathrm{eff}}
=\zeta t^{1/2}q^{-1/4}.
\eea

\paragraph{Combinatorial support of the flux sum}
Only fluxes that form reverse plane partitions contribute
by the same argument as in \cite[App.~D.1]{Crew:2020psc}.

For a box $\mathsf{a}$ in $\lambda_A^{(\alpha)}$, write
\bea
\mathsf{a}=(A,\alpha;i,j),
\qquad
A=a b\in\{12,23,31\},
\qquad
k_{A,\alpha}(i,j)=k_\mathsf{a}.
\eea
For the corner box $\mathsf{a}_0=(A,\alpha;1,1)$, we have
$s_{\mathsf{a}_0}=v_{A4,\alpha}$.
Its framing contribution contains
\bea
\left(
\frac{v_{A4,\alpha}}{s_{\mathsf{a}_0}};q_4
\right)_{-k_{\mathsf{a}_0}}
=
(1;q_4)_{-k_{\mathsf{a}_0}},
\eea
which vanishes for $k_{\mathsf{a}_0}<0$.
Hence $k_{\mathsf{a}_0}\geq0$.

\begin{figure}[t]
    \centering
    \includegraphics[width=0.35\linewidth]{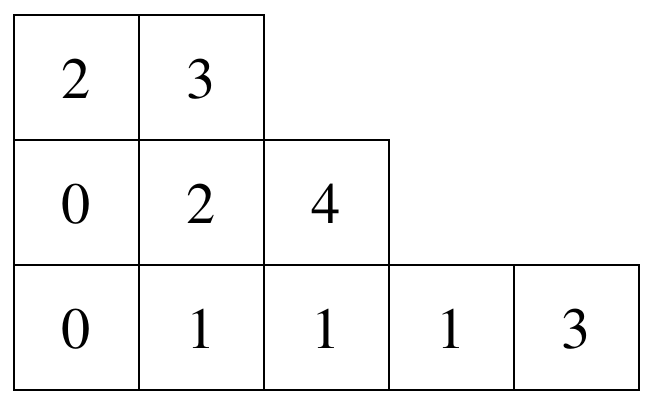}
    \caption{An example of a reverse plane partition.}
    \label{fig:reverse-plane-partition}
\end{figure}

Consider next two adjacent boxes
$\mathsf{a}=(A,\alpha;i,j)$ and $\mathsf{b}=(A,\alpha;i+1,j)$.
Their fixed-point weights satisfy
\bea
s_\mathsf{b}=q_{a}^{-1}s_\mathsf{a}.
\eea
The corresponding adjoint factor in the triality-invariant form \eqref{eq:exceptional-vortex-triality} is
\bea
\left(
q_{a}^{-1}\frac{s_\mathsf{a}}{s_\mathsf{b}};q_4
\right)'_{k_\mathsf{a}-k_\mathsf{b}}
\propto
(1;q_4)_{k_\mathsf{a}-k_\mathsf{b}},
\eea
which vanishes whenever $k_\mathsf{a}>k_\mathsf{b}$.
The same argument for $\mathsf{b}=(A,\alpha;i,j+1)$ gives
$k_\mathsf{a}\leq k_\mathsf{b}$.
Consequently, a nonzero contribution must satisfy
\bea\label{eq:colored-rpp-condition}
k_{A,\alpha}(i,j)\geq0,
\qquad
k_{A,\alpha}(i,j)\leq k_{A,\alpha}(i+1,j),
\qquad
k_{A,\alpha}(i,j)\leq k_{A,\alpha}(i,j+1),
\eea
whenever the indicated boxes belong to the diagram.
Thus the flux sum is supported on collections of
\textit{reverse plane partitions} with bases
$\underline{\vec\lambda}$ (see Figure~\ref{fig:reverse-plane-partition}),
\bea
\underline{\vec k}\in
\operatorname{RPP}(\underline{\vec\lambda})
:=
\prod_{A\in\three}
\prod_{\alpha=1}^{N_{A4}}
\operatorname{RPP}\bigl(\lambda_A^{(\alpha)}\bigr).
\eea

For generic framing fugacities, factors involving boxes
in different Young diagrams impose no further inequalities.
The vortex sum may therefore be restricted to
$\operatorname{RPP}(\underline{\vec\lambda})$, with degree
\bea
|\underline{\vec k}|
=
\sum_{\mathsf{a}\in\underline{\vec\lambda}}k_\mathsf{a}.
\eea

\paragraph{Contour integral representation with a Neumann vector multiplet}
We propose a contour integral representation of the normalized vortex partition function obtained above, motivated by~\cite{Piazzalunga:2023qik,Cao:2023lon}. Consider the spiked quiver with $N_4$ D2$_4$-branes and
$N_{A4}$ D6$_{A4}$-branes for $A\in\{12,23,31\}$.
We use the one-loop determinants associated with Neumann boundary
conditions for the vector multiplet, the adjoint chirals
$\Phi_{1,2,3}$ and the framing chirals $p_i$, and Dirichlet boundary
conditions for $\tilde p_i$.
With the $R$-charge assignments in \eqref{eq:flavor_table_sym},
the contour block is
\bea
\mathcal B
&=\frac{{(q_{4};q_{4})'_{\infty}}^{N_{4}}}{N_4!}\oint
\prod_{I=1}^{N_4}\frac{dy_I}{2\pi i y_I}\, \prod_{I=1}^{N_{4}}y_I^\xi
\prod_{I=1}^{N_4}
\prod_{A\in\{12,23,31\}}
\prod_{\alpha=1}^{N_{A4}}
\frac{
(q_{A4}q_4^{1/2}v_{A4,\alpha}/y_I;q_4)'_\infty
}{
(q_4^{1/2}v_{A4,\alpha}/y_I;q_4)'_\infty
}\\
&\qquad\times
\frac{
\prod_{I\neq J}(y_I/y_J;q_4)'_\infty
}{
\prod_{I,J}\prod_{s=1}^{3}
(q_s^{-1}y_I/y_J;q_4)'_\infty
},
\eea
where $q_{A4}=q_{a}q_{b}q_4=q_{c}^{-1}$ for
$A={ab}$ and $\{{a},{b},{c}\}=\{1,2,3\}$.

For the Higgs vacuum labeled by $\underline{\vec\lambda}$,
we label the integration variables by its boxes,
$y_I=y(\Bbox)=y_{A,\alpha}(i,j)$ for
$\Bbox\in\lambda_A^{(\alpha)}$.
The corresponding labeled contour block is
\bea\label{eq:halfindex-integral}
\mathcal B_{\underline{\vec\lambda}}
 &={(q_{4};q_{4})'_{\infty}}^{N_{4}}\oint_{\Gamma_{\underline{\vec\lambda}}}
\prod_{\Abox\in\underline{\vec\lambda}}
\frac{dy(\Bbox)}{2\pi i y(\Bbox)}\,y(\Bbox)^\xi
\prod_{\Abox\in\underline{\vec\lambda}}
\prod_{A\in\{12,23,31\}}
\prod_{\alpha=1}^{N_{A4}}
\frac{
(q_{A4}q_4^{1/2}v_{A4,\alpha}/y(\Bbox);q_4)'_\infty
}{
(q_4^{1/2}v_{A4,\alpha}/y(\Bbox);q_4)'_\infty
}\\
&\qquad\times
\frac{
\prod_{\Abox\neq\AboxF}
(y(\Bbox)/y(\BboxF);q_4)'_\infty
}{
\prod_{\Abox,\AboxF\in\underline{\vec\lambda}}
\prod_{s=1}^{3}
(q_s^{-1}y(\Bbox)/y(\BboxF);q_4)'_\infty
}.
\eea
The labeling of the integration variables by boxes absorbs the Weyl multiplicity. 

We specify $\Gamma_{\underline{\vec\lambda}}$ in \eqref{eq:halfindex-integral} by the multiplicative analogue of the contour prescription in~\cite[Proposition~7.12]{Cao:2023lon}, as in the K-theoretic contour integral of~\cite{Piazzalunga:2023qik}. For each nonempty $\lambda_A^{(\alpha)}$, the contour encloses the corner poles
\bea
y_{A,\alpha}(1,1)=q_4^{1/2}v_{A4,\alpha}q_4^k,
\qquad k\in\mathbb Z_{\geq0}.
\eea
For $A=ab$, we then add the remaining boxes one at a time, each after its neighbors at $(i-1,j)$ and $(i,j-1)$, and take the pole families
\bea
y_{A,\alpha}(i+1,j)&=q_a^{-1}y_{A,\alpha}(i,j)q_4^k,
\qquad k\in\mathbb Z_{\geq0},\\
y_{A,\alpha}(i,j+1)&=q_b^{-1}y_{A,\alpha}(i,j)q_4^k,
\qquad k\in\mathbb Z_{\geq0},
\eea
whenever the indicated boxes belong to the diagram. Along each edge, $k$ is the flux of the target box minus that of the source box. The pole locations are therefore $y_\mathsf{a}=q_4^{k_\mathsf{a}}\tilde s_\mathsf{a}$, with $\underline{\vec k}\in\operatorname{RPP}(\underline{\vec\lambda})$ as in \eqref{eq:colored-rpp-condition}.

Let $F_{\underline{\vec\lambda}}(y)$ denote the integrand of \eqref{eq:halfindex-integral} with the logarithmic measure omitted. In the formal fixed-point expansion of the Neumann block, we denote the contribution of the sector $k_\mathsf{a}=0$ for all boxes by $\mathcal B_{\underline{\vec\lambda},0}$. Factoring out this zero-vortex contribution, we write
\bea
\mathcal B_{\underline{\vec\lambda}}
=
\mathcal B_{\underline{\vec\lambda},0}\,
\mathcal Z_{\mathrm{vortex}}^{\underline{\vec\lambda}}.
\eea
To compute the weight of each flux sector relative to $\mathcal B_{\underline{\vec\lambda},0}$, we form the ratio of the shifted and unshifted integrands at generic $y_\mathsf{a}$. Following the treatment of singular factors in~\cite[Remark~7.11]{Cao:2023lon}, we cancel common factors before setting $y_\mathsf{a}=\tilde s_\mathsf{a}$. By \eqref{eq:symm-q-Pochhammer}, the one-loop determinants reduce to the finite products in \eqref{eq:exceptional-vortex-triality}, while the FI term gives $\mathfrak q^{|\underline{\vec k}|}$. Thus
\bea
\mathcal Z_{\mathrm{vortex}}^{\underline{\vec\lambda}}
=
\sum_{\underline{\vec k}\in\operatorname{RPP}(\underline{\vec\lambda})}
\left.
\frac{F_{\underline{\vec\lambda}}(\{q_4^{k_\mathsf{a}}y_\mathsf{a}\})}
     {F_{\underline{\vec\lambda}}(\{y_\mathsf{a}\})}
\right|_{y_\mathsf{a}=\tilde s_\mathsf{a}}.
\eea

For $\mathcal B_{\underline{\vec\lambda},0}$, we set $y_\mathsf{a}=\tilde s_\mathsf{a}$ at the level of the one-loop character. Since the character of a single hemisphere can contain a trivial weight, we take the index only after combining the north- and south-pole characters, as described in Appendix~\ref{app:edge-comparison}.

\paragraph{Geometric aspects of the half-index}

The mathematical formulation of vortex partition functions uses quasimaps from $\mathbb{P}^1$ to the target space, typically a GIT quotient~\cite{Okounkov:2015spn}. The 3d vortex partition function is identified with the K-theoretic vertex function, the generating function of equivariant Euler characteristics of the twisted virtual structure sheaf on the quasimap moduli spaces. Quasimaps from $\mathbb{P}^1$ to Nakajima varieties of affine quivers correspond to stable pairs on CY threefolds, so the associated vertex functions agree with one-leg PT vertices~\cite{Liu:2020jus}. The basic example is quasimaps $\mathbb{P}^1\to\operatorname{Hilb}(\mathbb{C}^2)$ to the Hilbert scheme of points on $\mathbb{C}^2$, which give the one-leg PT vertices of $\mathbb{C}^3$. Cao and Zhao extended this correspondence to higher dimensions by showing that quasimaps to $\operatorname{Hilb}(\mathbb{C}^3)$ give the one-leg PT vertices of $\mathbb{C}^4$~\cite{Cao:2023lon}; see also~\cite{Piazzalunga:2023qik} for the K-theoretic version. Since $\operatorname{Hilb}(\mathbb{C}^3)$ is realized by a quiver with potential, it is not a Nakajima variety in the usual sense.

Our setup is closer to this second case. The vortex partition function of the spiked quiver should be the vertex function of quasimaps to moduli spaces of spiked instantons on $\mathbb{C}^3$, which are also realized by a quiver with potential~\cite{Rapcak:2018nsl,Rapcak:2020ueh}. We leave the mathematical foundation of these quasimaps to future work.

For the spiked quiver, the Dirichlet flux sum and the Neumann contour integral considered above give the same vortex series after normalization by their respective zero-vortex contributions. In 3d $\mathcal{N}=4$ theories, the change of basis between half-indices associated with exceptional Dirichlet and enriched Neumann boundary conditions is described by elliptic stable envelopes~\cite{Aganagic:2016aa,BullimoreZhang:2021Elliptic,Dedushenko:2021mds,Dedushenko:2023qjq,Crew:2023tky}, which have a close connection with the capping operators appearing in the context of vertex functions~\cite{Okounkov:2016sya}. It would be interesting to determine whether an analogous elliptic stable envelope governs the dependence of the full spiked-quiver half-index on the choice of boundary condition.

\subsection{One-leg Pandharipande--Thomas counting and vortex partition function}\label{sec:DTPT-vortexQM}
 Donaldson--Thomas (DT) and Pandharipande--Thomas (PT) theories
originate from \cite{MNOP} and
\cite{Pandharipande:2007kc,Pandharipande:2007sq}, respectively, and
their relation was established in
\cite{pandharipande2010stable,Bridgeland:HallDTPT}. The K-theoretic refinement relevant
here was conjectured in \cite{Nekrasov:2014nea} and proven at the
vertex level in \cite{Kuhn:2023msy} by wall-crossing; our
gauge-theoretic realization in this paper follows
\cite{Kimura:2025lfo,Kimura:2025lig}.

For a single D6-brane, the vortex
partition function of the ADHM theory agrees with the
one-leg PT vertex \cite{Crew:2020psc}. The magnetic fluxes form a reverse plane partition over the Young diagram $\lambda$ specifying the Higgs vacuum at the boundary (see \eqref{eq:colored-rpp-condition}). We take this correspondence as our starting point and ask what replaces the one-leg PT vertex when several D6-branes are present.

The D0-brane interpretation of the vortex number follows from the Wess--Zumino coupling on the D2-branes,
\bea
\mu_2(2\pi\alpha')\int_{\mathbb{S}^{1}\times\mathbb C_4} C_1\wedge\operatorname{Tr}F . 
\eea
A vortex with magnetic flux $k=\frac{1}{2\pi}\int_{\mathbb C_4}\operatorname{Tr}F$ therefore carries $k$ units of D0-brane charge. At a fixed Higgs vacuum, the D2-branes prescribe the asymptotic leg, while the D0-branes count finite excitations above it. This leads us to consider the supersymmetric quantum mechanics of D0-branes bound to the D6--D2 system.

\paragraph{Supersymmetric quantum mechanics of PT$_{3}$ counting}
We first review the generating function of PT$_3$ invariants\footnote{We use $\mathrm{PT}_3$ for PT counting on the complex
three-dimensional space wrapped by a D6-brane, and $\mathrm{PT}_4$ for
PT counting on $\mathbb C^4$. The D2--D6 system considered here gives
a one-leg $\mathrm{PT}_3$ vertex. Replacing the D6-branes by D8-branes
wrapping $\mathbb C^4$ gives its one-leg
$\mathrm{PT}_4$ counterpart \cite{Piazzalunga:2023qik}.} in the language of quiver quantum mechanics, following \cite{Kimura:2025lfo,Kimura:2025lig}.

We start with the one-leg PT$_{3}$ quantum mechanics coming from a D6-brane on $\mathbb{C}^{3}_{123}$ with D2-branes on $\mathbb{C}_{3}$. We denote the associated one-leg PT$_{3}$ vertex by $\mathsf{PT}_{\bar{4};\varnothing\varnothing\lambda}[\fq;q_{1},q_{2},q_{3},q_{4}]$. For the D6-brane on $\mathbb C^3_{ab4}$ and D2-branes on $\mathbb{C}_{4}$ in our configuration \eqref{eq:DbraneADHMflavor}, with
$c=\overline{A4}$ and $A=ab$, we relabel the four directions by
$(1,2,3,4)\mapsto(a,b,4,c)$, so that the single-stack vertex is
\bea
\mathsf{PT}_{\bar c;\varnothing\varnothing\lambda_A}
[\mathfrak q;q_1,q_2,q_3,q_4]
=
\mathsf{PT}_{\bar 4;\varnothing\varnothing\lambda_A}
[\mathfrak q;q_a,q_b,q_4,q_c].
\label{eq:one-leg-coordinate-relabeling}
\eea

For a single D6-brane on $\mathbb{C}^{3}_{123}$ with D2-branes on $\mathbb{C}_{3}$ and Higgs vacuum $\lambda$, the D0-brane theory is the 1d $\mathcal{N}=4$ quiver quantum mechanics
\bea
\adjustbox{valign=c}{\begin{tikzpicture}[decoration={markings,mark=at position \arrowHeadPosition with {\arrow{latex}}}]
 \tikzset{
        box/.style={draw, minimum width=0.6cm, minimum height=0.6cm, text centered,thick},
        ->-/.style={decoration={
        markings,mark=at position #1 with {\arrow[scale=1.5]{>}}},postaction={decorate},line width=0.5mm},
        -<-/.style={decoration={
        markings,
        mark=at position #1 with {\arrow[scale=1.5]{<}}},postaction={decorate},line width=0.5mm}    
    }
\begin{scope}[xshift=4cm]
    \draw[black,thick,scale=1.3, postaction={decorate}] (-0.75,0) arc(90:450:0.2 and 0.45) ;
     \draw[black,thick,scale=1.3, postaction={decorate}] (-0.75,0) arc(0:360:0.45 and 0.2) ;
      \draw[black,thick,scale=1.3, postaction={decorate}] (-0.75,0) arc(180:540:0.45 and 0.2) ;
      \node[right] at (0.2,0){$\mathsf{B}_{3}$};
      \node[left] at (-2.2,0){$\mathsf{B}_{2}$};
      \node[below] at (-0.95,-1.25){$\mathsf{B}_{1}$};
    \draw[fill=white,thick](-0.95,0) circle(0.3cm);
    \node at (-0.95,0) {$k$};
    \node[fill=white,box] at (-0.95,1.5){$1$};
    \draw[thick,postaction={decorate}] (-1.05,1.2)--(-1.05,0.3);
   \draw[thick,postaction={decorate}] (-0.85,0.3)--(-0.85,1.2);
    \node[right] at (-0.85,0.75){$\mathsf{J}_{w}$};
     \node[left] at (-1.05,0.75){$\mathsf{I}_{z}$};
\end{scope}
\end{tikzpicture}}\qquad \begin{array}{c|c}
\text{$F$-term}\,\,&\mathsf{W}^{\D6\tbar\D2\tbar\D0}=\mathsf{W}^{\D6\tbar\D0}+\sum_{w}\delta \mathsf{W}_{w}\\
&\mathsf{W}^{\D6\tbar\D0}=\Tr\,\mathsf{B}_{3}[\mathsf{B}_{1},\mathsf{B}_{2}]\\
\text{$D$-term}\,\,& \sum_{a=1}^{3}[\mathsf{B}_{a},\mathsf{B}_{a}^{\dagger}]+\sum_{z}\mathsf{I}_{z}\mathsf{I}_{z}^{\dagger}-\sum_{w}\mathsf{J}_{w}^{\dagger}\mathsf{J}_{w}=\zeta_{\text{1d}} \mathbf{1}_{k}
\end{array}\label{eq:one-leg-PTquiver}
\eea
where we use 1d $\mathcal{N}=4$ notation for the quantum-mechanical fields. The moduli space is
\bea
\mathcal{M}^{\PT}_{k,\lambda}=\left\{\text{$F$-term, $D$-term of \eqref{eq:one-leg-PTquiver}\text{ with $\zeta_{\text{1d}}<0$}}\right\}/U(k).
\eea

Compared with the D6--D0 quiver, there are additional fundamental chirals $\mathsf{I}_{z}$ and antifundamental chirals $\mathsf{J}_{w}$, together with deformations $\delta\mathsf{W}_w$ of the superpotential, all determined by the Young diagram $\lambda$: the $\mathsf{I}_{z}$ correspond to the addable boxes $z=(z_{1},z_{2})\in A(\lambda)$, and the $\mathsf{J}_{w}$ to the removable boxes $w=(w_{1},w_{2})\in R(\lambda)$. The explicit formulas are given in Appendix~\ref{app:one-leg-detail}.

The rotations of the planes $\mathbb{C}_{a}$ $(a=1,2,3)$ act on the chiral fields as
\bea
\mathsf{B}_{1}\rightarrow q_{1}^{-1}\mathsf{B}_{1},\quad \mathsf{B}_{2}\rightarrow q_{2}^{-1}\mathsf{B}_{2},\quad \mathsf{B}_{3}\rightarrow q_{3}^{-1}\mathsf{B}_{3},\\
\mathsf{I}_{z}\rightarrow q_{1}^{-z_{1}+1}q_{2}^{-z_{2}+1}\mathsf{I}_{z},\quad \mathsf{J}_{w}\rightarrow q_{4}q_{1}^{w_{1}}q_{2}^{w_{2}}\mathsf{J}_{w}.
\eea

As shown in \cite{Kimura:2025lfo}, the chamber $\zeta_{\text{1d}}>0$ gives DT counting, while the chamber $\zeta_{\text{1d}}<0$ gives the PT counting we are interested in. The two chambers correspond to the reference vectors of the Jeffrey--Kirwan (JK) residue prescription
\bea\label{eq:DTPT-etavector}
\eta_{0}=(1,\ldots,1),\quad \tilde{\eta}_{0}=(-1,\ldots,-1),
\eea
respectively. The equivariant PT$_{3}$ one-leg vertex is then defined by
\bea
\mathsf{PT}_{\bar{4};\varnothing\varnothing\lambda}[\fq;q_{1,2,3,4}]=\sum_{k=0}^{\infty}\fq^{k}\mathsf{PT}_{\bar{4};\varnothing\varnothing\lambda}[k]
\eea
where
\bea
\mathsf{PT}_{\bar{4};\varnothing\varnothing\lambda}[k]&=\frac{1}{k!}\left(\frac{\sh(\eps_{14,24,34})}{\sh(\eps_{1,2,3,4})}\right)^{k}\oint_{\tilde{\eta}_{0}}\prod_{I=1}^{k}\left[\frac{d\phi_I}{2\pi i}\mathcal{Z}^{\D6\tbar\D2\tbar\D0}_{\bar{4};\varnothing\varnothing\lambda}(\fra,\phi_I)\right]\prod_{I<J}\mathcal{Z}^{\D0\tbar\D0}(\phi_I,\phi_J),
\eea
and 
\bea
\mathcal{Z}^{\D6\tbar\D2\tbar\D0}_{\bar{4};\varnothing\varnothing\lambda}(\fra,\phi_I)&=\prod_{(i,j)\in A(\lambda)}\frac{\sh(\phi_I-\fra+(i-1)\eps_{1}+(j-1)\eps_2+\eps_4)}{\sh(\phi_I-\fra+(i-1)\eps_{1}+(j-1)\eps_2)}\\
&\times \prod_{(i,j)\in R(\lambda)}\frac{\sh(\fra-\phi_I-i\eps_{1}-j\eps_2)}{\sh(\fra-\phi_I-(i-1)\eps_{1}-(j-1)\eps_2+\eps_3)},\\
\mathcal{Z}^{\D0\tbar\D0}(\phi_I,\phi_J)&=\frac{\sh(\phi_I-\phi_J)\sh(\phi_I-\phi_J+\eps_{14,24,34})}{\sh(\phi_I-\phi_J+\eps_{1,2,3,4})}\frac{\sh(\phi_J-\phi_I)\sh(\phi_J-\phi_I+\eps_{14,24,34})}{\sh(\phi_J-\phi_I+\eps_{1,2,3,4})}
\eea
and $\sh(x)=2\sinh(x/2)$. The contour $\oint_{\tilde\eta_0}$ is the JK prescription \eqref{eq:JK-full}.

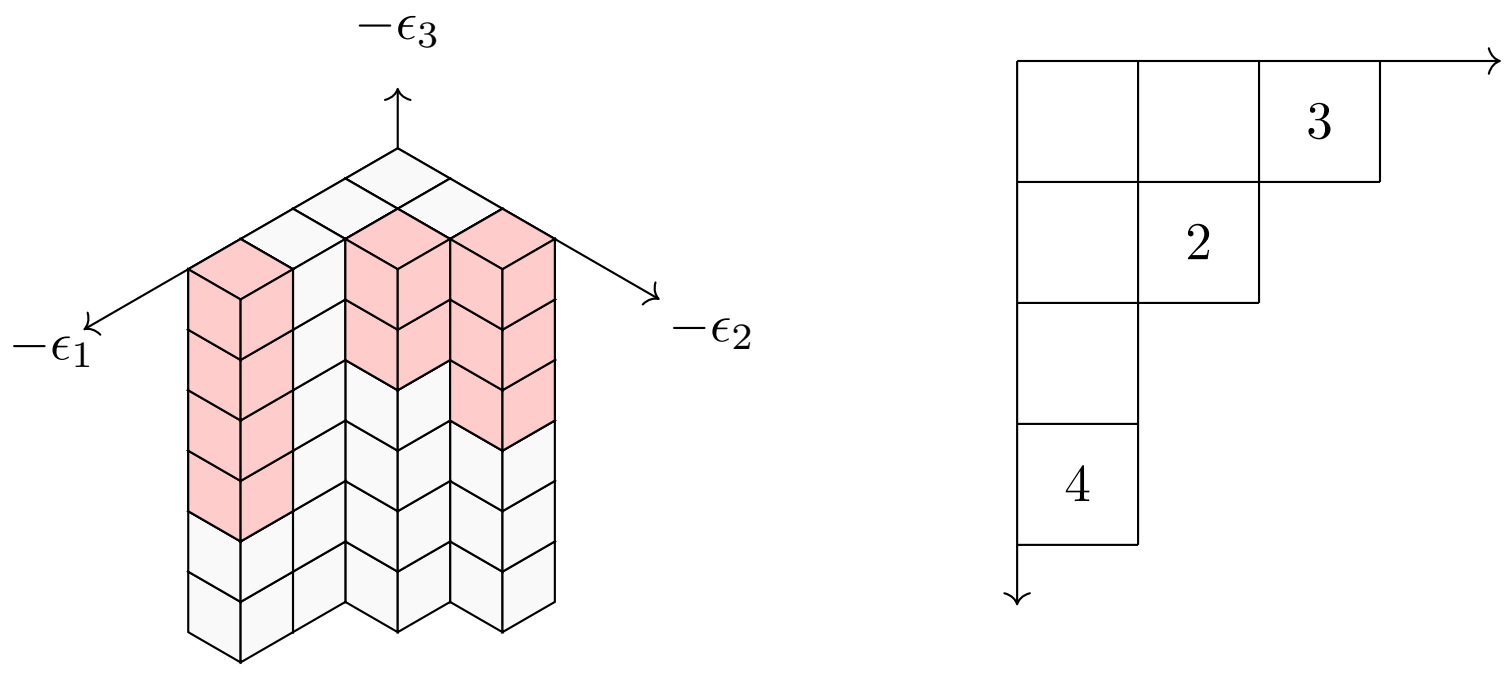
\begin{figure}
\centering
\adjustbox{valign=c}{\begin{tikzpicture}[scale=0.4]

\draw[->] (0,6)--+(-30:5); 
    \draw[->] (0,6)--+(210:6);
    \draw[->] (0,0)--(90:7);
    \node[yshift=2.4cm] at (210:6.6){$-\eps_1$};
    \node[yshift=2.4cm] at (-30:6){$-\eps_2$};
    \node[] at (90:8){$-\eps_3$};
    
    \planepartition{{6,6,6},{6,6},{6},{6}}
     \redxcube{1}{3}{4}
     \redxcube{1}{3}{5}
     \redxcube{1}{3}{6}

    \redxcube{2}{2}{5}
     \redxcube{2}{2}{6}

     \redxcube{4}{1}{3}
     \redxcube{4}{1}{4}
     \redxcube{4}{1}{5}
     \redxcube{4}{1}{6}
 
\end{tikzpicture}}\qquad \qquad  \adjustbox{valign=c}{ \begin{tikzpicture}[scale=0.8]
    \draw[->] (0,0)--(4,0);
     \draw[->] (0,0)--(0,-4.5);
     \draw (0,-1)--(3,-1);
     \draw (0,-2)--(2,-2);
     \draw (0,-3)--(1,-3);
     \draw (0,-4)--(1,-4);
     \draw (3,0)--(3,-1);
     \draw (2,0)--(2,-2);
     \draw (1,0)--(1,-4);
      \node at (2.5,-0.5){$3$};
     \node at (1.5,-1.5){$2$};
     \node at (0.5,-3.5){$4$};
\end{tikzpicture}}
\caption{One-leg PT counting and reverse plane partition.}
\label{fig:one-leg-PT}
\end{figure}

Evaluating the JK residues gives a combinatorial description of PT configurations, which is simple for a single leg. The Young diagram $\lambda$ forms the base of an
infinite leg extending in the negative direction. PT configurations are obtained by stacking finitely many boxes inside this infinite leg, regarded as an empty tube, where each box must be supported
as if gravity pointed toward the direction $(1,1,1)$. The resulting configurations are described by reverse plane
partitions with base $\lambda$, the same set that supports the vortex flux sum in \eqref{eq:colored-rpp-condition}; the number of added boxes is the
D0-brane charge. Figure~\ref{fig:one-leg-PT} shows a one-leg PT configuration and its reverse plane partition.

We now turn to the higher-rank generalizations, which were also proposed on the PT side in \cite{Kimura:2025lfo}. D6-branes with different orientations break half of the supersymmetry, and the D0-brane theory becomes a 1d $\mathcal{N}=2$ quantum mechanics. Its quiver and $J$- and $E$-terms are lengthy (see Appendix~\ref{app:one-leg-detail}), and the partition function can be written directly in terms of the rank-one framing factors. We label the three families of legs by
$A=12,23,31$, with $c=\overline{A4}$, and write their asymptotic data as
$\underline{\vec\lambda}=\{\lambda_A^{(\alpha)}\}$.
Each framing factor is obtained from the reference
$\mathrm{D6}_{\bar4}$ one-leg factor by the ordered relabeling
$(1,2,3,4)\mapsto(a,b,4,c)$ for $A=ab$, as in \eqref{eq:one-leg-coordinate-relabeling}. The resulting one-leg PT vertex is
\bea\label{eq:higherrank-PT-vertex}
\mathsf{PT}_{\underline{\vec{\lambda}}}[\fq,q_{1,2,3,4}]&=\sum_{k=0}^{\infty}\fq^{k}\mathsf{PT}_{\underline{\vec{\lambda}}}[k],\\
\mathsf{PT}_{\underline{\vec{\lambda}}}[k]&=\frac{1}{k!}\left(\frac{\sh(\eps_{14,24,34})}{\sh(\eps_{1,2,3,4})}\right)^{k}\oint_{\tilde{\eta}_{0}}\prod_{I=1}^{k}\left[\frac{d\phi_I}{2\pi i}\prod_{a=1}^{3}\prod_{\alpha=1}^{N_{\bar{a}}}\mathcal{Z}^{\D6\tbar\D2\tbar\D0}_{\bar{a};\varnothing\varnothing\lambda_{\overline{a4}}^{(\alpha)}}(\fra_{\bar{a},\alpha},\phi_I)\right]\prod_{I<J}\mathcal{Z}^{\D0\tbar\D0}(\phi_I,\phi_J)
\eea
where $(N_{234},N_{134},N_{124})=(L,M,N)$. For a single stack of D6-branes, this is the higher-rank one-leg PT vertex.

Tetrahedron instantons describe D0-branes bound to intersecting
D6-branes supported on the coordinate three-planes of
$\mathbb{C}^4$
\cite{Pomoni:2021hkn,Pomoni:2023nlf}.
Their interpretation in Donaldson--Thomas theory was developed
in \cite{Fasola:2023ypx}.
The construction above gives the PT counterpart of this
counting problem, with fixed one-leg asymptotic data
$\underline{\vec{\lambda}}$ on the nonempty D6-brane stacks.
When several stacks are nonempty, we therefore refer to \eqref{eq:higherrank-PT-vertex} as the one-leg
\emph{tetrahedron PT vertex}.
The different stacks share the same D0 gauge node, so this
vertex includes their interactions.

We give a gluing prescription for these vertices in
Appendix~\ref{app:tetrahedronDTPTvertex}, leaving its mathematical foundations for
future work.

\paragraph{Proposal}
Fix a Higgs vacuum $\underline{\vec\lambda}$ of the 3d spiked quiver, with
$|\underline{\vec\lambda}|=N_4$. The D6--D2--D0 quantum mechanics with this prescribed
leg defines the one-leg PT vertex
$\mathsf{PT}_{\underline{\vec\lambda}}$ of \eqref{eq:higherrank-PT-vertex}. For the boundary conditions in \eqref{eq:boundary-conditions} and \eqref{eq:halfindex-integral}, the flux-dependent signs discussed in Section~\ref{sec:halfindex-def} cancel in each flux sector. With this choice, we propose that the normalized vortex partition function agrees with the
one-leg PT generating function at the same vacuum:
\bea
\mathcal{Z}_{\mathrm{vortex}}^{\underline{\vec\lambda}}
(\mathfrak q;q_1,q_2,q_3,q_4)
\,{=}\,
\mathsf{PT}_{\underline{\vec\lambda}}
[\mathfrak q;q_1,q_2,q_3,q_4].
\label{eq:higher-rank-vortex-PT-proposal}
\eea
On the left, $\mathfrak q$ counts the total magnetic flux on the D2-branes;
on the right, it counts the rank of the D0 gauge node. The collection of Young diagrams $\underline{\vec\lambda}$ is fixed on both sides.

We further propose that the charge-$k$ vortex moduli space
$\mathcal M^{\mathrm{vort}}_{k,\underline{\vec\lambda}}$ is equivariantly
identified with the moduli space
$\mathcal M^{\mathrm{PT}}_{k,\underline{\vec\lambda}}$ of PT-stable
D6--D2--D0 quiver representations with leg $\underline{\vec\lambda}$. Note that for a single D6-brane, this is the known one-leg correspondence \cite{Crew:2020psc}.

\paragraph{Comparison with vortex partition functions}
We have confirmed \eqref{eq:higher-rank-vortex-PT-proposal} by direct computation in various cases. Below, we present three representative examples in which the vortex and PT series can be compared explicitly in every flux sector.

\begin{itemize}
    \item $(L,M,N)=(0,0,1)$ with $\Yboxdim{7pt}\lambda_{12}=\yng(1)$.
    On the PT side, the framing node contribution is
    \bea\Yboxdim{5pt}
\mathcal{Z}^{\D6\tbar\D2\tbar\D0}_{\bar{3};\varnothing\varnothing\,\yng(1)}(\fra,\phi_I)=\frac{\sh(\phi_I-\fra+\eps_{13,23})}{\sh(\phi_I-\fra+\eps_{1,2})}\frac{\sh(\fra-\phi_I+\eps_{34})}{\sh(\fra-\phi_I+\eps_{4})}. \label{eq:framing-oneleg-onebox}
\eea
With the reference vector $\tilde{\eta}_0$, the poles with nonzero residue are
\bea\label{eq:PT-oneleg-poles}
\phi_I=\fra+\eps_{4}I,\quad I=1,\ldots,k
\eea
and we have{\Yboxdim{5pt}
\bea
\mathsf{PT}_{\bar{3};\varnothing\varnothing\,\yng(1)}[\fq;q_{1,2,3,4}]&=\sum_{k=0}^{\infty}\fq^{k} \mathsf{PT}_{\bar{3};\varnothing\varnothing\,\yng(1)}[k],\quad 
\mathsf{PT}_{\bar{3};\varnothing\varnothing\,\yng(1)}[k]=\prod_{i=1}^{k}\frac{\sh(\eps_3+i\eps_4)}{\sh(i\eps_{4})}.
\label{eq:PT-oneleg-onebox}
\eea}

On the vortex side, we have
\bea
\mathcal{Z}^{\underline{\vec{\lambda}}}_{\text{vortex}}&=\sum_{k\in\mathbb{Z}_{\geq 0}}\fq^{k} q_{3}^{\frac{k}{2}}\frac{(1;q_{4})_{-k}}{(q_{3}^{-1};q_{4})_{-k}}=\sum_{k=0}^{\infty}\fq^{k} q_{3}^{k/2}\prod_{i=1}^{k}\frac{1-q_{3}^{-1}q_{4}^{-i}}{1-q_{4}^{-i}}
\eea  
and the two sides agree.

\item $(L,M,N)=(0,0,2)$ with $\Yboxdim{7pt}\vec{\lambda}_{12}=(\yng(1),\varnothing)$. On the PT side, the framing node contribution is
\bea\Yboxdim{5pt}
\mathcal{Z}^{\D6\tbar\D2\tbar\D0}_{\bar{3};\varnothing\varnothing\,\yng(1)}(\fra_{\bar3,1},\phi_I)\mathcal{Z}^{\D6\tbar\D2\tbar\D0}_{\bar{3};\varnothing\varnothing\varnothing}(\fra_{\bar3,2},\phi_I)&=\frac{\sh(\phi_I-\fra_{\bar3,1}+\eps_{13,23})}{\sh(\phi_I-\fra_{\bar3,1}+\eps_{1,2})}\frac{\sh(\fra_{\bar3,1}-\phi_I+\eps_{34})}{\sh(\fra_{\bar3,1}-\phi_I+\eps_{4})}\\
&\times \frac{\sh(\phi_I-\fra_{\bar3,2}+\eps_{3})}{\sh(\phi_I-\fra_{\bar3,2})}.
\eea
With the reference vector $\tilde{\eta}_0$, the poles are the same as in the previous example, $\phi_I=\fra_{\bar3,1}+\eps_4 I$ $(I=1,\ldots,k)$, but the second D6-brane contributes additional matter factors. The partition function is
\bea
\mathsf{PT}_{\underline{\vec{\lambda}}}[\fq;q_{1,2,3,4}]&=\sum_{k=0}^{\infty}\fq^{k}\prod_{i=1}^{k}\left(\frac{\sh(\eps_{3}+i\eps_4)}{\sh(i\eps_4)}\frac{\sh(\fra_{\bar3,1}-\fra_{\bar3,2}+\eps_3+i\eps_4)}{\sh(\fra_{\bar3,1}-\fra_{\bar3,2}+i\eps_{4})}\right).
\eea

On the vortex side, we find
\bea
\mathcal{Z}_{\mathrm{vortex}}^{\underline{\vec{\lambda}}}
&=\sum_{k\geq 0}(\fq q_3)^k
\frac{(1;q_4)_{-k}\,(v_{\bar3,2}/v_{\bar3,1};q_4)_{-k}}
     {(q_3^{-1};q_4)_{-k}\,(q_3^{-1}v_{\bar3,2}/v_{\bar3,1};q_4)_{-k}}\\
     &=\sum_{k\geq 0}\fq^k\prod_{i=1}^{k}
\left[
\frac{q_3(1-q_3^{-1}q_4^{-i})}{1-q_4^{-i}}\,
\frac{1-q_3^{-1}q_4^{-i}v_{\bar3,2}/v_{\bar3,1}}{1-q_4^{-i}v_{\bar3,2}/v_{\bar3,1}}
\right].
\eea
The two sides agree.

\item $(L,M,N)=(1,1,0)$ with ${\lambda}_{23}={\lambda}_{31}={\Yboxdim{7pt}\yng(1)}$. On the PT side, the framing node contribution is{\Yboxdim{5pt}
\bea
&\mathcal{Z}^{\D6\tbar\D2\tbar\D0}_{\bar{1};\varnothing\varnothing\,\yng(1)}(\fra_{\bar1},\phi_I)\mathcal{Z}^{\D6\tbar\D2\tbar\D0}_{\bar{2};\varnothing\varnothing\,\yng(1)}(\fra_{\bar2},\phi_I)\\
=&\frac{\sh(\phi_I-\fra_{\bar1}+\eps_{12,13})}{\sh(\phi_I-\fra_{\bar1}+\eps_{2,3})}\frac{\sh(\fra_{\bar1}-\phi_I+\eps_{14})}{\sh(\fra_{\bar1}-\phi_I+\eps_{4})}\frac{\sh(\phi_I-\fra_{\bar2}+\eps_{12,23})}{\sh(\phi_I-\fra_{\bar2}+\eps_{1,3})}\frac{\sh(\fra_{\bar2}-\phi_I+\eps_{24})}{\sh(\fra_{\bar2}-\phi_I+\eps_{4})}
\eea}
For the reference vector $\tilde{\eta}_0$, the poles form two strings of lengths $k_{23}$ and $k_{31}$, with $k_{23}+k_{31}=k$. Up to a permutation of the integration variables, their positions are
\bea
\{\phi_I\}_{I=1}^{k}
&= \{\mathfrak{a}_{\bar1}+i\epsilon_4 \mid 1\leq i\leq k_{23}\}
\cup \{\fra_{\bar2}+j\epsilon_4 \mid 1\leq j\leq k_{31}\}.
\eea
The partition function is then
\bea
 \mathsf{PT}_{\underline{\vec{\lambda}}}[\fq,q_{1,2,3,4}]&=\sum_{k_{23},k_{31}}\fq^{k_{23}+k_{31}}
R_{k_{23},k_{31}},\\
R_{k_{23},k_{31}}
&=
\prod_{i=1}^{k_{23}}
\frac{\sh(\epsilon_1+i\epsilon_4)}{\sh(i\epsilon_4)}
\frac{\sh(\Delta+\epsilon_2+i\epsilon_4)}
     {\sh(\Delta+i\epsilon_4)}
\prod_{j=1}^{k_{31}}
\frac{\sh(\epsilon_2+j\epsilon_4)}{\sh(j\epsilon_4)}
\frac{\sh(-\Delta+\epsilon_1+j\epsilon_4)}
     {\sh(-\Delta+j\epsilon_4)}
\\
&\times
\prod_{\ell=0}^{k_{23}-k_{31}-1}
\frac{\sh(\Delta+(\ell+1)\epsilon_4)
      \prod_{a=1}^{3}\sh(\Delta+\ell\epsilon_4-\epsilon_a)}
     {\sh(\Delta+\ell\epsilon_4)
      \prod_{a<b}\sh(\Delta+\ell\epsilon_4-\epsilon_{ab})},
\eea
where $\Delta=\fra_{\bar1}-\fra_{\bar2}$. For $k_{23}<k_{31}$, the last product denotes the inverse product
over $\ell=k_{23}-k_{31},\ldots,-1$.

On the vortex side, we find
\bea
\mathcal{Z}_{\mathrm{vortex}}^{\underline{\vec{\lambda}}}
&=\sum_{k_{23},k_{31}\geq0}
(\fq q_{12}^{1/2})^{k_{23}+k_{31}}\,
F_{23}(k_{23})F_{31}(k_{31})
H(x,k_{23}-k_{31}),\\
F_{23}(k)
&=
\frac{(1;q_4)_{-k}(x^{-1};q_4)_{-k}}
     {(q_1^{-1};q_4)_{-k}(q_2^{-1}x^{-1};q_4)_{-k}},\quad
F_{31}(k)
=
\frac{(x;q_4)_{-k}(1;q_4)_{-k}}
     {(q_1^{-1}x;q_4)_{-k}(q_2^{-1};q_4)_{-k}},\\
H(x,n)
&=
\frac{(q_1^{-1}x;q_4)_n(q_2^{-1}x;q_4)_n
      (q_3^{-1}x;q_4)_n(q_4x;q_4)_n}
     {(x;q_4)_n(q_{12}^{-1}x;q_4)_n
      (q_{13}^{-1}x;q_4)_n(q_{23}^{-1}x;q_4)_n},\qquad x=v_{\bar1}/v_{\bar2}
\eea
which agrees with the PT computation.

\end{itemize}

\subsection{BPS/CFT correspondence}\label{sec:BPSCFT}
The BPS/CFT correspondence relates supersymmetric partition functions of $\Omega$-deformed gauge theories to conformal blocks and Ward identities of 2d conformal field theories and their $q$-deformations \cite{Nekrasov:2015wsu,Nekrasov:2016ydq,Nekrasov:2017rqy,Nekrasov:2017gzb}. The $qq$-characters and the associated quiver W-algebras play a central role. Regularity of the appropriate $qq$-character expectation values encodes the non-perturbative Dyson--Schwinger equations obeyed by the partition function \cite{Kimura:2016ebq,Kimura:2019hnw,Kimura:2022zsm}. In the gauge-origami setting, each stack of D-branes is assigned a free-field vertex operator, and the sum over torus fixed points is generated by acting with screening charges \cite{Kimura:2023bxy,Kimura:2024xpr,Kimura:2024osv,Kimura:2025lfo}.

Here we apply this construction to the half-index of the spiked quiver. The D2$_{4}$- and D6$_{\bar{a}}$-branes of \eqref{eq:DbraneADHMflavor} are represented by vertex operators $\mathsf{S}_{4}$ and $\mathsf{W}_{\bar{a}}$, the Higgs vacuum determines which operators are inserted and where, and the screening charge $\mathscr{Q}_{4}$ generates the sum over vortex fluxes. The result is a free-field realization of the vortex partition functions for a single D6-brane stack and for the rank-$(L,M,N)$ spiked quiver. Since the same vertex operators generate the DT$_3$ and PT$_3$ $qq$-characters of \cite{Kimura:2024osv,Kimura:2025lfo}, this realization is the algebraic counterpart of the relation between vortex counting and PT counting discussed in the previous section. A related analysis of non-perturbative Dyson--Schwinger equations directly in three dimensions may also be found in \cite{Haouzi:2020bso}.

Following \cite{Kimura:2023bxy}, we introduce the D-brane vertex operators associated with the D2- and D6-branes of \eqref{eq:DbraneADHMflavor} as
\bea
\mathsf{S}_{a}(x)=\mathsf{s}_{a,0}(x):\exp\left(\sum_{n\neq 0}\mathsf{s}_{a,n}x^{-n}\right):,\quad \mathsf{W}_{\bar{a}}(x)=\mathsf{w}_{\bar{a},0}(x):\exp\left(\sum_{n\neq 0}\mathsf{w}_{\bar{a},n}x^{-n}\right):,\quad a\in\four,
\eea
where
\bea
\mathsf{s}_{a,n}=\frac{\mathsf{a}_{n}}{\bfP_{a}^{[-n]}},\quad \mathsf{w}_{\bar{a},n}=\frac{\mathsf{a}_{n}}{\bfP_{\bar{a}}^{[-n]}},\quad [\mathsf{a}_{n},\mathsf{a}_{m}]=-\frac{1}{n}\delta_{n+m,0}\bfP_{\four}^{[n]}
\eea
and $\mathsf{s}_{a,0}(x),\mathsf{w}_{\bar{a},0}(x)$ are zero-modes. Here, $:\mathcal{O}(x):$ denotes normal ordering and $\bfP_{\mathcal S}^{[n]}=\prod_{a\in\mathcal S}(1-q_a^n)$ is the $n$-th Adams operation applied to $\bfP_{\mathcal S}$. For later use, an operator such as $\mathcal{O}(x)^{-1}$ means $:\mathcal{O}(x)^{-1}:$.

We fix the zero-mode contributions by requiring that the operator products among the D2$_{4}$ and D6$_{\bar{1},\bar{2},\bar{3}}$ vertex operators take the form
\bea\label{eq:screening-framing-check}
\mathsf{S}_{4}(x)\mathsf{S}_{4}(x')&=\frac{(x'/x;q_{4})'_{\infty}(q_{14,24,34}x'/x;q_{4})'_{\infty}}{(q_{1,2,3}^{-1}x'/x;q_{4})'_{\infty}(q_{4}x'/x;q_{4})'_{\infty}}:\mathsf{S}_{4}(x)\mathsf{S}_{4}(x'):,\\
\mathsf S_4(q_4x)\mathsf W_{\bar i}(q_i^{-1}v)^{-1}
&=\frac{(q_i^{-1}v/x;q_4)'_\infty}{(v/x;q_4)'_\infty}
:\mathsf S_4(q_4x)\mathsf W_{\bar i}(q_i^{-1}v)^{-1}:
\eea
where we assumed $|q_4|<1$ and $(x;q_{4})'_{\infty}$ is defined in \eqref{eq:symm-q-Pochhammer}.

We also define the screening charge as
\bea
\mathscr{Q}_{4}(x)=\sum_{k\in\mathbb{Z}}\mathfrak{q}^{k}\mathsf{S}_{4}(xq_{4}^{k}).
\eea
For later use, we renormalize the screening charge as
\bea
\widetilde{\mathscr{Q}}_{4}(x)=\left(q_{4}^{-1/2}x\right)^{\xi}\mathscr{Q}_{4}(x)
\eea
where $\xi$ is the FI parameter \eqref{eq:3dFI}.

\paragraph{Rank-one ADHM theory}
Consider first the rank-one 3d $\mathcal{N}=4$ $U(N_{4})$ ADHM theory, realized by $N_{4}$ D2$_{4}$-branes and a single D6$_{124}$-brane. Its Higgs branch vacua are labeled by a single Young diagram $\lambda$ with $|\lambda|=N_{4}$ boxes, and we define the operator-valued block by
\bea
\mathbb B_{\lambda}&=\prod_{\mathsf{a}\in\lambda}^{\succ}\left[\frac{(1;q_{4})'_{\infty}(q_{12}^{-1};q_{4})'_{\infty}}{(q_{1}^{-1};q_{4})'_{\infty}(q_{2}^{-1};q_{4})'_{\infty}}\widetilde{\mathscr{Q}}_{4}(q_{4}s_{\mathsf{a}})\right]\mathsf{W}_{\bar{3}}(q_{3}^{-1}v)^{-1},\quad s_{\mathsf{a}}=vq_{1}^{-i+1}q_{2}^{-j+1},\quad (i,j)\in\lambda,
\eea
where the product is specified as follows. 

We order the boxes by choosing a sequence of Young diagrams
\bea
    \varnothing=\lambda^{(0)}
    \subset\lambda^{(1)}
    \subset\cdots
    \subset\lambda^{(N_4)}=\lambda,
    \qquad
    \lambda^{(j)}\setminus\lambda^{(j-1)}=\{\mathsf{a}_j\}.
\eea
Thus \(\mathsf{a}_j\) is the box added at step \(j\), and the corresponding
weights are ordered as \(s_{\mathsf{a}_1},\ldots,s_{\mathsf{a}_{N_4}}\). We place
each newly added screening factor to the left of the preceding
ones. With this convention,
\bea
    \prod_{\mathsf{a}\in\lambda}^{\succ}\mathcal O_\mathsf{a}
    :=\mathcal O_{\mathsf{a}_{N_4}}\cdots\mathcal O_{\mathsf{a}_1}.
\eea
The framing operator stands to the right of the screening factors. The factor \((1;q_4)'_\infty\) in each diagonal prefactor
vanishes on its own. We cancel these factors against the
corresponding poles in the ordered contractions before
evaluation. The resulting product is well defined for
generic equivariant parameters.

The half-index is then computed as
\bea
\left\langle\mathbb B_\lambda\right\rangle
&=
\left(\prod_{\mathsf{a}\in\lambda}\widetilde s_{\mathsf{a}}^{\,\xi}\right)
\sum_{\boldsymbol k\in\mathbb{Z}^{N_{4}}}
\fq^{\sum_{\mathsf{a}\in\lambda} k_{\mathsf{a}}}
\Biggl[
\left(
\frac{(1;q_4)'_\infty(q_{12}^{-1};q_4)'_\infty}
     {(q_1^{-1};q_4)'_\infty(q_2^{-1};q_4)'_\infty}
\right)^{N_4}
\prod_{\mathsf{a}\in\lambda}
\frac{(q_3^{-1}q_4^{-k_{\mathsf{a}}}v/s_{\mathsf{a}};q_4)'_\infty}
     {(q_4^{-k_{\mathsf{a}}}v/s_{\mathsf{a}};q_4)'_\infty}\\
&\qquad\times
\prod_{\mathsf{b}\succ\mathsf{a}}
\frac{
    (q_4^{k_{\mathsf{a}}-k_{\mathsf{b}}}s_{\mathsf{a}}/s_{\mathsf{b}};q_4)'_\infty
    \displaystyle\prod_{1\le i<j\le 3}
    (q_{ij}^{-1}q_4^{k_{\mathsf{a}}-k_{\mathsf{b}}}s_{\mathsf{a}}/s_{\mathsf{b}};q_4)'_\infty
}{
    (q_4^{1+k_{\mathsf{a}}-k_{\mathsf{b}}}s_{\mathsf{a}}/s_{\mathsf{b}};q_4)'_\infty
    \displaystyle\prod_{i=1}^{3}
    (q_i^{-1}q_4^{k_{\mathsf{a}}-k_{\mathsf{b}}}s_{\mathsf{a}}/s_{\mathsf{b}};q_4)'_\infty
}
\Biggr]_{\mathrm{reg}},
\label{eq:freefield-block-vev}
\eea
where $\left\langle \mathcal{O}\right \rangle =\bra{0}\mathcal{O}\ket{0}$. The subscript \(\mathrm{reg}\) means that common factors are
canceled before the product is evaluated at the fixed-point
weights. After this cancellation, only configurations in which $\{k_{\mathsf{a}}\}_{\mathsf{a}\in\lambda}$ forms a reverse plane partition contribute.

Dividing by the vacuum expectation value of the zero-flux sector $\mathbb{B}_{\lambda,0}$ ($\boldsymbol{k}=\boldsymbol{0}$) of the block, we obtain
\bea\label{eq:freefield-vortex-forms}
\frac{\left\langle \mathbb B_\lambda\right\rangle}{\left\langle \mathbb B_{\lambda,0}\right\rangle}
&=\sum_{\{k_{\mathsf{a}}\in\mathbb{Z}\}_{\mathsf{a}\in\lambda}}
\mathfrak{q}_{\text{eff}}^{\sum_{\mathsf{a}}k_{\mathsf{a}}}
\prod_{\mathsf{a}\in\lambda}
\frac{(v/s_{\mathsf{a}};q_{4})_{-k_{\mathsf{a}}}}
     {(q_{3}^{-1}v/s_{\mathsf{a}};q_{4})_{-k_{\mathsf{a}}}}
\prod_{\mathsf{b}\succ\mathsf{a}}
\frac{
(q_{1,2,3}^{-1}s_{\mathsf{a}}/s_{\mathsf{b}};q_{4})_{k_{\mathsf{a}}-k_{\mathsf{b}}}
(q_{4}s_{\mathsf{a}}/s_{\mathsf{b}};q_{4})_{k_{\mathsf{a}}-k_{\mathsf{b}}}
}{
(s_{\mathsf{a}}/s_{\mathsf{b}};q_{4})_{k_{\mathsf{a}}-k_{\mathsf{b}}}
(q_{12,23,13}^{-1}s_{\mathsf{a}}/s_{\mathsf{b}};q_{4})_{k_{\mathsf{a}}-k_{\mathsf{b}}}
},
\eea
where the effective topological term is \eqref{eq:effective-topological}. After using the \(q_4\)-Pochhammer identities, one can rewrite the
product over \(\mathsf b\succ  \mathsf a\) as a product over \(\mathsf a\ne  \mathsf  b\) and obtain the equality with
\eqref{eq:exceptional-vortex-triality} or \eqref{eq:exceptional-vortex-2}:
\bea
\frac{\left\langle \mathbb B_\lambda\right \rangle}{\left\langle \mathbb B_{\lambda,0}\right \rangle}=&\mathcal{Z}^{\lambda}_{\text{vortex}}.
\eea

\paragraph{Rank-\texorpdfstring{$N$}{N} ADHM theory}
Next, consider the rank-$N$ ADHM theory with $N_{4}$ D2$_{4}$-branes and $N$ D6$_{\bar{3}}$-branes. Its Higgs branch vacua are labeled by $N$-tuples of Young diagrams $\vec{\lambda}=(\lambda^{(\alpha)})_{\alpha=1}^{N}$ with $|\vec{\lambda}|=N_{4}$ boxes in total, and the half-index is computed from
\bea\label{eq:freefield-rankN}
\mathbb B_{\vec{\lambda}}=\prod_{\mathsf a\in\vec{\lambda}}^{\succ}\left[\frac{(1;q_{4})'_{\infty}(q_{12}^{-1};q_{4})'_{\infty}}{(q_{1}^{-1};q_{4})'_{\infty}(q_{2}^{-1};q_{4})'_{\infty}}\widetilde{\mathscr{Q}}_{4}(q_{4}s_{\mathsf{a}})\right]:\prod_{\alpha=1}^{N}\mathsf{W}_{\bar{3}}(q_{3}^{-1}v_{\alpha})^{-1}:,
\eea
where
\bea
s_{\mathsf{a}}=v_{\alpha}q_{1}^{-i+1}q_{2}^{-j+1},\quad(i,j)\in\lambda^{(\alpha)}.
\eea
For higher rank, we extend the order \(\succ\) to the boxes of
\(\vec\lambda\) by building the tuple from the empty tuple.
At each step, we add one box to one component, keeping every
component a Young diagram. We write \(\mathsf b\succ \mathsf a\) if \(\mathsf b\) is
added after \(\mathsf a\), and place the screening factor for \(\mathsf b\)
to the left of that for \(\mathsf a\).

The half-index is again the vacuum expectation value of the block,
\bea
\left\langle \mathbb{B}_{\vec{\lambda}}\right \rangle.
\eea
Denoting by \(\mathbb{B}_{\vec{\lambda},0}\) the zero-flux term and dividing by it, we obtain
\bea
\frac{\langle \mathbb{B}_{\vec{\lambda}} \rangle}
     {\langle \mathbb{B}_{\vec{\lambda},0} \rangle}
&= \mathcal{Z}^{\vec{\lambda}}_{\mathrm{vortex}} .
\eea

\paragraph{Rank-\texorpdfstring{$(L,M,N)$}{(L,M,N)} spiked quiver theory}
With $(L,M,N)$ D6$_{\bar{1},\bar{2},\bar{3}}$-branes and $N_{4}$ D2$_{4}$-branes, the Higgs branch vacua are labeled by the collection of Young diagrams $\underline{\vec\lambda}$ as defined in \eqref{eq:spiked-alldiagrams} with $|\underline{\vec{\lambda}}|=N_{4}$ boxes in total.

We define the operator-valued block as
\bea\label{eq:freefield-tetra}
\mathbb{B}_{\underline{\vec{\lambda}}}&=\prod_{\mathsf a\in\underline{\vec{\lambda}}}^{\succ}\left[\frac{(1;q_{4})'_{\infty}(q_{c d}^{-1};q_{4})'_{\infty}}{(q_{c}^{-1};q_{4})'_{\infty}(q_{d}^{-1};q_{4})'_{\infty}}\widetilde{\mathscr{Q}}_{4}(q_{4}s_{\mathsf{a}})\right]\\
&\quad \times :\prod_{\alpha=1}^{L}\mathsf{W}_{\bar{1}}(q_{1}^{-1}v_{234,\alpha})^{-1}\prod_{\beta=1}^{M}\mathsf{W}_{\bar{2}}(q_{2}^{-1}v_{314,\beta})^{-1}\prod_{\gamma=1}^{N}\mathsf{W}_{\bar{3}}(q_{3}^{-1}v_{124,\gamma})^{-1}:
\eea
where
\bea
s_{\mathsf{a}}=v_{A4,\alpha}q_{c}^{-i+1}q_{d}^{-j+1},\quad (i,j)\in\lambda_{A}^{(\alpha)},\quad A=cd\in\{12,23,31\}.
\eea
For each box \(\mathsf{a}\in\lambda_A^{(\alpha)}\), the indices
\(c,d\) in the diagonal factor are determined
by its plane \(A=cd\). We also use the same ordering prescription for \(\prod^\succ\) as in the rank-\(N\) ADHM case.

The half-index is again the vacuum expectation value of this block, and dividing by the zero-flux contribution $\mathbb{B}_{\underline{\vec{\lambda}},0}$ gives the vortex partition function,
\bea
\frac{\langle \mathbb{B}_{\underline{\vec{\lambda}}} \rangle}
     {\langle \mathbb{B}_{\underline{\vec{\lambda}},0} \rangle}
&=\mathcal{Z}^{\underline{\vec{\lambda}}}_{\text{vortex}}.
\eea

\section{Twisted indices and gluing of PT vertices}\label{sec:twisted-gluing}

In this section, we study the relation between the refined twisted index of the 3d spiked quiver theory \eqref{spiked-instanton-quiver} on $\mathbb{S}^1\times \mathbb{S}^2$ and the gluing of PT vertices on a local Calabi--Yau fourfold. We consider the local curve
\bea
Z=\operatorname{Tot}\left(
\mathcal{O}(-r_1)\oplus
\mathcal{O}(-r_2)\oplus
\mathcal{O}(-r_3)
\longrightarrow \mathbb{P}^1
\right),
\qquad r_1+r_2+r_3=2,
\eea
where the integers $r_i$ are identified with the $R$-charges of the adjoint chiral multiplets. The D2-branes wrap the compact curve $\mathbb{P}^1$, while the D6-branes wrap threefolds obtained by restricting to two of the three normal directions. In each local chart of $\mathbb{P}^1$, this configuration reduces to the D2--D6 system studied in Sections~\ref{sec:3dADHMflavor} and~\ref{sec:halfindex}. The gluing formalism for the one-leg PT vertices is reviewed in Appendix~\ref{app:gluing}.

The refined twisted index factorizes into contributions from the north and south poles of $\mathbb{S}^2$, labeled by a common Higgs branch vacuum. Under the vortex--PT correspondence of Section~\ref{sec:DTPT-vortexQM}, this factorization takes the form of the gluing of two one-leg PT vertices along the compact curve. The Higgs branch vacuum determines the asymptotic Young diagrams carried by the common edge, while the vortex partition functions give the vertex contributions. The perturbative and classical factors account for the edge factor and the D0-brane weight, respectively.

For a class of 3d $\mathcal N=4$ gauge theories, the twisted index on $\mathbb{S}^1\times\Sigma$ admits a geometric interpretation as the virtual Euler characteristic of a moduli space of twisted quasimaps to the Higgs branch \cite{BullimoreFerrariKim:TwistedQuasimaps}. In the ADHM theory with a single fundamental hypermultiplet, the twisted index on $\mathbb{S}^1\times \mathbb{S}^2$ factorizes into hemisphere blocks, and its generating function over the gauge rank reproduces the PT partition function of the resolved conifold \cite{Crew:2020psc}. We extend this correspondence to stacks of parallel D6-branes on more general local curves and to intersecting D6-brane stacks preserving $\mathcal N=2$ supersymmetry.

We first introduce the refined twisted index of the spiked quiver theory in Section~\ref{sec:twisted-spiked}. In Section~\ref{sec:rankone-twistedindex}, we consider a single stack of $N$ D6-branes and propose its correspondence with the rank-$N$ PT partition function of a local curve. Section~\ref{sec:tetrahedron-twisted} extends the analysis to intersecting D6-brane stacks and the gluing of tetrahedron PT vertices. Finally, in Section~\ref{sec:D2-D8}, we consider the D2--D8 system and relate its twisted index to the gluing of PT$_{4}$ vertices on the local Calabi--Yau fourfold, with the D6-supported PT$_3$ gluing recovered through tachyon condensation.

\subsection{Spiked quiver and the refined twisted index}\label{sec:twisted-spiked}
We first review the 3d $\mathcal{N}=2$ twisted index on $\mathbb{S}^{1}\times\mathbb{S}^{2}$, refined by the angular momentum on $\mathbb{S}^{2}$, obtained via supersymmetric localization \cite{Benini:2015noa,Closset:2016arn} (see also \cite{Crew:2020jyf}), and then apply it to the spiked quiver.

The twisted index counts the graded dimensions of the Hilbert space on $\mathbb{S}^{2}$ in the topologically twisted theory, in which the background $U(1)_{R}$ gauge field is identified with the $U(1)$ spin connection of $\mathbb{S}^{2}$. It is defined as
\begin{equation}\label{eq:3dN=2TwistedIndex}
    \mathcal{I} = \operatorname{Tr}_{\mathcal{H}_{\mathbb{S}^2}}(-1)^F \prod_i t_i^{F_i} q^{2J+R}\fq^{\mathfrak{J}},
\end{equation}
where $\mathcal{H}_{\mathbb{S}^2}$ is the twisted Hilbert space, $F_i$ is the $i$-th flavor charge, $J$ is the angular momentum on $\mathbb{S}^{2}$, and $R$ is the $R$-charge. The fugacity $\fq$ weights the topological charge $\mathfrak{J}$. Only states annihilated by the scalar supercharge contribute.


\paragraph{The contour formula}
Upon localization onto BPS configurations, the path-integral representation of the twisted index \eqref{eq:3dN=2TwistedIndex} reduces to
\bea\label{eq:twisted-general-JK}
    \mathcal{I}= \frac{1}{|W|}\sum_{\mathfrak{m} \in \Gamma_{G^\vee} } \oint_{\text{JK}}  \prod^{\text{rk}(G)}_{I=1} \frac{\mathrm{d}x_I}{2 \pi i x_I}~ \mathcal{I}_{\text{cl}}(x)\, \mathcal{I}_{\text{chi}}(x)\, \mathcal{I}_{\text{vec}}(x),
\eea
where $|W|$ is the order of the Weyl group of the gauge group $G$, and $\Gamma_{G^{\vee}}$ is the cocharacter lattice of $G$. The factors $\mathcal{I}_{\text{cl}},\mathcal{I}_{\text{chi}},\mathcal{I}_{\text{vec}}$ are the classical contribution and the one-loop determinants of the chiral and vector multiplets, respectively.
The contour is specified by the JK residue prescription \eqref{eq:JK-full}, reviewed in Appendix~\ref{app:JK}.

\paragraph{Classical contributions}
For unitary gauge groups, one can also introduce an FI term, which contributes a factor $\mathfrak{q}^{\mathfrak{m}}$ to the classical term $\mathcal{I}_{\text{cl}}$. Since we will not need the Chern--Simons term in what follows, we omit it here.

\paragraph{One-loop determinants}
The one-loop determinants for the chiral multiplets and vector multiplets are given as follows.
\begin{itemize}
    \item Chiral multiplet
    \begin{equation}\label{eq:twisted-chiral}
        \mathcal{I}_{\text{chi}}
        = \prod_{\rho \in \mathfrak{R}}
        \frac{(x^{\rho})^{\frac{B}{2}}}
        {(x^{\rho}q^{(1-B)};q^2)_B},
        \qquad
        B=\rho(\mathfrak{m})-r_{\rho}+1,
    \end{equation}
    where $\mathfrak{R}$ is the representation of the chiral multiplets under the gauge (and flavor) group, $\mathfrak{m}$ is the magnetic flux of the (background) gauge field through $\mathbb{S}^{2}$, and $r_{\rho}$ is the $U(1)_{R}$ charge of the chiral multiplet. It is convenient to abbreviate a single weight of \eqref{eq:twisted-chiral} as
    \bea\label{eq:twisted-chiral-D}
    \mathscr{D}(x,B):=\frac{x^{B/2}}{(xq^{1-B};q^{2})_{B}} .
    \eea
    The integer $B$ is the Euler characteristic of the line bundle $\mathcal{O}(\rho(\mathfrak{m})-r_{\rho})$ on $\mathbb{S}^{2}$ in which the chiral multiplet takes values after the twist. For $B>0$, each of the $B$ holomorphic modes, with angular momenta $-\frac{B-1}{2},\ldots,\frac{B-1}{2}$, contributes a factor to the denominator of \eqref{eq:twisted-chiral-D}. For $B<0$, the first cohomology gives $|B|$ factors in the numerator by the negative-index convention \eqref{q-Pochhammer}. For $B=0$, $\mathscr{D}=1$. In particular,
    \bea\label{eq:twisted-D-small}
    \mathscr{D}(x,1)=\frac{x^{1/2}}{1-x}=\frac{1}{[x^{-1}]},\qquad
    \mathscr{D}(x,-1)=x^{-1/2}(1-x)=[x^{-1}]=-[x],
    \eea
    where $[x]=x^{1/2}-x^{-1/2}$.

    \item Vector multiplet
    \begin{equation}\label{eq:twisted-vector}
        \mathcal{I}_{\text{vec}}
        =(-q)^{-\sum_{\alpha\in\Delta_+}|\alpha(\mathfrak{m})|}
        \prod_{\alpha\in\Delta}
        \left(1-x^{\alpha}q^{|\alpha(\mathfrak{m})|}\right)
        =
        \prod_{\alpha\in\Delta}
        \frac{(x^{\alpha/2})^{\alpha(\mathfrak{m})-1}}
        {(x^{\alpha}q^{2-\alpha(\mathfrak{m})};q^2)_{\alpha(\mathfrak{m})-1}},
    \end{equation}
    where $\Delta$ and $\Delta_+$ denote the sets of roots and positive roots of the gauge group $G$, respectively. The second form is \eqref{eq:twisted-chiral} at $r_{\rho}=2$; that is, the vector multiplet contributes like an adjoint chiral multiplet of $R$-charge two. For an abelian gauge group, $\Delta=\varnothing$ and $\mathcal{I}_{\text{vec}}=1$.
\end{itemize}

\paragraph{The spiked-quiver twisted index}
We now specialize the contour formula \eqref{eq:twisted-general-JK} to the spiked quiver \eqref{spiked-instanton-quiver}. Throughout, the gauge group is $U(N_{4})$, with fugacities $y_{\mathsf{a}}$ and fluxes $\mathfrak{m}_{\mathsf{a}}$, $\mathsf{a}=1,\ldots,N_{4}$, and the $\alpha$-th brane of the stack D6$_{\bar{s}}$ carries the framing fugacity $v_{\bar{s},\alpha}$, $\alpha=1,\ldots,N_{\bar{s}}$.

To compare with the half-index, recall the rotation convention in \eqref{eq:qi-def}: the fugacity conjugate to $J$ is $q^{2}$, and it is identified with the rotation parameter of the plane $\mathbb{C}_{4}$ of the brane configuration \eqref{eq:DbraneADHMflavor} as
\bea\label{eq:twisted-q-dictionary}
q^{2}=q_{4}.
\eea
The localization formula is written in terms of the flavor fugacities $t_{1,2,3}$ of \eqref{eq:flavor_table_sym}, which obey $t_{1}t_{2}t_{3}=1$ and carry no $R$-charge, together with $q$. On the other hand, the enumerative formulas of Section~\ref{sec:halfindex} use the mixed parameters $q_{1,2,3,4}$ obeying $q_{1}q_{2}q_{3}q_{4}=1$. The dictionary between the two sets of parameters is \eqref{eq:qi-def}:
\bea\label{eq:twisted-tq-dictionary}
q_{i}=t_{i}q^{-r_{i}}=t_{i}q_{4}^{-r_{i}/2},\qquad i=1,2,3,
\eea
with $r_{1}+r_{2}+r_{3}=2$. In these variables, the $q$-Pochhammer symbols of \eqref{eq:twisted-chiral} have base $q_{4}$, and the single-weight contribution \eqref{eq:twisted-chiral-D} reads
\bea\label{eq:twisted-chiral-D-q4}
\mathscr{D}(x,B)=\frac{x^{B/2}}{\big(x\,q_{4}^{\frac{1-B}{2}};q_{4}\big)_{B}} .
\eea

Using the $R$-charge assignments in \eqref{eq:flavor_table_sym}, the twisted index of the spiked quiver \eqref{spiked-instanton-quiver} is
\begin{equation}\label{eq:spiked-twisted-master}
\begin{aligned}
\mathcal{I}^{(L,M,N)}_{N_{4}}={}&\sum_{\mathfrak{m}\in\mathbb{Z}^{N_4}}\frac{(-1)^{N_4}}{N_4!}\oint_{\mathrm{JK}}\prod_{\mathsf{a}=1}^{N_4}\frac{\mathrm{d}y_{\mathsf{a}}}{2\pi i\,y_{\mathsf{a}}}\;
\big((-1)^{\sum_{s}N_{\bar{s}}}\mathfrak{q}\big)^{\sum_{\mathsf{a}}\mathfrak{m}_{\mathsf{a}}}
\\
&\times\prod_{s=1}^{3}\prod_{\mathsf{a}=1}^{N_4}\prod_{\alpha=1}^{N_{\bar{s}}}
\frac{(y_{\mathsf{a}}v_{\bar{s},\alpha})^{\frac{\mathfrak{m}_{\mathsf{a}}}{2}}}{\big(y_{\mathsf{a}}v_{\bar{s},\alpha}\,q_{4}^{\frac{1-\mathfrak{m}_{\mathsf{a}}}{2}};q_{4}\big)_{\mathfrak{m}_{\mathsf{a}}}}\;
\frac{\Big(\frac{q_{s}q_{4}^{r_{s}/2}}{y_{\mathsf{a}}v_{\bar{s},\alpha}}\Big)^{\frac{-\mathfrak{m}_{\mathsf{a}}+r_{s}}{2}}}{\Big(\frac{q_{s}}{y_{\mathsf{a}}v_{\bar{s},\alpha}}\,q_{4}^{\frac{\mathfrak{m}_{\mathsf{a}}+1}{2}};q_{4}\Big)_{-\mathfrak{m}_{\mathsf{a}}+r_{s}}}
\\
&\times\prod_{\mathsf{a}\neq\mathsf{b}}\frac{(y_{\mathsf{a}}/y_{\mathsf{b}})^{\frac{1}{2}(\mathfrak{m}_{\mathsf{a}}-\mathfrak{m}_{\mathsf{b}}-1)}}{\Big(\frac{y_{\mathsf{a}}}{y_{\mathsf{b}}}\,q_{4}^{\frac{\mathfrak{m}_{\mathsf{b}}-\mathfrak{m}_{\mathsf{a}}}{2}+1};q_{4}\Big)_{\mathfrak{m}_{\mathsf{a}}-\mathfrak{m}_{\mathsf{b}}-1}}
\prod_{s=1}^{3}\prod_{\mathsf{a},\mathsf{b}=1}^{N_4}
\frac{\big(q_{s}^{-1}q_{4}^{-r_{s}/2}\,y_{\mathsf{a}}/y_{\mathsf{b}}\big)^{\frac{1}{2}(\mathfrak{m}_{\mathsf{a}}-\mathfrak{m}_{\mathsf{b}}+1-r_{s})}}{\Big(\frac{y_{\mathsf{a}}}{q_{s}y_{\mathsf{b}}}\,q_{4}^{\frac{\mathfrak{m}_{\mathsf{b}}-\mathfrak{m}_{\mathsf{a}}}{2}};q_{4}\Big)_{\mathfrak{m}_{\mathsf{a}}-\mathfrak{m}_{\mathsf{b}}+1-r_{s}}} ,
\end{aligned}
\end{equation}
where we normalized $\mathcal{I}_{N_{4}=0}=1$. In the notation \eqref{eq:twisted-chiral-D-q4}, the contributions of the multiplets are
\bea\label{eq:twisted-framing-pair}
\text{vector fields:}&\qquad \mathscr{D}(y_{\mathsf{a}}/y_{\mathsf{b}},\,\mathfrak{m}_{\mathsf{a}}-\mathfrak{m}_{\mathsf{b}}-1),\\
\text{adjoint chiral fields:}&\qquad \mathscr{D}(q_{s}^{-1}q_{4}^{-r_{s}/2}\,y_{\mathsf{a}}/y_{\mathsf{b}},\,\mathfrak{m}_{\mathsf{a}}-\mathfrak{m}_{\mathsf{b}}+1-r_{s}),\\
\text{framing chiral fields:}&\qquad \mathscr{D}\!\left(y_{\mathsf{a}}v_{\bar{s},\alpha},\,\mathfrak{m}_{\mathsf{a}}\right)\,
\mathscr{D}\!\left(\frac{q_{s}q_{4}^{r_{s}/2}}{y_{\mathsf{a}}v_{\bar{s},\alpha}},\,-\mathfrak{m}_{\mathsf{a}}+r_{s}\right).
\eea

When $\sum_{s}N_{\bar{s}}r_{s}$ is odd, the framing prefactors give half-integer powers $\prod_{\mathsf{a}}y_{\mathsf{a}}^{\sum_{s}N_{\bar{s}}(\mathfrak{m}_{\mathsf{a}}-r_{s}/2)}$ of the gauge variables.
The fractional part is independent of $\mathfrak{m}_{\mathsf{a}}$ and is present even at zero gauge flux.
A mixed gauge--$R$ Chern--Simons contact term can shift these powers through the background $R$-symmetry flux~\cite{Benini:2015noa}.
In the following calculations, we restrict to even $\sum_{s}N_{\bar{s}}r_{s}$ and set this contact term to zero.

\subsection{Flavor rank \texorpdfstring{$(0,0,N)$}{(0,0,N)} twisted index and gluing of rank \texorpdfstring{$N$}{N} PT vertices}\label{sec:rankone-twistedindex}

In this subsection, we consider $N$ parallel D6-branes on $\mathbb{C}^{3}_{124}$ only, with $R$-charges $(r_{1},r_{2},r_{3})=(r_{1},r_{2},0)$ and $r_{1}+r_{2}=2$. With this choice, the parity anomaly cancels without Chern--Simons terms.

We argue below that, after summing the twisted index over the D2-brane number $N_{4}$, its factorization into half-index blocks takes the form of
the rank-$N$ PT gluing formula: the vortex factors give the PT
vertices, and the classical and perturbative factors give the
edge factor. We therefore propose that the resulting partition function is the rank-$N$
PT partition function of
$X=\operatorname{Tot}\bigl(\mathcal O(-r_1)\oplus
\mathcal O(-r_2)\to\mathbb P^1\bigr)$.

Geometrically, in each local chart, the D2-branes wrap $\mathbb{C}_4\times \mathbb{S}^1$
and the D6-branes wrap $\mathbb{C}^{3}_{124}\times \mathbb{S}^1$.
In the glued geometry, $\mathbb{C}_4$ becomes a chart of
$\mathbb{P}^1$, while $\mathbb{C}_1$ and $\mathbb{C}_2$ are its two
normal directions. Their degrees are $-r_1$ and $-r_2$, fixed by
the $R$-charges of the adjoint chirals $\Phi_1$ and $\Phi_2$.

\paragraph{Pole structure}
The twisted index with $N$ D6$_{124}$-branes and $r_{3}=0$ simplifies to
\bea
\mathcal I^{(0,0,N)}_{N_4}
={}&\frac{(-1)^{N_4}}{N_4!}
\sum_{\mathfrak{m}\in\mathbb Z^{N_4}}
\oint_{\mathrm{JK}}
\prod_{\mathsf{a}=1}^{N_4}\frac{dy_{\mathsf{a}}}{2\pi i y_{\mathsf{a}}}\,
\bigl((-1)^N\mathfrak q\bigr)^{\sum_{\mathsf{a}} \mathfrak{m}_{\mathsf{a}}}\prod_{\mathsf{a}=1}^{N_4}\prod_{\alpha=1}^{N}
\mathscr{D}(y_{\mathsf{a}}v_{\bar3,\alpha},\mathfrak{m}_{\mathsf{a}})\,
\mathscr{D}\!\left(\frac{q_3}{y_{\mathsf{a}}v_{\bar3,\alpha}},-\mathfrak{m}_{\mathsf{a}}\right)\\
&\times
\prod_{\mathsf{a}\ne\mathsf{b}}
\mathscr{D}\!\left(\frac{y_{\mathsf{a}}}{y_{\mathsf{b}}},\mathfrak{m}_{\mathsf{a}}-\mathfrak{m}_{\mathsf{b}}-1\right)\prod_{\mathsf{a},\mathsf{b}=1}^{N_4}
\mathscr{D}\!\left(q_3^{-1}\frac{y_{\mathsf{a}}}{y_{\mathsf{b}}},
        \mathfrak{m}_{\mathsf{a}}-\mathfrak{m}_{\mathsf{b}}+1\right)\\
&\times
\prod_{\mathsf{a},\mathsf{b}=1}^{N_4}\mathscr{D}\!\left(q_1^{-1}q_4^{-r_1/2}\frac{y_{\mathsf{a}}}{y_{\mathsf{b}}},
        \mathfrak{m}_{\mathsf{a}}-\mathfrak{m}_{\mathsf{b}}+1-r_1\right)
\mathscr{D}\!\left(q_2^{-1}q_4^{-r_2/2}\frac{y_{\mathsf{a}}}{y_{\mathsf{b}}},
        \mathfrak{m}_{\mathsf{a}}-\mathfrak{m}_{\mathsf{b}}+1-r_2\right),
\eea
where $r_{1}+r_{2}=2$.

To evaluate this integral in the positive FI chamber, we take the reference vector $\eta$ to be a generic small perturbation of $(1,\ldots,1)$. By \eqref{eq:twisted-chiral-D-q4}, for $B>0$ the contribution $\mathscr{D}(x,B)$ has $B$ simple poles, at
\bea\label{eq:twisted-poles}
x=q_{4}^{\frac{B-1}{2}-\ell},\qquad \ell=0,1,\ldots,B-1,
\eea
while for $B\leq0$ it is $x^{B/2}$ times a polynomial in $x$ and has no poles in $\mathbb C^\times$.

Three families of singular hyperplanes therefore occur in the $y$-plane. The fundamental $p_{3}$ gives, for every $\mathsf{a}$ with $\mathfrak{m}_{\mathsf{a}}>0$, the hyperplanes
\bea\label{eq:twisted-seed-hyperplane}
y_{\mathsf{a}}=v_{\bar{3},\alpha}^{-1}\,q_{4}^{\frac{\mathfrak{m}_{\mathsf{a}}-1}{2}-\ell},\qquad 0\leq\ell\leq\mathfrak{m}_{\mathsf{a}}-1,
\eea
with charge vector $e_{\mathsf{a}}$; the antifundamental $\tilde{p}_{3}$ gives, for every $\mathsf{a}$ with $\mathfrak{m}_{\mathsf{a}}<r_{3}=0$, hyperplanes with charge vector $-e_{\mathsf{a}}$; and the adjoint $\Phi_{s}$ gives, for every pair $\mathsf{a}\neq\mathsf{b}$ with $\mathfrak{m}_{\mathsf{a}}-\mathfrak{m}_{\mathsf{b}}\geq r_{s}$, the hyperplanes
\bea\label{eq:twisted-transport-hyperplane}
y_{\mathsf{a}}=y_{\mathsf{b}}\,q_{s}\,q_{4}^{\frac{\mathfrak{m}_{\mathsf{a}}-\mathfrak{m}_{\mathsf{b}}}{2}-\ell},\qquad 0\leq\ell\leq\mathfrak{m}_{\mathsf{a}}-\mathfrak{m}_{\mathsf{b}}-r_{s},
\eea
with charge vector $e_{\mathsf{a}}-e_{\mathsf{b}}$.
When the pole positions are expressed in terms of $q_s$, the only explicit dependence on $r_s$ is through the upper bound on $\ell$. In the chosen chamber, only the cones generated by $\{e_{\mathsf{a}}\}$ and $\{e_{\mathsf{a}}-e_{\mathsf{b}}\}$ are selected: each gauge variable is either seeded on a fundamental hyperplane \eqref{eq:twisted-seed-hyperplane} or transported from a previously fixed one along an adjoint hyperplane \eqref{eq:twisted-transport-hyperplane}.

Following a chain of transports back to its seed, every selected pole takes the form
\bea\label{eq:twisted-pole-general}
y_{\mathsf{a}}=v_{\bar{3},\alpha}^{-1}\,q_{1}^{x_{1}-1}q_{2}^{x_{2}-1}q_{3}^{x_{3}-1}\,q_{4}^{\frac{\mathfrak{m}_{\mathsf{a}}-1}{2}-\rho_{\mathsf{a}}},
\qquad
\tilde{\rho}_{\mathsf{a}}:=\mathfrak{m}_{\mathsf{a}}-1-\rho_{\mathsf{a}}-\sum_{c=1}^{3}(x_{c}-1)r_{c}\geq0 ,
\eea
where $x_{c}-1\geq0$ counts the transports in the direction $c$ along the chain, and $\rho_{\mathsf{a}}\geq0$ is the sum of the integers $\ell$ picked up at the seed and at each transport. 

When all gauge variables carry distinct tuples $(\alpha;x_{1},x_{2},x_{3})$ in \eqref{eq:twisted-pole-general}, the zeros from the antifundamental and adjoint multiplets eliminate all configurations except those labeled by Young diagrams in the $12$-plane:
\begin{itemize}
    \item For a pole \eqref{eq:twisted-pole-general} at $(x_{1},x_{2},x_{3})=(1,1,2)$ transported from $(1,1,1)$ by the adjoint $\Phi_{3}$, we have
        \begin{equation}
            y_{\mathsf{a}}=v_{\bar{3},\alpha}^{-1}\,q_{3}\,q_{4}^{\frac{1}{2}(\tilde{\rho}_{\mathsf{a}}-\rho_{\mathsf{a}})}, \qquad \mathfrak{m}_{\mathsf{a}}=\rho_{\mathsf{a}}+\tilde{\rho}_{\mathsf{a}}+1.
        \end{equation}
        The antifundamental factor $\mathscr{D}\!\left(q_{3}(y_{\mathsf{a}}v_{\bar{3},\alpha})^{-1},-\mathfrak{m}_{\mathsf{a}}\right)$ then vanishes at $y_{\mathsf{a}}$.
    \item For a two-step transport $\mathsf{c}\to\mathsf{b}\to\mathsf{a}$ along two distinct directions $a\neq b$ with $\{a,b,c\}=\{1,2,3\}$, applying \eqref{eq:twisted-pole-general} together with $q_{1}q_{2}q_{3}q_{4}=1$ and $r_{1}+r_{2}+r_{3}=2$ gives
        \begin{equation}
            \frac{y_{\mathsf{c}}}{y_{\mathsf{a}}}=q_{c}\,q_{4}^{\frac{1}{2}\bigl(r_{c}+(\rho_{\mathsf{a}}-\rho_{\mathsf{c}})-(\tilde{\rho}_{\mathsf{a}}-\tilde{\rho}_{\mathsf{c}})\bigr)},
            \qquad
            \mathfrak{m}_{\mathsf{a}}-\mathfrak{m}_{\mathsf{c}}=(\rho_{\mathsf{a}}-\rho_{\mathsf{c}})+(\tilde{\rho}_{\mathsf{a}}-\tilde{\rho}_{\mathsf{c}})+2-r_{c}.
        \end{equation}
        Since $\rho_{\mathsf{a}}-\rho_{\mathsf{c}}\geq 0$ and $\tilde{\rho}_{\mathsf{a}}-\tilde{\rho}_{\mathsf{c}}\geq 0$, the adjoint factor $\mathscr{D}\!\left(q_{c}^{-1}q_{4}^{-r_{c}/2}y_{\mathsf{c}}/y_{\mathsf{a}},\mathfrak{m}_{\mathsf{c}}-\mathfrak{m}_{\mathsf{a}}+1-r_{c}\right)$ vanishes at $y_{\mathsf{a}}$.
\end{itemize}

Finally, for degenerate poles where two gauge variables $y_{\mathsf{a}},y_{\mathsf{b}}$ share the same tuple $(\alpha;x_{1},x_{2},x_{3})$, we have $\mathfrak{m}_{\mathsf{a}}-\mathfrak{m}_{\mathsf{b}}=\Delta\rho+\Delta\tilde{\rho}$ and $y_{\mathsf{a}}/y_{\mathsf{b}}=q_{4}^{(\Delta\tilde{\rho}-\Delta\rho)/2}$, where $\Delta\rho=\rho_{\mathsf{a}}-\rho_{\mathsf{b}}$ and $\Delta\tilde{\rho}=\tilde{\rho}_{\mathsf{a}}-\tilde{\rho}_{\mathsf{b}}$. The JK residue at such points is given by \eqref{eq:JK-deg}, and the vector multiplet contribution \eqref{eq:twisted-vector} for the pair $(\mathsf{a},\mathsf{b})$ evaluates to
\begin{equation}\label{eq:vec-colliding-sign}
    \begin{aligned}
        \mathcal{I}_{\mathrm{vec}}^{(\mathsf{a},\mathsf{b})}
        &=
        \operatorname{sgn}(\Delta\rho\,\Delta\tilde{\rho})\,
        (-q_4^{1/2})^{-(|\Delta\rho|+|\Delta\tilde{\rho}|)}
        (1-q_4^{|\Delta\rho|})(1-q_4^{|\Delta\tilde{\rho}|}).
    \end{aligned}
\end{equation}
Thus $\mathcal{I}_{\mathrm{vec}}^{(\mathsf{a},\mathsf{b})}$ need not vanish in a fixed flux sector when $\Delta\rho\neq 0$ and $\Delta\tilde{\rho}\neq 0$. Exchanging $\tilde{\rho}_{\mathsf{a}}\leftrightarrow\tilde{\rho}_{\mathsf{b}}$ at fixed $(\rho_{\mathsf{a}},\rho_{\mathsf{b}})$ preserves the total flux $\mathfrak{m}_{\mathsf{a}}+\mathfrak{m}_{\mathsf{b}}$ and all chiral and classical factors, while reversing the sign $\operatorname{sgn}(\Delta\rho\,\Delta\tilde{\rho})$ in \eqref{eq:vec-colliding-sign}. We observe that, after summing over the magnetic fluxes $\mathfrak{m}\in\mathbb{Z}^{N_{4}}$, the contributions from such poles cancel among themselves.

Consequently, the nontrivial poles are labeled by Young diagrams $\lambda_{12}^{(\alpha)}$ with $\sum_{\alpha}|\lambda_{12}^{(\alpha)}|=N_{4}$,
\begin{equation}\label{eq:twisted-pole-rankN}
    y_{\mathsf{a}}=v_{\bar{3},\alpha}^{-1}q_{1}^{i-1}q_{2}^{j-1}q_{4}^{\frac{\mathfrak{m}_{\mathsf{a}}-1}{2}-\pi^{(\alpha)}_{12,\mathsf{a}}},\qquad
    \mathfrak{m}_{\mathsf{a}}=\tilde{\pi}_{12,\mathsf{a}}^{(\alpha)}+\pi_{12,\mathsf{a}}^{(\alpha)}+r_{1}(i-1)+r_{2}(j-1)+1,
\end{equation}
where the non-negative integers $\pi^{(\alpha)}_{12,\mathsf{a}}=\rho_{\mathsf{a}}$ and $\tilde{\pi}^{(\alpha)}_{12,\mathsf{a}}=\tilde{\rho}_{\mathsf{a}}$ attached to the boxes of $\lambda^{(\alpha)}_{12}$ form reverse plane partitions, $\pi^{(\alpha)}_{12},\tilde{\pi}^{(\alpha)}_{12}\in\mathrm{RPP}(\lambda^{(\alpha)}_{12})$.

\paragraph{Factorization}
The identity
\bea\label{eq:twisted-block-split}
\mathscr{D}(x,B)=x^{B/2}\,\frac{\big(x\,q_{4}^{\frac{1+B}{2}};q_{4}\big)_{\infty}}{\big(x\,q_{4}^{\frac{1-B}{2}};q_{4}\big)_{\infty}}
=\frac{x^{B/2}}{\big(x\,q_{4}^{\frac{1-B}{2}};q_{4}\big)_{\infty}\,\big(x\,q_{4}^{\frac{B-1}{2}};q_{4}^{-1}\big)_{\infty}}
\eea
splits each contribution \eqref{eq:twisted-chiral-D-q4} into two infinite products with bases $q_{4}$ and $q_{4}^{-1}$. The second equality uses $(z;q_{4})_{\infty}=1/(zq_{4}^{-1};q_{4}^{-1})_{\infty}$, understood by meromorphic continuation.
 
After taking the residue at \eqref{eq:twisted-pole-rankN} associated with reverse plane partitions $\pi_{12}^{(\alpha)}$ with the Young diagram bases $\underline{\vec{\lambda}}=(\vec{\lambda}_{12})$, where $\vec{\lambda}_{12}=(\lambda_{12}^{(\alpha)})_{\alpha=1}^{N}$, the twisted index factorizes into
\bea\label{eq:rankN-block-decomp}
\mathcal{I}^{(0,0,N)}_{N_{4}}=\sum_{|\vec{\lambda}_{12}|=N_{4}}  \fq^{\chi(\vec{\lambda}_{12},(r_{1},r_{2}))}  \mathcal{I}_{\text{edge}}\,\mathcal{Z}^{\vec{\lambda}_{12}}_{\text{vortex}}\,\widetilde{\mathcal{Z}}^{\vec{\lambda}_{12}}_{\text{vortex}}
\eea
where
\bea\label{eq:rankN-block-equiv-param}
&\mathcal{Z}^{\vec{\lambda}_{12}}_{\text{vortex}}=\mathcal{Z}^{\vec{\lambda}_{12}}_{\text{vortex}}(\fq,q_{1},q_{2},q_{3},q_{4}),\quad \widetilde{\mathcal{Z}}^{\vec{\lambda}_{12}}_{\text{vortex}}=\mathcal{Z}^{\vec{\lambda}_{12}}_{\text{vortex}}(\fq,\tilde{q}_{1},\tilde{q}_{2},\tilde{q}_{3},\tilde{q}_{4}),\\
&\tilde{q}_{1}=q_{1}q_{4}^{r_{1}},\quad \tilde{q}_{2}=q_{2}q_{4}^{r_{2}},\quad \tilde{q}_{3}=q_{3},\quad \tilde{q}_{4}=q_{4}^{-1}
\eea
and
\bea
\chi(\lambda,(r_{1},r_{2}))=\sum_{(i,j)\in\lambda}(r_{1}(i-1)+r_{2}(j-1)+1),\quad 
\chi(\vec{\lambda},(r_{1},r_{2}))=\sum_{\alpha=1}^{N}\chi(\lambda^{(\alpha)},(r_{1},r_{2})).
\eea
This is the standard block factorization of a twisted index
\cite{Beem:2012mb,Bullimore:2016nji,Closset:2016arn}. The factor $\mathcal{I}_{\mathrm{edge}}$ here is the contribution at zero vortex number at both poles, with the classical factor $\fq^\chi$ factored out. The extra topological term $\chi(\vec{\lambda}_{12},(r_{1},r_{2}))$ comes from
\bea
\sum_{\mathsf{a}}\mathfrak{m}_{\mathsf{a}}=\sum_{\alpha=1}^{N}|\pi_{12}^{(\alpha)}|+\sum_{\alpha=1}^{N}|\tilde{\pi}_{12}^{(\alpha)}|+\chi(\vec{\lambda}_{12},(r_{1},r_{2}))
\eea
where the first and second terms are absorbed in the vortex partition functions and the last term is the extra topological term coming from the base Young diagrams.

\begin{figure}[t]
\centering
\begin{tikzpicture}[
    line width=0.8pt,
    every node/.style={font=\small}
]

\node at (-2,0)
    {$\displaystyle\sum_{\vec{\lambda}_{12}}Q^{|\vec{\lambda}_{12}|}$};

\begin{scope}[xshift=0.1cm]
    \draw (0,0.9)
        arc[start angle=90,end angle=270,radius=0.9];

    \draw (0,-0.9)
        arc[start angle=270,end angle=450,
            x radius=0.18,y radius=0.9];
    \draw[dashed] (0,0.9)
        arc[start angle=90,end angle=270,
            x radius=0.18,y radius=0.9];

    \fill (-0.9,0) circle (1.4pt);
    \node at (-0.48,0) {$\mathrm N$};
    \node[above] at (-0.45,1.05) {$q_{1,2,3,4}$};
    \node[below] at (0,-1.05) {$\vec{\lambda}_{12}$};
\end{scope}

\node at (0.85,0) {$\times$};

\begin{scope}[xshift=1.7cm]
    \draw (0,-0.9)
        arc[start angle=270,end angle=450,radius=0.9];

    \draw (0,-0.9)
        arc[start angle=270,end angle=450,
            x radius=0.18,y radius=0.9];
    \draw[dashed] (0,0.9)
        arc[start angle=90,end angle=270,
            x radius=0.18,y radius=0.9];

    \fill (0.9,0) circle (1.4pt);
    \node at (0.48,0) {$\mathrm S$};
    \node[above] at (0.45,1.05) {$\tilde q_{1,2,3,4}$};
    \node[below] at (0,-1.05) {$\vec{\lambda}_{12}$};
\end{scope}

\node at (4.4,0) {$\longleftrightarrow$};

\begin{scope}[xshift=8.5cm]
    \coordinate (L) at (-0.9,0);
    \coordinate (R) at ( 0.9,0);

    \draw (L) -- ++(-1, 0.9);
    \draw (L) -- ++(-1,-0.9);
    \draw (R) -- ++( 1, 0.9);
    \draw (R) -- ++( 1,-0.9);

    \draw (-0.85,0)--(0.85,0);
    \fill (L) circle (1.5pt);
    \fill (R) circle (1.5pt);

    \node at (0,-0.5) {$\mathbb P^1$};
    \node[above] at (0,0.3) {$\vec{\lambda}_{12}$};

    \node[above ] at (-1.9,0.9){$q_{1}$};
    \node[above ] at (1.9,0.9){$\tilde{q}_{1}$};
    \node[below left] at (-1.9,-0.9){$q_{2}$};
    \node[below right] at (1.9,-0.9){$\tilde{q}_{2}$};
    
    \node[left,align=center] at (-1.3,0.0)
        {$\mathsf{PT}_{\vec{\lambda}_{12}}$};
    \node[right,align=center] at (1.3,0.0)
        {$\widetilde{\mathsf{PT}}_{\vec{\lambda}_{12}}$};

    \node[below,align=center] at (0,-1)
        {$Q^{|\vec{\lambda}_{12}|}
          \fq^{\chi(\vec{\lambda}_{12},(r_1,r_2))}$
         $\mathsf{E}_{\bar3;\vec{\lambda}_{12}}$};
\end{scope}

\end{tikzpicture}
\caption{
Block factorization of the twisted index and PT gluing.
Left: the twisted index of the 3d $\mathcal{N}=4$ ADHM theory
factorizes into two hemisphere blocks.
Right: D2-branes wrap the compact $\mathbb{P}^{1}$ of the local
curve, while $N$ flavor D6-branes wrap the entire Calabi--Yau
threefold.
The two hemisphere blocks correspond to the two PT$_{3}$ vertices,
joined along a common leg labeled by $\vec{\lambda}_{12}$.
}
\label{fig:twisted-PT-gluing}
\end{figure}
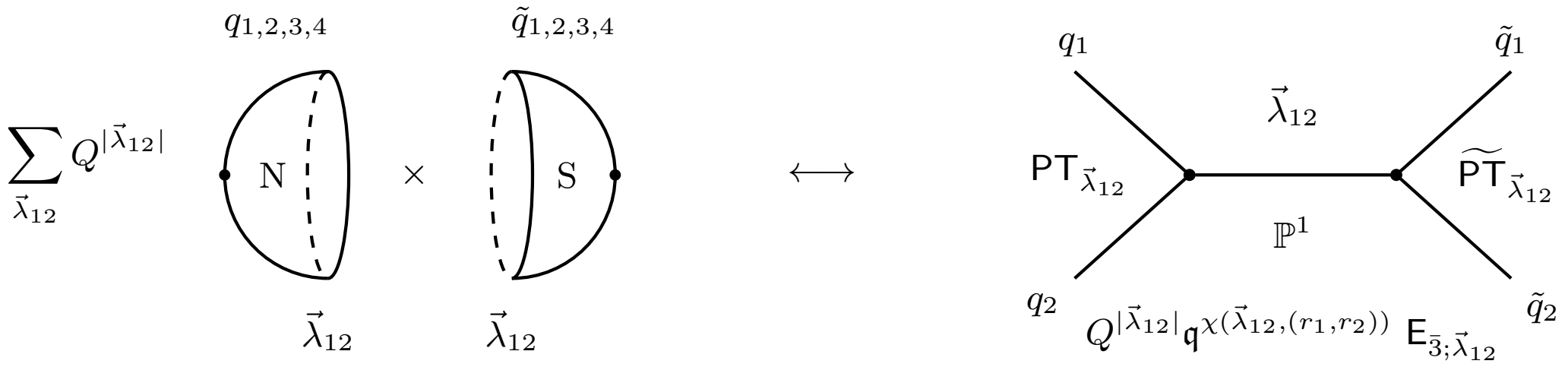

The two vortex factors correspond to the north and south poles of $\mathbb S^2$ (see Figure~\ref{fig:twisted-PT-gluing}). The two blocks are exchanged by
$q_4\to q_4^{-1}$ at fixed global flavor fugacities $t_i$. Thus the north-pole parameters are $q_i=t_iq_4^{-r_i/2}$, while the south-pole parameters are
$\tilde q_i=t_iq_4^{r_i/2}=q_iq_4^{r_i}$, as in
\eqref{eq:rankN-block-equiv-param}. The factor $\fq^{\chi(\vec{\lambda}_{12},(r_{1},r_{2}))}\mathcal{I}_{\text{edge}}$ is identified with the classical and one-loop part of the holomorphic blocks of the north and south hemispheres, up to additional \textit{signs} depending on the choice of the boundary condition. In particular, the FI-term coming from the two blocks gives the topological term as
\bea
\exp\left[\frac{\log \fq}{\log q_{4}}\sum_{\mathsf{a}\in\vec{\lambda}}\log \tilde{s}_{\mathsf{a}}|_{q_{1,2,3,4}}+\frac{\log \fq}{\log \tilde{q}_{4}}\sum_{\mathsf{a}\in\vec{\lambda}}\log 
\tilde{s}_{\mathsf{a}}|_{\tilde{q}_{1,2,3,4}}\right]=\fq^{\chi(\vec{\lambda},(r_{1},r_{2}))}
\eea
and we have
\bea
\fq^{\chi(\vec{\lambda},(r_1,r_2))}
\mathcal I_{\mathrm{edge}}^{\vec\lambda}
&=
\begin{dcases}
\varepsilon_{\mathrm D}
\left(\mathcal Z^{\vec{\lambda}}_{\mathrm{cl}}
      \mathcal Z^{\vec{\lambda}}_{\text{1-loop}}\right)
\left(\widetilde{\mathcal Z}^{\vec{\lambda}}_{\mathrm{cl}}
      \widetilde{\mathcal Z}^{\vec{\lambda}}_{\text{1-loop}}\right),
& \text{for \eqref{eq:exceptional-half-index}},\\
\varepsilon_{\mathrm N}\,
\mathcal B_{\vec{\lambda},0}
\widetilde{\mathcal B}_{\vec{\lambda},0},
& \text{for \eqref{eq:halfindex-integral}},\\
\varepsilon_{\mathrm{CFT}}\,
\left\langle\mathbb B_{\vec{\lambda},0}\right\rangle
\left\langle\widetilde{\mathbb B}_{\vec{\lambda},0}\right\rangle,
& \text{for \eqref{eq:freefield-rankN}}.
\end{dcases}
\eea
The tilded factors here are evaluated at the south-pole equivariant parameters and $\varepsilon_{\mathrm D}$,
$\varepsilon_{\mathrm N}$, and $\varepsilon_{\mathrm{CFT}}$ are extra signs taking values in $\{\pm 1\}$, compared with the twisted-index contribution. For the explicit sign rule, see Appendix~\ref{app:edge-comparison}; for the exceptional Dirichlet boundary condition of Appendix~\ref{app:rankNADHM_bcd}, \(\varepsilon_{\mathrm D}=(-1)^{r_1N_4}\) by \eqref{eq:parallel-d6-edge-sign}. We note that the relative signs can depend on the boundary condition,
the Young diagrams $\vec{\lambda}_{12}$, and the twisting
parameters $(r_1,r_2)$.

\paragraph{Relation to the gluing of PT$_{3}$ vertices}
The block factorization of the
ADHM twisted index and the PT gluing for the resolved conifold were identified in \cite{Crew:2020psc}. Their $R$-charge assignment $(r_1,r_2,r_3)=(1,1,0)$ fixes the normal bundle to
$\mathcal O(-1)\oplus\mathcal O(-1)$. We propose that the same correspondence holds for arbitrary integral
$(r_1,r_2,0)$ with $r_1+r_2=2$ and for $N$ D6-branes.
More precisely, the generating function of the twisted index,
\bea
\mathcal{I}^{(0,0,N)}_{\text{twist}}
=\sum_{N_4=0}^{\infty}Q^{N_4}\mathcal{I}^{(0,0,N)}_{N_4},
\eea
equals the rank-$N$ PT partition function of the local curve
\bea
X=\operatorname{Tot}\left(
\mathcal O(-r_1)\oplus\mathcal O(-r_2)
\longrightarrow\mathbb P^1\right).
\eea

Recall that the PT partition function of $X$ is obtained by gluing
the one-leg vertices at its two torus fixed points.
The common leg is labeled by an $N$-tuple of Young diagrams
$\vec{\lambda}_{12}$. Summing over this leg with the edge factor
and the topological weight gives\footnote{The gluing of PT vertices requires a consistent identification of the equivariant parameters and of the orientations of the asymptotic Young diagrams in the two local charts. These details matter in explicit computations, but we suppress them here and describe the gluing schematically.  }
\bea
\mathsf{PT}[X;N]
={}&\sum_{\vec{\lambda}_{12}}
Q^{|\vec{\lambda}_{12}|}
\fq^{\chi(\vec{\lambda}_{12},(r_1,r_2))}
\mathsf{E}_{\bar{3};\vec{\lambda}_{12}}
(q_1,q_2\mid\tilde q_1,\tilde q_2)
\mathsf{PT}_{\vec{\lambda}_{12}}[\fq,q_{1,2,3,4}]
\mathsf{PT}_{\vec{\lambda}_{12}}[\fq,\tilde q_{1,2,3,4}],
\eea
as reviewed in Appendix~\ref{app:rankNDTPTvertex}.
The local equivariant parameters are related as in
\eqref{eq:rankN-block-equiv-param}, and the explicit edge factor
is given in \eqref{eq:rankN-PT-edge}.

Using the vortex--PT correspondence of Section~\ref{sec:DTPT-vortexQM}, we identify
\bea
\mathcal{Z}^{\vec{\lambda}_{12}}_{\text{vortex}}
(\fq,q_1,q_2,q_3,q_4)
&=\mathsf{PT}_{\vec{\lambda}_{12}}[\fq,q_{1,2,3,4}],\quad \mathcal{Z}^{\vec{\lambda}_{12}}_{\text{vortex}}
(\fq,\tilde q_1,\tilde q_2,\tilde q_3,\tilde q_4)
=\mathsf{PT}_{\vec{\lambda}_{12}}[\fq,\tilde q_{1,2,3,4}].
\eea
The factor $\mathcal{I}_{\text{edge}}$ is identified with the edge factor
\bea
\mathcal{I}_{\text{edge}}
=\mathsf{E}_{\bar{3};\vec{\lambda}_{12}}
(q_1,q_2\mid\tilde q_1,\tilde q_2).
\eea
Since $|\vec{\lambda}_{12}|=N_4$, the block factorization
reproduces the PT gluing formula, yielding
\bea
\mathcal{I}^{(0,0,N)}_{\text{twist}}=\mathsf{PT}[X;N].
\eea

\paragraph{Plethystic exponential formulas}
For $N_{4}=1$, the twisted index is
\bea
\mathcal{I}_{1}^{(0,0,N)}&=(-1)\sum_{\mathfrak{m}=1}^{\infty}\left((-1)^{N}\fq\right)^{\mathfrak{m}}\prod_{i=1}^{3}\mathscr{D}\left(q_{i}^{-1}q_{4}^{-\frac{r_{i}}{2}},1-r_{i}\right)\\
&\qquad \times\oint_{\text{JK}}\frac{dy}{2\pi i y }\prod_{\alpha=1}^{N}\mathscr{D}(y v_{\bar{3},\alpha},\mathfrak{m})\mathscr{D}\left(\frac{q_{3}}{ yv_{\bar{3},\alpha}},-\mathfrak{m}\right)
\eea
where the JK prescription restricts the flux to $\mathfrak{m}>0$. The integral is then 
\bea\label{eq:rankonetwist-comp}
&\oint_{\text{JK}}\frac{dy}{2\pi i y }\prod_{\alpha=1}^{N}\mathscr{D}(y v_{\bar{3},\alpha},\mathfrak{m})\mathscr{D}\left(\frac{q_{3}}{ yv_{\bar{3},\alpha}},-\mathfrak{m}\right)\\
=&(-1)^{N\mathfrak{m}}\oint_{\text{JK}}\frac{dy}{2\pi i y } \prod_{\alpha=1}^{N}\prod_{k=0}^{\mathfrak{m}-1}q_{3}^{1/2}\frac{1-q_{3}^{-1}yv_{\bar{3},\alpha}q_{4}^{\frac{1-\mathfrak{m}}{2}}q_{4}^{k}}{1-yv_{\bar{3},\alpha}q_{4}^{\frac{1-\mathfrak{m}}{2}}q_{4}^{k}}\\
=&(-1)^{N\mathfrak{m}}\left(q_{3}^{N\mathfrak{m}/2}-q_{3}^{-N\mathfrak{m}/2}\right)
\eea
where the JK residue at a pole $y=v_{\bar{3},\alpha}^{-1} q_{4}^{\frac{\mathfrak{m}-1}{2}}q_{4}^{-k}$ is
\bea
\lim_{y\rightarrow v_{\bar{3},\alpha}^{-1} q_{4}^{\frac{\mathfrak{m}-1}{2}}q_{4}^{-k}}\left(1-yv_{\bar{3},\alpha}q_{4}^{\frac{1-\mathfrak{m}}{2}}q_{4}^{k}\right)\prod_{\beta=1}^{N}\mathscr{D}(y v_{\bar{3},\beta},\mathfrak{m})\mathscr{D}\left(\frac{q_{3}}{ yv_{\bar{3},\beta}},-\mathfrak{m}\right)
\eea
in our notation. In the last line, we used the global residue theorem and evaluated the residues at $y=0$ and $y=\infty$. We thus have
\bea
\mathcal{I}_{1}^{(0,0,N)}&=-\prod_{i=1}^{3}\mathscr{D}\left(q_{i}^{-1}q_{4}^{-\frac{r_{i}}{2}},1-r_{i}\right)\sum_{\mathfrak{m}=1}^{\infty}\fq^{\mathfrak{m}}\left(q_{3}^{N\mathfrak{m}/2}-q_{3}^{-N\mathfrak{m}/2}\right)\\
&=\prod_{i=1}^{3}\mathscr{D}\left(q_{i}^{-1}q_{4}^{-\frac{r_{i}}{2}},1-r_{i}\right)\frac{[q_{3}^{N}]}{[\fq q_{3}^{N/2}][\fq^{-1} q_{3}^{N/2}]},\qquad [x]=x^{1/2}-x^{-1/2}.
\eea
For any $(r_1,r_2,0)$ with $r_1+r_2=2$ and any $N$, the $N_{4}=1$ twisted index is independent of the D6-brane flavor fugacities.

For $(r_1,r_2,r_{3})=(1,1,0)$ and $(2,0,0)$, we compare the full generating function $\mathcal I_{\text{twist}}$ with the plethystic expressions collected in Appendix~\ref{app:PE-local}. For $(r_1,r_2,r_{3})=(1,1,0)$, the proposed identity is
\bea\label{eq:conifoldPE-rankN}
\mathcal{I}^{(0,0,N)}_{\text{twist}}
=
\mathsf{PT}\left[
\mathcal O(-1)\oplus\mathcal O(-1)
\rightarrow\mathbb P^1;N\right]
=
\operatorname{PE}\left[
\frac{Q[q_3^N]}
{[q_3][q_3^{N/2}\fq][q_3^{N/2}\fq^{-1}]}
\right],
\eea
where the plethystic exponential is defined by
\begin{equation}\label{plethystic}
\mathrm{PE}\!\left[f(x,y,\ldots)\right] 
= \exp\!\left[\sum_{d=1}^\infty \frac{1}{d} f(x^d,y^d,\ldots)\right].
\end{equation}

For $(r_1,r_2,r_{3})=(2,0,0)$, the proposed identity is
\bea\label{eq:Z2orbifoldPE-rankN}
\mathcal{I}^{(0,0,N)}_{\text{twist}}
=
\mathsf{PT}\left[
\mathcal O(-2)\oplus\mathcal O
\rightarrow\mathbb P^1;N\right]
=
\operatorname{PE}\left[
\frac{[q_3^N][q_{23}]}{[q_3][q_2]}
\frac{Q}
{[\fq q_3^{N/2}][\fq q_3^{-N/2}]}
\right].
\eea
Both plethystic expressions are independent of the D6-brane flavor fugacities. To first order in $Q$, the rank-one gauge calculation gives
\bea
\mathcal{I}_{1}^{(0,0,N)}&=\begin{dcases}
    \frac{1}{[q_{3}]}\frac{[q_{3}^{N}]}{[\fq q_{3}^{N/2}][\fq^{-1} q_{3}^{N/2}]},\quad (r_{1},r_{2},r_{3})=(1,1,0)\\
    \frac{[q_{23}]}{[q_{3}][q_{2}]} \frac{[q_{3}^{N}]}{[\fq q_{3}^{N/2}][\fq q_{3}^{-N/2}]},\quad (r_{1},r_{2},r_{3})=(2,0,0)
\end{dcases}
\eea
where we used
 \bea\label{eq:Dfunct-Rcharge}
    \mathscr{D}(q_{i}^{-1}q_{4}^{-r_{i}/2},1-r_i)&=\begin{dcases}
        \frac{1}{[q_i]},\qquad r_i=0,\\
        1,\qquad \quad  r_i=1,\\
        [q_{i}q_{4}],\qquad r_{i}=2.
    \end{dcases}
    \eea

\paragraph{Membrane index interpretation}
The plethystic formulas \eqref{eq:conifoldPE-rankN} and \eqref{eq:Z2orbifoldPE-rankN} admit a common interpretation in terms of an M2-brane wrapping the zero section $C=\mathbb P^1$ of the local threefold $X$. The relation between membrane counting on Calabi--Yau fivefolds and K-theoretic stable-pair invariants has been further explored in~\cite{GiacchettoPandharipandeSchuler:2026,HolmesSchuler:2026}, building on the membrane/sheaf correspondence~\cite{Nekrasov:2014nea}. To see how the rank enters the geometry, consider $N$ D6-branes wrapping $X\times \mathbb{S}^1$. T-duality along the worldvolume circle gives $N$ D5-branes wrapping $X$, which become $N$ NS5-branes under type IIB S-duality. T-duality along the dual circle, now transverse to the NS5-branes, gives a Taub--NUT space of charge $N$~\cite{Dijkgraaf:2006um}. Near coincident centers, the transverse surface is $S_N=\mathbb C^2/\mathbb Z_N$. Appendix~\ref{app:orb-6dTFT} describes how the two circle reductions of the same M-theory background relate this orbifold to the stack of $N$ D6-branes. 

Let us compute the membrane index on the smooth fivefold $Z_N=X\times\widetilde S_N$, where $\widetilde S_N$ is the minimal resolution of $S_N$. For a torus-fixed membrane $C_p=C\times\{p\}$, the single-particle contribution is determined by the equivariant cohomology of its normal bundle~\cite{Nekrasov:2014nea}. If the deformation space $H^0(C_p,N_{C_p/Z_N})$ has weights $a_i$ and the obstruction space $H^1(C_p,N_{C_p/Z_N})$ has weights $b_j$, the contribution is
\bea
Q\,\frac{\prod_j[b_j]}{\prod_i[a_i]} .
\eea
Here $Q$ is the fugacity for the curve class $[C]$. The product geometry gives the splitting
\bea
N_{C_p/Z_N}=N_{C/X}\oplus\bigl(\mathcal O_C\otimes T_p\widetilde S_N\bigr).
\eea
Since $H^0(C,\mathcal O_C)=\mathbb C$ and $H^1(C,\mathcal O_C)=0$, the two tangent weights of $\widetilde S_N$ at $p$ both enter the denominator. Summing over the $N$ fixed points gives the transverse factor common to both local curves,
\bea
\mathcal K_N
=\sum_{p\in\widetilde S_N^T}\frac{1}{[w_{p,1}][w_{p,2}]}
=q_3^{1/2}\operatorname{Hilb}_T(S_N)
=\frac{[q_3^N]}{[q_3][q_3^{N/2}\mathfrak q][q_3^{N/2}\mathfrak q^{-1}]} .
\eea
The factor $q_3^{1/2}$ comes from symmetrization, since $w_{p,1}w_{p,2}=q_3$ at every fixed point. Equivalently, the Hilbert series follows from the equation $uv=z_3^N$, with $u,v,z_3$ of weights $q_3^{N/2}\mathfrak q$, $q_3^{N/2}\mathfrak q^{-1}$, and $q_3$, respectively. At $k=N$, the Molien formula \eqref{eq:Zk-gauging} gives $\operatorname{Hilb}_T(S_N)$. Multiplying by $q_3^{1/2}$ reproduces the transverse character $\mathcal K_N$ above.

For the resolved conifold, $N_{C/X}=\mathcal O(-1)\oplus\mathcal O(-1)$ has vanishing $H^0$ and $H^1$. The single-particle index is therefore $Q\mathcal K_N$, and its plethystic exponential gives \eqref{eq:conifoldPE-rankN}.

For $X=\operatorname{Tot}(\mathcal O(-2)\oplus\mathcal O\to\mathbb P^1)$, the trivial summand gives one deformation of weight $q_2$, corresponding to motion of the wrapped curve. The $\mathcal O(-2)$ summand gives one obstruction of weight $q_1q_4$. The single-particle index is thus
\bea
Q\mathcal K_N\,\frac{[q_1q_4]}{[q_2]}
=
\frac{Q[q_3^N][q_2q_3]}
{[q_3][q_2][\mathfrak q q_3^{N/2}][\mathfrak q q_3^{-N/2}]} ,
\eea
where we used $q_1q_4=(q_2q_3)^{-1}$ and $[x^{-1}]=-[x]$. Its plethystic exponential gives \eqref{eq:Z2orbifoldPE-rankN}. In this case, the fully equivariant index retains the weight $q_2$ of the moving curve. At rank one, the limit $q_2\to\infty$ recovers the refined topological string expression \eqref{eq:Z2orbifold-refinedlimit}.

\subsection{Flavor rank \texorpdfstring{$(L,M,N)$}{(L,M,N)} twisted index and gluing of tetrahedron PT vertices}\label{sec:tetrahedron-twisted}
We now turn to intersecting D6-branes. To cancel the parity anomaly, we impose
\bea\label{eq:tetra-parity-anomaly}
Lr_{1}+Mr_{2}+Nr_{3}\in 2\mathbb{Z}
\eea
so that no Chern--Simons terms are needed.

As in Section~\ref{sec:rankone-twistedindex}, the twisted index factorizes into half-index blocks. Under the proposed vortex--PT correspondence, the vortex factors give the tetrahedron PT vertices, and the classical and perturbative factors give the edge contributions of Appendix~\ref{app:tetrahedronDTPTvertex}. The D2-branes wrap the compact $\mathbb{P}^{1}$ of the local curve $Z$. Each D6-brane stack wraps the total space of two of the three normal line bundles of $\mathbb P^1$, $\mathcal{O}(-r_{a})\oplus \mathcal{O}(-r_{b})\to\mathbb P^1$.

\paragraph{Pole structure and factorization}
The twisted index of the spiked quiver is given in \eqref{eq:spiked-twisted-master}. We again take the reference vector $\eta$ to be a generic small perturbation of $(1,\ldots,1)$. For generic $(L,M,N)$, all the fundamentals $p_{s}$ contribute. As in Section~\ref{sec:rankone-twistedindex}, the poles selected by the JK prescription are
\bea\label{eq:tetra-twist-pole}
y_{\mathsf{a}}=v_{\bar{s},\alpha}^{-1}\,q_{1}^{x_{1}-1}q_{2}^{x_{2}-1}q_{3}^{x_{3}-1}\,q_{4}^{\frac{\mathfrak{m}_{\mathsf{a}}-1}{2}-\rho_{\mathsf{a}}},
\eea
where
\bea\label{eq:tetra-flux}
\mathfrak{m}_{\mathsf{a}}=\tilde{\rho}_{\mathsf{a}}+\rho_{\mathsf{a}}+\sum_{c=1}^{3}(x_{c}-1)r_{c}+1 .
\eea
After summing over the magnetic fluxes, contributions cancel as in Section~\ref{sec:rankone-twistedindex}, and the remaining poles are labeled by Young diagrams $\lambda_{23}^{(\alpha)}\,(\alpha=1,\ldots,L)$, $\lambda_{31}^{(\beta)}\,(\beta=1,\ldots,M)$, $\lambda_{12}^{(\gamma)}\,(\gamma=1,\ldots, N)$ with the condition $\sum_{\alpha=1}^{L}|\lambda_{23}^{(\alpha)}|+\sum_{\beta=1}^{M}|\lambda_{31}^{(\beta)}|+\sum_{\gamma=1}^{N}|\lambda_{12}^{(\gamma)}|=N_{4}$:
\bea
y_{\mathsf{a}}&=v_{\bar{1},\alpha}^{-1}q_{2}^{i-1}q_{3}^{j-1}q_{4}^{\frac{\mathfrak{m}_{\mathsf{a}}-1}
{2}-\pi^{(\alpha)}_{23,\mathsf{a}}},\quad \mathfrak{m}_{\mathsf{a}}=\tilde{\pi}_{23,\mathsf{a}}^{(\alpha)}+{\pi}_{23,\mathsf{a}}^{(\alpha)}+r_{2}(i-1)+r_{3}(j-1)+1,\\
y_{\mathsf{a}}&=v_{\bar{2},\beta}^{-1}q_{3}^{i-1}q_{1}^{j-1}q_{4}^{\frac{\mathfrak{m}_{\mathsf{a}}-1}
{2}-\pi^{(\beta)}_{31,\mathsf{a}}},\quad \mathfrak{m}_{\mathsf{a}}=\tilde{\pi}_{31,\mathsf{a}}^{(\beta)}+{\pi}_{31,\mathsf{a}}^{(\beta)}+r_{3}(i-1)+r_{1}(j-1)+1,\\
y_{\mathsf{a}}&=v_{\bar{3},\gamma}^{-1}q_{1}^{i-1}q_{2}^{j-1}q_{4}^{\frac{\mathfrak{m}_{\mathsf{a}}-1}
{2}-\pi^{(\gamma)}_{12,\mathsf{a}}},\quad \mathfrak{m}_{\mathsf{a}}=\tilde{\pi}_{12,\mathsf{a}}^{(\gamma)}+{\pi}_{12,\mathsf{a}}^{(\gamma)}+r_{1}(i-1)+r_{2}(j-1)+1,
\eea
where $\pi^{(\alpha)}_{23},\tilde{\pi}_{23}^{(\alpha)}\in\operatorname{RPP}(\lambda_{23}^{(\alpha)})$, $\pi^{(\beta)}_{31},\tilde{\pi}_{31}^{(\beta)}\in\operatorname{RPP}(\lambda_{31}^{(\beta)})$, and $\pi^{(\gamma)}_{12},\tilde{\pi}_{12}^{(\gamma)}\in\operatorname{RPP}(\lambda_{12}^{(\gamma)})$.

After taking the residue associated with $\underline{\vec{\lambda}}$, the twisted index \eqref{eq:spiked-twisted-master} factorizes into
\bea
\mathcal{I}_{N_{4}}^{(L,M,N)}=\sum_{|\underline{\vec{\lambda}}|=N_{4}}\fq^{\chi(\underline{\vec{\lambda}},(r_{1},r_{2},r_{3}))}\mathcal{I}_{\text{edge}}\,\mathcal{Z}_{\text{vortex}}^{\underline{\vec{\lambda}}}\,\widetilde{\mathcal{Z}}_{\text{vortex}}^{\underline{\vec{\lambda}}}
\eea
where the vortex factors are \eqref{eq:exceptional-vortex} evaluated at the north- and south-pole equivariant parameters,
\bea
\mathcal{Z}_{\text{vortex}}^{\underline{\vec{\lambda}}}&=\mathcal{Z}_{\text{vortex}}^{\underline{\vec{\lambda}}}(\fq,q_{1},q_{2},q_{3},q_{4}),\quad \widetilde{\mathcal{Z}}_{\text{vortex}}^{\underline{\vec{\lambda}}}=\mathcal{Z}_{\text{vortex}}^{\underline{\vec{\lambda}}}(\fq,\tilde{q}_{1},\tilde{q}_{2},\tilde{q}_{3},\tilde{q}_{4}),\\
\tilde{q}_{1}=q_{1}q_{4}^{r_{1}},&\quad \tilde{q}_{2}=q_{2}q_{4}^{r_{2}},\quad \tilde{q}_{3}=q_{3}q_{4}^{r_{3}},\quad \tilde{q}_{4}=q_{4}^{-1}
\eea
and $\mathcal{I}_{\text{edge}}$ is the contribution at zero vortex number. The topological terms are
\bea
\chi(\underline{\vec{\lambda}},(r_{1},r_{2},r_{3})) &=\chi(\vec{\lambda}_{23},(r_{2},r_{3}))+\chi(\vec{\lambda}_{31},(r_{3},r_{1}))+\chi(\vec{\lambda}_{12},(r_{1},r_{2})),
\eea
which come from
\bea
\sum_{\mathsf{a}}\mathfrak{m}_{\mathsf{a}}&=\sum_{\alpha=1}^{L}|\pi_{23}^{(\alpha)}|+\sum_{\alpha=1}^{L}|\tilde{\pi}_{23}^{(\alpha)}|+\chi(\vec{\lambda}_{23},(r_{2},r_{3}))\\
&+\sum_{\beta=1}^{M}|\pi_{31}^{(\beta)}|+\sum_{\beta=1}^{M}|\tilde{\pi}_{31}^{(\beta)}|+\chi(\vec{\lambda}_{31},(r_{3},r_{1}))\\
&+\sum_{\gamma=1}^{N}|\pi_{12}^{(\gamma)}|+\sum_{\gamma=1}^{N}|\tilde{\pi}_{12}^{(\gamma)}|+\chi(\vec{\lambda}_{12},(r_{1},r_{2})).
\eea

As in Section~\ref{sec:rankone-twistedindex}, the topological term and the factor $\mathcal{I}_{\text{edge}}$ are identified with the classical and one-loop factors of the two blocks up to extra signs depending on the boundary condition,
the Young diagrams $\underline{\vec{\lambda}}$, and the twisting
parameters $(r_1,r_2,r_{3})$:
\bea
\fq^{\chi(\underline{\vec{\lambda}},(r_1,r_2,r_3))}
\mathcal I_{\mathrm{edge}}^{\underline{\vec{\lambda}}}
&=
\begin{dcases}
\varepsilon_{\mathrm D}
\left(\mathcal Z^{\underline{\vec{\lambda}}}_{\mathrm{cl}}
      \mathcal Z^{\underline{\vec{\lambda}}}_{\text{1-loop}}\right)
\left(\widetilde{\mathcal Z}^{\underline{\vec{\lambda}}}_{\mathrm{cl}}
      \widetilde{\mathcal Z}^{\underline{\vec{\lambda}}}_{\text{1-loop}}\right),
& \text{for \eqref{eq:exceptional-half-index}},\\
\varepsilon_{\mathrm N}\,
\mathcal B_{\underline{\vec{\lambda}},0}
\widetilde{\mathcal B}_{\underline{\vec{\lambda}},0},
& \text{for \eqref{eq:halfindex-integral}},\\
\varepsilon_{\mathrm{CFT}}\,
\left\langle\mathbb B_{\underline{\vec{\lambda}},0}\right\rangle
\left\langle\widetilde{\mathbb B}_{\underline{\vec{\lambda}},0}\right\rangle,
& \text{for \eqref{eq:freefield-tetra}}.
\end{dcases}
\eea
See Appendix~\ref{app:edge-comparison} for the explicit formulas.

\begin{figure}[t]
\centering
\begin{tikzpicture}[every node/.style={font=\small},
    line cap=round,line join=round,
    rim/.style={line width=.55pt},
    ray/.style={draw=black!57,line width=.42pt},
    boundary/.style={draw=black!72,line width=.6pt}]
\definecolor{tetraL}{RGB}{25,125,103}
\definecolor{tetraM}{RGB}{179,73,55}
\definecolor{tetraN}{RGB}{45,90,162}

\begin{scope}[shift={(-5.83,.23)},scale=1.14,shift={(5.65,-.43)},
    every node/.append style={font=\normalsize}]
    \node[inner sep=0pt,outer sep=0pt] at (-5.36,.43)
        {$\displaystyle\sum_{\underline{\vec{\lambda}}}
          Q^{|\underline{\vec{\lambda}}|}$};
    \foreach \x/\mirror/\pole/\weight in {-3.74/1/N/q,-2.46/-1/S/{\tilde{q}}}{
        \begin{scope}[shift={(\x,.43)},xscale=\mirror]
            \draw[boundary] (0,.70)
                arc[start angle=90,end angle=270,radius=.70];
            \draw[boundary] (0,-.70)
                arc[start angle=270,end angle=450,x radius=.13,y radius=.70];
            \draw[boundary,densely dashed] (0,.70)
                arc[start angle=90,end angle=270,x radius=.13,y radius=.70];
            \fill[black!85] (-.70,0) circle[radius=1.3pt];
            \node at (-.41,0) {$\mathrm{\pole}$};
            \node at (-.34,.95) {$\weight_{1,2,3,4}$};
            \node at (-.34,-1.01) {$\underline{\vec{\lambda}}$};
        \end{scope}
    }
    \node at (-3.10,.43) {$\times$};
\end{scope}
\node at (-.60,.33) {$\longleftrightarrow$};

\begin{scope}[xshift=.17cm]
    \begin{scope}[shift={(2.56,.43)},scale=.63]
        \coordinate (N0) at (-3.18,-.32);
        \coordinate (S0) at ( 3.18, .32);
        \foreach \pole/\x/\y in {N/-3.18/-.32,S/3.18/.32}{
            \begin{scope}[shift={(\x,\y)}]
                \coordinate (\pole1) at (-.72, 1.49);
                \coordinate (\pole2) at ( 1.06, .37);
                \coordinate (\pole3) at (-.69,-1.41);
            \end{scope}
        }

        \fill[tetraM,opacity=.11] (N3)--(N1)--(S1)--(S3)--cycle;

        \foreach \a/\b/\shade in {3/1/tetraM,2/3/tetraL,1/2/tetraN}{
            \fill[\shade,opacity=.25] (S0)--(S\a)--(S\b)--cycle;
        }

        \fill[tetraL,opacity=.17] (N2)--(N3)--(S3)--(S2)--cycle;
        \fill[tetraN,opacity=.16] (N1)--(N2)--(S2)--(S1)--cycle;
        \draw[rim,draw=tetraN!82!black] (N1)--(S1)--(S2);
        \draw[rim,draw=tetraL!82!black] (N3)--(S3)--(S2);
        \draw[rim,draw=black!50] (N2)--(S2);
        \draw[rim,draw=tetraM!82!black,densely dashed] (S3)--(S1);

        \foreach \a/\b/\shade in {3/1/tetraM,2/3/tetraL,1/2/tetraN}{
            \fill[\shade!33] (N0)--(N\a)--(N\b)--cycle;
            \draw[rim,draw=\shade!80!black] (N\a)--(N\b);
        }
        \foreach \i in {1,2,3}{
            \draw[ray] (N0)--(N\i);
            \draw[ray] (S0)--(S\i);
        }
        \draw[black!91,line width=1.15pt] (N0)--(S0);
        \fill[black!91] (N0) circle[radius=1.65pt];
        \fill[black!91] (S0) circle[radius=1.65pt];
        \node[above left,inner sep=2pt] at (N1) {$q_1$};
        \node[above right,inner sep=1.5pt] at (N2) {$q_2$};
        \node[below left,inner sep=2pt] at (N3) {$q_3$};
        \node[above,inner sep=2pt] at (S1) {$\tilde q_1$};
        \node[right,inner sep=2.5pt] at (S2) {$\tilde q_2$};
        \node[below,inner sep=2pt] at (S3) {$\tilde q_3$};

        \path (N0)--(S0)
            node[pos=.17,below=3pt] {$q_4$}
            node[pos=.50,below=3pt] {$\mathbb P^1$}
            node[pos=.83,below=3pt] {$\tilde q_4$};
    \end{scope}
    \node at (.73,-1.02) {$\mathsf{PT}_{\underline{\vec{\lambda}}}$};
    \node at (4.61,-1.02) {$\widetilde{\mathsf{PT}}_{\underline{\vec{\lambda}}}$};
    \node at (2.60,-1.67)
        {$Q^{|\underline{\vec{\lambda}}|}
          \mathfrak q^{\chi(\underline{\vec{\lambda}},(r_1,r_2,r_3))}$};
    \node at (2.60,-2.20)
        {$(-1)^{\sigma(\underline{\vec{\lambda}})}
          \mathsf E_{\underline{\vec{\lambda}}}
          (q_{1,2,3}\mid\tilde q_{1,2,3})$};
\end{scope}
\end{tikzpicture}
\caption{
The correspondence between the factorization of the spiked-quiver
twisted index and the gluing of two tetrahedron PT vertices.
The two hemispheres on the left share the boundary data
$\underline{\vec{\lambda}}
=(\vec{\lambda}_{23},\vec{\lambda}_{31},\vec{\lambda}_{12})$.
On the right, the green, red, and blue prisms represent the
$L$ D6$_{234}$-, $M$ D6$_{134}$-, and $N$ D6$_{124}$-brane supports,
respectively. The common interior edge represents the compact
$\mathbb P^1$ wrapped by the D2-branes.
}
\label{fig:tetrahedron-half-index-gluing}
\end{figure}
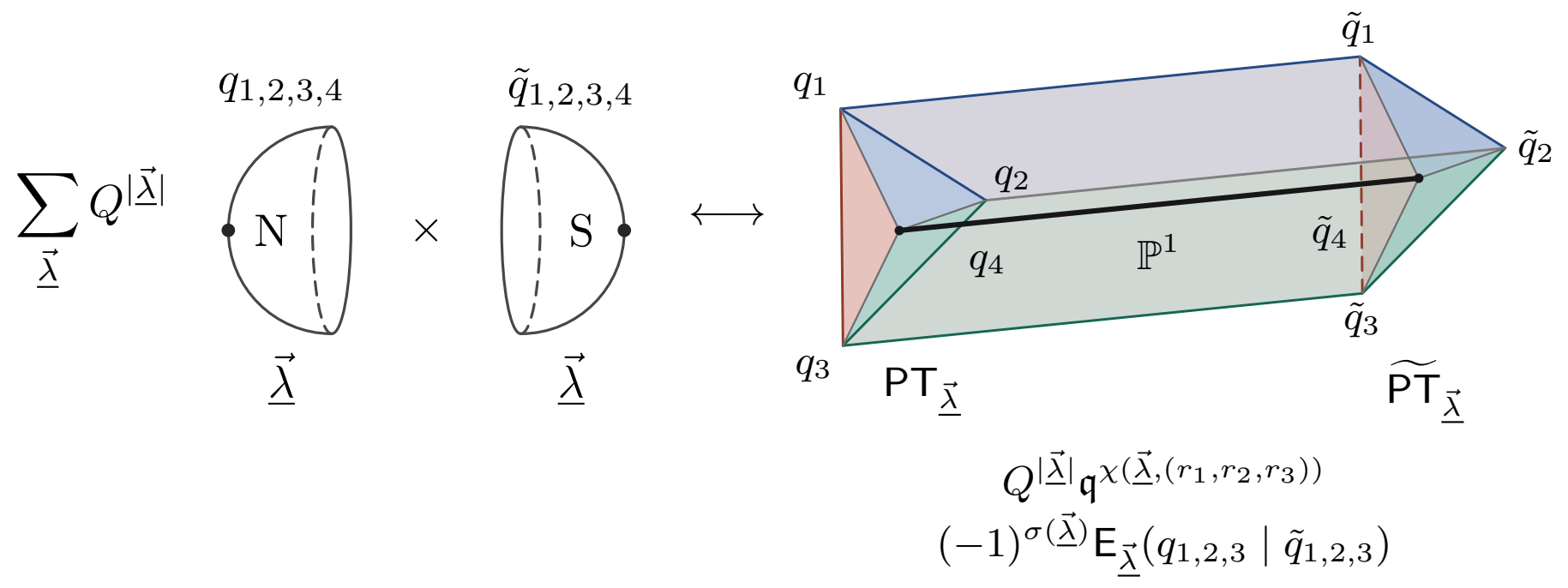

\paragraph{Relation to the gluing of tetrahedron PT vertices}
As in Section~\ref{sec:rankone-twistedindex}, we claim that the generating function of the twisted index
\bea
\mathcal{I}_{\text{twist}}^{(L,M,N)}=\sum_{N_{4}=0}^{\infty}Q^{N_{4}}\mathcal{I}_{N_{4}}^{(L,M,N)}
\eea
equals the rank-$(L,M,N)$ tetrahedron PT partition function of the local curve $Z$.

We wrap the $(L,M,N)$ D6-brane stacks on the threefolds obtained by setting the first, second, and
third fiber coordinates to zero, respectively. The two tetrahedron PT vertices at the torus fixed points
are glued along their common leg data
$\underline{\vec{\lambda}}=(\vec{\lambda}_{23},
\vec{\lambda}_{31},\vec{\lambda}_{12})$:
\bea
\mathsf{PT}[Z;(L,M,N)]
&=
\sum_{\underline{\vec{\lambda}}}
Q^{|\underline{\vec{\lambda}}|}
\fq^{\chi(\underline{\vec{\lambda}},(r_1,r_2,r_3))}(-1)^{\sigma(\underline{\vec{\lambda}})}
\mathsf{E}_{\underline{\vec{\lambda}}}
(q_{1,2,3}\mid\tilde q_{1,2,3})
\mathsf{PT}_{\underline{\vec{\lambda}}}[\fq,q_{1,2,3,4}]
\mathsf{PT}_{\underline{\vec{\lambda}}}[\fq,\tilde q_{1,2,3,4}].
\eea
Here, $\mathsf{E}_{\underline{\vec{\lambda}}}$ and $(-1)^{\sigma(\underline{\vec{\lambda}})}$ are the edge factor and the sign defined in Appendix~\ref{app:tetrahedronDTPTvertex}.

By the proposal \eqref{eq:higher-rank-vortex-PT-proposal}, the vortex blocks give the two tetrahedron PT vertices, and the factor $\mathcal{I}_{\text{edge}}$ is identified as
\bea
\mathcal{I}_{\text{edge}}=(-1)^{\sigma(\underline{\vec{\lambda}})}
\mathsf{E}_{\underline{\vec{\lambda}}}
(q_{1,2,3}\mid\tilde q_{1,2,3}).
\eea
Since $|\underline{\vec{\lambda}}|=N_4$, summing over the gauge rank gives
\bea
\mathcal I_{\text{twist}}^{(L,M,N)}
=
\sum_{N_4=0}^{\infty}
Q^{N_4}\mathcal I_{N_4}^{(L,M,N)}
=
\mathsf{PT}[Z;(L,M,N)].
\eea
Figure~\ref{fig:tetrahedron-half-index-gluing} schematically illustrates the correspondence.

For $(r_{1},r_{2},r_{3})=(2,0,0)$ and $(L,M,N)=(0,M,N)$, the condition \eqref{eq:tetra-parity-anomaly} is satisfied, and we propose the plethystic formula
\bea\label{eq:local-tetra-PE}
\mathcal{I}_{\text{twist}}^{(0,M,N)}&=\PE\left[\frac{[q_{2}^{M}q_{3}^{N}][q_{32}]}{[q_{3}][q_{2}]}\frac{Q}{[\fq q_{2}^{M/2}q_{3}^{N/2}][\fq q_{2}^{-M/2}q_{3}^{-N/2}]}\right]
\eea
which is independent of the individual D6-brane flavor fugacities. The specialization in Section~\ref{sec:D2-D8} checks the coefficient of $Q$ (see also Appendix~\ref{app:PE-local}).

\paragraph{Membrane index interpretation}
The plethystic formula \eqref{eq:local-tetra-PE} admits a membrane interpretation similar to that for parallel D6-branes. Consider the two D6-brane stacks supported on $D_2=\{z_2=0\}$ and $D_3=\{z_3=0\}$ with multiplicities $M$ and $N$, respectively. For a general twist, adjunction gives $K_{D_i}\cong\pi^*\mathcal O_{\mathbb P^1}(-r_i)$ as line bundles, where $\pi$ denotes the projection to $\mathbb P^1$. Thus both stacks wrap Calabi--Yau threefolds precisely when $r_2=r_3=0$. The condition $r_1+r_2+r_3=2$ then fixes $r_1=2$. In this case, the two stacks share the surface $T^*\mathbb P^1$, and the transverse coordinates $z_2,z_3$ belong to trivial line bundles over $\mathbb P^1$.

As in the parallel D6-brane case, we expect the M-theory circle to degenerate along the D6-brane locus. The two stacks define the divisor $z_2^M z_3^N=0$, which suggests the local fivefold
\bea
T^*\mathbb P^1\times Y_{M,N},
\qquad
Y_{M,N}=\{uv=z_2^M z_3^N\}\subset\mathbb C^4,
\eea
where the transverse threefold $Y_{M,N}$ is the generalized conifold. For $M=N=1$, we recover the ordinary conifold $uv=z_2z_3$. Setting $M=0$ instead gives $Y_{0,N}=\mathbb C_{z_2}\times(\mathbb C^2/\mathbb Z_N)$, in agreement with the transverse geometry of a single stack.

We compute the contribution of an M2-brane wrapping the zero section $C=\mathbb P^1\subset T^*\mathbb P^1$. Choose a toric crepant resolution $\widetilde Y_{M,N}$ of $Y_{M,N}$, and let $w_{p,1},w_{p,2},w_{p,3}$ be the tangent weights at a torus fixed point $p\in\widetilde Y_{M,N}$. The normal bundle of the membrane $C_p=C\times\{p\}$ is
\bea
N_{C_p}=\mathcal O_C(-2)\oplus\bigl(\mathcal O_C\otimes T_p\widetilde Y_{M,N}\bigr).
\eea
The second summand gives three deformations of weights $w_{p,1},w_{p,2},w_{p,3}$, while $H^1(C,\mathcal O_C(-2))$ gives one obstruction of weight $q_1q_4$. The membrane localization formula~\cite{Nekrasov:2014nea} therefore gives the single-particle index
\bea
Q\sum_{p\in\widetilde Y_{M,N}^T}
\frac{[q_1q_4]}{[w_{p,1}][w_{p,2}][w_{p,3}]} .
\eea

We assign $z_2,z_3$ the weights $q_2,q_3$ and $u,v$ the weights $q_2^{M/2}q_3^{N/2}\mathfrak q$ and $q_2^{M/2}q_3^{N/2}\mathfrak q^{-1}$, respectively. Since the defining equation has weight $q_2^M q_3^N$, the holomorphic volume form on $Y_{M,N}$ has weight $q_2q_3$. The resolution is crepant, so $w_{p,1}w_{p,2}w_{p,3}=q_2q_3$ at every fixed point. Using $q_1q_4=(q_2q_3)^{-1}$, we obtain
\bea
Q\sum_{p\in\widetilde Y_{M,N}^T}
\frac{[q_1q_4]}{\prod_{i=1}^3[w_{p,i}]}
=-Q(1-q_2q_3)\sum_{p\in\widetilde Y_{M,N}^T}
\frac{1}{\prod_{i=1}^3(1-w_{p,i})}
=-Q(1-q_2q_3)\operatorname{Hilb}_T(Y_{M,N}).
\eea
Here the equivariant Hilbert series follows directly from the four generators and the single defining relation of the generalized conifold,
\bea
\operatorname{Hilb}_T(Y_{M,N})
=
\frac{1-q_2^M q_3^N}
{(1-q_2)(1-q_3)(1-q_2^{M/2}q_3^{N/2}\mathfrak q)(1-q_2^{M/2}q_3^{N/2}\mathfrak q^{-1})}.
\eea
Taking the plethystic exponential, we recover \eqref{eq:local-tetra-PE} in the form
\bea
\mathcal I_{\mathrm{twist}}^{(0,M,N)}
=
\operatorname{PE}\!\left[-Q(1-q_2q_3)\operatorname{Hilb}_T(Y_{M,N})\right].
\eea

\subsection{D2--D8 system, twisted index and gluing of \texorpdfstring{PT$_{4}$}{PT4} vertices}\label{sec:D2-D8}
We now extend the correspondence between twisted indices and
PT gluing to Calabi--Yau fourfolds. We refer to PT counting on
these geometries as PT$_{4}$ counting.
The relevant brane system consists of D2-branes wrapping the
compact curve and D8-branes wrapping the Calabi--Yau fourfold.
We claim that the twisted index of the corresponding 3d gauge
theory factorizes into blocks that reproduce the PT$_4$ vertices,
with the remaining factors giving their gluing, and that its generating function is the PT$_{4}$ partition function of the local curve $Z$.

We wrap $N_{4}$ D2-branes on $\mathbb{C}_{4}\times\mathbb{S}^{1}$ and $N_{\four}$ D8--$\overline{\D8}$ pairs on $\mathbb{C}^{4}_{1234}\times\mathbb{S}^{1}$:
\bea\label{eq:D2-D8}
\renewcommand{\arraystretch}{1.05}
\begin{tabular}{|c|c|c|c|c|c|c|c|c|c|c|}
\hline
& \multicolumn{2}{c|}{$\mathbb{C}_{1}$} & \multicolumn{2}{c|}{$\mathbb{C}_{2}$} & \multicolumn{2}{c|}{$\mathbb{C}_{3}$} & \multicolumn{2}{c|}{$\mathbb{C}_{4}$} & \multicolumn{2}{c|}{$\mathbb{R}^{1}\times \mathbb{S}^{1}$} \\
\cline{2-11} \raisebox{-0mm}{}  & 1 & 2 & 3 & 4& 5 & 6 & 7 & 8 & 9& 0\\[-2pt]
\hline
    \raisebox{0mm}{$\textcolor{red!60!yellow}{\D2_{4}}$} & $\bullet$ & $\bullet$ & $\bullet$ & $\bullet$ & $\bullet$ & $\bullet$ & $-$ & $-$ & $\bullet$ & $-$ \\
\hline
\raisebox{-0.5mm}{$\textcolor{black}{{\D8}_{1234}} $}& $-$ & $-$& $-$ & $-$  & $-$ & $-$ & $-$ & $-$ & $\bullet$ & $-$ \\
\hline
\raisebox{-0.5mm}{$\textcolor{black}{{\overline{\D8}}_{1234}} $}& $-$ & $-$  & $-$ & $-$ & $-$ & $-$& $-$ & $-$ & $\bullet$ & $-$ \\
\hline
\end{tabular}
\eea
The low-energy theory on the D2-branes is a 3d $\mathcal{N}=2$ $U(N_{4})$ gauge theory with quiver and superpotential
\bea\label{eq:D2D8-quiver}
\adjustbox{valign=c}{\begin{tikzpicture}[decoration={markings,mark=at position \arrowHeadPosition with {\arrow{latex}}}]
 \tikzset{
        box/.style={draw, minimum width=0.6cm, minimum height=0.6cm, text centered,thick},
        ->-/.style={decoration={
        markings,mark=at position #1 with {\arrow[scale=1.5]{>}}},postaction={decorate},line width=0.5mm},
        -<-/.style={decoration={
        markings,
        mark=at position #1 with {\arrow[scale=1.5]{<}}},postaction={decorate},line width=0.5mm}    
    }
    
\begin{scope}[xshift=0cm]
\node[black,anchor=east] at (-0.3,0.65) {$p$};
\node[black,anchor=west] at ( 0.3,0.65) {$\widetilde p$};



    \draw[red!60!yellow,thick,scale=1.3,postaction={decorate}] (0,0) arc(90:450:0.13 and 0.5) ;
    \draw[red!60!yellow,thick,scale=1.3,postaction={decorate}, rotate=-90] (0,0) arc(90:450:0.13 and 0.5) ;
    \draw[red!60!yellow,thick,scale=1.3,postaction={decorate},rotate=90] (0,0) arc(90:450:0.13 and 0.5) ;
   
    \draw[red!60!yellow,thick,postaction={decorate}] (-1.02+0.50,1.2)--(-1.02+0.95,0.3);
  \draw[red!60!yellow,thick,postaction={decorate}](-0.88+0.95,0.3)--(-0.88+0.95+0.5,1.2);



     \draw[red!60!yellow,fill=white,thick](0,0) circle(0.3cm);
    \node[red!60!yellow] at (0,0) {$N_{4}$};

    \node[black,fill=white,box] at (-0.5,1.5){$N_{\four}$};
     \node[black,fill=white,box] at (+0.5,1.5){$N_{\four}$};

     \node[right] at (1.1,0.2){$\textcolor{blue}{\Phi_{3}}$};
      \node[left] at (-1.1,0.3){$\textcolor{red}{\Phi_{2}}$};
      \node[below] at (0,-1.25){$\textcolor{green!50!black}{\Phi_{1}}$};
    
\end{scope}
\end{tikzpicture}}\qquad {W}=\Tr\Phi_{3}[\Phi_{1},\Phi_{2}]
\eea
whose $F$-term relations are
\bea
\relax[\Phi_{1},\Phi_{2}]=[\Phi_{2},\Phi_{3}]=[\Phi_{3},\Phi_{1}]=0.
\eea

The Higgs branch at vanishing real
masses is
\bea
\mathcal{M}^{\D2\tbar\D8}_{ H}
=
\left\{
(\Phi_1,\Phi_2,\Phi_3,p,\widetilde{p})
\ \middle|\
\begin{aligned}
&[\Phi_a,\Phi_b]=0,\quad 1\leq a<b\leq 3,\\
&\sum_{a=1}^{3}[\Phi_a,\Phi_a^\dagger]
+pp^\dagger-\widetilde{p}^{\dagger}\widetilde{p}
=\xi\,\mathbf{1}_{N_4}
\end{aligned}
\right\}\big/U(N_4).
\eea

We specialize to $N_{\four}=1$ and choose a basis vector $e$ of the one-dimensional D8 framing space. For generic equivariant parameters, the torus-fixed vacua are labeled by plane partitions $\pi$ with $|\pi|=N_4$. Here a plane partition is a finite set of boxes $\pi\subset\mathbb Z_{\geq 1}^{3}$ such that decreasing any coordinate of a box by one gives another box of $\pi$ whenever that coordinate is greater than one. At the vacuum labeled by $\pi$, the gauge space has the basis
\bea
\left\{ \Phi_1^{i-1}\Phi_2^{j-1}\Phi_3^{k-1}p(e) \;\middle|\; (i,j,k)\in\pi \right\}.
\eea
Thus, the plane partition $\pi$ plays the same role as the
Young diagram $\lambda_{12}$ in the D2--D6 system. For generic $N_{\four}$, the Higgs vacua are specified by $N_{\four}$-tuples of plane partitions: $\vec{\pi}=(\pi^{(\alpha)})_{\alpha=1}^{N_{\four}}$ with $\sum_{\alpha=1}^{N_{\four}}|\pi^{(\alpha)}|=N_{4}$.

The half-index associated with each vacuum was studied in
\cite{Piazzalunga:2023qik}. The boundary condition used there is closer to the formula \eqref{eq:halfindex-integral}. We do not construct explicit Dirichlet boundary conditions for the vector multiplet in this case. With the equivariant parameters and normalization matched, the vortex partition function is known to reproduce the one-leg PT$_4$ vertex:
\bea\label{eq:halfindex-PT4-relation}
\mathcal{Z}_{\text{vortex}}^\pi
(\fq,\mu;q_1,q_2,q_3,q_4)
=
\mathsf{PT4}_{\varnothing\varnothing\varnothing\pi}
[\fq,\mu;q_1,q_2,q_3,q_4].
\eea
Here, $\pi$ labels the Higgs vacuum on the gauge-theory side
and the asymptotic plane partition along the fourth direction
on the PT$_4$ side. We use this identification below to relate the factorization of the twisted index to PT$_{4}$ gluing.

\paragraph{Twisted index and block factorization}
In the D2--D8 system, since the superpotential does not involve the framing fields, we choose both framing chiral multiplets to have $R$-charge one:
 \bea\label{eq:D2D8-flavor_table_sym}
 \begin{tabular}{c|c c c }
  &   $U(1)_{t_{1}} $  & $U(1)_{t_{2}}$  & $U(1)_{R,q}$ \\\hline
    $p$  & $0$& $0$ & $1$\\
    $\tilde{p}$  & $0$ & $0$ & $1$\\
  
 \end{tabular}
 \eea

For the $\alpha$-th D8--$\overline{\D8}$ pair, let $v_{\alpha}$ and $\mu_{\alpha}^{-1}v_{\alpha}$ be the brane and antibrane fugacities, respectively. We take the background fluxes to vanish and the masses $\mu_{\alpha}$ to be generic. The twisted index is then
\begin{equation}\label{eq:D2-D8-twisted-master}
\begin{aligned}
\mathcal{I}^{\text{D2-D8}}_{N_{4}}(\{\mu_{\alpha}\})=&\sum_{\mathfrak{m}\in\mathbb{Z}^{N_4}}\frac{(-1)^{N_4}}{N_4!}\oint_{\mathrm{JK}}\prod_{\mathsf{a}=1}^{N_4}\frac{\mathrm{d}y_{\mathsf{a}}}{2\pi i\,y_{\mathsf{a}}}\;
\big((-1)^{N_{\four}} \mathfrak{q}\big)^{\sum_{\mathsf{a}}\mathfrak{m}_{\mathsf{a}}}
\\
&\times\prod_{\mathsf{a}=1}^{N_4}\prod_{\alpha=1}^{N_{\four}}
\frac{(y_{\mathsf{a}}v_{\alpha})^{\frac{\mathfrak{m}_{\mathsf{a}}}{2}}}{\big(y_{\mathsf{a}}v_{\alpha}\,q_{4}^{\frac{1-\mathfrak{m}_{\mathsf{a}}}{2}};q_{4}\big)_{\mathfrak{m}_{\mathsf{a}}}}\;
\frac{\Big(\frac{\mu_{\alpha}}{y_{\mathsf{a}}v_{\alpha}}\Big)^{-\frac{\mathfrak{m}_{\mathsf{a}}}{2}}}{\Big(\frac{\mu_{\alpha}}{y_{\mathsf{a}}v_{\alpha}}\,q_{4}^{\frac{\mathfrak{m}_{\mathsf{a}}+1}{2}};q_{4}\Big)_{-\mathfrak{m}_{\mathsf{a}}}}
\\
&\times\prod_{\mathsf{a}\neq\mathsf{b}}\frac{(y_{\mathsf{a}}/y_{\mathsf{b}})^{\frac{1}{2}(\mathfrak{m}_{\mathsf{a}}-\mathfrak{m}_{\mathsf{b}}-1)}}{\Big(\frac{y_{\mathsf{a}}}{y_{\mathsf{b}}}\,q_{4}^{\frac{\mathfrak{m}_{\mathsf{b}}-\mathfrak{m}_{\mathsf{a}}}{2}+1};q_{4}\Big)_{\mathfrak{m}_{\mathsf{a}}-\mathfrak{m}_{\mathsf{b}}-1}}
\prod_{s=1}^{3}\prod_{\mathsf{a},\mathsf{b}=1}^{N_4}
\frac{\big(q_{s}^{-1}q_{4}^{-r_{s}/2}\,y_{\mathsf{a}}/y_{\mathsf{b}}\big)^{\frac{1}{2}(\mathfrak{m}_{\mathsf{a}}-\mathfrak{m}_{\mathsf{b}}+1-r_{s})}}{\Big(\frac{y_{\mathsf{a}}}{q_{s}y_{\mathsf{b}}}\,q_{4}^{\frac{\mathfrak{m}_{\mathsf{b}}-\mathfrak{m}_{\mathsf{a}}}{2}};q_{4}\Big)_{\mathfrak{m}_{\mathsf{a}}-\mathfrak{m}_{\mathsf{b}}+1-r_{s}}} ,
\end{aligned}
\end{equation}
normalized by $\mathcal{I}^{\text{D2-D8}}_{N_{4}=0}=1$. Each framing pair in the second line is
\bea\label{eq:D2-D8-framing-pair}
\mathscr{D}(y_{\mathsf{a}}v_{\alpha},\,\mathfrak{m}_{\mathsf{a}})\,\mathscr{D}\!\left(\frac{\mu_{\alpha}}{y_{\mathsf{a}}v_{\alpha}},\,-\mathfrak{m}_{\mathsf{a}}\right),
\eea
and the remaining factors, the topological fugacity, the vector multiplet and the three adjoint contributions, are the same as in \eqref{eq:spiked-twisted-master}. In this case, there is no parity anomaly for any flavor rank and $R$-charges $r_{1,2,3}$.

We specialize to a single framing pair, $N_{\four}=1$, and write $v=v_1$ for its flavor fugacity. As in Section~\ref{sec:rankone-twistedindex}, each chain of poles in the positive FI chamber starts at a fundamental pole of $p$, which fixes $y_{\mathsf b}$ for $\mathfrak m_{\mathsf b}>0$. An adjoint chiral $\Phi_c$ can then give a pole in $y_{\mathsf a}/y_{\mathsf b}$ whenever $\mathfrak m_{\mathsf a}-\mathfrak m_{\mathsf b}\geq r_c$. Iterating these relations gives the pole positions of \eqref{eq:twisted-pole-general},
\bea\label{eq:twisted-pole-D2D8}
y_{\mathsf a}&=v^{-1}q_1^{x_1-1}q_2^{x_2-1}q_3^{x_3-1}q_4^{\frac{\mathfrak m_{\mathsf a}-1}{2}-\rho_{\mathsf a}},\\
\mathfrak m_{\mathsf a}&=1+\rho_{\mathsf a}+\tilde\rho_{\mathsf a}+\sum_{c=1}^{3}(x_c-1)r_c ,
\eea
where $\rho_{\mathsf a},\tilde\rho_{\mathsf a}\in\mathbb Z_{\geq 0}$ and $x_c-1$ counts the steps in the $\Phi_c$ direction from the fundamental pole. For generic framing mass $\mu$, the zeros of the antifundamental $\tilde p$ do not remove any of these poles, so the base extends in all three adjoint directions. The base of a contributing configuration is therefore a plane partition $\pi$ with $|\pi|=N_4$, with $\rho_{\mathsf a}$ and $\tilde\rho_{\mathsf a}$ attached to each box $\mathsf a\in\pi$.

The non-negative integers $\rho_{\mathsf{a}}$ and $\tilde\rho_{\mathsf{a}}$ define
two reverse solid partitions with base $\pi$. A reverse solid partition with base $\pi$ is a map $\rho\colon\pi\to\mathbb Z_{\geq 0}$ satisfying
\bea
\rho(\mathbf{x}) &\leq \rho(\mathbf{x}+\mathbf{e}_c)
\qquad \text{whenever } \mathbf{x},\,\mathbf{x}+\mathbf{e}_c\in\pi,
\eea
for each $c=1,2,3$, where $\mathbf{e}_c$ is the unit vector in the $c$-th coordinate direction. This is the three-dimensional analogue of a reverse plane
partition. We denote the set of such assignments by
$\operatorname{RSP}(\pi)$, so that
$\rho,\tilde\rho\in\operatorname{RSP}(\pi)$.
The two assignments describe the vortex configurations at
the north and south poles, respectively.

After taking the residue at \eqref{eq:twisted-pole-D2D8}, the twisted index also factorizes into
\bea
\mathcal{I}^{\text{D2-D8}}_{N_{4}}(\mu)&=\sum_{|\pi|=N_{4}}\fq^{\chi(\pi,(r_{1},r_{2},r_{3}))}\mathcal{I}_{\text{edge}}\mathcal{Z}_{\text{vortex}}^{\pi}\widetilde{\mathcal{Z}}^{\pi}_{\text{vortex}},
\eea
where
\bea\label{eq:D2D8-block-param}
&\mathcal{Z}^{\pi}_{\text{vortex}}=\mathcal{Z}^{\pi}_{\text{vortex}}(\fq,\mu;q_{1},q_{2},q_{3},q_{4}),\quad \widetilde{\mathcal{Z}}^{\pi}_{\text{vortex}}=\mathcal{Z}^{\pi}_{\text{vortex}}(\fq,\mu;\tilde{q}_{1},\tilde{q}_{2},\tilde{q}_{3},\tilde{q}_{4}),\\
&\tilde{q}_{1}=q_{1}q_{4}^{r_{1}},\quad \tilde{q}_{2}=q_{2}q_{4}^{r_{2}},\quad \tilde{q}_{3}=q_{3}q_{4}^{r_{3}},\quad \tilde{q}_{4}=q_{4}^{-1}
\eea
and
\bea
\chi(\pi,(r_{1},r_{2},r_{3}))=\sum_{(i,j,k)\in\pi}(r_{1}(i-1)+r_{2}(j-1)+r_{3}(k-1)+1).
\eea
As before, the factor $\fq^{\chi(\pi,(r_{1},r_{2},r_{3}))}\mathcal{I}_{\text{edge}}$ is identified with the classical and one-loop factors of the north and south blocks up to extra signs depending on the boundary conditions.

\paragraph{Relation to the gluing of PT$_{4}$ vertices}
 We briefly recall the PT$_{4}$ gluing prescription reviewed in
Appendix~\ref{app:PT4vertexformalism}. Consider the local curve $Z$ above, with
D8-branes wrapping the entire fourfold.
The two one-leg PT$_4$ vertices are glued along their common
leg, labeled by a plane partition $\pi$:
\bea
\mathsf{PT4}[Z]
&=
\sum_{\pi}
Q^{|\pi|}
\fq^{\chi(\pi,(r_1,r_2,r_3))}
(-1)^{\sigma_a(\pi)}
\mathsf{E}^{(a)}_{\pi}
(q_{1,2,3}\mid\tilde q_{1,2,3})\\
&\qquad \times 
\mathsf{PT4}_{\varnothing\varnothing\varnothing\pi}
[\fq,\mu;q_{1,2,3,4}]
\mathsf{PT4}_{\varnothing\varnothing\varnothing\pi}
[\fq,\mu;\tilde q_{1,2,3,4}].
\eea
The edge factor depends on the orientation choice, represented
here by $a=1,2,3$ (see \eqref{eq:PT4-edgefactor}, \eqref{eq:PT4-edge-sign}). This dependence is compensated by
$(-1)^{\sigma_a(\pi)}$, so the signed edge contribution is
independent of $a$.

Comparing with the PT$_{4}$ vertex formalism, the vortex blocks reproduce the two PT$_4$ vertices, and the factor $\mathcal{I}_{\text{edge}}$ is identified with the edge contribution with this sign:
\bea
\mathcal I_{\text{edge}}
=
(-1)^{\sigma_a(\pi)}
\mathsf{E}^{(a)}_{\pi}
(q_{1,2,3}\mid\tilde q_{1,2,3}).
\eea
Thus the twisted index reproduces both the vertex gluing
and its orientation sign, giving
\bea
\mathcal I_{\text{twist}}^{\D2\tbar\D8}(\mu)
=
\mathsf{PT4}[Z],\quad \mathcal{I}_{\text{twist}}^{\D2\tbar\D8}(\mu)=\sum_{N_{4}=0}^{\infty}Q^{N_{4}}\mathcal{I}_{N_{4}}^{\D2\tbar\D8}(\mu).
\eea

\begin{figure}[t]
\centering
\begin{tikzpicture}[
    line width=0.8pt,
    every node/.style={font=\small}
]

\node at (-2,0)
    {$\displaystyle\sum_{\vec{\pi}}Q^{|\vec{\pi}|}$};

\begin{scope}[xshift=0.1cm]
    \draw (0,0.9)
        arc[start angle=90,end angle=270,radius=0.9];

    \draw (0,-0.9)
        arc[start angle=270,end angle=450,
            x radius=0.18,y radius=0.9];
    \draw[dashed] (0,0.9)
        arc[start angle=90,end angle=270,
            x radius=0.18,y radius=0.9];

    \fill (-0.9,0) circle (1.4pt);
    \node at (-0.48,0) {$\mathrm N$};
    \node[above] at (-0.45,1.05) {$q_{1,2,3,4}$};
    \node[below] at (0,-1.05) {$\vec{\pi}$};
\end{scope}

\node at (0.85,0) {$\times$};

\begin{scope}[xshift=1.7cm]
    \draw (0,-0.9)
        arc[start angle=270,end angle=450,radius=0.9];

    \draw (0,-0.9)
        arc[start angle=270,end angle=450,
            x radius=0.18,y radius=0.9];
    \draw[dashed] (0,0.9)
        arc[start angle=90,end angle=270,
            x radius=0.18,y radius=0.9];

    \fill (0.9,0) circle (1.4pt);
    \node at (0.48,0) {$\mathrm S$};
    \node[above] at (0.45,1.05) {$\tilde q_{1,2,3,4}$};
    \node[below] at (0,-1.05) {$\vec{\pi}$};
\end{scope}

\node at (4.4,0) {$\longleftrightarrow$};

\begin{scope}[xshift=8.5cm]
    \coordinate (L) at (-0.9,0);
    \coordinate (R) at ( 0.9,0);

    \draw (L) -- ++(-1, 0.9);
    \draw (L) -- ++(-1,-0.9);
     \draw (L) -- ++(-1.4,+0.3);
    \draw (R) -- ++( 1, 0.9);
    \draw (R) -- ++( 1,-0.9);
     \draw (R) -- ++(1.4,+0.3);

    \draw (-0.85,0)--(0.85,0);
    \fill (L) circle (1.5pt);
    \fill (R) circle (1.5pt);

    \node at (0,-0.5) {$\mathbb P^1$};
    \node[above] at (0,0.3) {$\vec{\pi}$};

    \node[above ] at (-1.9,0.9){$q_{1}$};
    \node[above ] at (1.9,0.9){$\tilde{q}_{1}$};
     \node[left] at (-2.2,0.3){$q_{2}$};
     \node[right] at (2.2,0.3){$\tilde{q}_{2}$};
    \node[below left] at (-1.9,-0.9){$q_{3}$};
    \node[below right] at (1.9,-0.9){$\tilde{q}_{3}$};
    
    \node[left,align=center] at (-1.3,-0.2)
        {$\mathsf{PT4}_{\varnothing\varnothing\varnothing\vec{\pi}}$};
    \node[right,align=center] at (1.3,-0.2)
        {$\widetilde{\mathsf{PT4}}_{\varnothing\varnothing\varnothing\vec{\pi}}$};

    \node[below,align=center] at (0,-1.5)        {$Q^{|\vec{\pi}|}\fq^{\chi(\vec{\pi},(r_{1},r_{2},r_{3}))}$};
    \node[below,align=center] at (0,-2){$ (-1)^{\sigma_{a}(\vec{\pi})}\mathsf{E}^{(a)}_{\vec{\pi}}(q_{1},q_{2},q_{3}\mid\tilde{q}_{1},\tilde{q}_{2},\tilde{q}_{3})$};
    
\end{scope}

\end{tikzpicture}
\caption{Block factorization of the D2--D8 twisted index and PT$_4$ gluing
on a local Calabi--Yau fourfold. On the left, the north- and
south-pole blocks are evaluated at $q_{1,2,3,4}$ and
$\tilde q_{1,2,3,4}$, respectively. On the right, the corresponding
one-leg PT$_4$ vertices are glued along the compact $\mathbb{P}^1$.
The sum runs over the common tuple of plane partitions
$\vec{\pi}$, with curve weight $Q^{|\vec{\pi}|}$, the classical
weight, and the signed PT$_4$ edge factor.
}
\label{fig:twisted-PT4-gluing}
\end{figure}
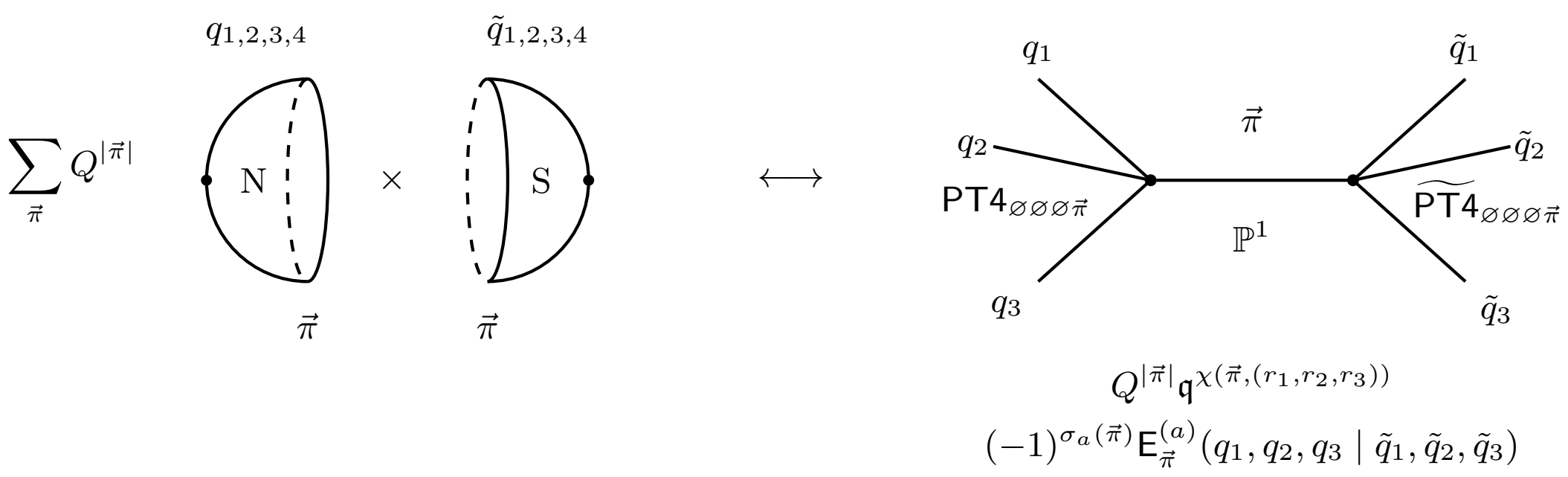

For several framing pairs, the vortex partition function labeled by the tuple of plane partitions $\vec{\pi}=(\pi^{(\alpha)})_{\alpha=1}^{N_{\four}}$ coincides with the rank-$N_{\four}$ PT$_{4}$ vertex:
\bea
\mathcal{Z}_{\text{vortex}}^{\vec{\pi}}
(\fq,\{\mu_{\alpha}\};q_1,q_2,q_3,q_4)
=
\mathsf{PT4}_{\varnothing\varnothing\varnothing\vec{\pi}}
[\fq,\{\mu_{\alpha}\};q_1,q_2,q_3,q_4].
\eea
The twisted index also factorizes and the generating function of the twisted index matches the rank-$N_{\four}$ PT$_{4}$ partition function:
\bea
\mathcal{I}^{\D2\tbar\D8}_{\text{twist}}(\{\mu_{\alpha}\})=\mathsf{PT4}[Z;N_{\four}].
\eea
The correspondence is schematically illustrated in Figure~\ref{fig:twisted-PT4-gluing}.

\paragraph{Plethystic exponential formulas}
For $N_{4}=1$, the twisted index is
\bea
\mathcal{I}^{\D2\tbar\D8}_{1}(\{\mu_{\alpha}\})&=-\prod_{s=1}^{3}\mathscr{D}(q_{s}^{-1}q_{4}^{-r_{s}/2},1-r_{s})
\sum_{\mathfrak{m}=1}^{\infty}((-1)^{N_{\four}}\fq)^{\mathfrak{m}}\\
&\qquad \times\oint_{\text{JK}}\frac{dy}{2\pi i y}\prod_{\alpha=1}^{N_{\four}}\mathscr{D}(yv_{\alpha},\mathfrak{m})\mathscr{D}\left(\frac{\mu_{\alpha}}{yv_{\alpha}},-\mathfrak{m}\right).
\eea
The integral is evaluated similarly to the computation in \eqref{eq:rankonetwist-comp}:
\bea
\oint_{\text{JK}}\frac{dy}{2\pi i y}\prod_{\alpha=1}^{N_{\four}}\mathscr{D}(yv_{\alpha},\mathfrak{m})\mathscr{D}\left(\frac{\mu_{\alpha}}{yv_{\alpha}},-\mathfrak{m}\right)&=(-1)^{N_{\four}\mathfrak{m}}\left(\mu^{\mathfrak{m}/2}-\mu^{-\mathfrak{m}/2}\right)
\eea
where
\bea\label{eq:mu-product}
\mu=\prod_{\alpha=1}^{N_{\four}}\mu_{\alpha}.
\eea
Summing over $\mathfrak{m}\in\mathbb{Z}_{>0}$ gives
\bea\label{eq:D2D8-rankone-formula}
\mathcal{I}^{\D2\tbar\D8}_{1}(\{\mu_{\alpha}\})&=\prod_{s=1}^{3}\mathscr{D}(q_{s}^{-1}q_{4}^{-r_{s}/2},1-r_{s})\frac{[\mu]}{[\fq\mu^{1/2}][\fq^{-1}\mu^{1/2}]}.
\eea
Thus the $N_{4}=1$ twisted index depends on the flavor fugacities of the D8--$\overline{\D8}$ pairs only through the product $\mu$ in \eqref{eq:mu-product}.

For $(r_{1},r_{2},r_{3})=(1,1,0)$ and $(2,0,0)$, the proposed plethystic formulas imply that the full twisted-index generating function depends only on the product \eqref{eq:mu-product},
\bea
\mathcal{I}_{\text{twist}}^{\D2\tbar\D8}(\{\mu_{\alpha}\}_{\alpha=1}^{N_{\four}})=\mathcal{I}_{\text{twist}}^{\D2\tbar\D8}(\mu).
\eea
For $(r_{1},r_{2},r_{3})=(1,1,0)$, it admits a plethystic expression (see Appendix~\ref{app:PE-local}):
\bea
\mathcal{I}_{\text{twist}}^{\D2\tbar\D8}(\{\mu_{\alpha}\})=\mathsf{PT4}\left[\mathcal{O}(-1)\oplus\mathcal{O}(-1)\oplus\mathcal{O}\rightarrow \mathbb{P}^{1};N_{\four}\right]&=\PE\left[\frac{Q[\mu]}{[q_{3}][\mu^{1/2}\fq][\mu^{1/2}\fq^{-1}]}\right].
\eea
Similarly, for $(r_{1},r_{2},r_{3})=(2,0,0)$, we have
\bea
\mathcal{I}_{\text{twist}}^{\D2\tbar\D8}(\{\mu_{\alpha}\})=\mathsf{PT4}\left[\mathcal{O}(-2)\oplus\mathcal{O}_{2}\oplus\mathcal{O}_{3}\rightarrow \mathbb{P}^{1};N_{\four}\right]&=\PE\left[\frac{[\mu][q_{2}q_{3}]}{[q_{2}][q_{3}]}\frac{Q}{[\mu^{1/2}\fq][\mu^{-1/2}\fq]}\right].
\eea
For both twists, substituting \eqref{eq:Dfunct-Rcharge} into \eqref{eq:D2D8-rankone-formula} gives the coefficient of $Q$ in the corresponding plethystic formula, with $\fq$ kept exact.

\paragraph{Relation with the D2--D6 system and tachyon condensation}
We now specialize the D2--D8 index to recover the D2--D6
systems discussed in the previous sections. We consider cases in which $r_s=0$ for some $s$ and place
D6-branes only on the corresponding stacks. The framing $R$-charges then agree in the two theories, and setting $\mu_{\alpha}=q_s=t_{s}$ identifies their
framing contributions:
\bea
\left.
\mathscr{D}(y_{\mathsf{a}} v_{\bar{s},\alpha},\mathfrak{m}_{\mathsf{a}})
\mathscr{D}\left(
\frac{\mu_{\alpha}}{y_{\mathsf{a}} v_{\bar{s},\alpha}},
-\mathfrak{m}_{\mathsf{a}}
\right)
\right|_{\mu_{\alpha}=q_s}
=
\mathscr{D}(y_{\mathsf{a}} v_{\bar{s},\alpha},\mathfrak{m}_{\mathsf{a}})
\mathscr{D}\left(
\frac{q_s}{y_{\mathsf{a}} v_{\bar{s},\alpha}},-\mathfrak{m}_{\mathsf{a}}
\right).
\eea
Geometrically, this specialization corresponds to tachyon
condensation of the D8--$\overline{\mathrm{D8}}$ pairs,
leaving D6-branes supported on the corresponding threefolds.
The PT$_{4}$ gluing formula therefore reduces to PT$_{3}$ gluing. For $r_s\neq0$, recovering the D2--D6 index also requires
changing the framing $R$-charges from $(1,1)$ to $(1,1-r_s)$.

We now consider the two $R$-charge assignments $(r_1,r_2,r_3)=(1,1,0)$ and $(2,0,0)$. For these two geometries, the generating function depends on
the framing masses only through their product $\mu$.
We can therefore use the single-pair expression with effective
mass $\mu$ to describe the following specializations.

For $(r_1,r_2,r_3)=(1,1,0)$, we take $N$ pairs with $\mu_\alpha=q_3=t_3$, so that $\mu=q_3^N$.
After tachyon condensation, the $N$ D8--$\overline{\D8}$ pairs
leave $N$ D6-branes wrapping
$\operatorname{Tot}(\mathcal{O}(-1)\oplus\mathcal{O}(-1)
\to\mathbb{P}^1)$.
The generating function then reduces to the rank-$N$
PT$_3$ partition function:
\bea
\mathcal{I}^{\D2\tbar\D8}_{\text{twist}}(\mu=q_{3}^{N})&=\PE\left[\frac{Q[q_{3}^{N}]}{[q_{3}][q_{3}^{N/2}\fq][q_{3}^{N/2}\fq^{-1}]}\right]\\
&=\mathsf{PT}[\mathcal{O}(-1)\oplus\mathcal{O}(-1)\rightarrow \mathbb{P}^{1};N]\\
&=\mathcal{I}^{(0,0,N)}_{\text{twist}}.
\eea

Similarly, for $(r_1,r_2,r_3)=(2,0,0)$, we take $M+N$ pairs, $M$ of them with $\mu_\alpha=q_2$ and $N$ of them with $\mu_\alpha=q_3$, so that
$\mu=q_2^M q_3^N$.
After tachyon condensation, these D8--$\overline{\D8}$ pairs
leave $M$ D6-branes wrapping
$\operatorname{Tot}(\mathcal{O}(-2)\oplus\mathcal{O}_3
\to\mathbb{P}^1)$ and $N$ D6-branes wrapping
$\operatorname{Tot}(\mathcal{O}(-2)\oplus\mathcal{O}_2
\to\mathbb{P}^1)$, where $\mathcal{O}_2$ and $\mathcal{O}_3$ denote the two
trivial line bundles associated with the second and third
fiber directions, respectively:
\bea
\mathcal{I}^{\D2\tbar\D8}_{\text{twist}}(\mu=q_{2}^{M}q_{3}^{N})&=\PE\left[\frac{[q_{2}^{M}q_{3}^{N}][q_{32}]}{[q_{3}][q_{2}]}\frac{Q}{[\fq q_{2}^{M/2}q_{3}^{N/2}][\fq q_{2}^{-M/2}q_{3}^{-N/2}]}\right]\\
&=\mathsf{PT}[\mathcal{O}(-2)\oplus \mathcal{O}_{2}\oplus \mathcal{O}_{3}\rightarrow \mathbb{P}^{1};(0,M,N)]\\
&=\mathcal{I}^{(0,M,N)}_{\text{twist}}.
\eea

\section{Conclusion and outlook}\label{sec:conclusion}
We have proposed that the normalized vortex contribution to the
half-index of the 3d $\mathcal{N}=2$ spiked quiver is the one-leg tetrahedron PT vertex, and checked this in examples. At a Higgs vacuum $\underline{\vec{\lambda}}$, the proposal reads 
\bea\label{eq:summary-chain}
\mathcal{Z}_{\text{vortex}}^{\underline{\vec{\lambda}}}
=
\mathsf{PT}_{\underline{\vec{\lambda}}}.
\eea
The Higgs vacuum specifies the asymptotic Young diagrams of the
vertex, and the vortex number counts the D0-brane charge relative to the prescribed leg. For parallel D6-branes, the right-hand side is the higher-rank one-leg PT vertex, and for a single D6-brane it is the ordinary one-leg PT vertex, where the identity is known \cite{Crew:2020psc}.

The factorization of the twisted index on
$\mathbb{S}^1\times\mathbb{S}^2$ connects this local correspondence to PT counting on local curves. The vortex factors at the two poles give the two vertices, and summing over their common Higgs vacuum implements gluing along the compact curve. The contribution at zero vortex number at both poles supplies the edge factor and the classical weight. The adjoint $R$-charges determine the degrees of the normal line bundles, while the fugacity for the D2 gauge
rank becomes the curve-counting parameter.

For intersecting D6-brane stacks, we use this gluing construction to formulate tetrahedron PT invariants of local curves. We test the resulting relations between twisted-index and PT generating functions in explicit calculations for several local curves and brane configurations. We also extend the comparison to the D2--D8 system, where the twisted index is related to the gluing of fourfold PT vertices. When $r_s=0$ for each nonempty D6-brane stack, the specialization $\mu_{\alpha}=t_s$ reduces the D2--D8 twisted index to the corresponding D2--D6 twisted index.

\paragraph{Boundary conditions}
Exceptional Dirichlet boundary conditions give a realization of
vacuum-labeled holomorphic blocks in 3d
$\mathcal{N}=4$ gauge theories
\cite{Bullimore:2016nji,Crew:2020psc,Bullimore:2020jdq,Crew:2023tky}.
Supersymmetric interfaces and enriched Neumann boundary conditions
give physical realizations of elliptic stable envelopes
\cite{Aganagic:2016aa,Dedushenko:2021mds,BullimoreZhang:2021Elliptic}.
The cigar half-index also relates vortex counting to
K-theoretic quasimap vertex functions \cite{Dedushenko:2023qjq}.

The construction for parallel D6-branes in
Appendix~\ref{app:rankNADHM_bcd} provides a starting point for
the spiked quiver. For intersecting stacks, it would be interesting
to classify the vacuum boundary conditions compatible with the
superpotential and boundary gauge anomaly cancellation
\cite{Dimofte:2017tpi}, and to determine which of them reproduce
the tetrahedron PT vortex blocks and the corresponding edge factors
after gluing. Elliptic stable envelopes for certain
$\mathcal{N}=2$ quivers and critical stable envelopes for quivers
with potentials suggest ways to investigate the dependence of
these boundary conditions on the mass chamber
\cite{Ishtiaque:2023acr,CaoOkounkovZhouZhou:2025Critical}.

\paragraph{Mirror symmetry}
In the $\mathcal{N}=4$ ADHM theory, mirror symmetry exchanges
the $A$- and $B$-twisted indices \cite{Crew:2020psc}.
Mirror dualities also occur in 3d $\mathcal{N}=2$
theories and can be tested at the level of holomorphic blocks
\cite{Aharony:1997bx,Beem:2012mb}.
It would be interesting to identify a mirror description of the
spiked quiver with intersecting D6-brane stacks and determine
how the mirror map acts on the vacuum boundary conditions and
the vortex fugacity. Since the $R$-charge assignments determine
the normal-bundle degrees in our gluing construction, a further
question is whether the mirror map on the background fields
has a geometric interpretation in tetrahedron PT theory.

\paragraph{Vortex origami and the three-leg PT vertex}
A single D6-brane wrapping a coordinate $\mathbb{C}^{3}$ admits
three types of D2-branes, one along each complex coordinate line
inside its worldvolume. With all three types present, the Higgs
vacua of the three D2 sectors specify the three asymptotic
partitions. The coupled vortex partition function is then
expected to reproduce the three-leg K-theoretic PT vertex
\cite{Pandharipande:2007sq,Kimura:2025lfo}.

More generally, one can include D6-brane stacks in all four
coordinate orientations in $\mathbb{C}^{4}$, with D2-branes
along the coordinate lines contained in their supports. The
resulting system of coupled vortex sectors is what we call
\emph{vortex origami}. It would be
interesting to determine the couplings at the intersections
and to identify the PT vertices computed by this system.

\paragraph{Gopakumar--Vafa-type invariants for tetrahedron PT theory}
For a Calabi--Yau threefold, the Gopakumar--Vafa (GV) formula expresses the topological string free energy in terms of the BPS indices of wrapped M2-branes through a one-loop Schwinger computation~\cite{Gopakumar:1998ii,Gopakumar:1998jq,Dedushenko:2014nya}. The relation between stable-pair invariants and BPS indices gives an enumerative counterpart to this description~\cite{pandharipande2010stable,Choi:2012jz}, while direct quantization of wrapped M2-branes provides an independent approach to the BPS spectrum~\cite{Dedushenko:2014nya,Huang:2026mwe}.

The plethystic formulas obtained in this work admit a related interpretation in terms of the membrane index~\cite{Nekrasov:2014nea}. For $N$ coincident parallel D6-branes, the transverse geometry near the stack is $\mathbb C^2/\mathbb Z_N$~\cite{Sen:1997KK}. For both local threefolds considered in \eqref{eq:conifoldPE-rankN} and \eqref{eq:Z2orbifoldPE-rankN}, the single-particle index of an M2-brane wrapping the zero section $\mathbb P^1$ reproduces the argument of the plethystic exponential. In the resolved-conifold case, the same rank dependence follows from the orbifold projection in Appendix~\ref{app:orbifold-topstring}.

For two intersecting stacks of ranks $(0,M,N)$ and twist $(r_1,r_2,r_3)=(2,0,0)$, the proposed fivefold is $T^*\mathbb P^1\times Y_{M,N}$, where $Y_{M,N}=\{uv=z_2^M z_3^N\}$ is the generalized conifold. The two D6-brane ranks enter through its defining equation. The index of an M2-brane wrapping the zero section of $T^*\mathbb P^1$ is $-Q(1-q_2q_3)\operatorname{Hilb}_T(Y_{M,N})$, whose plethystic exponential reproduces \eqref{eq:local-tetra-PE}. It would be interesting to determine whether the tetrahedron PT partition function admits a BPS expansion for more general ranks and twists, and how its single-particle contributions are organized by curve class and spin. Such an expansion would give a definition of GV-type invariants for tetrahedron PT theory and allow a comparison with the GV-type invariants of Calabi--Yau fourfolds~\cite{KlemmPandharipande:2007CY4,CaoMaulikToda:2019PTGV}.

\paragraph{D2-brane probes of toric Calabi--Yau threefolds}
A further direction is to consider D2-branes wrapping the
$\mathbb{C}$ factor in $X\times\mathbb{C}$ and probing a
general toric Calabi--Yau threefold $X$. Brane tilings and
the D2-brane realization of 3d
$\mathcal{N}=2$ quivers provide a starting point for
identifying the probe theory
\cite{Franco:2005rj,Aganagic:2009M2}.
One can then ask which boundary conditions give half-indices
whose vortex contributions compute PT vertices of $X$, and
how the quiver and its superpotential encode the gluing
along torus-invariant curves.

\acknowledgments
GN thanks Nicol\`o Piazzalunga, Yutong Huang, Qinjian Lou, Yi-Nan Wang, and Hao Zhang for useful discussions.
This work is supported by the Shanghai Municipal Science and Technology Major Project (No. 24ZR1403900). 
The work of TK is supported by EIPHI Graduate School (No.~ANR-17-EURE-0002) and the Bourgogne-Franche-Comté region.
The work of GN is supported by the National Natural Science Foundation of China under Grant No. 12422503. The work of JJ is supported by the Institute for Basic Science under the project IBS-R003-Y3.

\paragraph{Statement on the use of AI} We acknowledge the use of OpenAI Codex, Google Gemini and Anthropic Claude Code for assistance with
drafting, editing, consistency checks, and manuscript organization. The authors checked all results and take full responsibility for the content.

\appendix
\section{Notation and definitions}\label{app:notations}
We summarize the notation and definitions used throughout the paper. The Young-diagram conventions apply both to the vacuum labels of Section~\ref{sec:halfindex-def} and to the asymptotic data of Section~\ref{sec:DTPT-vortexQM}.

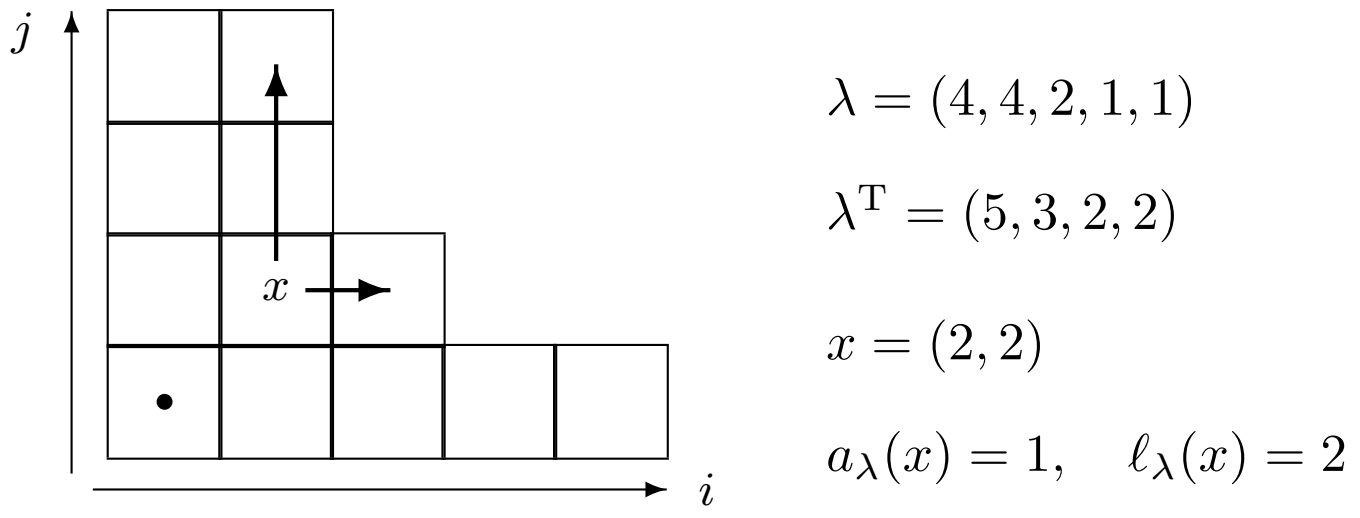
\begin{figure}[htbp]
\centering
\begingroup
\setlength{\unitlength}{1pt}
\begin{picture}(312,109)
  \thinlines
  \multiput(23,16)(21,0){5}{\framebox(21,21){}}
  \multiput(23,37)(21,0){3}{\framebox(21,21){}}
  \multiput(23,58)(21,0){2}{\framebox(21,21){}}
  \multiput(23,79)(21,0){2}{\framebox(21,21){}}
  \put(20,10){\vector(1,0){108}}
  \put(134,10){\makebox(0,0)[l]{\small \(i\)}}
  \put(16,13){\vector(0,1){87}}
  \put(9,96){\makebox(0,0)[r]{\small \(j\)}}
  \put(33.5,26.5){\circle*{3.2}}
  \put(54.5,47.5){\makebox(0,0){\small \(x\)}}
  \thicklines
  \put(60,47.5){\vector(1,0){16}}
  \put(54.5,53){\vector(0,1){37}}
  \put(158,83){\makebox(0,0)[l]{\(\lambda=(4,4,2,1,1)\)}}
  \put(158,62){\makebox(0,0)[l]{\(\lambda^\rmT=(5,3,2,2)\)}}
  \put(158,37){\makebox(0,0)[l]{\(x=(2,2)\)}}
  \put(158,16){\makebox(0,0)[l]{\(a_\lambda(x)=1,\quad\ell_\lambda(x)=2\)}}
\end{picture}
\endgroup
\caption{The dot marks \((1,1)\). The arrows from \(x\) run along its arm and leg.}
\label{fig:adhm-young-convention}
\end{figure}

\paragraph{Young diagrams}
For each ordered plane \(A=ab\), with \(ab=12,23,31\), we draw the \(a\)-axis to the right and the \(b\)-axis upwards (see Figure~\ref{fig:adhm-young-convention}). A box \(x=(i,j)\) has indices starting at \((1,1)\) in the lower-left corner. A step in the \(i\)-direction acts by \(\Phi_a\), and a step in the \(j\)-direction acts by \(\Phi_b\). Thus the box vector is \(\Phi_a^{i-1}\Phi_b^{j-1}p_{A4}e_{A}^{(\alpha)}\), with inverse gauge weight \(s_x=v_{A4,\alpha}q_a^{1-i}q_b^{1-j}\) before the common half-index shift.

 We identify the partition with its set of boxes by
\bea
\lambda&=(\lambda_1\geq\lambda_2\geq\cdots\geq0),\qquad
(i,j)\in\lambda\quad\Longleftrightarrow\quad i,j\geq1,\quad j\leq\lambda_i .
\eea
Only finitely many parts are nonzero. We do not distinguish a partition from its Young diagram.

The transpose has parts \(\lambda_j^\rmT=\#\{i:\lambda_i\geq j\}\), so \(\lambda_j^\rmT\) is the length of row \(j\) and \((i,j)\in\lambda\) is equivalent to \((j,i)\in\lambda^\rmT\). The size of the diagram is
\bea
|\lambda|&=\sum_{i\geq1}\lambda_i=\sum_{j\geq1}\lambda_j^\rmT .
\eea

For \(x=(i,j)\in\lambda\), the arm counts boxes strictly to the right and the leg counts boxes strictly above \(x\):
\[
a_\lambda(i,j)=\lambda_j^\rmT-i,\qquad
\ell_\lambda(i,j)=\lambda_i-j .
\]
The rightmost box of each nonempty row has zero arm. These are precisely the boxes in \((\lambda)_{\mathrm B}=\{(i,j)\in\lambda:i=\lambda_j^\rmT\}\) used in the boundary conditions. For a box of another diagram we use the same formulas, extending the parts of \(\lambda\) and \(\lambda^\rmT\) by zero. The resulting relative arm and leg lengths in a mixed framing block can be negative.

For framings of a single orientation \(A=ab\), we write \(\vec{\lambda}\), or \(\vec{\lambda}_A\) when specifying the plane. With \(N_{A4}\) framings of this orientation,
\bea
\vec{\lambda}_A&=(\lambda_A^{(1)},\ldots,\lambda_A^{(N_{A4})}),\qquad
|\vec{\lambda}_A|=\sum_{\alpha=1}^{N_{A4}}|\lambda_A^{(\alpha)}| .
\eea
For framings of different orientations, an underline denotes the collection of these tuples. In the cyclic order of the planes,
\bea
\underline{\vec{\lambda}}&=(\vec{\lambda}_{23},\vec{\lambda}_{31},\vec{\lambda}_{12}),\qquad
|\underline{\vec{\lambda}}|=\sum_{A\in\{12,23,31\}}|\vec{\lambda}_A|=N_4 .
\eea
A plane without D6-branes contributes the empty tuple.

\section{Boundary conditions for rank \texorpdfstring{$N$}{N} 3d \texorpdfstring{$\mathcal{N}=4$}{N=4} ADHM theory}\label{app:rankNADHM_bcd}
The half-index in Section~\ref{sec:halfindex-def} is computed with the boundary condition \eqref{eq:boundary-conditions}. This boundary condition is motivated by the exceptional Dirichlet boundary condition of Crew--Dorey--Zhang \cite{Crew:2020psc}, but it is not itself an exceptional Dirichlet boundary condition. In terms of the criteria of Appendix~\ref{app:rankN-bc-criteria}, an exceptional Dirichlet boundary condition satisfies criteria 1--4, whereas \eqref{eq:boundary-conditions} satisfies criteria 1, 2, 5, and 6. The failure of criterion~3 is the issue noted in the footnote of Section~\ref{sec:halfindex-def}: with several framings, the boundary locus of \eqref{eq:boundary-conditions} can contain directions along the complex gauge orbit of the vacuum, even when all D6-branes are parallel. One can also confirm that \eqref{eq:boundary-conditions} do not satisfy the criterion~4.

In this appendix, we remove these directions for the \((L,M,N)=(0,0,N)\) theory, i.e.\ the rank-\(N\) \(3\mathrm d\ \mathcal N=4\) ADHM theory with gauge group \(U(N_4)\) on \(N\) parallel \(\mathrm{D6}_{124}\)-branes, whose twisted index is studied in Section~\ref{sec:rankone-twistedindex}. We do so by adding tree constraints in the matrix blocks that connect different framings. The result is a rank-\(N\) generalization of the exceptional Dirichlet boundary condition of \cite{Crew:2020psc}: it satisfies criteria 1--4, and we show that it also satisfies criteria 5 and 6. The analogous construction for intersecting D6-branes is left for future work (see Section~\ref{sec:conclusion}).

The main results of this appendix are as follows.
\begin{itemize}
\item The normalized vortex factor is the same as for \eqref{eq:boundary-conditions}, namely \eqref{eq:exceptional-vortex-triality} with \(\underline{\vec\lambda}=\vec\lambda_{12}\). Hence the vortex--PT proposal \eqref{eq:higher-rank-vortex-PT-proposal} applies unchanged (Appendix~\ref{app:rankN-bc-halfindex}).
\item The boundary locus contains no complex gauge directions, and its image in the Higgs branch has the attracting tangent space \eqref{eq:parallel-d6-attracting-tangent} at the vacuum (Appendix~\ref{app:rankN-bc-tangent}).
\item The zero-flux factors of the two hemispheres glue to the rank-\(N\) PT edge factor \eqref{eq:rankN-PT-edge} up to the sign \((-1)^{r_1N_4}\) (Appendix~\ref{app:rankN-bc-edge}). This justifies the identification \(\mathcal I_{\mathrm{edge}}=\mathsf E_{\bar 3;\vec\lambda_{12}}\) used in Section~\ref{sec:rankone-twistedindex}.
\end{itemize}

\subsection{Criteria for the boundary condition}\label{app:rankN-bc-criteria}
We consider boundary conditions associated with a torus-fixed Higgs vacuum, and we judge them by the following six criteria. The first four concern supersymmetry and the local geometry at the chosen vacuum; they are the properties of an exceptional Dirichlet boundary condition \cite{Crew:2020psc}. The last two test the proposed relation to PT counting. For each criterion, the subsection in which it is verified is given in parentheses.

\begin{enumerate}
    \item \textbf{Compatibility with the vacuum.}
The boundary condition should admit the chosen fixed-point configuration and preserve the combined torus action that fixes it, so that the half-index can be graded by the corresponding fugacities. (Appendix~\ref{app:rankN-bc-def}.)

    \item \textbf{Boundary superpotential.}
The bulk superpotential should vanish on the allowed boundary field configurations, and so should its variation along the allowed boundary directions. Then the superpotential gives no boundary contribution under the preserved supersymmetry \cite{Yoshida:2014ssa}. (Appendix~\ref{app:rankN-bc-def}.)

    \item \textbf{Absence of residual complex gauge directions.}
At the fixed point, the tangent space of the allowed boundary locus should intersect the tangent space of the complex gauge orbit trivially. Otherwise, gauge-orbit directions would appear as additional boundary modes \cite{Crew:2020psc}. (Appendix~\ref{app:rankN-bc-tangent}.)

    \item \textbf{Attracting tangent space.}
After imposing the bulk vacuum equations and projecting to the Higgs branch, the allowed boundary deformations should give the attracting tangent space in the chosen mass chamber. This geometric criterion comes from the relation between exceptional Dirichlet boundary conditions and attracting sets \cite{Crew:2020psc,Ishtiaque:2023acr}. (Appendix~\ref{app:rankN-bc-tangent}.)

    \item \textbf{Vortex factor.}
After division by the zero-flux contribution, the half-index should reproduce the PT vertex with the same asymptotic data, as proposed in \eqref{eq:higher-rank-vortex-PT-proposal}. The sign of each flux sector in this comparison is fixed by the chosen fugacity convention. (Appendix~\ref{app:rankN-bc-halfindex}.)

    \item \textbf{Glued edge factor.}
Gluing the zero-flux factors at the two ends of a compact edge should reproduce the PT edge contribution, as in the block factorization \eqref{eq:rankN-block-decomp}. The relative sign is tracked in the chosen orientation convention. (Appendix~\ref{app:rankN-bc-edge}.)
\end{enumerate}

\subsection{Fixed points and equivariant weights}\label{app:rankN-bc-fixed}
For \((L,M,N)=(0,0,N)\), only the plane \(A=12\) is occupied, with transverse direction \(c=3\). The fields of the spiked quiver reduce to the ADHM data \((\Phi_1,\Phi_2,p_3,\tilde p_3)\) together with the adjoint \(\Phi_3\), where \(p_3\equiv p_{124}\) as in \eqref{eq:spiked-Adictionary}. The superpotential is \(W=\operatorname{Tr}(\Phi_3\mu_{\mathbb C})\) with the complex moment map
\bea
\mu_{\mathbb C}=[\Phi_1,\Phi_2]+p_{3}\tilde{p}_{3}.
\eea
For a positive real FI parameter, the Higgs branch \(\mathcal M_H\) is the moduli space of rank-\(N\) framed instantons of charge \(N_4\) on \(\mathbb C^2_{12}\). By Section~\ref{sec:3dADHMflavor}, its torus fixed points are labeled by \(\vec\lambda_{12}=(\lambda_{12}^{(1)},\ldots,\lambda_{12}^{(N)})\) with \(|\vec\lambda_{12}|=N_4\).

We use the following notation, which specializes \eqref{eq:spiked-Kbasis} to this case. Let \(V=\bigoplus_\alpha V_\alpha\) be the gauge space, with basis \(e_{\alpha;i,j}\) for \((i,j)\in\lambda_{12}^{(\alpha)}\), and let \(W_\alpha=\mathbb C f_\alpha\) be the framing line of the \(\alpha\)-th D6-brane. A box is written as \(x=(\alpha;i_x,j_x)\); it is the box \((12,\alpha;i_x,j_x)\) of Section~\ref{sec:halfindex-def}. When we want to display the framing label, we write \(e_{\beta,x}\) for the basis vector of a box \(x\in\lambda_{12}^{(\beta)}\). By \eqref{eq:spiked-fixedpoint-components} and \eqref{eq:spiked-fixedpoint-framing}, the fixed-point configuration is
\bea
\begin{aligned}
 \Phi_1 e_{\alpha;i,j}&=e_{\alpha;i+1,j},&
 \Phi_2 e_{\alpha;i,j}&=e_{\alpha;i,j+1},\\
 p_{{3}}f_\alpha&=e_{\alpha;1,1},&
 \tilde{p}_{{3}}&=\Phi_3=0 ,
\end{aligned}
 \label{eq:parallel-d6-fixed-point}
\eea
where a basis vector outside the diagram is understood to be zero. In particular, the fixed-point data are block diagonal with respect to the framing labels.

We draw increasing \(i\) to the right and increasing \(j\) upwards, as in Appendix~\ref{app:notations}. The weights of \eqref{eq:boundary-fugacities} become
\bea
 s_x&=v_{\bar{3},\alpha}q_1^{1-i_x}q_2^{1-j_x},&
 \tilde s_x&=q_4^{1/2}s_x,&
 \mathbf K&=\sum_x\tilde s_x,&
 \mathbf N_{124}&=q_4^{1/2}\sum_{\alpha=1}^N v_{\bar{3},\alpha}.
 \label{eq:parallel-d6-weights}
\eea
As in Section~\ref{sec:halfindex-def}, a step in direction \(d\) multiplies the inverse gauge fugacity \(y_x\), which equals \(\tilde s_x\) at zero flux, by \(q_d^{-1}\). We also write \(q_{12}=q_1q_2\).

For matrix blocks, we use the convention that the block \(\alpha\beta\) of a map has source \(\beta\) and target \(\alpha\). Thus \(\Phi_d^{\alpha\beta}:V_\beta\to V_\alpha\), \(p_3^{\alpha\beta}:W_\beta\to V_\alpha\), and \(\tilde p_3^{\alpha\beta}:V_\beta\to W_\alpha\). The hypermultiplets form the symplectic pairs
\bea
\bigl((\Phi_1)_{xy},(\Phi_2)_{yx}\bigr),\qquad
\bigl((p_{3})_{x\alpha},(\tilde{p}_{3})_{\alpha x}\bigr).
\label{eq:parallel-d6-pairs}
\eea
Note that the symplectic form pairs the block \(\alpha\beta\) with the block \(\beta\alpha\).

\subsection{The boundary condition}\label{app:rankN-bc-def}
We write \(\mathrm N\) for Neumann and \(\mathrm D\) for Dirichlet. A Dirichlet component is fixed to its value in \eqref{eq:parallel-d6-fixed-point}, and we write \(\mathrm D_0\) when this value is zero.

\paragraph{Vector multiplet and hypermultiplets}
We impose the \(\mathcal N=(2,2)\) Dirichlet boundary condition on the \(3\mathrm d\ \mathcal N=4\) vector multiplet. In \(\mathcal N=2\) notation, the vector multiplet is Dirichlet and \(\Phi_3=0\) at the boundary \cite{Crew:2023tky}. As in Section~\ref{sec:halfindex-def}, the gauge symmetry then becomes a boundary global symmetry. We choose its background so that the nonzero Dirichlet values fixed below, namely the entries of \(p_3\) and the adjoints that equal one in \eqref{eq:parallel-d6-fixed-point}, are invariant. This sets the gauge fugacities to \(y_x=\tilde s_x\), as in \eqref{eq:boundary-fugacities}.

Each hypermultiplet pair in \eqref{eq:parallel-d6-pairs} has one Dirichlet component and one Neumann component. We give the assignments for blocks within and between framings below.

\paragraph{Same framing}
Within each block \(\alpha\alpha\), we use the prescription of \cite{Crew:2020psc}, which is also the same-framing part of \eqref{eq:boundary-conditions}. Recall the set \((\lambda)_{\mathrm B}=\{(i,j)\in\lambda:i=\lambda^\rmT_j\}\) of Appendix~\ref{app:notations}. In our drawing, these are the rightmost boxes of the rows. (They are the bottom boxes in the convention of \cite{Crew:2020psc}, where \(i\) increases downwards.) For boxes \(x,y\) in the same diagram, the condition \(\mathcal C(x,y)\) of Section~\ref{sec:halfindex-def} holds if \(y\) is not the rightmost box of its row, or if it is the rightmost box and \(i_x>i_y\). The pair \(((\Phi_1)_{xy},(\Phi_2)_{yx})\) is assigned \((\mathrm D,\mathrm N)\) if \(\mathcal C(x,y)\) holds and \((\mathrm N,\mathrm D)\) otherwise. The framing pair \((p_3^{\alpha\alpha},\tilde p_3^{\alpha\alpha})\) is assigned \((\mathrm D,\mathrm N)\).

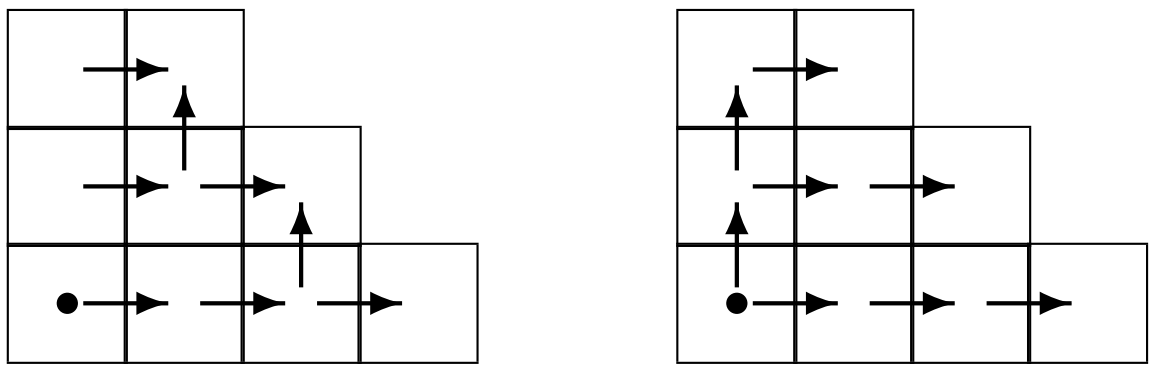
\begin{figure}[tbp]
\centering
\begingroup
\setlength{\unitlength}{1pt}
\def\adhmgrid{\thinlines
  \multiput(8,8)(22,0){4}{\framebox(22,22){}}
  \multiput(8,30)(22,0){3}{\framebox(22,22){}}
  \multiput(8,52)(22,0){2}{\framebox(22,22){}}
  \put(19,19){\circle*{4}}
  \thicklines
  \multiput(22,19)(22,0){3}{\vector(1,0){16}}
  \multiput(22,41)(22,0){2}{\vector(1,0){16}}
  \put(22,63){\vector(1,0){16}}}
\begin{picture}(232,95)
  \put(4,15){\begin{picture}(100,76)\adhmgrid
    \put(63,22){\vector(0,1){16}}\put(41,44){\vector(0,1){16}}
  \end{picture}}
  \put(130,15){\begin{picture}(100,76)\adhmgrid
    \multiput(19,22)(0,22){2}{\vector(0,1){16}}
  \end{picture}}
  \put(55,5){\makebox(0,0){\small Same framing: \(T_\alpha\)}}
  \put(181,5){\makebox(0,0){\small Mixed block: \(S_\beta\)}}
\end{picture}
\endgroup
\caption{The two trees used in the boundary condition, drawn for \(\lambda=(3,3,2,1)\). Horizontal and vertical arrows correspond to \(\Phi_1\) and \(\Phi_2\), and the dot is \((1,1)\). Left: the tree \(T_\alpha\) of \cite{Crew:2020psc}; an arrow marks a same-framing entry fixed to one. Right: the directed tree \(S_\beta\) used in the mixed blocks. An arrow \(x\xrightarrow{d}y\) selects the entire column \(\Phi_d^{\alpha\beta}e_{\beta,x}\) to be fixed to zero for \(\alpha>\beta\).}
\label{fig:parallel-d6-tree}
\end{figure}

In addition, let \(T_\alpha\) be the spanning tree of \(\lambda_{12}^{(\alpha)}\) formed by all horizontal \(\Phi_1\) edges together with the vertical \(\Phi_2\) edges that end on the rightmost box of a row (left of Figure~\ref{fig:parallel-d6-tree}). Exactly as in Section~\ref{sec:halfindex-def}, the adjoint entries along \(T_\alpha\) are fixed to their fixed-point value one, overriding the assignment above.

\paragraph{Different framings: initial assignment}
Between distinct framings, the assignment depends on a choice of mass chamber. We choose a chamber in which every mixed variation in a block \(\alpha\beta\) with \(\alpha>\beta\) has positive weight under the compensated torus, i.e.\ the equivariant torus combined with the gauge transformation that keeps \eqref{eq:parallel-d6-fixed-point} fixed. Concretely, on the mass torus we take
\bea
q_1=t^{-1},\qquad q_2=t,\qquad v_{\bar 3,\alpha}=t^{-K\alpha},\qquad t\to0,
\label{eq:parallel-d6-chamber}
\eea
with \(K\) large, for example \(K=4N_4+4\), and regard positive powers of \(t\) as attracting. Every weight in the block \(\alpha\beta\) contains the framing ratio \(v_{\bar3,\beta}/v_{\bar3,\alpha}=t^{K(\alpha-\beta)}\), so for this choice of \(K\) the framing gaps dominate the box-dependent shifts. Within each diagram, this is the chamber used in \cite{Crew:2020psc}. The cocharacter \eqref{eq:parallel-d6-chamber} only selects the chamber; the index itself is kept fully equivariant.

The initial assignment makes the attracting (positive) mixed blocks Neumann and the repelling ones Dirichlet. Together with the same-framing prescription above, it reads
\bea
\begin{array}{c|l|c}
\text{component or pair}&\text{condition}&\text{boundary condition}\\
\hline
U(N_4)\text{ vector}&\text{all components}&\mathrm D\\
\Phi_3&\text{all components}&\mathrm D_0\\
\bigl((\Phi_1)_{xy},(\Phi_2)_{yx}\bigr)
 &x,y\text{ in the same diagram and }\mathcal C(x,y)
 &(\mathrm D,\mathrm N)\\
\bigl((\Phi_1)_{xy},(\Phi_2)_{yx}\bigr)
 &x,y\text{ in the same diagram and }\neg\mathcal C(x,y)
 &(\mathrm N,\mathrm D)\\
\bigl(p_{{3}}^{\alpha\alpha},\tilde{p}_{{3}}^{\alpha\alpha}\bigr)
 &\text{same framing}&(\mathrm D,\mathrm N)\\
 \Phi_d^{\alpha\beta},\ p_{{3}}^{\alpha\beta},\ \tilde{p}_{{3}}^{\alpha\beta}
 &\alpha>\beta&\mathrm N\\
 \Phi_d^{\beta\alpha},\ p_{{3}}^{\beta\alpha},\ \tilde{p}_{{3}}^{\beta\alpha}
 &\alpha>\beta&\mathrm D_0
\end{array}
\label{eq:parallel-d6-bc}
\eea
where \(d=1,2\) in the last two rows. Since the symplectic pairing exchanges the blocks \(\alpha\beta\) and \(\beta\alpha\), each pair still contains exactly one Neumann component.

\paragraph{Different framings: tree constraints}
The initial assignment still violates criterion~3. For \(\alpha>\beta\), an infinitesimal complex gauge transformation \(\theta_{\alpha\beta}\in\operatorname{Hom}(V_\beta,V_\alpha)\) moves the fixed point only within the Neumann blocks \(\Phi_d^{\alpha\beta}\) and \(p_3^{\alpha\beta}\), so these gauge directions lie inside the boundary locus. To remove these directions, we freeze exactly \(\dim\operatorname{Hom}(V_\beta,V_\alpha)\) coordinates of the positive block, chosen along a tree.

For each nonempty diagram \(\lambda_{12}^{(\beta)}\), let \(S_\beta\) be the directed tree rooted at \((1,1)\) that contains every horizontal edge and every vertical edge on the first column \(i=1\) (right of Figure~\ref{fig:parallel-d6-tree}). In other words, the parent of \((i,j)\) is \((i-1,j)\) if \(i>1\), and \((1,j-1)\) if \(i=1\) and \(j>1\), so every box other than the root has exactly one incoming edge. For \(\alpha>\beta\) with \(\lambda_{12}^{(\beta)}\neq\varnothing\), we change
\bea
 p_{{3}}^{\alpha\beta},\qquad \Phi_d^{\alpha\beta}e_{\beta,x}
 \quad\text{for every edge }x\xrightarrow{d}y\text{ of }S_\beta
 \label{eq:parallel-d6-tree-change}
\eea
from \(\mathrm N\) to \(\mathrm D_0\), and their symplectic partners from \(\mathrm D_0\) to \(\mathrm N\). Here \(\Phi_d^{\alpha\beta}e_{\beta,x}\) denotes the whole column of \(\Phi_d^{\alpha\beta}\) with source \(x\); its partner is the row of \(\Phi_{3-d}^{\beta\alpha}\) with target \(x\). In particular, \(\tilde{p}_{{3}}^{\beta\alpha}\) becomes \(\mathrm N\). The frozen coordinates form \(|\lambda_{12}^{(\beta)}|\) columns in \(V_\alpha\), one for the root and one for each of the \(|\lambda_{12}^{(\beta)}|-1\) edges, so their number is \(|\lambda_{12}^{(\alpha)}||\lambda_{12}^{(\beta)}|=\dim\operatorname{Hom}(V_\beta,V_\alpha)\). When the source diagram is empty, the block has no gauge directions and nothing is changed.

\paragraph{Comparison with \eqref{eq:boundary-conditions}}
For parallel D6-branes, the final assignment agrees with \eqref{eq:boundary-conditions} except in the mixed blocks, where the two differ by exchanging \(\mathrm N\) and \(\mathrm D\) within whole symplectic pairs. The exchanged pairs are the tree columns of \eqref{eq:parallel-d6-tree-change} with their dual rows and, when \(\lambda_{12}^{(\beta)}=\varnothing\), the framing pair \((p_3^{\alpha\beta},\tilde p_3^{\beta\alpha})\).

\paragraph{Criteria 1 and 2}
All frozen mixed entries vanish at the fixed point, since the ADHM data \eqref{eq:parallel-d6-fixed-point} are block diagonal. The nonzero same-framing entries along \(T_\alpha\) and the root values of \(p_3^{\alpha\alpha}\) have weight one under the compensated torus. Hence the fixed point satisfies the boundary condition, and all coordinate boundary subspaces are invariant under this torus. Since the boundary condition fixes \(\Phi_3=0\), the superpotential \(W=\operatorname{Tr}(\Phi_3\mu_{\mathbb C})\) vanishes on the boundary. Its variation along the allowed boundary directions also vanishes, since \(\delta\Phi_3=0\) and the remaining term \(\operatorname{Tr}(\Phi_3\,\delta\mu_{\mathbb C})\) is zero.

\subsection{Half-index and vortex factor}\label{app:rankN-bc-halfindex}
Let \(\mathscr N\) and \(\mathscr D\) be the sets of chiral components that are Neumann and Dirichlet, respectively, in the final assignment. The half-index is then \eqref{eq:exceptional-half-index} with \(\underline{\vec\lambda}=\vec\lambda_{12}\) and these sets. For a flux \(k=(k_x)\in\mathbb Z^{N_4}\), the flux-shifted weights of the table in Section~\ref{sec:halfindex-def} become
\bea
\begin{array}{c|c}
\chi&w_{\chi,k}\\ \hline
 (\Phi_d)_{yx} & q_d^{-1}q_4^{k_x-k_y}s_x/s_y\\
 (p_{3})_{x\alpha}& q_4^{-k_x}v_{\bar{3},\alpha}/s_x\\
 (\tilde{p}_{3})_{\alpha x}&q_{12}^{-1}q_4^{k_x}s_x/v_{\bar{3},\alpha}
\end{array}
\eea
where \(d=1,2,3\). With \(\xi=\log\fq/\log q_4\), the half-index reads
\bea
\begin{aligned}
 \widetilde{\mathcal Z}_{\mathbb S^1\times D}^{\vec\lambda_{12}}
 ={}&\left(\prod_x\tilde s_x^{\,\xi}\right)
 \sum_{k\in\mathbb Z^{N_4}}\fq^{\sum_x k_x}
 \prod_{x,y}\frac{1}{(q_4^{1+k_x-k_y}s_x/s_y;q_4)'_\infty}
 \\
 &\times
 \prod_{\chi\in\mathscr N}\frac{1}{(w_{\chi,k};q_4)'_\infty}
 \prod_{\chi\in\mathscr D}(q_4 w_{\chi,k}^{-1};q_4)'_\infty .
\end{aligned}
 \label{eq:parallel-d6-half-index}
\eea

\paragraph{The vortex factor is unchanged}
We now check criterion~5. Let \(d_\chi\) be the flux shift of \(\chi\), defined by \(w_{\chi,k}=q_4^{d_\chi}w_\chi\) as in \eqref{eq:exceptional-vortex}. By \eqref{eq:exceptional-vortex}, a Neumann component contributes \((w_\chi;q_4)'_{d_\chi}\) to the normalized vortex factor, and a Dirichlet component contributes \(1/(q_4w_\chi^{-1};q_4)'_{-d_\chi}\). Using \eqref{eq:symm-q-Pochhammer}, one finds
\bea
\frac{1}{(q_4w^{-1};q_4)'_{-d}}=(-1)^{d}\,(w;q_4)'_{d},\qquad d\in\mathbb Z .
\label{eq:parallel-d6-ND-sign}
\eea
Thus switching a single component \(\chi\) between \(\mathrm N\) and \(\mathrm D\) multiplies its normalized contribution by \((-1)^{d_\chi}\). The two members of a symplectic pair \eqref{eq:parallel-d6-pairs} have opposite flux shifts, because their weights multiply to \(q_{12}^{-1}\) in every flux sector. Exchanging \(\mathrm N\) and \(\mathrm D\) within a whole pair therefore produces no sign. Since the present assignment differs from \eqref{eq:boundary-conditions} only by such exchanges (Appendix~\ref{app:rankN-bc-def}), the two give the same normalized vortex factor, namely \eqref{eq:exceptional-vortex-triality} with only the plane \(A=12\) occupied and \(q_{A4}=q_3^{-1}\):
\bea
\mathcal Z_{\mathrm{vortex}}^{\vec\lambda_{12}}
&=\sum_{k\in\mathbb Z^{N_4}}\fq^{\sum_x k_x}
\prod_{x,\alpha}\frac{(v_{\bar3,\alpha}/s_x;q_4)'_{-k_x}}{(q_3^{-1}v_{\bar3,\alpha}/s_x;q_4)'_{-k_x}}
\prod_{x\ne y}\frac{\prod_{d=1}^3(q_d^{-1}s_x/s_y;q_4)'_{k_x-k_y}}{(s_x/s_y;q_4)'_{k_x-k_y}} .
\label{eq:parallel-d6-vortex}
\eea
The products are evaluated with the fixed-point prescription of Section~\ref{sec:halfindex-def}, and by \eqref{eq:colored-rpp-condition} the flux sum is supported on reverse plane partitions with bases \(\vec\lambda_{12}\). Under the vortex--PT identification \eqref{eq:higher-rank-vortex-PT-proposal}, the sum gives the rank-\(N\) one-leg PT series \(\mathsf{PT}_{\vec\lambda_{12}}\).

\subsection{Gauge directions and the attracting tangent space}\label{app:rankN-bc-tangent}
We now check criteria~3 and~4. Let \(L_\partial\) be the space of Neumann variations of the hypermultiplets at the fixed point, with character
\bea
\mathcal L_\partial=\operatorname{ch}L_\partial =\sum_{\chi\in\mathscr N}w_{\chi,0}.
\eea
Each symplectic pair contains exactly one Neumann component, so \(L_\partial\) is Lagrangian. We write \(d_0\) for the infinitesimal complex gauge action and \(d_1=d\mu_{\mathbb C}\) for the linearized complex moment map, so that \(d_1d_0=0\). We treat the same-framing blocks, the positive mixed blocks \(\alpha\beta\) with \(\alpha>\beta\), and the reverse mixed blocks \(\beta\alpha\) in turn.

\paragraph{Same-framing blocks}
In each block \(\alpha\alpha\), the assignment, including all its Dirichlet entries in \eqref{eq:parallel-d6-bc}, is that of \cite{Crew:2020psc}. The corresponding gauge-orbit calculation is given in \cite[Appendix~D.2, eqs.~(D.28)--(D.38)]{Crew:2020psc}.

\paragraph{Positive mixed blocks}
Let \(\alpha>\beta\) with \(\lambda_{12}^{(\beta)}\neq\varnothing\), and let \(\theta_{\alpha\beta}:V_\beta\to V_\alpha\) be an infinitesimal complex gauge transformation. Since the fixed point is block diagonal, the variations it induces in the frozen coordinates \eqref{eq:parallel-d6-tree-change} are
\bea
 \delta p_{3}^{\alpha\beta}
 &=\theta_{\alpha\beta}e_{\beta;1,1},&
 (\delta\Phi_d^{\alpha\beta})e_{\beta,x}
 &=\theta_{\alpha\beta}e_{\beta,y}
   -\Phi_d^{\alpha\alpha}\theta_{\alpha\beta}e_{\beta,x},
 \qquad x\xrightarrow{d}y\text{ in }S_\beta .
 \label{eq:parallel-d6-tree-recursion}
\eea
We claim that this linear map from \(\operatorname{Hom}(V_\beta,V_\alpha)\) to the frozen coordinates is an isomorphism. The two spaces have the same dimension by the count in Appendix~\ref{app:rankN-bc-def}, and the inverse can be written down explicitly. Let \(h_y=\theta_{\alpha\beta}e_{\beta,y}\) denote the columns of \(\theta_{\alpha\beta}\). Given arbitrary values \(u_0\in V_\alpha\) for \(\delta p_3^{\alpha\beta}\) and \(u_y\in V_\alpha\) for the frozen column on the edge ending at each nonroot box \(y\), set \(h_{(1,1)}=u_0\) and, along the unique incoming edge \(x\xrightarrow{d}y\) of \(S_\beta\),
\[
h_y=u_y+\Phi_d^{\alpha\alpha}h_x .
\]
Because \(S_\beta\) is a directed tree rooted at \((1,1)\), this recursion determines exactly one column for every box of \(\lambda_{12}^{(\beta)}\), even when \(\Phi_d^{\alpha\alpha}\) has a kernel. Hence the projection of the infinitesimal orbit map to the frozen coordinates is an isomorphism. In particular, vanishing frozen variations imply \(\theta_{\alpha\beta}=0\), so no gauge direction of this block survives in the boundary locus.

The shape of the tree matters: each nonroot box must be the target of exactly one frozen column. This is visible already for a \(2\times2\) source diagram and a one-box target diagram. There \(\Phi_1^{\alpha\alpha}=\Phi_2^{\alpha\alpha}=0\), and the recursion reduces to \(h_y=u_y\). The vertical edge of \(S_\beta\) fixes \(h_{1,2}\), and its two horizontal edges fix \(h_{2,1}\) and \(h_{2,2}\). Had we used the tree \(T_\beta\) instead, whose vertical edge ends at \((2,2)\), two constraints would fix \(h_{2,2}\) and \(h_{1,2}\) would remain free. This is why the mixed blocks use \(S_\beta\) rather than the same-framing tree \(T_\beta\).

\paragraph{Reverse mixed blocks}
Let \(H_+\) and \(H_-\) be the spaces of hypermultiplet variations in the blocks \(\alpha\beta\) and \(\beta\alpha\), respectively. Let \(L_+\subset H_+\) be the kernel of the projection to the frozen coordinates, i.e.\ the allowed variations in the positive block. The isomorphism just proved gives the decomposition
\[
 H_+=d_0\operatorname{Hom}(V_\beta,V_\alpha)\oplus L_+ .
\]
Since \(L_\partial\) is Lagrangian, the symplectic pairing identifies the allowed reverse variations \(L_-\subset H_-\) with the annihilator of \(L_+\). For \(z\in L_-\), the quantities \(\operatorname{Tr}(\theta\,d_1z)\) are, up to a fixed sign, the pairings \(\omega(d_0\theta,z)\). Since \(L_-\) annihilates \(L_+\), its pairing with the first summand \(d_0\operatorname{Hom}(V_\beta,V_\alpha)\) is perfect. It follows that \(d_1\) restricts to an isomorphism from \(L_-\) onto the reverse moment-map space, and \(L_-\cap\ker d_1=0\). In particular, there is no allowed gauge direction in the reverse block either, since \(d_1d_0=0\) and \(d_0\) is injective: an endomorphism in its kernel kills the framing vectors and commutes with both adjoints, whose iterates on the framing vectors span the gauge space by the stability condition \eqref{eq:spiked-stability}. Similarly, in the positive block every class in \(\ker d_1/\operatorname{im}d_0\) has a unique representative in \(L_+\cap\ker d_1\).

\paragraph{The attracting tangent space}
Combining the three cases, \(L_\partial\) contains no infinitesimal gauge directions, and, by the symplectic duality above, the moment-map differential restricted to \(L_\partial\) is surjective. Consequently, the tangent space of the boundary image in \(\mathcal M_H\) is \(L_\partial\cap\ker d_1\), with character
\bea
 T^+_{\vec\lambda_{12}}
 =\mathcal L_\partial-q_{12}^{-1}\mathbf K^\vee\mathbf K .
 \label{eq:parallel-d6-attracting-tangent}
\eea
The last term is the character of the complex moment-map equations.

To identify this with the attracting tangent space, we compare it block by block with the full tangent character of \(\mathcal M_H\) at the fixed point. Set \(\mathbf K_\alpha=\sum_{x\in\lambda_{12}^{(\alpha)}}\tilde s_x\) and \(\mathbf N_\alpha=q_4^{1/2}v_{\bar{3},\alpha}\), so that \(\mathbf K=\sum_\alpha\mathbf K_\alpha\) and \(\mathbf N_{124}=\sum_\alpha\mathbf N_\alpha\). By the analysis of the mixed blocks, for \(\alpha>\beta\) the character \(T^+_{\vec\lambda_{12}}\) contains the full tangent character of the block \(\alpha\beta\),
\bea
\begin{aligned}
 \bigl(T^+_{\vec\lambda_{12}}\bigr)_{\alpha\beta}
 ={}&\mathbf K_\alpha^\vee\mathbf N_\beta
 +q_{12}^{-1}\mathbf N_\alpha^\vee\mathbf K_\beta
 \\
 &-(1-q_1^{-1})(1-q_2^{-1})
   \mathbf K_\alpha^\vee\mathbf K_\beta ,
\end{aligned}
 \label{eq:parallel-d6-mixed-tangent}
\eea
and nothing from the block \(\beta\alpha\). The subtracted term accounts for both the gauge variation and the linearized complex moment map in this block.

In terms of arms and legs, with the conventions of Appendix~\ref{app:notations} (where the arm and leg of a box relative to another diagram may be negative), the full tangent character in the block \(\alpha\beta\) is
\bea
\begin{aligned}
T_{\alpha\beta}
=\frac{v_{\bar{3},\beta}}{v_{\bar{3},\alpha}}\biggl(
&\sum_{x\in\lambda_{12}^{(\alpha)}}
 q_1^{a_{\lambda_{12}^{(\alpha)}}(x)}
 q_2^{-\ell_{\lambda_{12}^{(\beta)}}(x)-1}+\sum_{x\in\lambda_{12}^{(\beta)}}
 q_1^{-a_{\lambda_{12}^{(\beta)}}(x)-1}
 q_2^{\ell_{\lambda_{12}^{(\alpha)}}(x)}
\biggr),
\end{aligned}
\label{eq:parallel-d6-arm-leg}
\eea
and the tangent character of \(\mathcal M_H\) at the fixed point is \(\sum_{\alpha,\beta}T_{\alpha\beta}\). In the chamber \eqref{eq:parallel-d6-chamber}, the attracting part consists of the second sum in each diagonal block \(T_{\alpha\alpha}\) and of the full block \(T_{\alpha\beta}\) for \(\alpha>\beta\). The same-framing blocks of \(T^+_{\vec\lambda_{12}}\) give the former by \cite{Crew:2020psc}, and the mixed blocks give the latter by \eqref{eq:parallel-d6-mixed-tangent}. Therefore
\bea
T^+_{\vec\lambda_{12}}
=
\sum_{\alpha=1}^N\sum_{x\in\lambda_{12}^{(\alpha)}}
 q_1^{-a_{\lambda_{12}^{(\alpha)}}(x)-1}
 q_2^{\ell_{\lambda_{12}^{(\alpha)}}(x)}
+\sum_{\alpha>\beta}T_{\alpha\beta}
\label{eq:parallel-d6-attracting-arm-leg}
\eea
is the attracting tangent character, as required by criterion~4.

\subsection{Perturbative factor and the PT edge}\label{app:rankN-bc-edge}
Finally, we check criterion~6. As in Appendix~\ref{app:edge-comparison}, we write the zero-flux factor as the classical FI factor times \(\widehat a[\mathcal X^{\mathrm D}/\bfP_4^\vee]\), with \(\bfP_4^\vee=1-q_4^{-1}\), and compute the character \(\mathcal X^{\mathrm D}\) for the present boundary condition.

The two weights in a hypermultiplet symplectic pair multiply to \(q_{12}^{-1}\). Since \(q_4^{-1}q_{12}^{-1}=q_3\), a Neumann weight \(w\) and its Dirichlet partner contribute
\bea
 w^{-1}-q_4^{-1}q_{12}^{-1}w^{-1}=(1-q_3)w^{-1}
\eea
to \(\mathcal X^{\mathrm D}\). The Dirichlet vector multiplet and \(\Phi_3\) together contribute \((q_4^{-1}-q_{12})\mathbf K^\vee\mathbf K =-(1-q_3)q_{12}\mathbf K^\vee\mathbf K\). Comparing with \eqref{eq:parallel-d6-attracting-tangent}, we find that the full zero-flux factor, including the classical contribution, is
\bea
 \mathcal Z_{\mathrm{pert}}^{\vec\lambda_{12}}
 &=\left(\prod_x\tilde s_x^{\,\xi}\right)
   \widehat a\left[\frac{\mathcal X^{\mathrm D}}{1-q_4^{-1}}\right],
 &
 \mathcal X^{\mathrm D}
 &=(1-q_3)\bigl(T^+_{\vec\lambda_{12}}\bigr)^\vee .
 \label{eq:parallel-d6-perturbative-tangent}
\eea
Thus the zero-flux factor is determined by the attracting tangent space of Appendix~\ref{app:rankN-bc-tangent}.

\paragraph{Gluing}
We now glue the north and south hemispheres as in Section~\ref{sec:rankone-twistedindex}. Take \(r_3=0\) and \(r_1+r_2=2\), with the south-pole parameters \(\tilde q_4=q_4^{-1}\) and \(\tilde q_i=q_iq_4^{r_i}\) of \eqref{eq:rankN-block-equiv-param}. After removing the classical factor \(\fq^{\chi(\vec\lambda_{12},(r_1,r_2))}\), the glued perturbative factor is
\bea
\mathcal E_{\mathrm{edge}}^{\mathrm D}
=\widehat a \left[ \frac{\mathcal X^{\mathrm D}}{1-q_4^{-1}} +\frac{\widetilde{\mathcal X}^{\mathrm D}}{1-\tilde{q}_4^{-1}} \right],
\eea
where the tilde denotes evaluation at the south-pole parameters.

To compare with the PT edge factor, we start from the boundary condition \eqref{eq:boundary-conditions}, whose sign is given in Appendix~\ref{app:edge-comparison}. In the parallel configuration, every transverse direction is \(3\) and \(r_3=0\), so the exponent \(\eta_{\mathrm D}\) of the Dirichlet sign rule there reduces to \(r_1\sum_\alpha|\lambda_{12}^{(\alpha)}|=r_1N_4\). By Appendix~\ref{app:rankN-bc-def}, the present boundary condition differs from \eqref{eq:boundary-conditions} only by exchanges within mixed symplectic pairs. We now show that such an exchange does not change the glued determinant.

Denote the sum of the two patch characters by \(G(u)=u/(1-q_4^{-1})+\tilde u/(1-\tilde q_4^{-1})\). For a Neumann weight \(w\), put \(u=w^{-1}\), so that the pair contributes \((1-q_3)u\) to \(\mathcal X^{\mathrm D}\). Exchanging \(\mathrm N\) and \(\mathrm D\) within the pair makes the partner, whose weight is \(q_{12}^{-1}w^{-1}\), Neumann, and thus replaces \(u\) by \(q_{12}u^{-1}\) in the contribution \((1-q_3)u\). Using
\bea
q_{12}=q_4^{-1}q_3^{-1},\qquad \tilde q_3=q_3,\qquad G(q_4^{-1}u^{-1})=-G(u)^{\vee},
\eea
we find that the old glued character minus the new one is
\[
Y-Y^\vee,\qquad Y=(1-q_3)G(u).
\]
Every weight in \(Y\) carries a nontrivial ratio of distinct framing fugacities. Using \([z^{-1}]=-[z]\) in \eqref{eq:ahat-definition}, we obtain \(\widehat a[Y-Y^\vee]=(-1)^{\operatorname{rk}Y}=1\), where \(\operatorname{rk}Y=0\) because of the factor \(1-q_3\). The exchanges therefore leave the sign unchanged, and the glued determinant is related to the rank-\(N\) PT edge factor \eqref{eq:rankN-PT-edge} by the same sign as for \eqref{eq:boundary-conditions}:
\bea
 (-1)^{r_1N_4}\mathcal E_{\mathrm{edge}}^{\mathrm D}
 &=
 \mathsf E_{\bar 3;\vec\lambda_{12}}
 (q_1,q_2\mid\tilde q_1,\tilde q_2).
 \label{eq:parallel-d6-edge-sign}
\eea
In the notation of Section~\ref{sec:rankone-twistedindex}, the sign \(\varepsilon_{\mathrm D}\) is therefore \((-1)^{r_1N_4}\). Combined with the vortex factor \eqref{eq:parallel-d6-vortex} and the vortex--PT proposal \eqref{eq:higher-rank-vortex-PT-proposal}, the block factorization \eqref{eq:rankN-block-decomp} with the present boundary condition takes the form of the PT gluing formula for \(\mathsf{PT}[X;N]\) in Section~\ref{sec:rankone-twistedindex}.


\section{Review of Jeffrey--Kirwan residue integrals}\label{app:JK}
This appendix reviews the JK residue prescription, which defines the contour integrals of the D0 quantum mechanics in Section~\ref{sec:DTPT-vortexQM} and of the twisted index \eqref{eq:twisted-general-JK} in Section~\ref{sec:twisted-spiked}. For each integral, the charge vectors and the reference vector must be specified.


We follow the standard presentation \cite{jeffrey1995localization,vergne1996note,brion1999arrangement,szenes2003toric,%
Benini:2013nda,Benini:2013xpa}. 
The Witten index of a supersymmetric gauge theory with a compact gauge group $G$
is computed by a contour integral over the Cartan torus of $G$.
For a rank-$k$ gauge group $G$ (typically $U(k)$ in our context), the index takes the form
\bea\label{eq:Witten-index}
  Z
  =
  \oint_{\mathrm{JK}}
  \prod_{I=1}^{k}
  \frac{d\phi_I}{2\pi i}\;
  \mathcal{I}(\phi),
\eea
where $\phi_I$ are coordinates on the complexification $\mathfrak{h}_{\mathbb C}$ of the Cartan subalgebra $\mathfrak{h}\cong\mathbb{R}^k$, and $\mathcal{I}(\phi)$ is a meromorphic function determined by the field content.
In the integrals considered here, the signs from shifting the individual $\sh$ factors by a gauge period cancel, giving
\bea
  \mathcal{I}(\ldots,\phi_I,\ldots) = \mathcal{I}(\ldots,\phi_I+2\pi i,\ldots),
\eea
so $\phi$ takes values in $\mathfrak{h}_{\mathbb C}/(2\pi i\,\mathbb{Z}^k)$.
The JK prescription specifies the integration cycle through a reference vector $\eta\in\mathfrak{h}^{\ast}$, which selects the poles that contribute.

The reference vector $\eta$ is determined by the
FI parameters of the gauge theory:
different FI chambers correspond to different choices of $\eta$, and
the Witten index is independent of $\eta$ as long as $\eta$ remains in the same chamber.

\paragraph{Pole structure}
The poles of $\mathcal{I}$ arise from zeros of the denominator.
Each such zero defines a singular hyperplane in $\mathfrak{h}_{\mathbb C}$,
\begin{align}\label{eq:hyperplane}
  H_{j,n_j}:\quad
  Q_j(\phi) + f_j(a,\epsilon) = 2\pi i\, n_j,
  \qquad n_j\in\mathbb{Z},\quad j=1,2,\ldots,
\end{align}
where $Q_j\in\mathfrak{h}^{\ast}$ is the \emph{charge vector} of the $j$-th factor
and $f_j(a,\epsilon)$ encodes the dependence on mass and $\Omega$-background parameters.
Choose an integral basis of the gauge cocharacter lattice, so that $\phi\sim\phi+2\pi i\,m$ for $m\in\mathbb Z^k$. Collect the $k$ linearly independent integral charge vectors of a candidate pole as columns of
$\boldsymbol Q=(Q_{j_1}|\cdots|Q_{j_k})$, and set $f=(f_{j_1},\ldots,f_{j_k})^{\mathsf T}$. The hyperplane equations and their solutions are
\begin{align}
\boldsymbol Q^{\mathsf T}\phi+f&=2\pi i\,n,\\
\phi_n&=(\boldsymbol Q^{\mathsf T})^{-1}(2\pi i\,n-f)
\quad\bmod 2\pi i\,\mathbb Z^k.
\end{align}
A gauge-period shift sends $n$ to $n+\boldsymbol Q^{\mathsf T}m$. Thus distinct solutions are labeled by representatives of the finite lattice quotient
\begin{align}
n\in\mathbb Z^k/\boldsymbol Q^{\mathsf T}\mathbb Z^k,
\qquad \#\bigl(\mathbb Z^k/\boldsymbol Q^{\mathsf T}\mathbb Z^k\bigr)=|\det\boldsymbol Q|.
\end{align}
For example, for $k=1$ and $Q=2$, representatives $n=0,1$ give $\phi=-f/2$ and $\phi=-f/2+\pi i$, the two distinct solutions modulo $2\pi i$.

Let $\mathcal{M}_{\mathrm{sing}}=\bigcup_{j,n_j} H_{j,n_j}$ be the singular locus before quotienting by the gauge periods.
We denote by $\mathcal{M}^{\ast}_{\mathrm{sing}}$ the set of isolated points at which
at least $k$ linearly independent hyperplanes intersect.
A point $\phi_{\ast}\in\mathcal{M}^{\ast}_{\mathrm{sing}}$
at which exactly $k$ hyperplanes meet is called \emph{non-degenerate}.
If $n>k$ hyperplanes meet at $\phi_{\ast}$, we distinguish two subcases:
the intersection is \emph{degenerate} if the numerator of $\mathcal{I}$ vanishes
to order $n-k$ at $\phi_{\ast}$ (so that the net pole order is $k$),
and a \emph{higher-order pole} otherwise.

Suppose that hyperplanes $H_{j_1},\ldots,H_{j_n}$ meet at $\phi_{\ast}$ ($n\ge k$), and
let $Q_{\ast}$ denote the corresponding charge vectors.
Choose an ordered sequence of $k$ linearly independent vectors in
$Q_{\ast}$, relabeled $Q_{j_1},\ldots,Q_{j_k}$, and form the complete flag
\begin{align}\label{eq:flag}
  \{0\}\subset F_1\subset\cdots\subset F_k=\mathbb{R}^k,
  \qquad
  F_i=\mathrm{span}\{Q_{j_1},\ldots,Q_{j_i}\}.
\end{align}
Different sequences yielding the same flag are identified;
we fix a basis $\mathcal{B}(F,Q_{\ast})=(Q_{j_1},\ldots,Q_{j_k})$ for each flag.
From the flag and the full set $Q_{\ast}$, define the \emph{cumulative charge vectors}
\begin{align}\label{eq:kappa}
  \kappa(F,Q_{\ast})=(\kappa_1,\ldots,\kappa_k),
  \qquad
  \kappa_i = \sum_{\substack{Q\in Q_{\ast}\\Q\in F_i}} Q.
\end{align}
The JK residue at a degenerate or higher-order pole is then
\begin{align}\label{eq:JK-deg}
  \underset{\phi=\phi_{\ast}}{\mathrm{JK\text{-}Res}}
  \bigl(Q(\phi_{\ast}),\eta\bigr)\,\mathcal{I}(\phi)
  =
  \sum_{F}
  \delta(F,\eta)\;
  \frac{\mathrm{sign}\det\kappa(F,Q_{\ast})}{\det\mathcal{B}(F,Q_{\ast})}
  \underset{\delta_k=0}{\mathrm{Res}}
  \cdots
  \underset{\delta_1=0}{\mathrm{Res}}
  \mathcal{I}(\phi)
  \Big|_{Q_{j_i}(\phi)+f_{j_i}=2\pi i n_{j_i}+\delta_i},
\end{align}
where the sum runs over all flags,
$\delta(F,\eta)=1$ if $\eta$ lies in the closed cone spanned by $\kappa(F,Q_{\ast})$
and $\delta(F,\eta)=0$ otherwise,
and the iterated residues are taken in the order $\delta_1,\ldots,\delta_k$.

\paragraph{Full contour integral}
For a generic reference vector $\eta$
(one that does not lie on any wall between chambers),
the contour integral \eqref{eq:Witten-index} is evaluated as
\begin{align}\label{eq:JK-full}
  \oint_{\eta}
  \prod_{i=1}^{k}
  \frac{d\phi_i}{2\pi i}\;
  \mathcal{I}(\phi)
  =
  \sum_{\phi_{\ast}\in\mathcal{M}^{\ast}_{\mathrm{sing}}}
  \underset{\phi=\phi_{\ast}}{\mathrm{JK\text{-}Res}}
  \bigl(Q(\phi_{\ast}),\eta\bigr)\,\mathcal{I}(\phi).
\end{align}
The right-hand side is a finite sum: only those isolated singular points
$\phi_{\ast}$ whose associated cone contains $\eta$ contribute.

Across a wall between chambers, the selected residues can change, and in supersymmetric quantum mechanics the jump can receive a contribution from infinity on the Coulomb branch. In Section~\ref{sec:DTPT-vortexQM} the two choices in \eqref{eq:DTPT-etavector} select the DT and PT chambers, respectively.

\section{Superpotential of the \texorpdfstring{PT$_{3}$}{PT3} vertex}\label{app:one-leg-detail}
In this appendix, we write down the superpotential and the
$J$- and $E$-terms of the one-leg D6--D2--D0 quantum mechanics
used in Section~\ref{sec:DTPT-vortexQM}. See \cite{Kimura:2025lfo} and \cite{Galakhov:2021xum} for details.

We begin with a D6-brane on $\mathbb C^3_{123}$ and a D2-brane leg
along $\mathbb C_3$, specified by a Young diagram $\nu$ in the
$12$-plane. Following the main text, the PT configuration occupies the directions $123$ and the leg points along the $3$-direction, with the associated Young diagram $\nu$:
\bea
\adjustbox{valign=c}{%
\begin{tikzpicture}[scale=0.4]
  \planepartition{{6,6,6},{6,6},{6},{6}}

  \coordinate (top) at (0,6);

  \draw[->] (top) -- ++(210:6)
    node[pos=1.12] {$1$};
  \draw[->] (top) -- ++(-30:5)
    node[pos=1.16] {$2$};
  \draw[->] (top) -- ++(90:2)
    node[pos=1.16] {$3$};
\end{tikzpicture}%
}\qquad\qquad \qquad   \adjustbox{valign=c}{\begin{tikzpicture}[scale=0.8,rotate=90]
    \draw [thick,->] (0,0)--(4,0);
    \draw [thick,->] (0,0)--(0,-5);
    \node[above] at (4,0){$2$};
    \node[right] at (0,-5){$1$};
    \draw[thick] (3,0)--(3,-1)--(2,-1)--(2,-2)--(1,-2)--(1,-3)--(1,-4)--(0,-4);
    \draw[fill=red,xshift=3cm] (0,0)--(0.3,0)--(0.3,-0.3)--(0,-0.3)--(0,0);
    \draw[fill=red,xshift=2cm,yshift=-1cm] (0,0)--(0.3,0)--(0.3,-0.3)--(0,-0.3)--(0,0);
    \draw[fill=red,xshift=1cm,yshift=-2cm] (0,0)--(0.3,0)--(0.3,-0.3)--(0,-0.3)--(0,0);
    \draw[fill=red,xshift=0cm,yshift=-4cm] (0,0)--(0.3,0)--(0.3,-0.3)--(0,-0.3)--(0,0);

     \draw[fill=blue,xshift=3cm,yshift=-1cm] (0,0)--(0.3,0)--(0.3,-0.3)--(0,-0.3)--(0,0);
       \draw[fill=blue,xshift=2cm,yshift=-2cm] (0,0)--(0.3,0)--(0.3,-0.3)--(0,-0.3)--(0,0);
         \draw[fill=blue,xshift=1cm,yshift=-4cm] (0,0)--(0.3,0)--(0.3,-0.3)--(0,-0.3)--(0,0);

         \node[scale=2] at (1,-1){$\nu$};
    
\end{tikzpicture}}\eea
Decomposing $\nu$ into rectangles, we parametrize them as
\bea
A(\nu)&=\{(r_{i-1}+1,s_{i}+1)\mid i=1,\ldots,f+1\},\\
R(\nu)&=\{(r_{i},s_{i})\mid i=1,\ldots,f\},
\eea
where $r_{0}=0,s_{f+1}=0$ and $f$ is the number of rectangles.

As in the main text, we introduce a fundamental chiral multiplet $\mathsf{I}_{z}$ for each addable box $z\in A(\nu)$ and an antifundamental chiral multiplet $\mathsf{J}_{w}$ for each removable box $w\in R(\nu)$. To lighten the notation, we shift the labels by $(-1,-1)$ only for $\mathsf{I}_{z}$, and denote the chiral multiplets as
\bea
\mathsf{I}_{(r_{i-1},s_{i})},\quad (i=1,\ldots,f+1),\qquad \mathsf{J}_{(r_{i},s_{i})},\quad (i=1,\ldots,f).
\eea
The quiver is then written as
\bea
\adjustbox{valign=c}{\begin{tikzpicture}[decoration={markings,mark=at position \arrowHeadPosition with {\arrow{latex}}}]
 \tikzset{
        box/.style={draw, minimum width=0.6cm, minimum height=0.6cm, text centered,thick},
        ->-/.style={decoration={
        markings,mark=at position #1 with {\arrow[scale=1.5]{>}}},postaction={decorate},line width=0.5mm},
        -<-/.style={decoration={
        markings,
        mark=at position #1 with {\arrow[scale=1.5]{<}}},postaction={decorate},line width=0.5mm}    
    }
\begin{scope}[xshift=4cm]
    \draw[black,thick,scale=1.3, postaction={decorate}] (-0.75,0) arc(90:450:0.2 and 0.45) ;
     \draw[black,thick,scale=1.3, postaction={decorate}] (-0.75,0) arc(0:360:0.45 and 0.2) ;
      \draw[black,thick,scale=1.3, postaction={decorate}] (-0.75,0) arc(180:540:0.45 and 0.2) ;
      \node[right] at (0.2,0){$\mathsf{B}_{3}$};
      \node[left] at (-2.2,0){$\mathsf{B}_{2}$};
      \node[below] at (-0.95,-1.25){$\mathsf{B}_{1}$};
    \draw[fill=white,thick](-0.95,0) circle(0.3cm);
    \node at (-0.95,0) {$k$};
    \node[fill=white,box] at (-0.95,1.5){$1$};
    \draw[thick,postaction={decorate}] (-1.05,1.2)--(-1.05,0.3);
   \draw[thick,postaction={decorate}] (-0.85,0.3)--(-0.85,1.2);
    \node[right] at (-0.85,0.75){$\mathsf{J}_{(r_{i},s_{i})}$};
     \node[left] at (-1.05,0.75){$\mathsf{I}_{(r_{i-1},s_{i})}$};
\end{scope}
\end{tikzpicture}}
\eea
Following \cite{Galakhov:2021xum,Kimura:2025lfo}, the superpotential is
\bea\label{eq:DT3-oneleg-superpotential}
\mathsf{W}_{\varnothing\varnothing\nu}&=\mathsf{W}_{0}+\delta\mathsf{W}_{\varnothing\varnothing\nu},\\
\mathsf{W}_{0}&=\Tr \mathsf{B}_{1}[\mathsf{B}_{2},\mathsf{B}_{3}],\\
\delta\mathsf{W}_{\varnothing\varnothing\nu}&=\sum_{i=1}^{f}\Tr\,\mathsf{J}_{(r_{i},s_{i})}\left(\mathsf{B}_{1}^{r_{i}-r_{i-1}}\mathsf{I}_{(r_{i-1},s_{i})}-\mathsf{B}_{2}^{s_{i}-s_{i+1}}\mathsf{I}_{(r_{i},s_{i+1})}\right).
\eea
The flavor charges are determined from this superpotential as
\bea\label{eq:onelegDTflavorcharge}
[\mathsf{I}_{(r_{i-1},s_{i})}]=q_{1}^{-r_{i-1}}q_{2}^{-s_{i}},\quad [\mathsf{J}_{(r_{i},s_{i})}]=q_{4}q_{1}^{r_{i}}q_{2}^{s_{i}}.
\eea

\paragraph{\texorpdfstring{$\mathcal{N}=2$ $J$- and $E$-terms}{N=2 J- and E-terms}}
In 1d $\mathcal{N}=2$ language, each $\mathcal{N}=4$ chiral multiplet splits into an $\mathcal{N}=2$ chiral multiplet and an $\mathcal{N}=2$ Fermi multiplet, while the $\mathcal{N}=4$ vector multiplet splits into an $\mathcal{N}=2$ vector multiplet and an $\mathcal{N}=2$ adjoint chiral multiplet, which we denote by $\mathsf{B}_{4}$. The field content is then
\bea
\text{$\mathcal{N}=2$ chiral}:&\qquad \mathsf{B}_{1,2,3,4},\quad \mathsf{I}_{(r_{i-1},s_{i})},\quad \mathsf{J}_{(r_{i},s_{i})},\\
\text{$\mathcal{N}=2$ Fermi}:&\qquad \Lambda_{1,2,3},\quad \Lambda_{(r_{i-1},s_{i})},\quad \widetilde{\Lambda}_{(r_{i},s_{i})}
\eea
where $\Lambda_{1,2,3}$ are the Fermi multiplets associated with $\mathsf{B}_{1,2,3}$, and $\Lambda_{(r_{i-1},s_{i})},\widetilde{\Lambda}_{(r_{i},s_{i})}$ those associated with $\mathsf{I}_{(r_{i-1},s_{i})},\mathsf{J}_{(r_{i},s_{i})}$, respectively.

The $E$-terms are 
\bea\label{eq:DT-oneleg-Eterm}
\mathcal{E}_{i}&=[\mathsf{B}_{4},\mathsf{B}_{i}],\quad \mathcal{E}_{(r_{i-1},s_{i})}=\mathsf{B}_{4}\mathsf{I}_{(r_{i-1},s_{i})},\quad \widetilde{\mathcal{E}}_{(r_{i},s_{i})}=-\mathsf{J}_{(r_{i},s_{i})}\mathsf{B}_{4}.
\eea
The $J$-terms are computed as
\bea\label{eq:DT-oneleg-Jterm}
\mathcal{J}_{1}&=\frac{\partial \mathsf{W}}{\partial \mathsf{B}_{1}}=[\mathsf{B}_{2},\mathsf{B}_{3}]+\sum_{i=1}^{f}\sum_{a=1}^{r_{i}-r_{i-1}}\mathsf{B}_{1}^{r_{i}-r_{i-1}-a}\mathsf{I}_{(r_{i-1},s_{i})}\mathsf{J}_{(r_{i},s_{i})}\mathsf{B}_{1}^{a-1},\\
\mathcal{J}_{2}&=\frac{\partial \mathsf{W}}{\partial \mathsf{B}_{2}}=[\mathsf{B}_{3},\mathsf{B}_{1}]-\sum_{i=1}^{f}\sum_{a=1}^{s_{i}-s_{i+1}}\mathsf{B}_{2}^{s_{i}-s_{i+1}-a}\mathsf{I}_{(r_{i},s_{i+1})}\mathsf{J}_{(r_{i},s_{i})}\mathsf{B}_{2}^{a-1},\\
\mathcal{J}_{3}&=\frac{\partial \mathsf{W}}{\partial \mathsf{B}_{3}}=[\mathsf{B}_{1},\mathsf{B}_{2}],\\
\mathcal{J}_{(r_{i-1},s_{i})}&=\frac{\partial \mathsf{W}}{\partial \,\mathsf{I}_{(r_{i-1},s_{i})}}=\begin{cases}
    \mathsf{J}_{(r_{1},s_{1})}\mathsf{B}_{1}^{r_{1}},\quad i=1\\
\mathsf{J}_{(r_{i},s_{i})}\mathsf{B}_{1}^{r_{i}-r_{i-1}}-\mathsf{J}_{(r_{i-1},s_{i-1})}\mathsf{B}_{2}^{s_{i-1}-s_{i}},\quad i=2,\ldots,f\\
    -\mathsf{J}_{(r_{f},s_{f})}\mathsf{B}_{2}^{s_{f}},\quad i=f+1
\end{cases}\\
\widetilde{\mathcal{J}}_{(r_{i},s_{i})}&=\frac{\partial \mathsf{W}}{\partial\, \mathsf{J}_{(r_{i},s_{i})}}=\mathsf{B}_{1}^{r_{i}-r_{i-1}}\mathsf{I}_{(r_{i-1},s_{i})}-\mathsf{B}_{2}^{s_{i}-s_{i+1}}\mathsf{I}_{(r_{i},s_{i+1})},\quad (i=1,\ldots,f)
\eea

\paragraph{One-box case}
For the simplest case $\Yboxdim{6pt}\nu=\yng(1)$, we have $r_{1}=s_{1}=1$. The superpotential is 
\bea\Yboxdim{4.5pt}
\delta\mathsf{W}_{\varnothing\varnothing\,\yng(1)}=\Tr \,\mathsf{J}_{(1,1)}\left(\mathsf{B}_{1}\mathsf{I}_{(0,1)}-\mathsf{B}_{2}\mathsf{I}_{(1,0)}\right).
\eea
The $E$-terms are
\bea
\mathcal{E}_{1,2,3}&=[\mathsf{B}_{4},\mathsf{B}_{1,2,3}],\quad \mathcal{E}_{(0,1)}=\mathsf{B}_{4}\mathsf{I}_{(0,1)},\quad\mathcal{E}_{(1,0)}=\mathsf{B}_{4}\mathsf{I}_{(1,0)},\quad  \widetilde{\mathcal{E}}_{(1,1)}=-\mathsf{J}_{(1,1)}\mathsf{B}_{4}
\eea
and the $J$-terms are
\bea
\mathcal{J}_{1}&=\frac{\partial \mathsf{W}}{\partial \mathsf{B}_{1}}=[\mathsf{B}_{2},\mathsf{B}_{3}]+\mathsf{I}_{(0,1)}\mathsf{J}_{(1,1)},\quad 
\mathcal{J}_{2}=\frac{\partial \mathsf{W}}{\partial \mathsf{B}_{2}}=[\mathsf{B}_{3},\mathsf{B}_{1}]-\mathsf{I}_{(1,0)}\mathsf{J}_{(1,1)},\quad 
\mathcal{J}_{3}=\frac{\partial \mathsf{W}}{\partial \mathsf{B}_{3}}=[\mathsf{B}_{1},\mathsf{B}_{2}],\\
\mathcal{J}_{(0,1)}&=\frac{\partial \mathsf{W}}{\partial \,\mathsf{I}_{(0,1)}}=\mathsf{J}_{(1,1)}\mathsf{B}_{1},\quad
\mathcal{J}_{(1,0)}=\frac{\partial \mathsf{W}}{\partial \,\mathsf{I}_{(1,0)}}=-\mathsf{J}_{(1,1)}\mathsf{B}_{2},\quad
\widetilde{\mathcal{J}}_{(1,1)}=\frac{\partial \mathsf{W}}{\partial\, \mathsf{J}_{(1,1)}}=\mathsf{B}_{1}\mathsf{I}_{(0,1)}-\mathsf{B}_{2}\mathsf{I}_{(1,0)}.
\eea

\paragraph{Higher-rank case}
The construction generalizes to higher rank by including several D6-branes with D2-branes bound to them. For $n$ D6$_{\bar{4}}$-branes, the asymptotic data form $n$-tuples of Young diagrams $\vec{\nu}$, where $\vec{\nu}=(\nu^{(\alpha)})_{\alpha=1}^{n}$. This configuration still preserves four supercharges, and the superpotential is simply additive:
\bea
\delta\mathsf{W}_{\vec{\varnothing}\vec{\varnothing}\vec{\nu}}=\sum_{\alpha=1}^{n}\delta \mathsf{W}_{\varnothing\varnothing\nu^{(\alpha)}}.
\eea
In particular, for $(\vec{\varnothing},\vec{\varnothing},\vec{\nu})$ one introduces $\mathsf{I}_{(r^{(\alpha)}_{i-1},s^{(\alpha)}_{i})}$ and $\mathsf{J}_{(r_{i}^{(\alpha)},s_{i}^{(\alpha)})}$ for each $\alpha=1,\ldots,n$, and sums the contributions \eqref{eq:DT3-oneleg-superpotential}.

\paragraph{Tetrahedron instanton case}
So far, only D6$_{\bar{4}}$-branes were present, but we may equally include D6-branes wrapping the other three-planes $\mathbb{C}^{3}_{\bar{a}}$. Let $N_{\bar{a}}\,(a\in\four)$ be the number of D6$_{\bar{a}}$-branes. The asymptotic Young diagrams are collectively denoted by 
\bea
\underline{\vec{\nu}}=(\nu_{\bar{a}}^{(\alpha)})_{a\in\four}^{\alpha=1,\ldots,N_{\bar{a}}}
\eea

Such a configuration preserves only two supercharges, and the quiver diagram takes the form 
\bea\label{eq:tetrahedronDTquiver}
\adjustbox{valign=c}{
\begin{tikzpicture}[decoration={markings,mark=at position \arrowHeadPosition with {\arrow{latex}}}]
    \tikzset{
        cir/.style={circle, draw, minimum size=0.8cm, text centered, thick},
        point/.style={circle, fill, inner sep=1.5pt},
        box/.style={draw,minimum width=0.7cm, minimum height=0.7cm, text centered,thick},
        ->-/.style={decoration={
        markings,mark=at position #1 with {\arrow[scale=1.5]{>}}},postaction={decorate},line width=0.5mm},
        -<-/.style={decoration={
        markings,
        mark=at position #1 with {\arrow[scale=1.5]{<}}},postaction={decorate},line width=0.5mm}    
    }

    \node[cir] (i1) at ($(0,0)$) { };
    \begin{scope}[rotate=0]
    \node[box,rotate=0] (n13) at ($(0,-2)$) {\rotatebox{-0}
    {$\D6_{\bar{4};\vec{\varnothing}\vec{\varnothing}\vec{\nu}_{\bar{4}}}$}};
   
    \draw[postaction={decorate}, thick] ($(0.06,-1.65)$) -- ($(0.06,-0.4)$);
    \draw[postaction={decorate},red, thick] ($(-0.06,-1.65)$) -- ($(-0.06,-0.4)$);
   
    \end{scope}

    \begin{scope}[rotate=90]
    \node[box,rotate=90] (n12) at  ($(0,-2.6)$) {\rotatebox{-90}
    {$\D6_{\bar{1};\vec{\varnothing}\vec{\varnothing}\vec{\nu}_{\bar{1}}}$}};
  
    \draw[postaction={decorate}, thick] ($(0.06,-1.65)$) -- ($(0.06,-0.4)$);
    \draw[postaction={decorate},red, thick] ($(-0.06,-1.65)$) -- ($(-0.06,-0.4)$);
   
    \end{scope}
    
    \begin{scope}[rotate=180]
    \node[box,rotate=180] (n12) at  ($(0,-2)$) {\rotatebox{-180}{$\D6_{\bar{2};\vec{\varnothing}\vec{\varnothing}\vec{\nu}_{\bar{2}}}$}};
   
    \draw[postaction={decorate}, thick] ($(0.06,-1.65)$) -- ($(0.06,-0.4)$);
    \draw[postaction={decorate},red, thick] ($(-0.06,-1.65)$) -- ($(-0.06,-0.4)$);
    \end{scope}
    
    \begin{scope}[rotate=270]
    \node[box,rotate=270] (n23) at  ($(0,-2.6)$) {\rotatebox{-270}{$\D6_{\bar{3};\vec{\varnothing}\vec{\varnothing}\vec{\nu}_{\bar{3}}}$}};
   
    \draw[postaction={decorate}, thick] ($(0.06,-1.65)$) -- ($(0.06,-0.4)$);
    \draw[postaction={decorate},red, thick] ($(-0.06,-1.65)$) -- ($(-0.06,-0.4)$);
    \end{scope}

    \begin{scope}[rotate=45]
    \chiralarc[postaction={decorate},thick](0,0.55)(-45:225:0.22:0.65)
    \fermiarc[postaction={decorate},thick](0,0.5)(-45:225:0.1:0.5)
    \end{scope}
    \begin{scope}[rotate=135]
    \chiralarc[postaction={decorate},thick](0,0.55)(-45:225:0.22:0.65)
    \fermiarc[postaction={decorate},thick](0,0.5)(-45:225:0.1:0.5)
    \end{scope}
    \begin{scope}[rotate=225]
    \chiralarc[postaction={decorate},thick](0,0.55)(-45:225:0.22:0.65)
    \end{scope}
    \begin{scope}[rotate=315]
    \chiralarc[postaction={decorate},thick](0,0.55)(-45:225:0.22:0.65)
    \fermiarc[postaction={decorate},thick](0,0.5)(-45:225:0.1:0.5)
    \end{scope}    
\end{tikzpicture}} 
\eea
Each arrow joining a flavor node to the gauge node stands for several $\mathcal{N}=2$ fundamental and antifundamental chiral and Fermi multiplets, whose explicit form we omit.

Besides the quiver, we must specify the $J$- and $E$-terms. We first
consider a single flavor node and write the $\mathcal N=4$ theory in
$\mathcal N=2$ language. We then add the other flavor nodes and use
quadrality to determine the new terms. With the signs given below,
$\sum\Tr(\mathcal J\cdot\mathcal E)=0$ holds without imposing any
equations on $\mathsf B_a,\mathsf I,\mathsf J$. This condition fixes
the relative signs of the $(1,0)$ and $(0,1)$ terms. 

Since the main text only requires D6$_{\bar{1},\bar{2},\bar{3}}$-branes together with D2$_{4}$-branes, we specialize to a single brane of each type with $\Yboxdim{6pt}\nu_{\bar{1},\bar{2},\bar{3}}=\yng(1)$. The (anti)fundamental chiral fields are
\bea
\mathsf{I}_{\bar{a},(1,0)},\quad \mathsf{I}_{\bar{a},(0,1)},\quad \mathsf{J}_{\bar{a},(1,1)},\quad a=1,2,3
\eea
and the associated Fermi fields are
\bea
\Lambda_{\bar{a},(1,0)},\quad \Lambda_{\bar{a},(0,1)},\quad \widetilde{\Lambda}_{\bar{a},(1,1)},\quad a=1,2,3.
\eea
The labels $(1,0),(0,1),(1,1)$ are coordinates in the two-dimensional plane obtained from $\bar{a}$ by removing the $4$-axis; for $\mathsf{I}_{\bar{1},(1,0)}$, for instance, $(1,0)$ refers to the $(x_{2},x_{3})$-plane.

Using the quadrality symmetry, the $J$- and $E$-terms are determined to be
\bea
\begin{array}{c|cc}
  & \text{$E$-term}   &  \text{$J$-term}  \\\hline
 \Lambda_{1}  & [\mathsf{B}_{4},\mathsf{B}_{1}]-\mathsf{I}_{\bar{2},(1,0)}\mathsf{J}_{\bar{2},(1,1)}-\mathsf{I}_{\bar{3},(1,0)}\mathsf{J}_{\bar{3},(1,1)} &[\mathsf{B}_{2},\mathsf{B}_{3}]  \\
 \Lambda_{2}  & [\mathsf{B}_{4},\mathsf{B}_{2}]-\mathsf{I}_{\bar{1},(1,0)}\mathsf{J}_{\bar{1},(1,1)}-\mathsf{I}_{\bar{3},(0,1)}\mathsf{J}_{\bar{3},(1,1)}&[\mathsf{B}_{3},\mathsf{B}_{1}]\\
 \Lambda_3 &[\mathsf{B}_{4},\mathsf{B}_{3}]-\mathsf{I}_{\bar{1},(0,1)}\mathsf{J}_{\bar{1},(1,1)}-\mathsf{I}_{\bar{2},(0,1)}\mathsf{J}_{\bar{2},(1,1)} &[\mathsf{B}_{1},\mathsf{B}_{2}]\\
 \Lambda_{\bar{1},(1,0)} & \mathsf{B}_{1}\mathsf{I}_{\bar{1},(1,0)}  & \mathsf{J}_{\bar{1},(1,1)}\mathsf{B}_{3}   \\
 \Lambda_{\bar{1},(0,1)} & \mathsf{B}_{1}\mathsf{I}_{\bar{1},(0,1)} &-\mathsf{J}_{\bar{1},(1,1)}\mathsf{B}_{2}\\
 \widetilde{\Lambda}_{\bar{1},(1,1)}  & -\mathsf{J}_{\bar{1},(1,1)}\mathsf{B}_{1}& - \mathsf{B}_{2}\mathsf{I}_{\bar{1},(0,1)} +\mathsf{B}_{3} \mathsf{I}_{\bar{1},(1,0)}\\
 \Lambda_{\bar{2},(1,0)} &  \mathsf{B}_{2}\mathsf{I}_{\bar{2},(1,0)} &-\mathsf{J}_{\bar{2},(1,1)}\mathsf{B}_{3}\\
 \Lambda_{\bar{2},(0,1)} &\mathsf{B}_{2}\mathsf{I}_{\bar{2},(0,1)}&\mathsf{J}_{\bar{2},(1,1)}\mathsf{B}_{1}\\
 \widetilde{\Lambda}_{\bar{2},(1,1)} &-\mathsf{J}_{\bar{2},(1,1)}\mathsf{B}_{2} &\mathsf{B}_{1}\mathsf{I}_{\bar{2},(0,1)} -\mathsf{B}_{3} \mathsf{I}_{\bar{2},(1,0)} \\
 \Lambda_{\bar{3},(1,0)} &\mathsf{B}_{3}\mathsf{I}_{\bar{3},(1,0)}&\mathsf{J}_{\bar{3},(1,1)}\mathsf{B}_{2}\\
 \Lambda_{\bar{3},(0,1)} &\mathsf{B}_{3}\mathsf{I}_{\bar{3},(0,1)}&-\mathsf{J}_{\bar{3},(1,1)}\mathsf{B}_{1}\\
 \widetilde{\Lambda}_{\bar{3},(1,1)} &-\mathsf{J}_{\bar{3},(1,1)}\mathsf{B}_{3}& -\mathsf{B}_{1}\mathsf{I}_{\bar{3},(0,1)} +
 \mathsf{B}_{2} \mathsf{I}_{\bar{3},(1,0)}
\end{array}
\eea

\paragraph{Superpotential for PT$_3$ counting}
In addition to the superpotential terms, the moduli space of the quiver quantum mechanics is governed by the $D$-terms. As discussed in \cite{Kimura:2025lfo,Kimura:2025lig}, the PT counting is reproduced by the change of reference vector $(1,\ldots,1)\rightarrow (-1,\ldots,-1)$, or equivalently by flipping the FI parameter $\zeta\rightarrow -\zeta$. Alternatively, one may construct a different quiver quantum mechanics that reproduces the same PT counting, obtained simply by reversing the orientation of the arrows between the flavor and gauge nodes. The two descriptions are expected to be related by quiver mutation.

Here, we illustrate this second route for the one-leg, one-box PT counting $\Yboxdim{6pt}(\varnothing,\varnothing,\yng(1))$. As noted in the main text \eqref{eq:PT-oneleg-poles}, the surviving configurations are columns of boxes extending in a single direction. In the plane-partition language, this is a pit reduction with pits at the positions $(2,1,1)$ and $(1,2,1)$. In general, a pit at position $(l+1,m+1,n+1)$ is imposed by introducing an antifundamental chiral multiplet $\mathsf{J}$ and modifying the D6--D0 superpotential by
\bea
\delta\mathsf{W}=\Tr \,(\mathsf{J}_{(l,m,n)}\mathsf{B}_{1}^{l}\mathsf{B}_{2}^{m}\mathsf{B}_{3}^{n}\mathsf{I}).
\eea
For the one-leg one-box PT counting, we therefore introduce two multiplets $\mathsf{J}_{(1,0)},\mathsf{J}_{(0,1)}$, corresponding to the pits at $(2,1,1)$ and $(1,2,1)$, respectively, and the superpotential is modified by
\bea
\delta\mathsf{W}=\Tr \left[\left(\mathsf{J}_{(1,0)} \mathsf{B}_{1}+\mathsf{J}_{(0,1)}\mathsf{B}_{2}\right)\mathsf{I}\right].
\eea
Up to a sign and a relabeling of the fields, this is the superpotential \eqref{eq:DT3-oneleg-superpotential} with the roles of the fundamental and antifundamental multiplets exchanged. We note that this model was also studied in \cite[Sec.~3.5]{Rapcak:2021hdh}. 

\section{Gluing of one-leg PT vertices}\label{app:gluing}
In this appendix, we review the gluing procedure for the one-leg PT vertices constructed in Section~\ref{sec:DTPT-vortexQM}. The local vertex keeps the asymptotic data fixed; gluing sums over their common value and supplies the edge character and the classical charge of the reference configuration. We review rank-one and rank-$N$ PT$_{3}$ invariants in Appendices~\ref{app:rankoneDTPTvertex} and~\ref{app:rankNDTPTvertex}, propose their tetrahedron generalization in Appendix~\ref{app:tetrahedronDTPTvertex}, and review the PT$_{4}$ vertex formalism in Appendix~\ref{app:PT4vertexformalism}. Plethystic formulas for the PT invariants of local curves are collected in Appendix~\ref{app:PE-local}.

Following the main text, we introduce
\bea
\bfP_{a}=1-q_{a},\quad \bfP_{\mathcal{S}}=\prod_{a\in\mathcal{S}}\bfP_{a}
\eea
where $\mathcal{S}$ is a subset of the four-set $\four=\{1,2,3,4\}$.

For a finite virtual torus character $V=\sum_w n_w w$, define
\begin{equation}\label{eq:ahat-definition}
[w]=w^{1/2}-w^{-1/2},\qquad
\widehat a[V]=\prod_w[w]^{-n_w},\qquad
\widehat a[V-W]=\widehat a[V]/\widehat a[W].
\end{equation}
Square roots are fixed consistently on the double cover of the equivariant torus. This is the symmetrized multiplicative Euler factor used below. The characters $\bfN$ denote framing bundles, $\bfK$ tautological bundles, and $\mathbf e$ their asymptotic leg characters. A superscript $\vee$ dualizes all weights. The edge character is formed by combining the two local rational expressions before applying $\widehat a$; this cancels the infinite leg contribution. In the fourfold theory, a polarization and an orientation of the virtual determinant are additional data.

The bundles $\mathcal O(-r_i)$ are the three normal line bundles of the zero section $\mathbb P^1$. Their degrees add to $-2$, so $K_Z$ is trivial: for $E=\bigoplus_i\mathcal O(-r_i)$, $K_{\operatorname{Tot}E}=\pi^*(K_{\mathbb P^1}\otimes\det E^{-1})$. Equivariant lifts of these bundles give the transition factors in \eqref{eq:localcurveCY4}.

Let $Z$ be a toric Calabi--Yau fourfold and $\Delta(Z)$ its moment polytope, whose sets of vertices and edges we denote, abusing notation, by $\Delta_{0}$ and $\Delta_{1}$. Vertices $\mathsf{v}\in\Delta_{0}$ correspond to the $(\mathbb{C}^{\ast})^{4}$ fixed points of $Z$, while edges $\mathsf{e}\in\Delta_{1}$ correspond to $(\mathbb{C}^{\ast})^{4}$-invariant curves: a noncompact edge gives a line $\mathbb{C}\subset Z$, and a compact edge gives a $\mathbb{P}^{1}\subset Z$ with normal bundle
\bea
N_{\mathbb{P}^{1}/Z}\simeq \mathcal{O}(-r_{1})\oplus \mathcal{O}(-r_{2})\oplus \mathcal{O}(-r_{3}),\quad r_{1}+r_{2}+r_{3}=2.
\eea
At each vertex, we choose local equivariant parameters $q_{i}^{\mathsf{v}}$ for the $(\mathbb{C}^{\ast})^{4}$ action on the corresponding open chart. Along an edge, the two sets of local parameters are related by the transition rules
\bea\label{eq:localcurveCY4}
\adjustbox{valign=c}{
\begin{tikzpicture}
    \draw[thick, postaction={mid arrow}] (0,0)--(-1.3,1.3);
    \draw[thick, postaction={mid arrow}] (0,0)--(-1.7,-0.1);
    \draw[thick, postaction={mid arrow}] (0,0)--(0,-1.5);
    \draw[thick, postaction={mid arrow}] (0,0)--(1,0);
    \draw[thick, postaction={mid arrow}] (2,0)--(1,0);
    \draw[thick, postaction={mid arrow}] (2,0)--(3.3,0.1);
    \draw[thick, postaction={mid arrow}] (2,0)--(2.4, 1.3);
    \draw[thick, postaction={mid arrow}] (2,0)--(2.4,-1.5);
    \fill (0,0) circle (2pt);
    \node[below left] at (0,0){$\mathsf{v}_{1}$};
    \fill (2,0) circle (2pt);
    \node[below right] at (2,0){$\mathsf{v}_{2}$};
    \node[above] at (1,0){$\mathsf{e}$};
    \node[above right] at (-0.9,0.9){$q_{1}^{\mathsf{v_{1}}}$};
    \node[below left] at (0,-0.6){$q_{3}^{\mathsf{v_{1}}}$};
    \node[below left] at (-1.7,-0.1){$q_{2}^{\mathsf{v_{1}}}$};
    \node[below] at (0.5,-0.1){$q_{4}^{\mathsf{v_{1}}}$};
    \node[below] at (1.5,-0.1){$q_{4}^{\mathsf{v_{2}}}$};
    \node[right] at (2.3,0.8){$q_{1}^{\mathsf{v_{2}}}$};
     \node[right] at (3.3,0.1){$q_{2}^{\mathsf{v_{2}}}$};
    \node[right] at (2.3,-0.8){$q_{3}^{\mathsf{v_{2}}}$};
\end{tikzpicture}}\qquad \begin{cases}q_{1}^{\mathsf{v}_{2}}=(q_{4}^{\mathsf{v}_{1}})^{r_{1}}q_{1}^{\mathsf{v}_{1}},\\
    q_{2}^{\mathsf{v}_{2}}=(q_{4}^{\mathsf{v}_{1}})^{r_{2}}q_{2}^{\mathsf{v}_{1}},\\
     q_{3}^{\mathsf{v}_{2}}=(q_{4}^{\mathsf{v}_{1}})^{r_{3}}q_{3}^{\mathsf{v}_{1}},\\
    q_{4}^{\mathsf{v}_{2}}=(q_{4}^{\mathsf{v}_{1}})^{-1}.
    \end{cases}
\eea
We are mainly interested in the local curve $Z=\operatorname{Tot}\,(\mathcal{O}(-r_{1})\oplus \mathcal{O}(-r_{2})\oplus \mathcal{O}(-r_{3})\rightarrow \mathbb{P}^{1})$, whose polytope has just two vertices joined by a single edge.

\subsection{Rank-one \texorpdfstring{PT$_{3}$}{PT3} invariants}\label{app:rankoneDTPTvertex}
We set $r_{3}=0$ and take a single D6-brane wrapping the Calabi--Yau threefold $X=\operatorname{Tot}(\mathcal{O}(-r_{1})\oplus \mathcal{O}(-r_{2})\rightarrow \mathbb{P}^{1})$ with $r_{1}+r_{2}=2$. Its D6--D2--D0 bound states compute the rank-one equivariant PT$_{3}$ partition functions. Since the D6-brane does not extend along the direction with weight $q_{3}$, we omit that leg and draw the geometry as
\bea\label{eq:localcurveCY3}
\begin{tikzpicture}
    \draw[thick, postaction={mid arrow}] (0,0)--(-1.3,1.3);
    \draw[thick, postaction={mid arrow}] (0,0)--(0,-1.5);
    \draw[thick, postaction={mid arrow}] (0,0)--(1,0);
    \draw[thick, postaction={mid arrow}] (2,0)--(1,0);
    \draw[thick, postaction={mid arrow}] (2,0)--(2.4, 1.3);
    \draw[thick, postaction={mid arrow}] (2,0)--(2.4,-1.5);
    \fill (0,0) circle (2pt);
    \node[below left] at (0,0){$\mathsf{v}_{1}$};
    \fill (2,0) circle (2pt);
    \node[below right] at (2,0){$\mathsf{v}_{2}$};
    \node[above] at (1,0){$\mathsf{e}$};
    \node[above right] at (-0.9,0.9){$q_{1}$};
    \node[below left] at (0,-0.6){$q_{2}$};
    \node[below] at (0.5,-0.1){$q_{4}$};
    \node[below] at (1.5,-0.1){$\tilde{q}_{4}$};
    \node[right] at (2.3,0.8){$\tilde{q}_{1}$};
    \node[right] at (2.3,-0.8){$\tilde{q}_{2}$};
\end{tikzpicture}
\eea
with 
\bea
\tilde{q}_{1}=q_{1}q_{4}^{r_{1}},\quad \tilde{q}_{2}=q_{2}q_{4}^{r_{2}},\quad \tilde{q}_{3}=q_{3},\quad \tilde{q}_{4}=q_{4}^{-1}.
\eea

Given such a diagram, the PT partition function of the local geometry is assembled as follows. 
\begin{enumerate}
    \item We assign a Young diagram to each edge of the diagram. An edge carrying a vertex at both ends is assigned a Young diagram that is summed over at the end; an edge with a free end is assigned the trivial Young diagram.
    \bea\label{eq:localcurveCY3-YD}
\begin{tikzpicture}
    \draw[thick, postaction={mid arrow}] (0,0)--(-1.3,1.3);
    \draw[thick, postaction={mid arrow}] (0,0)--(0,-1.5);
    \draw[thick, postaction={mid arrow}] (0,0)--(1,0);
    \draw[thick, postaction={mid arrow}] (2,0)--(1,0);
    \draw[thick, postaction={mid arrow}] (2,0)--(2.4, 1.3);
    \draw[thick, postaction={mid arrow}] (2,0)--(2.4,-1.5);
    \fill (0,0) circle (2pt);
    \fill (2,0) circle (2pt);
    \node[above] at (1,0){\Large$\nu$};
    \node[above right] at (-0.6,0.6){$q_{1}$};
    \node[above left] at (-1.3,1.3){\Large{$\varnothing$}};
    \node[above right] at (2.4,1.3){\Large${\varnothing}$};
    \node[below right] at (2.4,-1.5){\Large${\varnothing}$};
    \node[below left] at (0,-1.5){\Large${\varnothing}$};
    \node[below left] at (0,-0.6){${q}_{2}$};
    \node[below] at (0.5,-0.1){$q_{4}$};
    \node[below] at (1.5,-0.1){$\tilde{q}_{4}$};
    \node[right] at (2.3,0.8){$\tilde{q}_{1}$};
    \node[right] at (2.3,-0.8){$\tilde{q}_{2}$};
\end{tikzpicture}
\eea

\item \textbf{Vertex term:} To each vertex, we assign the PT$_{3}$ vertex, evaluated on the local equivariant parameters of that vertex. For the left vertex, for instance,
\bea
\begin{tikzpicture}
    \draw[thick, postaction={mid arrow}] (0,0)--(-1.3,1.3);
    \draw[thick, postaction={mid arrow}] (0,0)--(0,-1.5);
    \draw[thick, postaction={mid arrow}] (0,0)--(1.3,0);
    \fill (0,0) circle (2pt);
    \node[above] at (1,0){\Large$\nu$};
    \node[above right] at (-0.6,0.6){$q_{1}$};
    \node[above left] at (-1.3,1.3){\Large{$\varnothing$}};
    \node[below left] at (0,-1.5){\Large${\varnothing}$};
    \node[below left] at (0,-0.6){${q}_{2}$};
    \node[below] at (0.5,-0.1){$q_{4}$};

    \node at (4,0){\large$\mathsf{PT}_{\bar{3};\varnothing\varnothing\nu}[\fq,q_{1},q_{2},q_{3},q_{4}]$};
\end{tikzpicture}
\eea
while for the right vertex we have
\bea
\begin{tikzpicture}
    \draw[thick, postaction={mid arrow}] (2,0)--(0.7,0);
    \draw[thick, postaction={mid arrow}] (2,0)--(2.4, 1.3);
    \draw[thick, postaction={mid arrow}] (2,0)--(2.4,-1.5);
    \fill (2,0) circle (2pt);
    \node[above] at (1,0){\Large$\nu$};
      \node[above right] at (2.4,1.3){\Large${\varnothing}$};
    \node[below right] at (2.4,-1.5){\Large${\varnothing}$};
   
\node[below] at (1.5,-0.1){$\tilde{q}_{4}$};
    \node[right] at (2.3,0.8){$\tilde{q}_{1}$};
    \node[right] at (2.3,-0.8){$\tilde{q}_{2}$};
    
    \node at (6,0){\large$\mathsf{PT}_{\bar{3};\varnothing\varnothing\nu}[\fq,\tilde{q}_{1},\tilde{q}_{2},\tilde{q}_{3},\tilde{q}_{4}]$};
\end{tikzpicture}
\eea

The same ordered transverse axes are used at both vertices. If a vertex convention swaps those axes, replace $\nu$ by $\nu^{\rm T}$ and swap the corresponding equivariant parameters at the same time. The inversion of the edge parameter alone does not transpose the transverse diagram. 

\item \textbf{Edge term:} For each vertex, we introduce the character
\bea
\mathbf{E}_{\bar{3};\varnothing\varnothing\nu}(q_{1},q_{2},q_{4})&=\bfP_{3}\bfN^{\vee}\mathbf{e}_{\varnothing\varnothing\nu}(q_{1},q_{2},q_{4})-\bfP_{124}\mathbf{e}_{\varnothing\varnothing\nu}^{\vee}(q_{1},q_{2},q_{4})\mathbf{e}_{\varnothing\varnothing\nu}(q_{1},q_{2},q_{4}),\\
\bfN=v,\quad &\mathbf{e}_{\varnothing\varnothing\nu}(q_{1},q_{2},q_{4})=\frac{\boldsymbol{\nu}}{1-q_{4}^{-1}} ,\quad \boldsymbol{\nu}=\sum_{(i,j)\in\nu}vq_{1}^{-i+1}q_{2}^{-j+1}
\eea
where we identified the Young diagram $\nu$ with its character.

Combining the contributions of the two vertices, the edge character is
\bea
\mathcal{\mathbf{E}}_{\bar{3};\nu}(q_{1},q_{2}\mid \tilde{q}_{1},\tilde{q}_{2})=\mathbf{E}_{\bar{3};\varnothing\varnothing\nu}(q_{1},q_{2},q_{4})+\mathbf{E}_{\bar{3};\varnothing\varnothing\nu}(\tilde{q}_{1},\tilde{q}_{2},\tilde{q}_{4}).
\eea
The edge factor is then
\bea
\mathsf{E}_{\bar{3};\nu}(q_{1},q_{2}\mid \tilde{q}_{1},\tilde{q}_{2})&=\widehat{a}[\,\mathcal{\mathbf{E}}_{\bar{3};\nu}(q_{1},q_{2}\mid \tilde{q}_{1},\tilde{q}_{2})]
\eea

\item \textbf{Topological term:} There are two topological terms, one counting D2-branes and the other counting D0-branes.

\begin{itemize}
    \item D2 topological term: For each edge $\mathsf{e}$, we introduce a counting parameter $Q_{\mathsf{e}}$ which counts the number of boxes
\bea
    Q_{\mathsf{e}}^{|\nu|}.
\eea
    They correspond to the K\"ahler parameters.

    \item D0 topological term: the edges also carry D0-brane contributions, counting the boxes of the leg $\nu$:
    \bea
   \left| \frac{\boldsymbol{\nu}}{1-q_{4}^{-1}}  \right|
    \eea
    Taken on its own this is infinite, but combined with the contribution from the vertex at the other end of the edge it becomes finite:
      \bea
\frac{\boldsymbol{\nu}}{1-q_{4}^{-1}}+\frac{\tilde{\boldsymbol{\nu}}}{1-\tilde{q}_{4}^{-1}}&=\sum_{(i,j)\in\nu}vq_{1}^{-i+1}q_{2}^{-j+1}\frac{1-q_{4}^{-r_{1}(i-1)-r_{2}(j-1)-1}}{1-q_{4}^{-1}}\\
&=\sum_{(i,j)\in\nu}vq_{1}^{-i+1}q_{2}^{-j+1}(1+q_{4}^{-1}+q_{4}^{-2}+\cdots +q_{4}^{-r_{1}(i-1)-r_{2}(j-1)})
\eea
where we used the parameter correspondence
\bea
\tilde{q}_{4}=q_{4}^{-1},\quad \tilde{q}_{2}=q_{4}^{r_{2}}q_{2},\quad \tilde{q}_{1}=q_{4}^{r_{1}}q_{1},\quad r_{1}+r_{2}=2.
\eea
When $r_1$ or $r_2$ is negative, the displayed finite sum is understood as the Laurent polynomial $(1-q_4^{-d})/(1-q_4^{-1})$ with $d=r_1(i-1)+r_2(j-1)+1$. For $d<0$, it is a negative character of first cohomology, not a positive number of boxes. Its value at $q_4=1$ is still $d$. The regularized size is then
\bea
\chi(\nu,(r_{1},r_{2}))&=\left| \frac{\boldsymbol{\nu}}{1-q_{4}^{-1}}+\frac{\tilde{\boldsymbol{\nu}}}{1-\tilde{q}_{4}^{-1}}  \right|=\left(\frac{\boldsymbol{\nu}}{1-q_{4}^{-1}}+\frac{\tilde{\boldsymbol{\nu}}}{1-\tilde{q}_{4}^{-1}}\right)|_{v,q_{1,2,3,4}\rightarrow 1}\\
&=\sum_{(i,j)\in\nu}(r_{1}(i-1)+r_{2}(j-1)+1)
\eea

The D0 topological term is then
\bea
\fq^{\chi(\nu,(r_{1},r_{2}))}
\eea
\end{itemize}

\item \textbf{PT$_{3}$ partition function:} Combining the vertex, edge, and topological terms, the PT$_{3}$ partition function is
\bea
\mathsf{PT}[X]&=\sum_{\nu}Q^{|\nu|}\fq^{\chi(\nu,(r_{1},r_{2}))}\mathsf{E}_{\bar{3};\nu}(q_{1},q_{2}\mid \tilde{q}_{1},\tilde{q}_{2})\\ &\qquad\times\mathsf{PT}_{\bar{3};\varnothing\varnothing\nu}[\fq,q_{1},q_{2},q_{3},q_{4}]\mathsf{PT}_{\bar{3};\varnothing\varnothing\nu}[\fq,\tilde{q}_{1},\tilde{q}_{2},\tilde{q}_{3},\tilde{q}_{4}].
\label{eq:PT3-gluing-rankone}
\eea

\end{enumerate}

\subsection{Rank \texorpdfstring{$N$}{N} \texorpdfstring{PT$_{3}$}{PT3} invariants}\label{app:rankNDTPTvertex}
The generalization to higher rank is straightforward. We again set $r_{3}=0$, and now wrap $N$ D6-branes on $\operatorname{Tot}(\mathcal{O}(-r_{1})\oplus \mathcal{O}(-r_{2})\rightarrow \mathbb{P}^{1})$. The edge now carries an $N$-tuple of Young diagrams $\vec{\nu}=(\nu^{(\alpha)})_{\alpha=1}^{N}$, and we introduce flavor fugacities $\vec{v}=(v_{\alpha})_{\alpha=1}^{N}$ for the D6-branes.

\paragraph{Vertex term}
The vertex term is simply replaced by its higher-rank counterpart
\bea
\mathsf{PT}_{\vec{\nu}}[\fq,q_{1,2,3,4}]
\eea
where the boxes extend along the $4$-direction over the base $\vec{\nu}$ in the $12$-plane. 

\paragraph{Edge term}
The edge contribution comes from the character
\bea
\mathcal{\mathbf{E}}_{\bar{3};\vec{\nu}}(q_{1},q_{2}\mid \tilde{q}_{1},\tilde{q}_{2})=\mathbf{E}_{\bar{3};\vec{\varnothing}\vec{\varnothing}\vec{\nu}}(q_{1},q_{2},q_{4})+\mathbf{E}_{\bar{3};\vec{\varnothing}\vec{\varnothing}\vec{\nu}}(\tilde{q}_{1},\tilde{q}_{2},\tilde{q}_{4}).
\eea
where
\bea
\mathbf{E}_{\bar{3};\vec{\varnothing}\vec{\varnothing}\vec{\nu}}(q_{1},q_{2},q_{4})&=\bfP_{3}\bfN^{\vee}\mathbf{e}_{\vec{\varnothing}\vec{\varnothing}\vec{\nu}}(q_{1},q_{2},q_{4})-\bfP_{124}\mathbf{e}_{\vec{\varnothing}\vec{\varnothing}\vec{\nu}}^{\vee}(q_{1},q_{2},q_{4})\mathbf{e}_{\vec{\varnothing}\vec{\varnothing}\vec{\nu}}(q_{1},q_{2},q_{4}),\\
\bfN=\sum_{\alpha=1}^{N}v_{\alpha},\quad &\mathbf{e}_{\vec{\varnothing}\vec{\varnothing}\vec{\nu}}(q_{1},q_{2},q_{4})=\frac{\vec{\boldsymbol{\nu}}}{1-q_{4}^{-1}} ,\quad \boldsymbol{\vec{\nu}}=\sum_{\alpha=1}^{N}\sum_{(i,j)\in\nu^{(\alpha)}}v_{\alpha}q_{1}^{-i+1}q_{2}^{-j+1}
\eea
Taking the index, the edge factor is
\bea\label{eq:rankN-PT-edge}
\mathsf{E}_{\bar{3};\vec{\nu}}(q_{1},q_{2}\mid \tilde{q}_{1},\tilde{q}_{2})&=\widehat{a}[\,\mathcal{\mathbf{E}}_{\bar{3};\vec{\nu}}(q_{1},q_{2}\mid \tilde{q}_{1},\tilde{q}_{2})]
\eea

\paragraph{Topological term}
The topological term is modified to
\bea
Q^{|\vec{\nu}|}=Q^{\sum_{\alpha=1}^{N}|\nu^{(\alpha)}|}
\eea
for the D2-branes, and
\bea
\fq^{\chi(\vec{\nu},(r_{1},r_{2}))}=\fq^{\sum_{\alpha=1}^{N}\chi(\nu^{(\alpha)},(r_{1},r_{2}))}
\eea
for the D0-branes.

\paragraph{Rank-\texorpdfstring{$N$}{N} \texorpdfstring{PT$_3$}{PT3} partition function}
Putting everything together, we have
\bea
\mathsf{PT}[X;N]&=\sum_{\vec{\nu}}Q^{|\vec{\nu}|}\fq^{\chi(\vec{\nu},(r_{1},r_{2}))}\mathsf{E}_{\bar{3};\vec{\nu}}(q_{1},q_{2}\mid \tilde{q}_{1},\tilde{q}_{2})\mathsf{PT}_{\vec{\nu}}[\fq,q_{1,2,3,4}]\mathsf{PT}_{\vec{\nu}}[\fq,\tilde{q}_{1,2,3,4}].
\eea

\subsection{Tetrahedron PT invariants}\label{app:tetrahedronDTPTvertex}
We extend the usual PT vertex gluing prescription to the
tetrahedron PT vertices introduced in Section~\ref{sec:DTPT-vortexQM}.
This gives a gluing formula for the intersecting D6-brane
configurations considered in this paper.
A geometric construction of the corresponding moduli spaces
and a mathematical derivation of this gluing formula are left
for future work.

Keeping $r_{1,2,3}$ general, we wrap $(L,M,N)$ D6-branes on $\mathcal{O}(-r_{a})\oplus\mathcal{O}(-r_{b})\rightarrow \mathbb{P}^{1}$ for $(a,b)=(2,3),(3,1),(1,2)$, respectively. The edge now carries $(L,M,N)$-tuples of Young diagrams, collectively denoted as in the main text by $\underline{\vec{\lambda}}=(\lambda_{A}^{(\alpha)})_{A\in\{12,23,31\}}$, and the flavor fugacities of the D6-branes are collected into $\underline{\vec{v}}=(v_{\bar{a},\alpha})^{a=1,2,3}_{\alpha=1,\ldots, N_{\bar{a}}}$.

\paragraph{Vertex term}
The vertex term is replaced by 
\bea
\mathsf{PT}_{\underline{\vec{\lambda}}}[\fq,q_{1,2,3,4}].
\eea

\paragraph{Edge term}
The one-leg edge contribution of the tetrahedron setup comes from the character
\bea\label{eq:PT3-tetrahedron-edge-character}
\mathbf{E}_{\underline{\vec{\varnothing}}\underline{\vec{\varnothing}}\underline{\vec{\lambda}}}&=\frac{1}{\bfP_4^{\vee}}\left(\sum_{a,b=1}^{3}\bfP_{a}\bfN_{\bar{a}}^{\vee}\boldsymbol{\lambda}_{\overline{b4}}-\sum_{a=1}^{3}\bfP_{\overline{a4}}\boldsymbol{\lambda}_{\overline{a4}}^{\vee}\boldsymbol{\lambda}_{\overline{a4}}-\sum_{1\leq a<b\leq 3}\bfP_{123}\boldsymbol{\lambda}_{\overline{a4}}^{\vee}\boldsymbol{\lambda}_{\overline{b4}}      \right)
\eea
where
\bea
\bfN_{\bar{a}}=\sum_{\alpha=1}^{N_{\bar{a}}}v_{\bar{a},\alpha},\quad \boldsymbol{\lambda}_{A}=\sum_{\alpha=1}^{N_{A4}}\sum_{(i,j)\in\lambda_{A}^{(\alpha)}}v_{A4,\alpha}q_{a}^{-i+1}q_{b}^{-j+1},\quad A=(ab).
\eea
After combining the two patches, we get
\bea
\mathbf{E}_{\underline{\vec{\lambda}}}(q_{1,2,3}\mid \tilde{q}_{1,2,3})&=\mathbf{E}_{\underline{\vec{\varnothing}}\underline{\vec{\varnothing}}\underline{\vec{\lambda}}}(q_{1,2,3})+\mathbf{E}_{\underline{\vec{\varnothing}}\underline{\vec{\varnothing}}\underline{\vec{\lambda}}}(\tilde{q}_{1,2,3})
\eea
The contribution to the partition function is then
\bea\label{eq:PT3-tetrahedron-edgefactor}
\mathsf{E}_{\underline{\vec{\lambda}}}(q_{1,2,3}\mid \tilde{q}_{1,2,3})=\widehat{a}\left[\,\mathbf{E}_{\underline{\vec{\lambda}}}(q_{1,2,3}\mid \tilde{q}_{1,2,3})\right].
\eea

\paragraph{Topological term}
The topological term is modified to
\bea
Q^{|\underline{\vec{\lambda}}|}=Q^{\sum_{A=12,23,31}|\vec{\lambda}_{A}|}
\eea
and
\bea
\fq^{\chi(\underline{\vec{\lambda}},(r_{1},r_{2},r_{3}))},\quad \chi(\underline{\vec{\lambda}},(r_{1},r_{2},r_{3})) &=\chi(\vec{\lambda}_{23},(r_{2},r_{3}))+\chi(\vec{\lambda}_{31},(r_{3},r_{1}))+\chi(\vec{\lambda}_{12},(r_{1},r_{2})).
\eea

\paragraph{Sign rule}
For general $r_{1,2,3}$, we conjecture the sign factor
\bea\label{eq:tetrahedronPT-sign}
(-1)^{\sigma(\underline{\vec{\lambda}})},\qquad \sigma(\underline{\vec{\lambda}})=\sum_{a=1}^{3}r_{a}|\vec{\lambda}_{\overline{a4}}|.
\eea
and leave a general analysis to future work. This sign factor depends on the choice of edge term; the rule applies to the choice made above. 

For $r_{3}=0$ with $N$ D6-branes wrapping $\mathcal{O}(-r_{1})\oplus \mathcal{O}(-r_{2})\rightarrow \mathbb{P}^{1}$, the sign is trivial and we recover the rules of Appendix~\ref{app:rankNDTPTvertex}.

Such sign issues are common in the computation of PT$_{4}$ invariants (see \eqref{eq:PT4-edge-sign}). The sign rule \eqref{eq:tetrahedronPT-sign} is compatible with \eqref{eq:PT4-edge-sign}. Consider a single D6-brane normal to direction \(a\), with \(\{a,b,c\}=\{1,2,3\}\). Its asymptotic Young diagram \(\lambda_{bc}\) defines a plane partition of thickness one in
direction \(a\),
\bea
    \pi_a(\lambda_{bc})
    =
    \bigl\{
        (x_a,x_b,x_c)\in\mathbb Z_{>0}^{3}
        \,\big|\,
        x_a=1,\ (x_b,x_c)\in\lambda_{bc}
    \bigr\}.
\eea
Choose \(a\) as the distinguished direction in the D8--D2 edge
polarization. In the convention \(\bfN=(1-\mu^{-1})v\), the
specialization \(\mu=q_a\) gives \(\bfN^\vee=(1-q_a)v^{-1}\).
Writing \(\bfK\) for the leg character of \(\boldsymbol{\pi}_a(\boldsymbol{\lambda}_{bc})\),
the edge character on each patch reduces as
\bea
    \bfN^\vee \bfK-\bfP_{bc4}\bfK^\vee \bfK
    \ \xrightarrow{\ \mu=q_a\ }\
    (1-q_a)v^{-1}\bfK-\bfP_{bc4}\bfK^\vee \bfK.
\eea
The right-hand side is the single-D6 edge character in \eqref{eq:PT3-tetrahedron-edge-character}. When $r_a=0$, we have $\tilde q_a=q_a$, so the same mass specialization applies at both north and south poles of $\mathbb{P}^1$, as in Section~\ref{sec:D2-D8}.

The D8--D2 edge sign is \((-1)^{\sigma_a(\pi)}\), where
\bea
    \sigma_a(\pi)
    =
    r_a|\pi|
    +
    \#\bigl\{
        (x_a,x_b,x_c)\in\pi
        \,\big|\,
        x_b=x_c<x_a
    \bigr\}.
\eea
Every box in \(\pi_a(\lambda_{bc})\) has \(x_a=1\), so the
second term vanishes. Hence
\bea
    (-1)^{\sigma_a(\pi_a(\lambda_{bc}))}
    =
    (-1)^{r_a|\lambda_{bc}|}.
\eea
Thus the specialization reproduces both the edge character and
the edge sign for a single D6-brane.

For several D6-branes, the signs come from the product of the single-brane signs as
\bea
\sum_{a=1}^{3}\sum_{\alpha=1}^{N_{\bar{a}}}\sigma_{a}(\pi_{a}(\lambda_{\overline{a4}}^{(\alpha)}))=\sigma(\underline{\vec{\lambda}}).
\eea
We note that no additional sign comes from the mixed terms between distinct D6-branes.

\paragraph{Tetrahedron PT partition function}
Gathering everything, we obtain
\bea
\mathsf{PT}[Z;(L,M,N)]&=\sum_{\underline{\vec{\lambda}}}Q^{|\underline{\vec{\lambda}}|}\fq^{\chi(\underline{\vec{\lambda}},(r_{1},r_{2},r_{3}))}  (-1)^{\sigma(\underline{\vec{\lambda}})}\mathsf{E}_{\underline{\vec{\lambda}}}(q_{1,2,3}\mid \tilde{q}_{1,2,3})\mathsf{PT}_{\underline{\vec{\lambda}}}[\fq,q_{1,2,3,4}]\mathsf{PT}_{\underline{\vec{\lambda}}}[\fq,\tilde{q}_{1,2,3,4}].
\eea

\subsection{\texorpdfstring{PT$_{4}$}{PT4} invariants}\label{app:PT4vertexformalism}
Donaldson--Thomas theory of Calabi--Yau fourfolds has been developed both on the mathematical and on the physical side. The vertex formalism for toric Calabi--Yau fourfolds that we use below, including the sign rule reviewed in this appendix, was set up in \cite{Monavari:2022rtf}, and extended to intersecting D6-branes in \cite{Fasola:2023ypx} and to more general gauge origami configurations in \cite{Monavari:2025xrs}. On the physics side, the corresponding D8--D0 and D8--D2--D0 counting problems are the Magnificent Four of \cite{Nekrasov:2017cih}, generalized to several stacks of branes in \cite{Nekrasov:2018xsb,Nekrasov:2023nai}, and to tetrahedron instantons in \cite{Pomoni:2021hkn,Pomoni:2023nlf}. The one-leg K-theoretic fourfold vertex, which is the case relevant to us, was obtained from a 3d gauge theory in \cite{Piazzalunga:2023qik}. For toric Calabi--Yau fourfolds, the DT/PT correspondence was conjectured in cohomology~\cite{Cao:2019tnw} and in equivariant K-theory~\cite{CaoKoolMonavari:2019}. Cohomological stable-pair invariants of local Calabi--Yau fourfolds were studied in~\cite{CaoKoolMonavari:2020}. For $Z=\operatorname{Tot}(K_X)$ over a smooth threefold $X$ satisfying the hypotheses of~\cite{Bojko:2025lyz}, the DT$_4$/PT$_4$ wall-crossing formula holds at the level of equivariant virtual fundamental classes. A general wall-crossing formula for equivariant DT$_4$ invariants was established in~\cite{BojkoKuhnLiuThimm:2026}. The conjectural K-theoretic local-curve generating functions used below are instead specified in \cite{CaoKoolMonavari:2019,CaoKoolMonavari:2023CRC}, with their status summarized in Appendix~\ref{app:PE-local}.

We may also consider a physical setup with D8-branes wrapping the whole Calabi--Yau fourfold $Z$, and compute their PT$_{4}$ invariants with the same vertex formalism. The differences are that plane partitions, rather than Young diagrams, are attached to the edges, that each vertex of the moment polytope carries a PT$_{4}$ vertex, and that one has to keep track of the signs arising from the choice of orientation of the moduli space. 

We focus on the local curve $Z=\operatorname{Tot}\,(\mathcal{O}(-r_{1})\oplus \mathcal{O}(-r_{2})\oplus \mathcal{O}(-r_{3})\rightarrow \mathbb{P}^{1})$ with a single D8--$\overline{\D8}$ pair, of flavor fugacities $v$ and $\mu^{-1}v$, respectively.\footnote{Thus the antibrane-to-brane weight ratio is $\mu^{-1}$, and the virtual framing character is $(1-\mu^{-1})v$. A convention using the ratio itself as the mass variable requires $\mu\mapsto\mu^{-1}$ in every framing and Euler factor.} The higher-rank generalization is straightforward.
\begin{enumerate}
    \item We assign a plane partition to each edge of the diagram. As before, an edge with a free end is assigned the empty plane partition. 
    \bea
\adjustbox{valign=c}{
\begin{tikzpicture}
    \draw[thick, postaction={mid arrow}] (0,0)--(-1.3,1.3);
    \draw[thick, postaction={mid arrow}] (0,0)--(-1.7,-0.1);
    \draw[thick, postaction={mid arrow}] (0,0)--(0,-1.5);
    \draw[thick, postaction={mid arrow}] (0,0)--(1,0);
    \draw[thick, postaction={mid arrow}] (2,0)--(1,0);
    \draw[thick, postaction={mid arrow}] (2,0)--(3.3,0.1);
    \draw[thick, postaction={mid arrow}] (2,0)--(2.4, 1.3);
    \draw[thick, postaction={mid arrow}] (2,0)--(2.4,-1.5);
    \fill (0,0) circle (2pt);
    \fill (2,0) circle (2pt);
    \node[above] at (1,0){\Large$\pi$};
    \node[above right] at (-0.9,0.9){$q_{1}$};
    \node[below left] at (0,-0.6){$q_{3}$};
    \node[below left] at (-0.5,-0.1){$q_{2}$}; 
    \node[above left] at (-1.3,1.3){\Large$\varnothing$};
    \node[below] at (0,-1.5){\Large$\varnothing$};
    \node[left] at (-1.7,-0.1){\Large$\varnothing$};
    \node[below] at (0.5,-0.1){$q_{4}$};
   \node[right] at (2.3,0.8){$\tilde{q}_{1}$};
     \node[below] at (2.5,0.){$\tilde{q}_{2}$};
     \node[below] at (1.8,-0.1){$\tilde{q}_{4}$};
    \node[right] at (2.3,-1){$\tilde{q}_{3}$};
    
    \node[right] at (2.4,1.3){\Large$\varnothing$};
     \node[right] at (3.3,0.1){\Large$\varnothing$};
    \node[below] at (2.3,-1.5){\Large$\varnothing$};
\end{tikzpicture}}
\eea
Here, $\pi$ is the plane partition, and $q_{1,2,3,4}$ ($\tilde{q}_{1,2,3,4}$) are the local equivariant parameters associated with each vertex.

\item \textbf{Vertex term:} To each vertex, we assign the PT$_{4}$ vertex, evaluated on the local equivariant parameters:
  \bea
\adjustbox{valign=c}{
\begin{tikzpicture}
    \draw[thick, postaction={mid arrow}] (0,0)--(-1.3,1.3);
    \draw[thick, postaction={mid arrow}] (0,0)--(-1.7,-0.1);
    \draw[thick, postaction={mid arrow}] (0,0)--(0,-1.5);
    \draw[thick, postaction={mid arrow}] (0,0)--(1.5,0);
    \fill (0,0) circle (2pt);
    \node[above] at (1,0){\Large$\pi$};
    \node[above right] at (-0.9,0.9){$q_{1}$};
    \node[below left] at (0,-0.6){$q_{3}$};
    \node[below left] at (-0.5,-0.1){$q_{2}$}; 
    \node[above left] at (-1.3,1.3){\Large$\varnothing$};
    \node[below] at (0,-1.5){\Large$\varnothing$};
    \node[left] at (-1.7,-0.1){\Large$\varnothing$};
    \node[below] at (0.5,-0.1){$q_{4}$};

        \node  at (4,-0){\large$\mathsf{PT4}_{\varnothing\varnothing\varnothing\pi}[\fq,\mu,q_{1},q_{2},q_{3},q_{4}]$};
    
\end{tikzpicture}}\\
\adjustbox{valign=c}{
\begin{tikzpicture}
    \draw[thick, postaction={mid arrow}] (2,0)--(0.5,0);
    \draw[thick, postaction={mid arrow}] (2,0)--(3.3,0.1);
    \draw[thick, postaction={mid arrow}] (2,0)--(2.4, 1.3);
    \draw[thick, postaction={mid arrow}] (2,0)--(2.4,-1.5);
    \fill (2,0) circle (2pt);
    \node[above] at (1,0){\Large$\pi$};
   \node[right] at (2.3,0.8){$\tilde{q}_{1}$};
     \node[below] at (2.5,0.){$\tilde{q}_{2}$};
     \node[below] at (1.8,-0.1){$\tilde{q}_{4}$};
    \node[right] at (2.3,-1){$\tilde{q}_{3}$};
    
    \node[right] at (2.4,1.3){\Large$\varnothing$};
     \node[right] at (3.3,0.1){\Large$\varnothing$};
    \node[below] at (2.3,-1.5){\Large$\varnothing$};
    
  \node  at (6.5,0.5){\large$\mathsf{PT4}_{\varnothing\varnothing\varnothing\pi}[\fq,\mu,\tilde{q}_{1},\tilde{q}_{2},\tilde{q}_{3},\tilde{q}_{4}]$};
\end{tikzpicture}}
\eea
For the explicit DT$_4$/PT$_4$ vertex formulas, see \cite{Nekrasov:2023nai,Liu:FourfoldPTVertex,Kimura:2025lig,CaoKoolMonavari:2019}. The transverse coordinate order is kept fixed at both ends of an edge, while the tangent weight is inverted by \eqref{eq:localcurveCY4}. Permuting two transverse axes requires the same permutation of the plane partition. This combinatorial convention is separate from the virtual-determinant orientation encoded in the sign rule below.

\item \textbf{Edge term:} The edge contribution comes from the character
\bea
\mathbf{E}^{(a)}_{\varnothing\varnothing\varnothing\pi}(q_{1},q_{2},q_{3},q_{4})
&=\bfN^{\vee}\mathbf{e}_{\varnothing\varnothing\varnothing\pi}
-\bfP_{bc4}\mathbf{e}_{\varnothing\varnothing\varnothing\pi}^{\vee}\mathbf{e}_{\varnothing\varnothing\varnothing\pi},
\eea
where $\{a,b,c\}=\{1,2,3\}$ and
\[
\bfN=(1-\mu^{-1})v,\qquad
\mathbf{e}_{\varnothing\varnothing\varnothing\pi}=\frac{\boldsymbol{\pi}}{1-q_{4}^{-1}},\qquad
\boldsymbol{\pi}=\sum_{(i,j,k)\in\pi}vq_{1}^{-i+1}q_{2}^{-j+1}q_{3}^{-k+1}.
\]
Combining the two vertex contributions gives
\bea
\mathbf{E}^{(a)}_{\pi}(q_{1},q_{2},q_{3}\mid\tilde{q}_{1},\tilde{q}_{2},\tilde{q}_{3})=\mathbf{E}^{(a)}_{\varnothing\varnothing\varnothing\pi}(q_{1},q_{2},q_{3},q_{4})+\mathbf{E}^{(a)}_{\varnothing\varnothing\varnothing\pi}(\tilde{q}_{1},\tilde{q}_{2},\tilde{q}_{3},\tilde{q}_{4}),\quad a=1,2,3
\eea
which is regularized by combining the two patches as a rational character before expansion. For negative bundle degrees, the corresponding character is virtual, with first cohomology counted negatively. The edge factor is its index:
\bea\label{eq:PT4-edgefactor}
\mathsf{E}^{(a)}_{\pi}(q_{1},q_{2},q_{3}\mid\tilde{q}_{1},\tilde{q}_{2},\tilde{q}_{3})=\widehat{a}[\mathbf{E}^{(a)}_{\pi}(q_{1},q_{2},q_{3}\mid\tilde{q}_{1},\tilde{q}_{2},\tilde{q}_{3})],\quad a=1,2,3
\eea

\item \textbf{Sign rule:} The edge character involves a choice of $a$, an ambiguity related to the orientation of the moduli space. The difference between the three choices is known to be compensated by suitable $\mathbb{Z}_{2}$ sign factors: for the choice $a$, the sign is $(-1)^{\sigma_{a}(\pi)}$ with
\bea
\sigma_{a}(\pi)=r_{a}|\pi|+\#\{(x_{1},x_{2},x_{3})\in\pi\mid x_{b}=x_{c}<x_{a}\},\quad \{a,b,c\}=\{1,2,3\}
\eea

Combining this with the edge term above, one forms
\bea\label{eq:PT4-edge-sign}
(-1)^{\sigma_{a}(\pi)}\mathsf{E}^{(a)}_{\pi}(q_{1},q_{2},q_{3}\mid\tilde{q}_{1},\tilde{q}_{2},\tilde{q}_{3})
\eea
which is indeed independent of the choice of $a=1,2,3$.

\item \textbf{Topological term:}
\begin{itemize}
    \item D2 topological term: It counts the number of D2-branes.
    \bea
    Q^{|\pi|}
    \eea
    Note that the orientation sign of the vertex part is already included in the PT$_4$ vertex, and is distinct from the edge sign $(-1)^{\sigma_a(\pi)}$ above.
    \item D0 topological term: write $d_{ijk}=1+r_1(i-1)+r_2(j-1)+r_3(k-1)$. As in the threefold case, this contribution comes from
    \bea
\frac{\boldsymbol{\pi}}{1-q_{4}^{-1}}+\frac{\widetilde{\boldsymbol{\pi}}}{1-\tilde{q}_{4}^{-1}}&=\sum_{(i,j,k)\in\pi}vq_{1}^{-i+1}q_{2}^{-j+1}q_{3}^{-k+1}\frac{1-q_4^{-d_{ijk}}}{1-q_4^{-1}}
    \eea
   and
   \bea
    \chi(\pi,(r_{1},r_{2},r_{3}))&=\left|\frac{\boldsymbol{\pi}}{1-q_{4}^{-1}}+\frac{\widetilde{\boldsymbol{\pi}}}{1-\tilde{q}_{4}^{-1}}\right|\\ &=\sum_{(i,j,k)\in\pi}(r_{1}(i-1)+r_{2}(j-1)+r_{3}(k-1)+1).
   \eea
   The D0 topological term is then
   \bea
    \fq^{\chi(\pi,(r_{1},r_{2},r_{3}))}.
   \eea
\end{itemize}

\item \textbf{PT$_{4}$ partition function:} The PT$_{4}$ partition function is
\bea\label{eq:PT4partfunct}
\mathsf{PT4}[Z]&=\sum_{\pi}Q^{|\pi|}\fq^{\chi(\pi,(r_{1},r_{2},r_{3}))}(-1)^{\sigma_{a}(\pi)}\mathsf{E}^{(a)}_{\pi}(q_{1},q_{2},q_{3}\mid\tilde{q}_{1},\tilde{q}_{2},\tilde{q}_{3})\\
&\times \mathsf{PT4}_{\varnothing\varnothing\varnothing\pi}[\fq,\mu,q_{1},q_{2},q_{3},q_{4}]\mathsf{PT4}_{\varnothing\varnothing\varnothing\pi}[\fq,\mu,\tilde{q}_{1},\tilde{q}_{2},\tilde{q}_{3},\tilde{q}_{4}].
\eea

\end{enumerate}

\subsection{Plethystic exponential formulas for local curves}\label{app:PE-local}
In this subsection, we collect the conjectural plethystic formulas for the generic-mass K-theoretic PT partition functions of the local curves $\operatorname{Tot}(\mathcal{O}(-r_{1})\oplus\mathcal{O}(-r_{2})\oplus \mathcal{O}(-r_{3})\rightarrow \mathbb{P}^{1})$ with $(r_{1},r_{2},r_{3})=(1,1,0),(2,0,0)$. 

\paragraph{The \texorpdfstring{$(r_{1},r_{2},r_{3})=(1,1,0)$}{(1,1,0)} case}
For this geometry, the plethystic formula of \cite[Conjecture~0.16]{CaoKoolMonavari:2019} reads, in the variables $(y,q,t_4)=(\mu,\fq,q_3)$,
\bea\label{eq:PT4-110} 
\mathsf{PT4}\left[(1,1,0);\mu,\fq,Q\right]=\PE\left[\frac{Q[\mu]}{[q_{3}][\mu^{1/2}\fq][\mu^{1/2}\fq^{-1}]}\right]
\eea
where the parameter $\mu$ controls the distance between the D8- and $\overline{\D8}$-branes. 

At $\mu=q_{3}$, the D8--$\overline{\D8}$ pair undergoes tachyon condensation, leaving a single D6-brane wrapping the resolved conifold:
\bea\label{eq:PT3-conifold}
\mathsf{PT3}\left[\mathcal{O}(-1)\oplus\mathcal{O}(-1)\rightarrow \mathbb{P}^{1};\fq,Q\right]=\PE\left[\frac{Q}{[q_{3}^{1/2}\fq][q_{3}^{1/2}\fq^{-1}]}\right].
\eea
The rank-one threefold conifold formula has an independent proof in \cite[Section~4]{KononovOkounkovOsinenko:2019}; its relation to the fourfold specialization, including the counting-variable sign, is explained in \cite[Appendix~B]{CaoKoolMonavari:2019}. With $\mathsf{q}=q_3^{1/2}\fq$ and $\mathsf{t}=q_3^{-1/2}\fq$, this agrees with the refined topological string partition function \eqref{eq:conifold-pf} of the resolved conifold~\cite{Iqbal:2007ii}. Taking in addition the limit $q_{3}\rightarrow 1$ gives its unrefined version \cite{AKMV}:
\bea
\mathsf{PT3}\left[\mathcal{O}(-1)\oplus\mathcal{O}(-1)\rightarrow \mathbb{P}^{1};\fq,Q\right]\xrightarrow{q_{3}\rightarrow 1} \PE\left[-\frac{Q}{[\fq]^{2}}\right].
\eea

If instead we take $N$ pairs with $\mu_\alpha=q_{3}$, so that $\mu=q_{3}^{N}$, the D8--$\overline{\D8}$ pairs turn into $N$ D6-branes wrapping the conifold, and we obtain the higher-rank PT$_{3}$ partition function 
\bea
    \label{eq:PT3-conifold-rankN}\mathsf{PT3}\left[\mathcal{O}(-1)\oplus\mathcal{O}(-1)\rightarrow \mathbb{P}^{1};N;\fq,Q\right]=\PE\left[\frac{Q[q_{3}^{N}]}{[q_{3}][q_{3}^{N/2}\fq][q_{3}^{N/2}\fq^{-1}]}\right].
\eea
Both routes, tachyon condensation from the PT$_{4}$ vertex and the higher-rank PT$_{3}$ vertex formalism of Appendix~\ref{app:rankNDTPTvertex}, give the same answer, and the result is independent of the flavor fugacities of the D6-branes. 

The plethystic expression \eqref{eq:PT3-conifold-rankN} agrees with the orbifold conifold series \eqref{eq:Zk_conifold_pf} at $k=N$ under $\mathsf{q}=q_{3}^{1/2}\fq^{1/N}$ and $\mathsf{t}=q_{3}^{-1/2}\fq^{1/N}$, with the same $Q$.
Thus the PT D0 fugacity is $\fq=(\mathsf q\mathsf t)^{N/2}$ in the orbifold variables.

\paragraph{The \texorpdfstring{$(r_{1},r_{2},r_{3})=(2,0,0)$}{(2,0,0)} case}
Here $Z$ is the minimal resolution of $\mathbb C^2/\mathbb Z_2$ times $\mathbb C^2$. The proposed formula is the $G=\mathbb Z_2\subset SU(2)$ case of \cite[Conjecture~4.3]{CaoKoolMonavari:2023CRC}. In that reference, set $y=\mu$, $q=\fq$ and identify the two trivial-direction weights $(t_3,t_4)$ with $(q_2,q_3)$; the plethystic formula is then
\bea\label{eq:PT4-200}
\mathsf{PT4}[(2,0,0);\mu,\fq,Q]=\PE\left[\frac{[\mu][q_{32}]}{[q_{3}][q_{2}]}\frac{Q}{[\fq\mu^{1/2}][\fq\mu^{-1/2}]}\right]
\eea

If we take $\mu=q_{3}$, this reduces to the PT$_{3}$ partition function of a D6-brane wrapping $\mathcal{O}(-2)\oplus\mathcal{O}_2\rightarrow \mathbb{P}^{1}$, where the subscripts on the trivial line bundles $\mathcal{O}_{2,3}$ record the associated equivariant parameters:
\bea
\mathsf{PT3}[\mathcal{O}(-2)\oplus\mathcal{O}_2\rightarrow \mathbb{P}^{1};\fq,Q]=\PE\left[\frac{[q_{32}]}{[q_{2}]}\frac{Q}{[\fq q_{3}^{1/2}][\fq q_{3}^{-1/2}]}\right]
\eea
which can also be obtained by considering the gluing of the rank-one PT$_{3}$ vertex. By the symmetry between the two trivial directions, taking $\mu=q_{2}$ instead yields the PT$_{3}$ partition function of a D6-brane wrapping 
$\mathcal{O}(-2)\oplus\mathcal{O}_3\rightarrow \mathbb{P}^{1}$:
\bea
\mathsf{PT3}[\mathcal{O}(-2)\oplus\mathcal{O}_3\rightarrow \mathbb{P}^{1};\fq,Q]=\PE\left[\frac{[q_{32}]}{[q_{3}]}\frac{Q}{[\fq q_{2}^{1/2}][\fq q_{2}^{-1/2}]}\right].
\eea

Returning to $\mu=q_{3}$ and taking the limit $q_{2}\rightarrow \infty$, we recover the refined topological string partition function of the same geometry, with refinement parameters $\mathsf{q}=\fq q_{3}^{1/2},\mathsf{t}=q_{3}^{-1/2}\fq$ \cite{Iqbal:2007ii}:
\bea\label{eq:Z2orbifold-refinedlimit}
\mathsf{PT3}[\mathcal{O}(-2)\oplus\mathcal{O}_2\rightarrow \mathbb{P}^{1};\fq,Q]\xrightarrow{q_{2}\rightarrow \infty}\PE\left[\frac{Qq_{3}^{1/2}}{[\fq q_{3}^{1/2}][\fq q_{3}^{-1/2}]}\right]=\PE\left[\frac{Q\sqrt{\mathsf{q}/\mathsf{t}}}{[\mathsf{q}][\mathsf{t}]}\right]
\eea
If we instead take the limit $q_{3}\rightarrow 1$, we obtain the unrefined topological string partition function~\cite{AKMV}:
\bea
\mathsf{PT3}[\mathcal{O}(-2)\oplus\mathcal{O}_2\rightarrow \mathbb{P}^{1};\fq,Q]\xrightarrow{q_{3}\rightarrow1}\PE\left[\frac{Q}{[\fq]^{2}}\right].
\eea

Imposing $\mu=q_{3}^{N}$, we obtain the rank-$N$ PT$_{3}$ partition function for $N$ D6-branes wrapping $\mathcal{O}(-2)\oplus\mathcal{O}_2\rightarrow \mathbb{P}^{1}$:
\bea
\mathsf{PT3}[\mathcal{O}(-2)\oplus\mathcal{O}_2\rightarrow \mathbb{P}^{1};N,\fq,Q]&=\PE\left[\frac{[q_{3}^{N}][q_{32}]}{[q_{3}][q_{2}]}\frac{Q}{[\fq q_{3}^{N/2}][\fq q_{3}^{-N/2}]}\right],
\eea
which can also be obtained from the higher-rank PT$_{3}$ vertex formalism of Appendix~\ref{app:rankNDTPTvertex}. Here too, the dependence on the D6-brane flavor fugacities cancels. 

Imposing $\mu=q_{2}^{M}q_{3}^{N}$, we obtain the rank-$(0,M,N)$ tetrahedron PT partition function of the geometry:
\bea
\mathsf{PT}[\mathcal{O}(-2)\oplus \mathcal{O}_{2}\oplus \mathcal{O}_{3}\rightarrow \mathbb{P}^{1};(0,M,N),\fq,Q]=\PE\left[\frac{[q_{2}^{M}q_{3}^{N}][q_{32}]}{[q_{3}][q_{2}]}\frac{Q}{[\fq q_{2}^{M/2}q_{3}^{N/2}][\fq q_{2}^{-M/2}q_{3}^{-N/2}]}\right]
\eea
where we have intersecting $M$ D6-branes wrapping $\mathcal{O}(-2)\oplus \mathcal{O}_{3}\rightarrow \mathbb{P}^{1}$ and $N$ D6-branes wrapping $\mathcal{O}(-2)\oplus \mathcal{O}_{2}\rightarrow \mathbb{P}^{1}$. This result can also be obtained from the tetrahedron DT/PT vertex formalism of Appendix~\ref{app:tetrahedronDTPTvertex}. Again, the D6-brane flavor fugacities drop out after gluing.

\section{Comparison of perturbative factors with PT edge factors}\label{app:edge-comparison}
In this appendix, we collect the one-loop factors of the half-indices of Section~\ref{sec:halfindex-def} and compare their gluing with the edge term of the PT vertex formalism.

Define the characters
\bea
\bfN_{A4}=\sum_{\alpha=1}^{N_{A4}}q_{4}^{1/2}v_{A4,\alpha},\quad \bfK=\sum_{\mathsf{a}\in\underline{\vec{\lambda}}}\tilde{s}_{\mathsf{a}}=\sum_{\mathsf{a}\in\underline{\vec{\lambda}}}q_{4}^{1/2}s_{\mathsf{a}}.
\eea
Using the $\widehat{a}$-index, the tower of modes along $\mathbb C_4$ is included by
\bea
\widehat a\left[\frac{\mathcal X}{\bfP_4^\vee}\right]
=\prod_w\prod_{n\geq0}[q_4^{-n}w]^{-n_w}=
\prod_w (w^{-1};q_4)'_\infty{}^{-n_w},
\eea
where we used the regularization of \eqref{eq:symm-q-Pochhammer}.

\paragraph{Half-index}
The classical part and one-loop perturbative part of the half-indices are generally written as
\bea
\left(\prod_{a\in\underline{\vec{\lambda}}}\tilde{s}_{a}^{\xi}\right)\widehat{a}\left[\frac{\mathcal{X}_{\underline{\vec{\lambda}}}}{\bfP_{4}^{\vee}}\right],
\eea
where $\mathcal{X}_{\underline{\vec{\lambda}}}$ is a character. For each theory, the characters are as follows. For the Neumann boundary condition, the single-hemisphere expression is kept formal whenever the fixed-point character contains a trivial weight, and $\widehat a$ is applied to the sum of the north- and south-pole characters.

\begin{itemize}
\item \textbf{Boundary condition with Dirichlet vector multiplet:}
\bea
\mathcal{X}^{\text{D}}_{\underline{\vec{\lambda}}}=q_{4}^{-1}\bfK^{\vee}\bfK+\sum_{\chi \in\mathscr{N}_{\underline{\vec{\lambda}}}}w_{\chi}^{-1}-q_{4}^{-1}\sum_{\chi \in\mathscr{D}_{\underline{\vec{\lambda}}}}w_{\chi}
\eea

\item \textbf{Boundary condition with Neumann vector multiplet \eqref{eq:halfindex-integral}:}
\bea
\mathcal{X}^{\text{N}}_{\underline{\vec{\lambda}}}&=\sum_{A\in\three}\bfP_{\overline{A4}}\bfN_{A4}^{\vee}\bfK-(1-q_{1}-q_{2}-q_{3})\bfK^{\vee}\bfK
\eea

\item \textbf{BPS/CFT boundary condition:}

\bea
\mathcal{X}^{\text{CFT}}_{\underline{\vec{\lambda}}}=\sum_{A\in\three}\bfP_{\overline{A4}}\bfN_{A4}^{\vee}\bfK-\bfP_{123}\left(\bfK^{\vee}\bfK\right)_{\succ}-\sum_{A\in\three}|\vec{\lambda}_{A}|\bfP_{A}
\eea
where we introduced
\bea
\left(\bfK^{\vee}\bfK\right)_{\succ}=\sum_{b\succ a}\frac{s_{b}}{s_{a}}
\eea
and $\succ$ is the ordering introduced in Section~\ref{sec:BPSCFT}. Note that we have the property
\bea
\bfK^{\vee}\bfK=\left(\bfK^{\vee}\bfK\right)_{\succ}+\left(\bfK^{\vee}\bfK\right)^{\vee}_{\succ}+|\underline{\vec{\lambda}}|\boldsymbol{1}
\eea

\end{itemize}

\paragraph{Comparison with the edge term}
Gluing the half-indices of the north and south hemispheres, the combined classical and one-loop contribution after removing the FI factor is
\bea
\mathcal{E}_{\text{edge}}=\widehat{a}\left[\frac{\mathcal{X}_{\underline{\vec{\lambda}}}}{\bfP_{4}^{\vee}}+\frac{\widetilde{\mathcal{X}}_{\underline{\vec{\lambda}}}}{\widetilde{\bfP}_{4}^{\vee}}\right]
\eea
This factor agrees with the edge term with the sign rule, $(-1)^{\sigma(\underline{\vec{\lambda}})}\mathsf{E}_{\underline{\vec{\lambda}}}(q_{1,2,3}\mid \tilde{q}_{1,2,3})$, up to a sign. 

\begin{itemize}
\item \textbf{Boundary condition with Dirichlet vector multiplet:} Generally, the sign difference is
\bea
(-1)^{\eta_{\text{D}}}\mathcal{E}_{\text{edge}}^{\text{D}}=(-1)^{\sigma(\underline{\vec{\lambda}})}\mathsf{E}_{\underline{\vec{\lambda}}}(q_{1,2,3}\mid \tilde{q}_{1,2,3})
\eea
where $\eta_{\text{D}}$ depends on the explicit boundary condition.

For the boundary conditions in \eqref{eq:boundary-conditions},
let $\alpha$ run over the ordered framing labels. Write
$A_\alpha=a_\alpha b_\alpha$ for the plane of framing $\alpha$,
$c_\alpha$ for its transverse direction, and
$n_\alpha=|\lambda_{A_\alpha}^{(\alpha)}|$. Under the
parity-anomaly cancellation condition \eqref{eq:tetra-parity-anomaly}, we find
\bea
\eta_{\mathrm D}
\equiv
\sum_\alpha r_{a_\alpha}n_\alpha
+\sum_{\alpha<\beta}r_{c_\alpha}n_\alpha n_\beta
\pmod{2}.
\eea

\item \textbf{Boundary condition with Neumann vector multiplet \eqref{eq:halfindex-integral}:} Let $\nu(\lambda)$ denote the number of $2\times2$ squares in
the Young diagram $\lambda$. Each square is specified by its
upper-right box, so
\bea
\nu(\lambda)
=
\#\bigl\{(i,j)\in\lambda \,\bigm|\, i\geq 2,\ j\geq 2\bigr\}.
\eea
For the tuple $\underline{\vec{\lambda}}$, we set
\bea
\nu(\underline{\vec{\lambda}})
=
\sum_{A,\alpha}\nu\bigl(\lambda_A^{(\alpha)}\bigr).
\eea
The sign difference is
\bea
(-1)^{|\underline{\vec{\lambda}}|+\nu(\underline{\vec{\lambda}})}
\mathcal{E}_{\text{edge}}^{\text{N}}
&=
(-1)^{\sigma(\underline{\vec{\lambda}})}
\mathsf{E}_{\underline{\vec{\lambda}}}
\bigl(q_{1,2,3}\mid \tilde{q}_{1,2,3}\bigr).
\eea

\item \textbf{BPS/CFT boundary condition:} For this case, the factor $\mathcal{E}_{\text{edge}}^{\text{CFT}}$ matches the edge term $\mathsf{E}_{\underline{\vec{\lambda}}}(q_{1,2,3}\mid \tilde{q}_{1,2,3})$ as
\bea
\mathcal{E}_{\text{edge}}^{\text{CFT}}=\mathsf{E}_{\underline{\vec{\lambda}}}(q_{1,2,3}\mid \tilde{q}_{1,2,3})
\eea
and so
\bea
(-1)^{\sigma(\underline{\vec{\lambda}})}\mathcal{E}_{\text{edge}}^{\text{CFT}}=(-1)^{\sigma(\underline{\vec{\lambda}})}\mathsf{E}_{\underline{\vec{\lambda}}}(q_{1,2,3}\mid \tilde{q}_{1,2,3}).
\eea

\end{itemize}

\section{Type IIA topological string on \texorpdfstring{$\mathbb{C}^2/\mathbb{Z}_k \times \text{CY}_{3}$}{C2/Zk times CY3}}\label{app:orbifold-topstring}
In this appendix, we obtain the rank dependence of the plethystic formula \eqref{eq:PT3-conifold-rankN} from a $\Z_k$ orbifold projection of the refined topological string. At $k=N$, the projected conifold partition function reproduces the rank-$N$ series \eqref{eq:PT3-conifold-rankN}, and the projection of single-particle states gives the transverse character $\mathcal K_N$ of the membrane index in Section~\ref{sec:rankone-twistedindex}.

Following~\cite{KimuraTaki:2013}, we start from M-theory on $\mathrm{TN}_k\times X\times\mathbb S^1$, where $X$ is a Calabi--Yau threefold and $\mathrm{TN}_k$ is the $k$-centered Taub--NUT space. Reducing on $\mathbb S^1$ gives type IIA string theory on $X$ with transverse space $\mathrm{TN}_k$, which becomes the orbifold $\C^2/\Z_k$ when the centers coincide and the Taub--NUT radius is large. Reducing instead on the circle fiber of $\mathrm{TN}_k$ gives $k$ D6-branes wrapping $X\times\mathbb S^1$. The exchange of the two reductions is the 9--11 flip~\cite{Dijkgraaf:2006um}, and it relates the orbifold order $k$ to the D6-brane rank. We state the projection rule in Appendix~\ref{app:orb-idea}, recall the M-theory setup in Appendix~\ref{sec:M-theory}, and describe the flip in Appendix~\ref{app:orb-6dTFT}. Appendices~\ref{app:orb-conifold} and~\ref{app:orb-CS} treat the resolved conifold and its geometric transition to Chern--Simons theory on $\mathbb{S}^{3}$.

\subsection{Main idea}\label{app:orb-idea}

The A-model topological string on a noncompact toric Calabi--Yau threefold $X$ geometrically engineers a 4d $\mathcal{N}=2$ gauge theory on $\R^{4}\cong\C^{2}$ in the self-dual $\Omega$-background~\cite{Katz:1996fh}. The full amplitude on $X$ is the partition function of the corresponding 5d gauge theory on $\C^2\times\mathbb S^1$, and the refined amplitude allows generic equivariant parameters $\epsilon_{1}$ and $\epsilon_{2}$~\cite{Awata:2005fa,Iqbal:2007ii}. In this construction, the gauge theory lives on the factor $\C^{2}$ of the spacetime $X\times\C^2$, so replacing $\C^{2}$ by another four-manifold $Y$ engineers the gauge theory on $Y$. We take $Y = \C^{2}/\Gamma$ with $\Gamma \subset SU(2)$ finite, and restrict to the type $A_{k-1}$ case $\Gamma = \Z_{k}$.

The refined partition function on $X \times \C^{2}$ is a meromorphic function of $(\mathsf{q}, \mathsf{t}) = (e^{\epsilon_1},e^{-\epsilon_2})$, the fugacities for the maximal torus $T = U(1) \times U(1) \subset SU(2) \times SU(2) \cong \mathrm{Spin}(4)$ acting on $(z_{1},z_{2}) \in \C^{2}$ as $T : (z_1, z_2) \longmapsto (\mathsf{q} z_1, \, \mathsf{t}^{-1} z_2)$.
The cyclic group acts by $\mathbb{Z}_k : (z_1, z_2) \mapsto (\omega z_1, \omega^{-1} z_2)$ with $\omega = e^{2\pi i/k}$, so the combined action is given by 
\begin{align}\label{eq:orb-embedding}
    T \times \mathbb{Z}_k : (z_1, z_2) \mapsto (\mathsf{q} \omega z_1, (\mathsf{t} \omega)^{-1} z_2) . 
\end{align}

The gauge theory on the orbifold $Y = \C^2/\Z_k$ is obtained from the $\Z_k$-invariant sector of the theory on the covering space $\C^{2}$. By analogy, we define the topological string amplitude on the orbifold by the $\Gamma$-invariant projection
\bea\label{eq:orb-projection}
  Z(X \times \mathbb{C}^2/\Gamma) = Z(X \times \mathbb{C}^2)^\Gamma ,
\eea
imposed on single-particle states, that is, on the argument of the plethystic exponential. By \eqref{eq:orb-embedding}, a monomial $\mathsf{q}^{a}\mathsf{t}^{b}$ carries $\Z_{k}$ charge $\omega^{a+b}$. When $Z(X\times\C^2)$ is written as an infinite product, \eqref{eq:orb-projection} therefore keeps exactly the factors whose exponents obey
\bea\label{eq:orb-selection}
 a + b \equiv 0 \pmod k \, .
\eea
Below, we test this rule against the rank-$N$ PT series of the resolved conifold and against refined Chern--Simons theory.
\subsection{M-theory description}\label{sec:M-theory}

The refined A-model topological string on a Calabi--Yau threefold $X$ is described in terms of M-theory on
\bea
 X \times \mathrm{TN}_{1} \times \mathbb{S}^{1} \, ,
\eea
where $\mathrm{TN}_k$ stands for the $k$-centered Taub--NUT space~\cite{Dijkgraaf:2006um}. The partition function of this setup is the M-theory index counting the BPS states of M2-branes,
\bea
 Z = \Tr \left[ (-1)^F \, \mathsf{q}^{s_1 - s_R} \, \mathsf{t}^{-s_2 + s_R} \right] \, ,
\eea
where $s_1$, $s_2$ generate the rotations of the two complex planes and $s_R$ the $R$-symmetry.

We may incorporate M5-branes in addition to M2-branes, wrapping a Lagrangian submanifold $L$ of $X$,
\bea
 L \times \C \times \mathbb{S}^{1} \, ,
\eea
where $\C$ is either of the complex planes for $z_1$ or $z_2$. Since these planes are inequivalent in the refined theory, the partition function depends on which one is chosen. For $X=T^*\!M$ and $L=M$, a stack of $N$ M5-branes gives rise to $SU(N)$ Chern--Simons theory on the three-manifold $M$; the existence of the $U(1)_R$ symmetry requires $M$ to be a Seifert manifold. Knot invariants are obtained in the same setup by adding M5-branes wrapping
\bea
 L_K \times \C \times \mathbb{S}^{1} \, ,
\eea
where $L_K \subset T^*\!M$ is a Lagrangian submanifold intersecting $M$ along a knot $K$,
\bea
 L_K \cap M = K \, .
\eea
These $L_K$ and $M$ intersect along an $\mathbb{S}^{1}$, on which the M2-branes ending on the two stacks of M5-branes are localized. The M-theory index is computed by turning on fugacities $U$ and $V$ for those M5-branes, interpreted as holonomies of the gauge fields on $M$ and $L_K$.

\subsection{(6+1)d topological field theory}\label{app:orb-6dTFT}

\paragraph{Rank-one case}
As shown in~\cite{Iqbal:2003ds}, the topological string for a toric Calabi--Yau threefold $X$ is equivalent to $U(1)$ maximally supersymmetric topologically twisted gauge theory in six dimensions.
One can understand this equivalence using the 9--11 flip~\cite{Dijkgraaf:2006um}.
We again start with M-theory on
\begin{equation}
 (\mbox{M}) \quad
 \mathrm{TN}_{k=1} \times X \times \mathbb{S}^{1} \, .
\end{equation}
There are two ways to obtain the IIA theory by dimensional reduction from this configuration.
The first reduction shrinks the extra $\mathbb{S}^{1}$:
\begin{equation}
 (\mbox{IIA}) \quad
  \mathrm{TN}_{k=1} \times X 
  \, .
\end{equation}
Near its center, the Taub--NUT space looks like flat Euclidean $\R^4 \cong \C^2$,
and in the large-radius limit the geometry becomes $\C^2 \times X$.

The second reduction uses the circle inside the Taub--NUT space:
\begin{equation}
(\mbox{IIA}) \quad
 \R^3 \times \underbrace{X \times \mathbb{S}^{1}}_{{\rm D}6}
 \, .
\end{equation}
This gives a D6-brane wrapping $X$ at the Taub--NUT center, where the circle shrinks to zero size.
We have a topologically twisted maximally supersymmetric $(6+1)$d $U(1)$ gauge theory as a worldvolume theory of the D6-brane.
By taking the zero radius limit for the extra $\mathbb{S}^{1}$, we then obtain a $U(1)$ topological field theory in six dimensions, whose partition function is given by the generating function of the equivariant Donaldson--Thomas invariants of $X$.
Keeping the radius of the extra $\mathbb{S}^{1}$ finite, we instead obtain the K-theoretic Donaldson--Thomas invariants.

\paragraph{Higher-rank case}

We then replace the Taub--NUT geometry with the multi-centered one~\cite{Pomoni:2023nlf}.
Starting with M-theory on
\begin{equation}
 (\mbox{M}) \quad
 \mathrm{TN}_{k} \times X \times \mathbb{S}^{1} \, ,
\end{equation}
we have the first reduction 
\begin{equation}
 (\mbox{IIA}) \quad
  \mathrm{TN}_{k} \times X 
  \, .
\end{equation}
In the large Taub--NUT radius limit, this configuration reduces to the topological string on the Calabi--Yau threefold $X$ with the transverse orbifold $\C^2/\Z_k$, used in the orbifold calculation below.

Another reduction of M-theory gives
\begin{equation}
(\mbox{IIA}) \quad
 \R^3 \times \underbrace{X \times \mathbb{S}^{1}}_{k\,{\rm D}6}
 \, .
\end{equation}
In this case, the geometry is replaced with $k$ D6-branes located at the centers of the Taub--NUT geometry.
The resulting $(6+1)$d topological field theory is then a maximally supersymmetric $U(k)$ gauge theory, and the partition function is given by the generating function of the rank-$k$ Donaldson--Thomas invariants of $X$.
When the positions of the centers are different from each other, the symmetry is generically broken to a subgroup of $U(k)$.

\paragraph{Asymptotics and colored partitions}
In six dimensions, the partition function is described combinatorially by plane partitions, in parallel with the two-dimensional Young diagrams of the 4d instanton partition function~\cite{Iqbal:2003ds}. For the $U(1)$ theory on $\C^{3}$ with coordinates $(z_1,z_2,z_3)$, one may impose a boundary condition in each direction: the asymptotic behavior of the gauge field as $z_1 \to \infty$ is the instanton configuration on the transverse plane $(z_2,z_3) \in \C^{2}$, labeled by a Young diagram $\lambda$. The three boundary conditions are thus a triple of partitions $(\lambda,\mu,\nu)$, which are precisely the asymptotic profiles of the plane partition along the three axes. For $U(k)$, the asymptotics are labeled by $k$-colored Young diagrams instead, which is exactly what the refined topological vertex of the orbifold theory produces~\cite{Chaimanowong:2018xyr}.

\subsection{Resolved conifold}\label{app:orb-conifold}

We illustrate the rule \eqref{eq:orb-selection} on the resolved conifold. For the K\"ahler parameter $Q = e^{-T}$, where $T$ is the complexified area of the $\mathbb{P}^{1}$ and is not to be confused with the refinement parameter $\mathsf{t}$, the refined topological string partition function is
\bea\label{eq:conifold-pf}
  Z(Q;\mathsf{q},\mathsf{t}) = \prod_{i,j=1}^\infty (1 - Q \mathsf{q}^{i-1/2} \mathsf{t}^{j-1/2}) = (Q \sqrt{\mathsf{q}\mathsf{t}};\mathsf{q},\mathsf{t})_\infty = \PE \left[ \frac{-Q}{[\mathsf{q}][\mathsf{t}]} \right] \, .
\eea
The factor labeled by $(i,j)$ carries $\Z_{k}$ charge $\omega^{(i-1/2)+(j-1/2)} = \omega^{i+j-1}$, so \eqref{eq:orb-selection} retains those with $i+j-1 \equiv 0$ and
\bea
  Z_{\Z_k}(Q;\mathsf{q},\mathsf{t}) = \prod_{\substack{i,j=1 \\ i+j-1 \equiv 0 \, (\mathrm{mod}\, k)}}^\infty (1 - Q \mathsf{q}^{i-1/2} \mathsf{t}^{j-1/2}) \, .
\eea
To evaluate this, solve the constraint as $i \equiv r$ and $j \equiv 1-r \pmod k$ for $r=1,\ldots,k$, that is,
\bea\label{eq:orb-conifold-param}
 i = r + k(m-1) \, , \qquad j = (k+1-r) + k(n-1) \, , \qquad m,n \geq 1 \, .
\eea
Since $\mathsf{q}^{i-1/2}\mathsf{t}^{j-1/2} = (\mathsf{q}/\mathsf{t})^{r-\frac{k+1}{2}} (\mathsf{q}^{k})^{m-1/2}(\mathsf{t}^{k})^{n-1/2}$, the product factorizes into $k$ copies of \eqref{eq:conifold-pf} with rescaled equivariant parameters and shifted K\"ahler moduli,
\bea\label{eq:Zk_conifold_pf}
  Z_{\Z_k}(Q;\mathsf{q},\mathsf{t}) = \prod_{r=1}^k Z((\mathsf{q}/\mathsf{t})^{r-\frac{k+1}{2}} Q;\mathsf{q}^k,\mathsf{t}^k) = \PE\left[ \frac{ - Q [(\mathsf{q}/\mathsf{t})^k]}{[\mathsf{q}^k][\mathsf{t}^k][\mathsf{q}/\mathsf{t}]} \right] \, ,
\eea
the last equality using $\sum_{r=1}^{k} x^{r-\frac{k+1}{2}} = [x^{k}]/[x]$ with $x = \mathsf{q}/\mathsf{t}$.

The plethystic form of \eqref{eq:Zk_conifold_pf} is the $\Z_k$-gauging formula
\bea\label{eq:Zk-gauging}
  \frac{1}{k} \sum_{r=0}^{k-1} \frac{1}{(1 - \omega^r \mathsf{q})(1 - \omega^{-r} \mathsf{t}^{-1})} = \frac{1 - (\mathsf{q}/\mathsf{t})^k}{(1 - \mathsf{q}^k)(1 - \mathsf{t}^{-k})(1 - \mathsf{q}/\mathsf{t})}.
\eea
The left-hand side is the Molien average of the Hilbert series of $\C[z_{1},z_{2}]$. The right-hand side is the Hilbert series of the invariant ring, generated by $X = z_1^{k}$, $Y = z_2^{k}$ and $Z = z_1 z_2$, of weights $\mathsf{q}^{k}$, $\mathsf{t}^{-k}$ and $\mathsf{q}/\mathsf{t}$ under \eqref{eq:orb-embedding}, subject to the single relation $XY = Z^{k}$ of weight $(\mathsf q/\mathsf t)^{k}$. This is the hypersurface realization of the $A_{k-1}$ singularity,
\bea
 \C^2/\Z_k \cong \{(X,Y,Z) \in \C^3 \mid XY - Z^k = 0\} \, ,
\eea
with $\C^2 \to \C^3$, $(z_1,z_2) \mapsto (z_1^k, z_2^k, z_1 z_2)$. We remark that in the unrefined case $\mathsf{q}=\mathsf{t}$ the shifts in \eqref{eq:Zk_conifold_pf} disappear and the orbifold partition function is simply $Z_{\Z_k}(Q;\mathsf{q},\mathsf{q}) = Z(Q;\mathsf{q}^k,\mathsf{q}^k)^k$.

At $k=N$, the change of variables given after \eqref{eq:PT3-conifold-rankN} identifies the orbifold conifold partition function \eqref{eq:Zk_conifold_pf} with the rank-$N$ twisted index \eqref{eq:conifoldPE-rankN} of Section~\ref{sec:rankone-twistedindex}.

\subsection{Geometric transition and orbifold Chern--Simons theory}\label{app:orb-CS}

The resolved conifold is related to the deformed conifold $T^*\!\mathbb{S}^3$ by the geometric transition~\cite{Gopakumar:1998ki}, so by the discussion of Section~\ref{sec:M-theory} the topological string on the resolved conifold is related to Chern--Simons theory on $\mathbb{S}^3$, and the two partition functions agree in the limit $N \to \infty$ once the K\"ahler parameter is identified with the 't Hooft coupling, $Q = \mathsf{q}^N$. The refinement of this duality leads to the refined Chern--Simons theory~\cite{Aganagic:2011sg,Aganagic:2012hs}. For a gauge group $G$ with dual Coxeter number $h^\vee$ and set of positive roots $R_+$, the partition function on $\mathbb{S}^3$ at level $\kappa$ is
\bea
  Z_{G}(\mathsf{q},\mathsf{t}) = \prod_{\alpha \in R_+} \frac{(\mathsf{t}^{\alpha};\mathsf{q})_\infty}{(\mathsf{t}^{\alpha+1};\mathsf{q})_\infty} \, , \qquad \mathsf{q} = \exp \left( \frac{2 \pi i}{\kappa + h^\vee} \right) \, ,
\eea
where $\mathsf{t}$ is the refinement parameter and $\mathsf{t}^{\alpha}$ is shorthand for $\mathsf{t}^{\operatorname{ht}(\alpha)}$, the height of the root; for $G = SU(N)$ and $\alpha = e_{i} - e_{j}$ one has $\operatorname{ht}(\alpha) = j-i$, so that%
\footnote{Setting $\mathsf{t} = \mathsf{q}^\beta$ with $\beta \in \mathbb{Z}_{> 0}$ reproduces the expression in~\cite{Aganagic:2011sg,Aganagic:2012hs},
\bea
  Z_{SU(N)}(\mathsf{q},\mathsf{q}^\beta) = \prod_{m=0}^{\beta - 1} \prod_{j=1}^{N-1} (1 - \mathsf{q}^{\beta j+m})^{N-j} \, .
\eea}
\bea\label{eq:ref_CS_pf}
  Z_{SU(N)}(\mathsf{q},\mathsf{t}) = \prod_{1 \le i < j \le N} \frac{(\mathsf{t}^{j-i};\mathsf{q})_\infty}{(\mathsf{t}^{j-i+1};\mathsf{q})_\infty} = \frac{(\mathsf{t};\mathsf{q})_\infty^N}{(\mathsf{t};\mathsf{q},\mathsf{t})_\infty} (\mathsf{t}^{N+1};\mathsf{q},\mathsf{t})_\infty \, .
\eea
Setting $\mathcal{Q} = \mathsf{t}^{N}$, the $\mathcal{Q}$-dependent part of the partition function is
\bea
  Z_{SU(N)}(\mathsf{q},\mathsf{t}) \sim (\mathcal{Q}\mathsf{t};\mathsf{q},\mathsf{t})_\infty = Z(\mathcal{Q} (\mathsf{t}/\mathsf{q})^{\frac{1}{2}}; \mathsf{q},\mathsf{t}) \, ,
\eea
so that $Q = \mathcal{Q}(\mathsf{t}/\mathsf{q})^{\frac{1}{2}}$ reproduces the refined topological string on the resolved conifold \eqref{eq:conifold-pf}.

We turn to the orbifold theory, defining $Z_{G}^\Gamma(\mathsf{q},\mathsf{t}) = Z_{G}(\mathsf{q},\mathsf{t})^\Gamma$ by the same projection \eqref{eq:orb-selection}, and taking $G = SU(N)$ with $N \in k \mathbb{Z}_{>0}$ for simplicity. In \eqref{eq:ref_CS_pf}, the factor $1 - \mathsf{t}^{m}\mathsf{q}^{n}$ carries charge $\omega^{m+n}$, so the surviving factors of $(\mathsf{t};\mathsf{q})_\infty$ are those with $n \equiv -1$, giving $(\mathsf{t}\mathsf{q}^{k-1};\mathsf{q}^{k})_\infty$, while in the double products the invariant factors organize into $k$ sectors labeled by $r=1,\ldots,k$ exactly as in \eqref{eq:orb-conifold-param}:
\bea\label{eq:orb_CS_pf}
  Z_{SU(N)}^{\Z_k}(\mathsf{q},\mathsf{t}) = (\mathsf{t}\mathsf{q}^{k-1};\mathsf{q}^k)_\infty^N \prod_{r=1}^k \frac{(\mathsf{t}^N (\mathsf{q}^{k-r}\mathsf{t}^r);\mathsf{q}^k,\mathsf{t}^k)_\infty}{((\mathsf{q}^{k-r}\mathsf{t}^r);\mathsf{q}^k,\mathsf{t}^k)_\infty} \, ,
\eea
which factorizes into $k$ sectors as in the case of the orbifold conifold partition function, and which reduces to \eqref{eq:ref_CS_pf} for $k=1$. Identifying $\mathsf{t}^N (\mathsf{t}/\mathsf{q})^{\frac{1}{2}} = Q$ as before, the $Q$-dependent part of \eqref{eq:orb_CS_pf} becomes $\prod_{r=1}^{k}(Q(\mathsf{q}/\mathsf{t})^{\frac{1}{2}}\mathsf{q}^{k-r}\mathsf{t}^{r};\mathsf{q}^{k},\mathsf{t}^{k})_\infty$, which is the orbifold conifold partition function~\eqref{eq:Zk_conifold_pf} after the relabeling $r \to k+1-r$.

\bibliographystyle{JHEP}
\bibliography{references}

\end{document}